\documentclass[a4paper,debug,notitlepage,nobib]{tufte-book}

\usepackage{graphicx}
\usepackage[svgnames]{xcolor}
\usepackage{hyperref}
\usepackage{epstopdf}
\usepackage{xspace} 
\usepackage{longtable} 
\usepackage{slashed}
\usepackage{amsfonts}
\usepackage[T1]{fontenc}
\usepackage{amsmath,amssymb}
\usepackage{slashed}
\usepackage{feynmp-auto}
\makeatletter
\let\ginnatwidth\Gin@nat@width
\let\ginnatheight\Gin@nat@height
\makeatother
\setkeys{Gin}{width=\linewidth,totalheight=\textheight,keepaspectratio}
\newenvironment{feynmandiagram}[1][]{\setkeys{Gin}{width=\ginnatwidth,totalheight=\ginnatheight}\begin{fmffile}{#1}
\begin{fmfgraph*}(100,70)\fmfpen{thick}}{\end{fmfgraph*}\end{fmffile}\setkeys{Gin}{width=\linewidth,totalheight=\textheight,keepaspectratio}}
\usepackage{subfig}
\usepackage{xhfill}

\setkeys{Gin}{width=\linewidth,totalheight=\textheight,keepaspectratio}
\usepackage{mathtools} 
\usepackage{booktabs} 
\usepackage{units}    
\usepackage{multicol} 
\usepackage{fancyvrb} 
\usepackage{fancyhdr}
\usepackage{refcount}
\usepackage{calc}
\usepackage{lastpage}
\IfFileExists{moderntex}{
  \usepackage[protrusion=true,expansion=true,tracking=true,kerning=true,spacing=true]{microtype}
}{}
\fvset{fontsize=\normalsize} 

\usepackage[marginpar]{todo} 
\let\nominalTodo\Todo
\renewcommand\Todo[1]{\nominalTodo{\normalfont\footnotesize\sffamily\bf #1}}

\usepackage{booktabs,colortbl, array}
\usepackage{rotating}

\makeatletter
\title{Dark Matter Benchmark Models for Early LHC Run-2 Searches:\\\noindent Report~of~the~ATLAS/CMS~Dark~Matter~Forum}
\author{ATLAS+CMS Dark Matter Forum}
\date{\today}
\usepackage{titling}

\newcommand{\MET}{\ensuremath{\slashed{E}_T}\xspace}
\newcommand{\chiDM}{\ensuremath{\chi}\xspace}

\newcommand{\mMed}{\ensuremath{M_{\rm{med}}}\xspace}
\newcommand{\MPhi}{\ensuremath{M_{\Phi}}\xspace}
\newcommand{\Mphi}{\ensuremath{M_{\phi}}\xspace}
\newcommand{\mmed}{\mMed}
\newcommand{\Mmed}{\mMed}
\newcommand{\gDM}{\ensuremath{g_{\chiDM}}\xspace}
\newcommand{\gdm}{\gDM}
\newcommand{\gq}{\ensuremath{g_{\rm q}}\xspace}

\newcommand{\mdm}{\ensuremath{m_{\chiDM}}\xspace}
\newcommand{\mDM}{\mdm}

\newcommand{\pT}{\ensuremath{p_{\rm T}}\xspace}
\newcommand{\kT}{\ensuremath{k_{\rm T}}\xspace}
\newcommand{\Qtr}{\ensuremath{Q_{\rm tr}}\xspace}
\newcommand{\Mstar}{\ensuremath{M_{*}}\xspace}
\newcommand{\gstar}{\ensuremath{g_{*}}\xspace}
\newcommand{\Mcut}{\ensuremath{M_{\text{cut}}}\xspace}

\newcommand{\Reft}{\ensuremath{R}\xspace}

\newcommand{\Zprime}{\ensuremath{Z^\prime}\xspace}
\newcommand{\Ztwo}{\ensuremath{\mathbb{Z}_2}\xspace}

\newcommand{\invfb}{\ensuremath{\mbox{\,fb}^{-1}}\xspace}
\newcommand{\ev}{\ensuremath{\mathrm{\,e\kern -0.1em V}}\xspace}
\newcommand{\kev}{\ensuremath{\mathrm{\,ke\kern -0.1em V}}\xspace}
\newcommand{\mev}{\ensuremath{\mathrm{\,Me\kern -0.1em V}}\xspace}
\newcommand{\mevc}{\ensuremath{{\mathrm{\,Me\kern -0.1em V\!/}c}}\xspace}
\newcommand{\mevcc}{\ensuremath{{\mathrm{\,Me\kern -0.1em V\!/}c^2}}\xspace}
\newcommand{\gev}{\ensuremath{\mathrm{\,Ge\kern -0.1em V}}\xspace}
\newcommand{\gevc}{\ensuremath{{\mathrm{\,Ge\kern -0.1em V\!/}c}}\xspace}
\newcommand{\gevcnospace}{\ensuremath{{\mathrm{\,Ge\kern -0.1em V\!/}c}}}
\newcommand{\gevcc}{\ensuremath{{\mathrm{\,Ge\kern -0.1em V\!/}c^2}}\xspace}
\newcommand{\tev}{\ensuremath{\mathrm{\,Te\kern -0.1em V}}\xspace}

\newcommand{\lag}{{\ensuremath{\cal L}}} 
\newcommand{\gmu}{\gamma^\mu}

\newcommand{\madgraph}{{\textsc{MadGraph5\_aMC@NLO}}\xspace}

\newcommand{\syscalc}{\textsc{SysCalc}\xspace}

\newcommand{\powheg}{{\namecaps{POWHEG}}\xspace}
\newcommand{\mcfm}{{\namecaps{MCFM}}\xspace}

\newcommand{\pythia}{{\namecaps{Pythia}}\xspace}
\newcommand{\pythiaEight}{{\namecaps{Pythia~8}}\xspace}

\newcommand{\bornsuppfact}{\texttt{bornsuppfact}\xspace}
\newcommand{\bornktmin}{\texttt{bornktmin}\xspace}
\newcommand{\runningwidth}{\texttt{runningwidth}\xspace}
\newcommand{\masslow}{\texttt{mass\_low}\xspace}
\newcommand{\masshigh}{\texttt{mass\_high}\xspace}
\newcommand{\ncallOne}{\texttt{ncall1}\xspace}
\newcommand{\ncallTwo}{\texttt{ncall2}\xspace}
\newcommand{\itmxOne}{\texttt{itmx1}\xspace}
\newcommand{\itmxTwo}{\texttt{itmx2}\xspace}
\newcommand{\foldsci}{\texttt{foldsci}\xspace}
\newcommand{\foldy}{\texttt{foldy}\xspace}
\newcommand{\foldphi}{\texttt{foldphi}\xspace}
\newcommand{\withnegweights}{\texttt{withnegweights}\xspace}

\newcommand{\modelDMV}{\texttt{DMV}\xspace}

\newcommand{\schannel}{\ensuremath{s\text{-channel}}\xspace}
\newcommand{\tchannel}{\ensuremath{t\text{-channel}}\xspace}
\newcommand{\spinzero}{\ensuremath{\text{spin-}0}\xspace}
\newcommand{\spinone}{\ensuremath{\text{spin-}1}\xspace}

\newcommand{\Spintwo}{\ensuremath{\text{Spin-}2}\xspace}

\newcommand{\complexi}{\ensuremath{i}\xspace}

\newcommand{\lsim}{ \mathop{}_{\textstyle \sim}^{\textstyle <} }

\newcommand{\beq}{\begin{equation}}
\newcommand{\eeq}{\end{equation}}
\newcommand{\be}{\begin{equation}}
\newcommand{\ee}{\end{equation}}
\newcommand{\bea}{\begin{eqnarray}}
\newcommand{\eea}{\end{eqnarray}}
 \def\be   {\begin{equation}}   \def\ee   {\end{equation}}
 \def\ba   {\begin{array}}      \def\ea   {\end{array}}
 \def\bea  {\begin{eqnarray}}   \def\eea  {\end{eqnarray}}
 \def\bean {\begin{eqnarray*}}  \def\eean {\end{eqnarray*}}

\newcommand{\bpm}{\begin{pmatrix}}
\newcommand{\epm}{\end{pmatrix}}

\def\bsp#1\esp{\begin{split}#1\end{split}}

\usepackage{etex}
\reserveinserts{20}
\usepackage[hyperref=true,url=false,backend=bibtex,style=alphabetic,backref=false,firstinits=true,doi=false,eprint=true,language=USenglish]{biblatex}
\DeclareFieldFormat{sentencecase}{\MakeSentenceCase{#1}}
\renewbibmacro*{title}{%
  \ifthenelse{\iffieldundef{title}\AND\iffieldundef{subtitle}}
    {}
    {\ifthenelse{\ifentrytype{article}\OR\ifentrytype{inbook}\OR\ifentrytype{report}%
      \OR\ifentrytype{incollection}\OR\ifentrytype{inproceedings}%
      \OR\ifentrytype{inreference}}
      {\printtext[title]{%
        \printfield[sentencecase]{title}%
        \setunit{\subtitlepunct}%
        \printfield[sentencecase]{subtitle}}}%
      {\printtext[title]{%
        \printfield[titlecase]{title}%
        \setunit{\subtitlepunct}%
        \printfield[titlecase]{subtitle}}}%
     \newunit}%
  \printfield{titleaddon}}
\DeclareFieldFormat[article]{journaltitle}{#1\isdot}
\DeclareFieldFormat[article]{journalsubtitle}{#1\isdot}
\DeclareFieldFormat[article]{volume}{\textbf{#1}\isdot}
\DeclareFieldFormat[article,inbook,incollection,inproceedings,patent,thesis,unpublished]
  {title}{\emph{#1\isdot}}
\renewbibmacro{in:}{}

\DefineBibliographyStrings{USenglish}{%
  page = {},
  pages = {}
}
\DefineBibliographyStrings{UKenglish}{%
  page = {},
  pages = {}
}
\begin{document}

\setcounter{secnumdepth}{3} 



\thispagestyle{empty}
\topskip0pt
\begin{fullwidth}
\sffamily
{
  \Large
  \fontsize{18}{24}\selectfont 
  \@title
}\\
\vspace{1\baselineskip}
{\Large 
\noindent
\@date\\
\vspace{1\baselineskip}
\noindent
\noindent\href{mailto:daniel.abercrombie@cern.ch}{Daniel Abercrombie} 
\emph{MIT, USA}\\
\noindent\href{mailto:nural.akchurin@cern.ch}{Nural Akchurin} 
\emph{Texas Tech University, USA}\\
\noindent\href{mailto:ece.akilli@cern.ch}{Ece Akilli} 
\emph{Universit\'e de Gen\`eve, DPNC, Switzerland}\\
\noindent\href{mailto:juan.alcaraz@cern.ch}{Juan Alcaraz Maestre} 
\emph{Centro de Investigaciones Energ\'eticas Medioambientales y Tecnol\'ogicas (CIEMAT), Spain}\\
\noindent\href{mailto:brandon.leigh.allen@cern.ch}{Brandon Allen} 
\emph{MIT, USA}\\
\noindent\href{mailto:barbara.alvarez.gonzalez@cern.ch}{Barbara Alvarez Gonzalez} 
\emph{CERN, Switzerland}\\
\noindent\href{mailto:Jeremy.Andrea@cern.ch}{Jeremy Andrea} 
\emph{Institut Pluridisciplinaire Hubert Curien/D\'epartement Recherches Subatomiques, Universit\'e de Strasbourg/CNRS-IN2P3, France}\\
\noindent\href{mailto:alexandre.arbey@ens-lyon.fr}{Alexandre Arbey} 
\emph{Universit\'e de Lyon and Centre de Recherche Astrophysique de Lyon, CNRS and Ecole Normale Sup\'erieure de Lyon, France and CERN Theory Division, Switzerland}\\
\noindent\href{mailto:georges.azuelos@cern.ch}{Georges Azuelos} 
\emph{University of Montreal and TRIUMF, Canada}\\
\noindent\href{mailto:Patrizia.Azzi@cern.ch}{Patrizia Azzi} 
\emph{INFN Padova, Italy}\\
\noindent\href{mailto:mihailo.backovic@uclouvain.be}{Mihailo Backovi\'{c}} 
\emph{Centre for Cosmology, Particle Physics and Phenomenology (CP3), Universit\'e catholique de Louvain, Belgium}\\
\noindent\href{mailto:yangbai@physics.wisc.edu}{Yang Bai} 
\emph{Department of Physics, University of Wisconsin-Madison, USA}\\
\noindent\href{mailto:swagato.banerjee@cern.ch}{Swagato Banerjee} 
\emph{University of Wisconsin-Madison, USA}\\
\noindent\href{mailto:j.beacham@cern.ch}{James Beacham} 
\emph{Ohio State University, USA}\\
\noindent\href{mailto:Alexander.Belyaev@cern.ch}{Alexander Belyaev} 
\emph{Rutherford Appleton Laboratory and University of Southampton, United Kingdom}\\
\noindent\href{mailto:antonio.boveia@cern.ch}{Antonio Boveia (editor)} 
\emph{CERN, Switzerland}\\
\noindent\href{mailto:amelia.jean.brennan@cern.ch}{Amelia Jean Brennan} 
\emph{The University of Melbourne, Australia}\\
\noindent\href{mailto:Oliver.Buchmueller@Cern.ch}{Oliver Buchmueller} 
\emph{Imperial College London, United Kingdom}\\
\noindent\href{mailto:mbuckley@physics.rutgers.edu}{Matthew R. Buckley} 
\emph{Department of Physics and Astronomy, Rutgers University, USA}\\
\noindent\href{mailto:giorgio.busoni@sissa.it}{Giorgio Busoni} 
\emph{SISSA and INFN, Sezione di Trieste, Italy}\\
\noindent\href{mailto:Michael.Buttignol@cern.ch}{Michael Buttignol} 
\emph{Institut Pluridisciplinaire Hubert Curien/D\'epartement Recherches Subatomiques, Universit\'e de Strasbourg/CNRS-IN2P3, France}\\
\noindent\href{mailto:cacciapa@ipnl.in2p3.fr}{Giacomo Cacciapaglia} 
\emph{Universit\'e de Lyon and Universit\'e Lyon 1, CNRS/IN2P3, UMR5822, IPNL, France}\\
\noindent\href{mailto:Regina.Caputo@cern.ch}{Regina Caputo} 
\emph{Santa Cruz Institute for Particle Physics, Department of Physics and Department of Astronomy and Astrophysics, University of California at Santa Cruz, USA}\\
\noindent\href{mailto:lmc@physics.osu.edu}{Linda Carpenter} 
\emph{Ohio State University, USA}\\
\noindent\href{mailto:nuno.castro@cern.ch}{Nuno Filipe Castro} 
\emph{LIP-Minho, Braga, and Departamento de F\'\i sica e Astronomia, Faculdade de Ci\^ encias da Universidade do Porto, Portugal}\\
\noindent\href{mailto:guillelmo.gomez.ceballos@cern.ch}{Guillelmo Gomez Ceballos} 
\emph{MIT, USA}\\
\noindent\href{mailto:Yangyang.Cheng@cern.ch}{Yangyang Cheng} 
\emph{University of Chicago, USA}\\
\noindent\href{mailto:john.paul.chou@cern.ch}{John Paul Chou} 
\emph{Rutgers University, USA}\\
\noindent\href{mailto:arely.cortes.gonzalez@cern.ch}{Arely Cortes Gonzalez} 
\emph{IFAE Barcelona, Spain}\\
\noindent\href{mailto:christopher.cowden@cern.ch}{Chris Cowden} 
\emph{Texas Tech University, USA}\\
\noindent\href{mailto:fraderamo@berkeley.edu}{Francesco D'Eramo} 
\emph{University of California and LBNL, Berkeley, USA}\\
\noindent\href{mailto:annapaola.de.cosa@cern.ch}{Annapaola De Cosa} 
\emph{University of Zurich, Switzerland}\\
\noindent\href{mailto:michele.degruttola@cern.ch}{Michele De Gruttola} 
\emph{CERN, Switzerland}\\
\noindent\href{mailto:deroeck@mail.cern.ch}{Albert De Roeck} 
\emph{CERN, Switzerland}\\
\noindent\href{mailto:andrea.desimone@sissa.it}{Andrea De Simone} 
\emph{SISSA and INFN, Sezione di Trieste, Italy}\\
\noindent\href{mailto:deandrea@ipnl.in2p3.fr}{Aldo Deandrea} 
\emph{Universit\'e de Lyon and Universit\'e Lyon 1, CNRS/IN2P3, UMR5822, IPNL, France}\\
\noindent\href{mailto:zeynep.demiragli@cern.ch}{Zeynep Demiragli} 
\emph{MIT, USA}\\
\noindent\href{mailto:adifranz@uci.edu}{Anthony DiFranzo} 
\emph{Department of Physics and Astronomy, University of California, Irvine and Theoretical Physics Department, Fermilab, USA}\\
\noindent\href{mailto:caterina.doglioni@cern.ch}{Caterina Doglioni (editor)} 
\emph{Lund University, Sweden}\\
\noindent\href{mailto:tdupree.cms@gmail.com}{Tristan du Pree} 
\emph{CERN, Switzerland}\\
\noindent\href{mailto:erbacher@physics.ucdavis.edu}{Robin Erbacher} 
\emph{University of California, Davis, USA}\\
\noindent\href{mailto:Johannes.Erdmann@cern.ch}{Johannes Erdmann} 
\emph{Institut f\"ur Experimentelle Physik IV, Technische Universit\"at Dortmund, Germany}\\
\noindent\href{mailto:cfischer@ifae.es}{Cora Fischer} 
\emph{IFAE Barcelona, Spain}\\
\noindent\href{mailto:Henning.Flacher@cern.ch}{Henning Flaecher} 
\emph{H.H. Wills Physics Laboratory, University of Bristol, United Kingdom}\\
\noindent\href{mailto:pjfox@fnal.gov}{Patrick J. Fox} 
\emph{Fermilab, USA}\\
\noindent\href{mailto:Benjamin.Fuks@cern.ch}{Benjamin Fuks} 
\emph{Institut Pluridisciplinaire Hubert Curien/D\'epartement Recherches Subatomiques, Universit\'e de Strasbourg/CNRS-IN2P3, France}\\
\noindent\href{mailto:genest@lpsc.in2p3.fr}{Marie-Helene Genest} 
\emph{LPSC, Université Grenoble-Alpes, CNRS/IN2P3, France}\\
\noindent\href{mailto:Bhawna.Gomber@cern.ch}{Bhawna Gomber} 
\emph{University of Wisconsin-Madison, USA}\\
\noindent\href{mailto:Andreas.Goudelis@oeaw.ac.at}{Andreas Goudelis} 
\emph{Institut f\"ur Hochenergiephysik, \"Osterreichische Akademie der Wissenschaften, Austria}\\
\noindent\href{mailto:johanna.gramling@cern.ch}{Johanna Gramling} 
\emph{Universit\'e de Gen\`eve, DPNC, Switzerland}\\
\noindent\href{mailto:gunion@physics.ucdavis.edu}{John Gunion} 
\emph{University of California, Davis, USA}\\
\noindent\href{mailto:kristian.hahn@cern.ch}{Kristian Hahn} 
\emph{Northwestern University, USA}\\
\noindent\href{mailto:ulrich.haisch@physics.ox.ac.uk}{Ulrich Haisch} 
\emph{Rudolf Peierls Centre for Theoretical Physics, University of Oxford, United Kingdom}\\
\noindent\href{mailto:roni@fnal.gov}{Roni Harnik} 
\emph{Theoretical Physics Department, Fermilab, USA}\\
\noindent\href{mailto:philip.coleman.harris@cern.ch}{Philip C. Harris} 
\emph{CERN, Switzerland}\\
\noindent\href{mailto:Kerstin.Hoepfner@cern.ch}{Kerstin Hoepfner} 
\emph{RWTH Aachen University, III. Physikalisches Institut A, Germany}\\
\noindent\href{mailto:siew.yan.hoh@cern.ch}{Siew Yan Hoh} 
\emph{National Centre for Particle Physics, Universiti Malaya, Malaysia}\\
\noindent\href{mailto:dylan.hsu@cern.ch}{Dylan George Hsu} 
\emph{MIT, USA}\\
\noindent\href{mailto:schsu@uw.edu}{Shih-Chieh Hsu} 
\emph{Physics, University of Washington, Seattle, USA}\\
\noindent\href{mailto:Yutaro.Iiyama@cern.ch}{Yutaro Iiyama} 
\emph{MIT, USA}\\
\noindent\href{mailto:valerio.ippolito@cern.ch}{Valerio Ippolito} 
\emph{Laboratory for Particle Physics and Cosmology, Harvard University, USA}\\
\noindent\href{mailto:thomas.jacques@unige.ch}{Thomas Jacques} 
\emph{Department of Theoretical Physics, University of Geneva, Switzerland}\\
\noindent\href{mailto:Xiangyang.Ju@cern.ch}{Xiangyang Ju} 
\emph{University of Wisconsin-Madison, USA}\\
\noindent\href{mailto:felix.kahlhoefer@desy.de}{Felix Kahlhoefer} 
\emph{DESY, Germany}\\
\noindent\href{mailto:Alexis.Kalogeropoulos@cern.ch}{Alexis Kalogeropoulos} 
\emph{Deutsches Elektronen-Synchrotron (DESY), Germany}\\
\noindent\href{mailto:laser.seymour.kaplan@cern.ch}{Laser Seymour Kaplan} 
\emph{University of Wisconsin-Madison, USA}\\
\noindent\href{mailto:Lashkar.Kashif@cern.ch}{Lashkar Kashif} 
\emph{University of Wisconsin-Madison, USA}\\
\noindent\href{mailto:valya.khoze@durham.ac.uk}{Valentin V. Khoze} 
\emph{Institute of Particle Physics Phenomenology, Durham University, United Kingdom}\\
\noindent\href{mailto:Raman.Khurana@cern.ch}{Raman Khurana} 
\emph{National Central University, Taiwan}\\
\noindent\href{mailto:khristian.kotov@cern.ch}{Khristian Kotov} 
\emph{The Ohio State University, USA}\\
\noindent\href{mailto:dmytro.kovalskyi@cern.ch}{Dmytro Kovalskyi} 
\emph{MIT, USA}\\
\noindent\href{mailto:suchita.kulkarni@oeaw.ac.at}{Suchita Kulkarni} 
\emph{Institut f\"ur Hochenergiephysik, \"Osterreichische Akademie der Wissenschaften, Austria}\\
\noindent\href{mailto:shuichi.kunori@ttu.edu}{Shuichi Kunori} 
\emph{Texas Tech University, USA}\\
\noindent\href{mailto:viktor.kutzner@physik.rwth-aachen.de}{Viktor Kutzner} 
\emph{RWTH Aachen University, III. Physikalisches Institut A, Germany}\\
\noindent\href{mailto:hminlee@cau.ac.kr}{Hyun Min Lee} 
\emph{Department of Physics, Chung-Ang University, Korea}\\
\noindent\href{mailto:sungwon.lee@cern.ch}{Sung-Won Lee} 
\emph{Texas Tech University, USA}\\
\noindent\href{mailto:spliew@lbl.gov}{Seng Pei Liew} 
\emph{Department of Physics, University of Tokyo, Japan}\\
\noindent\href{mailto:tongylin@gmail.com}{Tongyan Lin} 
\emph{Kavli Institute for Cosmological Physics, University of Chicago, USA}\\
\noindent\href{mailto:steven.lowette@cern.ch}{Steven Lowette (editor)} 
\emph{Vrije Universiteit Brussel - IIHE, Belgium}\\
\noindent\href{mailto:romain.madar@cern.ch}{Romain Madar} 
\emph{Laboratoire de Physique Corpusculaire, Clermont-Ferrand, France}\\
\noindent\href{mailto:sarah.malik@cern.ch}{Sarah Malik (editor)} 
\emph{Imperial College London, United Kingdom}\\
\noindent\href{mailto:Fabio.Maltoni@cern.ch}{Fabio Maltoni} 
\emph{Centre for Cosmology, Particle Physics and Phenomenology (CP3), Universit\'e catholique de Louvain, Belgium}\\
\noindent\href{mailto:mmp@ifae.es}{Mario Martinez Perez} 
\emph{IFAE Barcelona, Spain}\\
\noindent\href{mailto:o.p.c.mattelaer@durham.ac.uk}{Olivier Mattelaer} 
\emph{IPPP Durham, United Kingdom}\\
\noindent\href{mailto:kentarou.mawatari@vub.ac.be}{Kentarou Mawatari} 
\emph{Theoretische Natuurkunde and IIHE/ELEM, Vrije Universiteit Brussel, and International Solvay Institutes, Belgium}\\
\noindent\href{mailto:c.mccabe@uva.nl}{Christopher McCabe} 
\emph{GRAPPA, University of Amsterdam, Netherlands}\\
\noindent\href{mailto:megy.theo@gmail.com}{Th\'eo Megy} 
\emph{Laboratoire de Physique Corpusculaire, Clermont-Ferrand, France}\\
\noindent\href{mailto:enrico.morgante@unige.ch}{Enrico Morgante} 
\emph{Department of Theoretical Physics, University of Geneva, Switzerland}\\
\noindent\href{mailto:mrenna@fnal.gov}{Stephen Mrenna (editor)} 
\emph{FNAL, USA}\\
\noindent\href{mailto:siddharth.m.narayanan@cern.ch}{Siddharth M. Narayanan} 
\emph{MIT, USA}\\
\noindent\href{mailto:andrew.james.nelson@cern.ch}{Andy Nelson} 
\emph{University of California, Irvine, USA}\\
\noindent\href{mailto:Sergio.Novaes@cern.ch}{S\'ergio F. Novaes} 
\emph{Universidade Estadual Paulista, Brazil}\\
\noindent\href{mailto:klaas.ole.padeken@cern.ch}{Klaas Ole Padeken} 
\emph{RWTH Aachen University, III. Physikalisches Institut A, Aachen, Germany}\\
\noindent\href{mailto:Priscilla.Pani@cern.ch}{Priscilla Pani} 
\emph{Stockholm University, Sweden}\\
\noindent\href{mailto:mpapucci@lbl.gov}{Michele Papucci} 
\emph{Theoretical Physics Group, Lawrence Berkeley National Laboratory, and Berkeley Center for Theoretical Physics, University of California, Berkeley, USA}\\
\noindent\href{mailto:paulini@heps.phys.cmu.edu}{Manfred Paulini} 
\emph{Carnegie Mellon University, USA}\\
\noindent\href{mailto:paus@mit.edu}{Christoph Paus} 
\emph{MIT, USA}\\
\noindent\href{mailto:Jacopo.Pazzini@pd.infn.it}{Jacopo Pazzini} 
\emph{Universit\`a di Padova, Italy}\\
\noindent\href{mailto:Bjoern.Penning@cern.ch}{Bj\"orn Penning} 
\emph{Imperial College London, United Kingdom}\\
\noindent\href{mailto:mpeskin@slac.stanford.edu}{Michael E. Peskin} 
\emph{SLAC, Stanford University, USA}\\
\noindent\href{mailto:Deborah.Pinna@cern.ch}{Deborah Pinna} 
\emph{University of Zurich, Switzerland}\\
\noindent\href{mailto:massimiliano.procura@univie.ac.at}{Massimiliano Procura} 
\emph{Universit\"at Wien, Austria}\\
\noindent\href{mailto:shamona.fawad.qazi@cern.ch}{Shamona F. Qazi} 
\emph{National Centre for Physics, Quaid-i-Azam University, Pakistan}\\
\noindent\href{mailto:davide.racco@unige.ch}{Davide Racco} 
\emph{Department of Theoretical Physics, University of Geneva, Switzerland}\\
\noindent\href{mailto:emanuele.re@physics.ox.ac.uk}{Emanuele Re} 
\emph{Rudolf Peierls Centre for Theoretical Physics, University of Oxford, United Kingdom}\\
\noindent\href{mailto:antonio.riotto@unige.ch}{Antonio Riotto} 
\emph{Department of Theoretical Physics, University of Geneva, Switzerland}\\
\noindent\href{mailto:rizzo@slac.stanford.edu}{Thomas G. Rizzo} 
\emph{SLAC, USA}\\
\noindent\href{mailto:rainer.roehrig@cern.ch}{Rainer Roehrig} 
\emph{Max-Planck-Institut für Physik, Germany}\\
\noindent\href{mailto:David.Salek@cern.ch}{David Salek} 
\emph{Nikhef and GRAPPA, Netherlands}\\
\noindent\href{mailto:Arturo.Rodolfo.Sanchez.Pineda@cern.ch}{Arturo Sanchez Pineda} 
\emph{INFN Sezione di Napoli, and Dipartimento di Fisica, Universit\`a di Napoli, Italy}\\
\noindent\href{mailto:s.sarkar1@physics.ox.ac.uk}{Subir Sarkar} 
\emph{Rudolf Peierls Centre for Theoretical Physics, University of Oxford, United Kingdom, and Niels Bohr Institute, Copenhagen, Denmark}\\
\noindent\href{mailto:Alexander.Schmidt@cern.ch}{Alexander Schmidt} 
\emph{University of Hamburg, Germany}\\
\noindent\href{mailto:Steven.Schramm@cern.ch}{Steven Randolph Schramm} 
\emph{Universit\'e de Gen\`eve, DPNC, Switzerland}\\
\noindent\href{mailto:will.shepherd@gmail.com}{William Shepherd} 
\emph{University of California Santa Cruz Department of Physics and Santa Cruz Institute for Particle Physics, USA, and Niels Bohr International Academy, University of Copenhagen, Denmark}\\
\noindent\href{mailto:gurpreet.singh@cern.ch}{Gurpreet Singh} 
\emph{Chulalongkorn University, Thailand}\\
\noindent\href{mailto:Livia.Soffi@cern.ch}{Livia Soffi} 
\emph{Cornell University, USA}\\
\noindent\href{mailto:Norraphat.Srimanobhas@cern.ch}{Norraphat Srimanobhas} 
\emph{Chulalongkorn University, Faculty of Science, Department of Physics, Thailand}\\
\noindent\href{mailto:kevin.kai.hong.sung@cern.ch}{Kevin Sung} 
\emph{Northwestern University, USA}\\
\noindent\href{mailto:ttait@uci.edu}{Tim M. P. Tait} 
\emph{Department of Physics and Astronomy, University of California, Irvine, USA}\\
\noindent\href{mailto:Timothee.Theveneaux-Pelzer@cern.ch}{Timothee Theveneaux-Pelzer} 
\emph{Laboratoire de Physique Corpusculaire, Clermont Universit\'e and Universit\'e Blaise Pascal and CNRS/IN2P3, Clermont-Ferrand, France}\\
\noindent\href{mailto:M.C.Thomas@soton.ac.uk}{Marc Thomas} 
\emph{Southampton University, United Kingdom}\\
\noindent\href{mailto:mia.tosi@cern.ch}{Mia Tosi} 
\emph{University of Padova and INFN, Italy}\\
\noindent\href{mailto:Daniele.Trocino@cern.ch}{Daniele Trocino} 
\emph{Northeastern University, Boston, USA}\\
\noindent\href{mailto:sonaina.undleeb@ttu.edu}{Sonaina Undleeb} 
\emph{Texas Tech University, USA}\\
\noindent\href{mailto:alessandro.vichi@cern.ch}{Alessandro Vichi} 
\emph{Theory division, CERN, Switzerland }\\
\noindent\href{mailto:Fuquan.Wang@cern.ch}{Fuquan Wang} 
\emph{University of Wisconsin-Madison, USA}\\
\noindent\href{mailto:liantaow@uchicago.edu}{Lian-Tao Wang} 
\emph{Enrico Fermi Institute and Department of Physics and Kavli Institute for Cosmological Physics, University of Chicago, USA}\\
\noindent\href{mailto:renjie.wang@cern.ch}{Ren-Jie Wang} 
\emph{Department of Physics, Northeastern University, USA}\\
\noindent\href{mailto:alokin@uw.edu}{Nikola  Whallon} 
\emph{Physics, University of Washington, Seattle, USA}\\
\noindent\href{mailto:steven.worm@cern.ch}{Steven Worm} 
\emph{Particle Physics Department, Rutherford Appleton Laboratory, United Kingdom}\\
\noindent\href{mailto:mengqing.wu@cern.ch}{Mengqing Wu} 
\emph{Laboratoire de Physique Subatomique et de Cosmologie, Universit\'e Grenoble-Alpes, CNRS/IN2P3, France}\\
\noindent\href{mailto:sau.lan.wu@cern.ch}{Sau Lan Wu} 
\emph{University of Wisconsin-Madison, USA}\\
\noindent\href{mailto:Hongtao.Yang@cern.ch}{Hongtao Yang} 
\emph{University of Wisconsin-Madison, USA}\\
\noindent\href{mailto:Yong.Yang@cern.ch}{Yong Yang} 
\emph{Universit\"at Zurich, Switzerland}\\
\noindent\href{mailto:Shin-Shan.Yu@cern.ch}{Shin-Shan Yu} 
\emph{National Central University, Taiwan}\\
\noindent\href{mailto:bryan.zaldivar@ulb.ac.be}{Bryan Zaldivar} 
\emph{Universit\'e Libre de Bruxelles, Belgium}\\
\noindent\href{mailto:marco.zanetti@cern.ch}{Marco Zanetti} 
\emph{Universit\`a di Padova, Italy}\\
\noindent\href{mailto:zhangzq@lal.in2p3.fr}{Zhiqing Zhang} 
\emph{Laboratoire de l'Acc\'el\'erateur Lin\'eaire, Univ. Paris-Sud 11 et IN2P3/CNRS, France}\\
\noindent\href{mailto:a.zucchetta@cern.ch}{Alberto Zucchetta} 
\emph{Universit\`a di Padova, Italy}\\

~\\
\noindent Contact editors: \href{mailto:lhc-dmf-admin@cern.ch}{lhc-dmf-admin@cern.ch}
}\\
\end{fullwidth}
\vspace*{\fill}

\setcounter{tocdepth}{3}
\tableofcontents
 
\pagebreak

\chapter{Introduction}
\label{sec:Introduction}


Dark matter (DM) \sidenote{Many theories of physics beyond the Standard Model predict the existence
of stable, neutral, weakly-interacting and massive particles that are
putative Dark Matter candidates. In the following, we refer to such
matter as Dark Matter, even though the observation of such matter at a collider
could only establish that it is neutral, weakly-interactive, massive and stable
on the distance-scales of tens of meters.} has not yet been observed in particle physics experiments, and
there is not yet any evidence for non-gravitational interactions
between Dark Matter and Standard Model (SM) particles.  If such
interactions exist, particles of Dark Matter could be produced
at the LHC. Since Dark Matter particles themselves do not produce signals
in the LHC detectors, one way to observe them is when they are produced in association
with a visible SM particle X(=$g, q, \gamma, Z, W$, or $h$).
Such reactions, which are
observed at colliders as particles or jets recoiling against an invisible state, are
called ``mono-X'' or \MET{}+X reactions (see e.g 
Refs.~\cite{Birkedal:2004xn,Feng:2005gj,Petriello:2008pu,Beltran:2010ww,Bai:2010hh}), 
where \MET is the missing transverse momentum observable in the detector.

Early Tevatron and LHC Run-1 searches for \MET{}+X signatures at 
CDF~\cite{Aaltonen:2012jb}, 
ATLAS~\cite{Aad:2015zva,Aad:2014tda,ATLAS:2014wra,Aad:2014vka,Aad:2013oja,Aad:2014wza,Aad:2014vea,ATL-PHYS-PUB-2014-007}
and
CMS~\cite{Khachatryan:2014rra,Khachatryan:2014rwa,Khachatryan:2014tva,Khachatryan:2014uma,Khachatryan:2015nua,CMS-PAS-B2G-13-004,CMS-PAS-EXO-14-004},
employed a basis of contact interaction operators in effective field
theories (EFTs) \cite{Goodman:2010yf,Goodman:2010ku} to calculate the
possible signals. 
These EFTs assume that production of Dark Matter takes place through a
contact interaction involving a quark-antiquark pair, or two gluons,
and two Dark Matter particles.  In this case, the missing energy
distribution of the signal is determined by the nature and the mass of
the Dark Matter particles and the Lorentz structure of the
interaction. Only the overall production rate is a free parameter to
be constrained or measured.  Provided that the contact interaction
approximation holds, these EFTs provide a straightforward way to
compare the results from different collider searches with non-collider
searches for Dark Matter.  

The EFT describes the case when the mediator of the interaction between SM and DM particles are very heavy; 
if this is not the case, models that explicitly include these mediators are 
needed~\cite{Goodman:2010yf,Shoemaker:2011vi,Bai:2010hh,Kopp:2011eu,Fox:2011fx,Fox:2011pm,Shoemaker:2011vi,Busoni:2013lha}.
Some ``simplified models'' \cite{Alwall:2008ag,Goodman:2011jq,Alves:2011wf}
of Dark Matter production were constructed, including particles and interactions beyond the SM.
These models can be used consistently at LHC energies, and provide
an extension to the EFT approach. 
Many proposals for such models have emerged (see, for example
Refs. \cite{An:2012va,An:2012ue,Tait:2013,Buchmueller:2013dya,Bai:2013iqa,Bai:2014osa,An:2013xka,Yavin:14092893,Malik:2014ggr,Harris:2014hga,Buckley:2014fba,Haisch:2015ioa,Bai:2012xg,Carpenter:2012rg,Bell:2012rg,Petrov:2013nia,Carpenter:2013xra}). 
At the LHC, the kinematics of mono-X reactions occurring via a \tev-scale mediator can differ substantially from the prediction of the contact
interaction. The mediator may also produce qualitatively different signals, such as decays back into Standard Model particles. 
Thus, appropriate simplified models are an important component of the design, optimization, and interpretation of Dark Matter searches at ATLAS and CMS.
This has already been recognized in the CDF, ATLAS and CMS searches quoted above, where both EFT and selected simplified model
results are presented. 

\section{The ATLAS/CMS Dark Matter Forum}

To understand what signal models should be considered for the upcoming LHC Run-2, 
groups of experimenters from both ATLAS and CMS collaborations have held separate 
meetings with small groups of theorists, and discussed further at the DM@LHC 
workshop~\cite{Malik:2014ggr,Yavin:14092893,DMatLHCProceedings}. 
These discussions identified overlapping sets of simplified models as possible
benchmarks for early LHC Run-2 searches. 
Following the DM@LHC workshop, ATLAS and CMS organized a forum, called the~\textit{ATLAS-CMS Dark
Matter Forum}, to form a consensus on the use of these simplified models
and EFTs for early Run-2 searches with the participation of experts on
theories of Dark Matter. This is the final report of the ATLAS-CMS Dark Matter Forum.

One of the guiding principles of this report is to channel the efforts
of the ATLAS and CMS collaborations towards a minimal basis of dark
matter models that should influence the design of the early Run-2
searches. At the same time, a thorough survey of realistic collider
signals of Dark Matter is a crucial input to the overall design of the
search program.

The goal of this report is such a survey, though confined within some
broad assumptions and focused on benchmarks for kinematically-distinct
signals which are most urgently needed. As far as time and resources
have allowed, the assumptions have been carefully motivated by
theoretical consensus and comparisons of simulations. But, to achieve such a 
consensus in only a few months before the start of Run-2, it was
important to restrict the scope and timescale to the following:

\begin{enumerate}
\item The forum should propose a prioritized, compact set of benchmark
  simplified models that should be agreed upon by both collaborations for
  Run-2 searches. The values for the scan on the parameters of the models for which
  experimental results are provided should be specified, to facilitate theory reinterpretation 
  beyond the necessary model-independent limits that 
  should be provided by all LHC Dark Matter searches. 
\item The forum should recommend the use of the state of the art calculations
  for these benchmark models. Such a recommendation will aid the  
  standardization the event generator implementation
  of the simplified models and the harmonization of other common technical
  details as far as practical for early Run-2 LHC analyses. It
  would be desirable to have a common choice of leading order (LO) and 
  next-to-leading order (NLO) matrix elements corresponding to the state of the art calculations, 
  parton shower (PS) matching and merging, factorization and renormalization
  scales for each of the simplified models. This will also lead to a
  common set of theory uncertainties, which will facilitate the
  comparison of results between the two collaborations.
\item The forum should discuss how to apply the
  EFT formalism and present the results of EFT
  interpretations.
\item The forum should prepare a report summarizing these items,
  suitable both as a reference for the internal ATLAS and CMS
  audiences and as an explanation of early Run-2 LHC benchmark models for theory and non-collider
  readers. This report represents the views of its endorsers, as participants of the forum.
\end{enumerate}

\section{Grounding Assumptions}

We assume that interactions exist between Standard Model hadrons
and the particles that constitute cosmological Dark Matter. If this
is not the case, then proton collisions will not directly produce Dark Matter
particles, and Dark Matter will not scatter off nuclei in direct
detection experiments.

The Dark Matter itself is assumed to be a single particle, a Dirac
fermion WIMP, stable on collider timescales and non-interacting with
the detector.  
The former assumption is reductionistic.
The rich particle content of the Standard Model is circumstantial evidence that
the Dark Matter sector, which constitutes five times as much of the
mass of the universe, may be more complex than a single particle or a
single interaction. But, as was often the case in the discoveries of
the SM, here only one mediator and one search channel might play a
dominant role in the opening stages of an LHC discovery. The latter
assumption focuses our work on early LHC searches, where small
kinematic differences between models will not matter in a discovery
scenario, and with the imminent re-start of the LHC our report relies
heavily on a large body of existing theoretical work which assumed Dirac fermionic Dark Matter. 

Different spins of Dark Matter particles will typically
give similar results. Exceptions exist: For example, the choice of Majorana fermions forbids some
processes that are allowed for Dirac fermions~\cite{Goodman:2010yf}.
Aside from these, adjusting the choice of Dirac or Majorana fermions or scalars will produce only minor changes
in the kinematic distributions of the visible particle and is expected to have little effect
on cut-and-count\sidenote{Cut-and-count refers to an analysis
that applies a certain event selection and checks the inclusive number of events which pass. 
This is to be contrasted with a shape analysis, which compares the distribution of events.} analysis. Thus the choice of Dirac
fermion Dark Matter should be sufficient as benchmarks for the upcoming Run-2 searches. 


One advantage of collider experiments lies in their ability to study
and possibly characterize the mediator. A discovery of an anomalous
\MET signature at the LHC would not uniquely imply discovery of dark
matter, while at the same time e.g. discovery of an anomalous and
annually-modulated signal in a direct-detection experiment would leave
unanswered many questions about the nature of the interaction that
could be resolved by the simultaneous discovery of a new mediator
particle. Collider, direct, and indirect detection searches provide
complementary ways to approach this problem~\cite{Bauer:2013ihz}, and it is in this spirit
that much of our focus is on the mediator.

We systematically explore the basic possibilities for
mediators of various possible spins and couplings.
All models considered are assumed to produce a signature with pairs of Dark Matter particles.
Though more varied and
interesting possibilities are added to the literature almost daily,
these basic building blocks account for much of the physics studied at
hadron colliders in the past three decades.

We also assume that Minimal Flavor Violation (MFV) \cite{Chivukula:1987py,Hall:1990ac,Buras:2000dm,D'Ambrosio:2002ex} applies to the
models included in this report. This means that the flavor structure of the
couplings between Dark Matter and ordinary particles follows the same
structure as the Standard Model. This choice is simple, since no
additional theory of flavor is required, beyond what is already
present in the SM, and it provides a mechanism to ensure that the
models do not violate flavor constraints.  As a consequence, \spinzero
resonances must have couplings to fermions proportional to the SM Higgs couplings. 
Flavor-safe models can still be constructed beyond the MFV
assumption, for example ~\cite{Agrawal:2014aoa}, and deserve further study.
For a discussion of MFV in the context of the simplified models
included in this report, see Ref.~\cite{DMatLHCProceedings}.

In the parameter scan for the models considered in this report, we make the
assumption of a minimal decay width for the particles mediating the
interaction between SM and DM.  This means that only decays
strictly necessary for the self-consistency of the model (e.g.  to DM
and to quarks) are accounted for in the definition of the mediator
width. We forbid any further decays to other invisible particles of
the Dark Sector that may increase the width or produce striking, visible signatures. 
Studies within this report show that, for cut-and-count analyses, the kinematic distributions of
many models, and therefore the sensitivity of these searches, do not depend
significantly on the mediator width, as long as the width remains smaller
than the mass of the particle and that narrow mediators are sufficiently light.

The particle content of the models chosen as benchmarks is limited to
one single kind of DM whose self-interactions are not relevant for LHC
phenomenology, and to one type of SM/DM interaction at a time. These
assumptions only add a limited number of new particles and new interactions to the
SM. These simplified models, independently explored by different
experimental analyses, can be used as starting points to build more
complete theories. Even though this factorized picture does not always
lead to full theories and leaves out details that are necessary for
the self-consistency of single models (e.g. the mass generation for
mediator particles), it is a starting point to prepare a set of
distinct but complementary collider searches for Dark Matter, as
it leads to benchmarks that are easily comparable across channels.

\section{Choices of benchmarks considered in this report and parameter scans}

Contact interaction operators have been outlined as basis set of theoretical
building blocks representing possible types of interactions between SM and DM particles
in~\cite{Goodman:2010ku}. The approach followed by LHC searches (see e.g. Refs.~\cite{Khachatryan:2014rra,Aad:2015zva} 
for recent jet+\MET{} Run-1 searches with the 8 TeV dataset) 
so far has been to simulate only a prioritized set of the possible operators with distinct kinematics
for the interpretation of the constraints obtained, and provide results that may be reinterpreted in terms of the other operators.
This report intends to follow this strategy, firstly focusing on simplified models that allow the exploration 
of scenarios where the mediating scale is not as large.  In the limit of large mediator mass, the simplified models map onto
the EFT operators.
Secondly, this report considers specific EFT benchmarks 
whenever neither a simplified model completion 
nor other simplified models yielding similar kinematic distributions are available 
and implemented in one of the event generators used by both collaborations. 
This is the case for dimension-5 or dimension-7 operators with direct 
DM-electroweak boson couplings \sidenote{An example of a dimension-5 operator for scalar
DM is described in Appendix~\ref{app:EWSpecificModels_Appendix}. 
Dimension-7 operators of DM coupling to gauge bosons exist in the literature, but they require a larger particle spectrum
with respect to the models studied in this report.}.
Considering these models as separate experimental benchmarks 
will allow to target new signal regions and help validate the 
contact interaction limit of new simplified models 
developed to complete these specific operators. 
Results from these EFT benchmarks should include the condition that
the momentum transfer does not probe the scale of the interaction; whenever there is no model
that allows a direct mapping between these two quantities, various options should be tested to 
ensure a given fraction of events within the range of applicability of the EFT approach.
Experimental searches should in any case deliver 
results that are independent from the specific benchmark tested, such as fiducial cross-sections that
are excluded in a given signal region. 

When choosing the points to be scanned in the parameter space of the models,
this report does not quantitatively consider constraints that are 
external to the MET+X analyses. This is the case also for results from LHC experiments
searching for mediator decays. 
The main reason for not doing so in this report 
is the difficulty of incorporating these constraints in a rigorous quantitative way within
the timescale of the Forum. However, even if the parameter scans
and the searches are not optimized with those constraints in mind, 
we intend to make all information available to the community to exploit
the unique sensitivity of colliders to all possible DM signatures. 

\section{Structure of this report and dissemination of results}

The report provides a brief theoretical summary of the models considered, 
starting from the set of simplified models and contact interactions put forward 
in previous discussions and in the literature cited above. 
Its main body documents the studies 
done within this Forum to identify a kinematically distinct set of model parameters
to be simulated and used as benchmarks for early Run-2 searches. The implementation
of these studies according to the state of the art calculations is detailed,
including instructions on how to estimate theoretical uncertainties in the generators used
for these studies. The presentation of results for EFT benchmarks is also covered. 

Chapter \ref{subsec:MonojetLikeModels} of this report is dedicated to simplified
models with radiation of a hard object either from the initial state
or from the mediator. These models produce primarily monojet signatures, 
but should be considered for all \MET{}+X searches.
Chapter~\ref{subsec:EWSpecificModels} contains studies on the benchmark models
for final states specifically containing an electroweak 
boson ($W/Z/\gamma/H$). In this case, both 
simplified models leading to mono-boson signatures
and contact interaction operators are considered. 
Details of the state of the art calculations and on the implementation of the simplified models in
Monte Carlo generators are provided in
Chapter \ref{app:MonojetLikeModels_Appendix}.
Chapter \ref{sec:EFTValidity} is devoted to the treatment of the presentation of results for the benchmark
models from contact interaction operators. 
Chapter \ref{sec:TheoryUncertainties} prescribes how to estimate theoretical uncertainties on the simulation of these models. 
Chapter \ref{chapter:conclusions} concludes the report.

Further models that could be studied
beyond early searches and their implementation are described in Appendix~\ref{app:EWSpecificModels_Appendix}. 
For these models, either the implementation could not be fully developed by the time of this report,
or some of the grounding assumptions were not fully met.  
Some of these models have been used in previous ATLAS and CMS analyses and discussed thoroughly within the Forum. 
They are therefore worth considering for further studies and for Run-2 searches, since they lead to unique \MET{}+X signatures 
that are not shared by any other of the models included in this report. 
Appendix~\ref{app:Presentation_Of_Experimental_Results} contains the necessary elements that
should be included in the results of experimental searches to allow for further reinterpretation. 

It is crucial for the success of the work of this Forum that these studies can be employed as cross-check
and reference to the theoretical and experimental community interested in early Run-2 searches. 
For this reason, model files, parameter cards, and cross-sections for the models considered in these studies 
are publicly available. The SVN repository of the Forum~\cite{ForumSVN} contains the models and parameter files
necessary to reproduce the studies within this report. Details and cross-sections for these models, 
as a function of their parameters, will be published on HEPData~\cite{HEPData}.

\chapter{\texorpdfstring{Simplified models for all \MET+X analyses}{Simplified models for all MET+X analyses}}
\label{subsec:MonojetLikeModels}


In this Chapter we review 
models that yield $X$+\MET{} signatures,
where $X$ is a QCD parton or $\gamma, W, Z$ or $h$.

The primary simplified models for Dirac fermion DM studied and recommended by this Forum 
for early LHC Run-2 searches are detailed in this Chapter, 
comprising \spinzero and \spinone mediators. Section~\ref{sec:monojet_V} covers the
\schannel exchange of a vector mediator~\footnote{Colored vector mediators 
can be exchanged in the \tchannel, but there are no examples in literature so far.}, 
while we consider both \schannel and \tchannel exchange for scalar mediators in
Section~\ref{sec:monojet_scalar} and~\ref{sec:monojet_t_channel} respectively. 
\Spintwo mediators are briefly mentioned in Section~\ref{sec:spintwo}.
While these models are general and cover a broad set of signatures,
the discussion and studies are focused on the monojet final state. 
Details on final states with electroweak (EW) boson radiation and with heavy flavor quarks 
from diagrams arising within these models are also discussed in this Chapter.

A summary of the state of the art calculations and implementations for these models 
is provided in Table~
6.1. Section~\ref{app:MonojetLikeModels_Appendix}
details the implementation of these models that
have been used for the studies in this Chapter and that will be employed
for the simulation of early Run-2 benchmark models for LHC DM searches. 


\section{Vector and axial vector mediator, \schannel exchange}
\label{sec:monojet_V}

A simple extension of the Standard Model (SM) is an
additional $U(1)$ gauge symmetry, where a Dark Matter
candidate particle has charges only under this new group.
Assuming that some SM particles are also charged under
this group, a new gauge boson can mediate interactions
between the SM and DM.   

\begin{figure}[h!]
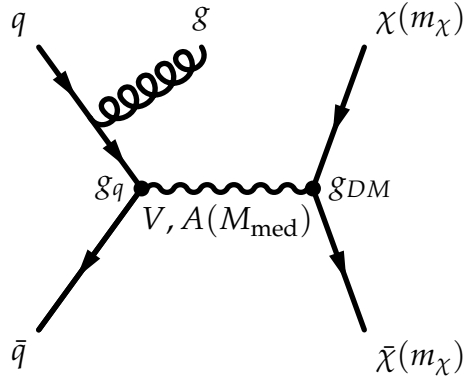

\centering
  \unitlength=0.005\textwidth
  \vspace{0.5\baselineskip}
  \begin{feynmandiagram}[modelVmonojetParameters]
    \fmfleft{i1,i2}
    \fmfright{o1,o2}
    \fmftop{isr}
    \fmfbottom{pisr}
    \fmf{wiggly,tension=0.6,label={\Large $V,,A(\mMed)$}}{v1,v2}
    \fmf{fermion}{o2,v2,o1}
    \fmf{fermion}{i2,visr,v1}
    \fmf{plain}{v1,pvisr,i1}
    \fmf{fermion,tension=0}{v1,i1}
    \fmfdot{v1,v2}
    \fmflabel{\Large ${g_q}$}{v1}
    \fmflabel{\Large ${g_{DM}}$}{v2}
    \fmflabel{\Large ${\bar{q}}$}{i1}
    \fmflabel{\Large ${q}$}{i2}
    \fmflabel{\Large ${\bar{\chiDM}(\mDM)}$}{o1}
    \fmflabel{\Large ${\chiDM(\mDM)}$}{o2}
    \fmf{gluon,tension=0}{visr,isr}
    \fmf{phantom,tension=0}{pvisr,pisr}
    \fmflabel{\Large ${g}$}{isr}
  \end{feynmandiagram}
\caption{Representative Feynman
diagram showing the pair production of Dark Matter particles in association with a parton from the initial state via a vector or axial-vector mediator.
The cross section and kinematics depend upon
the mediator and Dark Matter masses, and the mediator couplings to Dark Matter and quarks respectively: ($\mMed ,\, \mDM ,\, \gDM ,\, \gq)$. }
\label{fig:OP}
\setfloatalignment{t}
  \vspace{0.5\baselineskip}
\end{figure}

We consider the case of a DM particle \chiDM of mass \mdm that is a Dirac fermion and where the production 
proceeds via the exchange of a \spinone mediator of mass \mMed in
the \schannel, illustrated in Fig.~\ref{fig:OP}.

\begin{figure}
\centering
\includegraphics[width=0.95\textwidth]{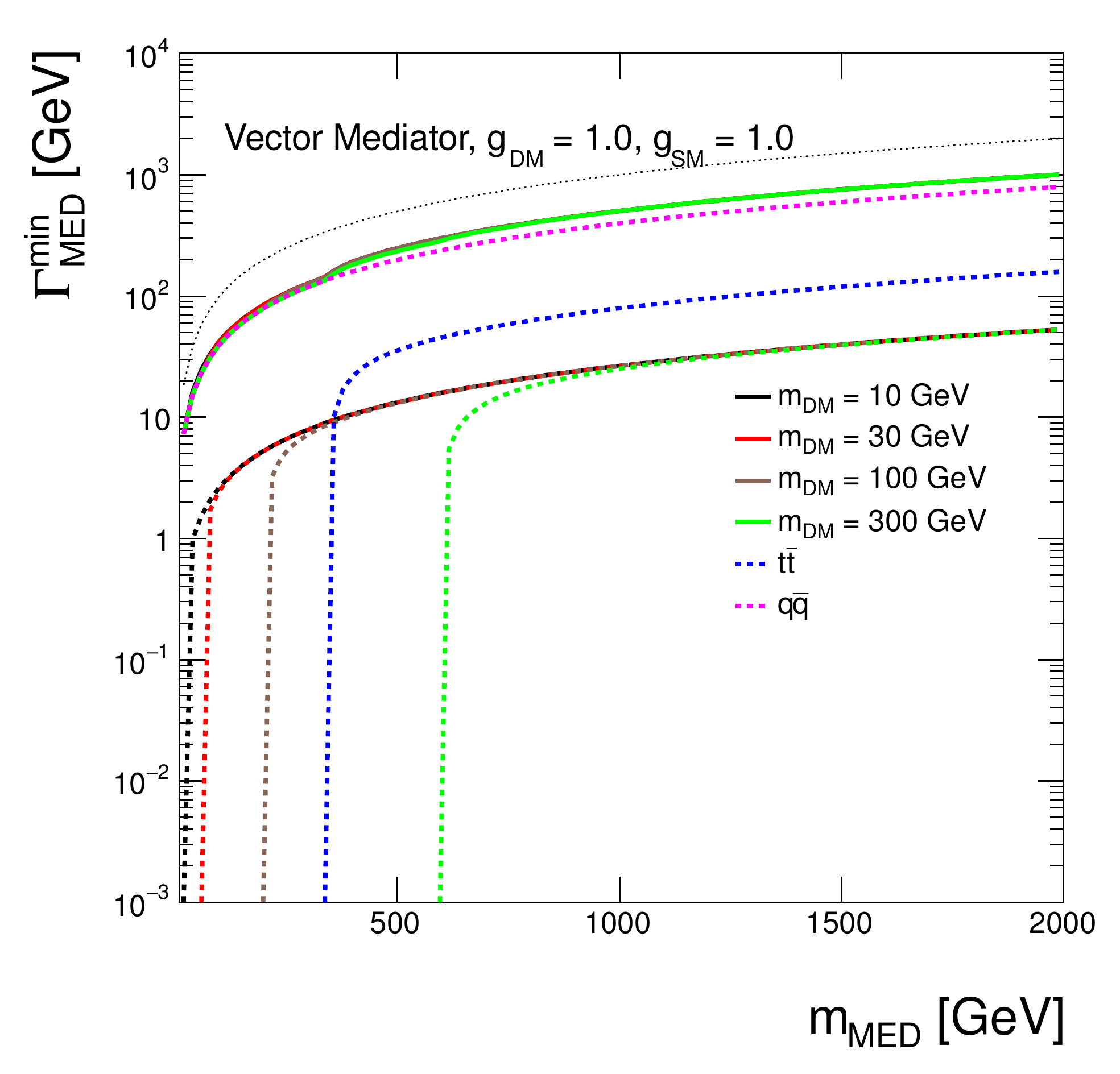}
\includegraphics[width=0.95\textwidth]{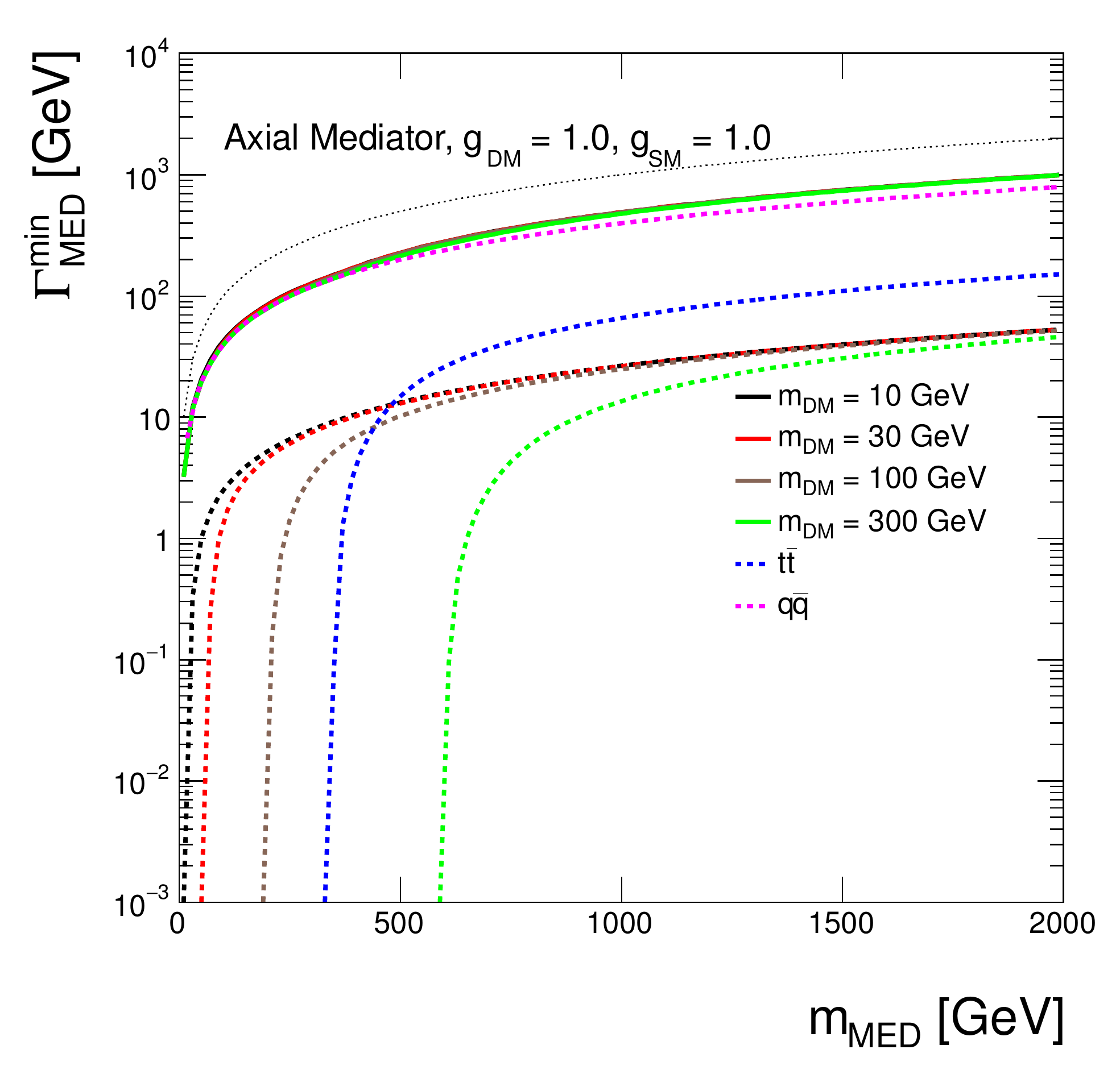}
\caption{Minimal width as a function of mediator mass for vector and axial-vector mediator assuming couplings of 1. The total width is shown as solid lines for Dark Matter masses of 10~\gev, 30~\gev, 100~\gev and 300~\gev in black, red, brown and green, respectively. The individual contributions from Dark Matter are indicated by dotted lines with the same colors. The contribution from all quarks but top is shown as magenta dotted line and the contribution from top quarks only is illustrated by the dotted blue line. The dotted black line shows the extreme case $\Gamma_{\rm{min}}=\mMed$.}
\label{fig:monojet_width_V}
\end{figure}

We consider two models with vector and axial-vector couplings
between the \spinone mediator $\Zprime$ and SM and DM fields, with
the corresponding interaction Lagrangians:

\begin{align}
\label{eq:AV} 
\mathcal{L}_{\mathrm{vector}} &= \gq \sum_{q={u,d,s,c,b,t}}  \Zprime_{\mu} \bar{q}\gamma^{\mu}q + \gDM \Zprime_{\mu} \bar{\chiDM}\gamma^{\mu}\chiDM \\
\mathcal{L}_{\rm{axial-vector}} &= \gq \sum_{q={u,d,s,c,b,t}}  \Zprime_{\mu} \bar{q}\gamma^{\mu}\gamma^5q + \gDM \Zprime_{\mu} \bar{\chiDM}\gamma^{\mu}\gamma^5\chiDM.
\end{align}
The coupling \gq is assumed to be universal to all quarks.
It is also possible to consider other models in which mixed vector and axial-vector couplings are considered, 
for instance the couplings to the quarks are axial-vector whereas those to DM are vector. 
As mentioned in the Introduction, when no additional visible or invisible decays contribute to the width of the mediator, 
the minimal width is fixed by the choices of couplings \gq and \gDM. The effect of larger 
widths is discussed in Section~\ref{paragraph:nonminimalwidth}. 
For the vector and axial-vector models, the minimal width is:

\begin{align}
\label{eq:monojet_min}
\Gamma^{\textrm{V}}_\textrm{min} &= 
\frac{\gDM^2 \mMed}{12\pi}\left(1+\frac{2 \mDM^2}{\mMed^2} \right)\beta_{DM} \theta(\mMed-2\mDM)\\\nonumber
 &+ \sum_q \frac{3 \gq^2 \mMed}{12\pi}\left(1+\frac{2 m_q^2}{\mMed^2} \right)\beta_q \theta(\mMed-2m_q),\\
\Gamma_{\textrm{min}}^{\rm{A}}&=\frac{\gDM^2 \mMed}{12\pi} \beta_{DM}^{3} \theta(\mMed-2\mDM)\\\nonumber
   &+ \sum_q \frac{3 \gq^2 \mMed}{12\pi}\beta_q^{3} \theta(\mMed-2m_q)\;.
\end{align}
$\theta(x)$ denotes the Heaviside step function, and
$\beta_f=\sqrt{1-\frac{4 m_f^2}{\mMed^2}}$ is the velocity of the
fermion $f$ with mass $m_f$  in the mediator rest frame.
Note the color factor 3 in the quark terms.
Figure\,\ref{fig:monojet_width_V} shows the minimal width as a function of mediator mass for both vector and axial-vector mediators assuming
the coupling choice $\gq=\gDM=1$. With this choice of the couplings, the dominant contribution to the minimal width comes from the quarks, due 
to the combined quark number and color factor enhancement. 
We specifically assume that the vector mediator does not couple to leptons.  If such a coupling were present, it would have a minor effect in increasing the mediator width, but it would also bring in constraints from measurements of the Drell-Yan process that would unnecessarily restrict the model space.

Therefore, the minimal set of parameters under consideration for these two models is
\bea
\left\{~\gq, ~ \gDM, ~\mDM,~\mMed,\right\} \,.
\eea
together with the spin structure of their couplings. 

A thorough discussion of these models and their parameters can also be found in~\cite{Buchmueller:2014yoa}.
 

These simplified models are known and available in event generators at NLO + PS accuracy, as detailed in Section~\ref{sec:monojet_implementation}. 
Results in this Section have been obtained using the model implementation within the \powheg generator (\textsc{v3359}) \cite{Haisch:2013ata},  interfaced to \pythiaEight~\cite{Sjostrand:2007gs} for the parton shower.

In addition, for the vector models considered, initial and final state radiation of a $Z'$ can occur which can appear as a narrow jet if it decays hadronically and may not be distinguishable from a QCD jet, thus accounting for some fraction of the monojet signal. The ISR and FSR of $Z'$ becomes more important at large values of the couplings~\cite{Bai:2015nfa}. 

\subsection{Parameter scan}
\label{sub:parameter_scan_monojet}

In order to determine an optimal choice of the parameter grid for the simulation of early Run-2 benchmark models, dependencies of the kinematic quantities and cross sections on the model parameters have been studied. Only points that are kinematically distinct will be fully simulated, while instructions on how to rescale the results according to models with different cross sections are presented in Section~\ref{sec:monojet_scaling}. The following paragraphs list the main observations from the scans over the parameters that support the final proposal for the benchmark signal grid.

\subsubsection{Scan over the couplings}

\begin{figure*}
	\centering
	\includegraphics[width=0.95\textwidth]{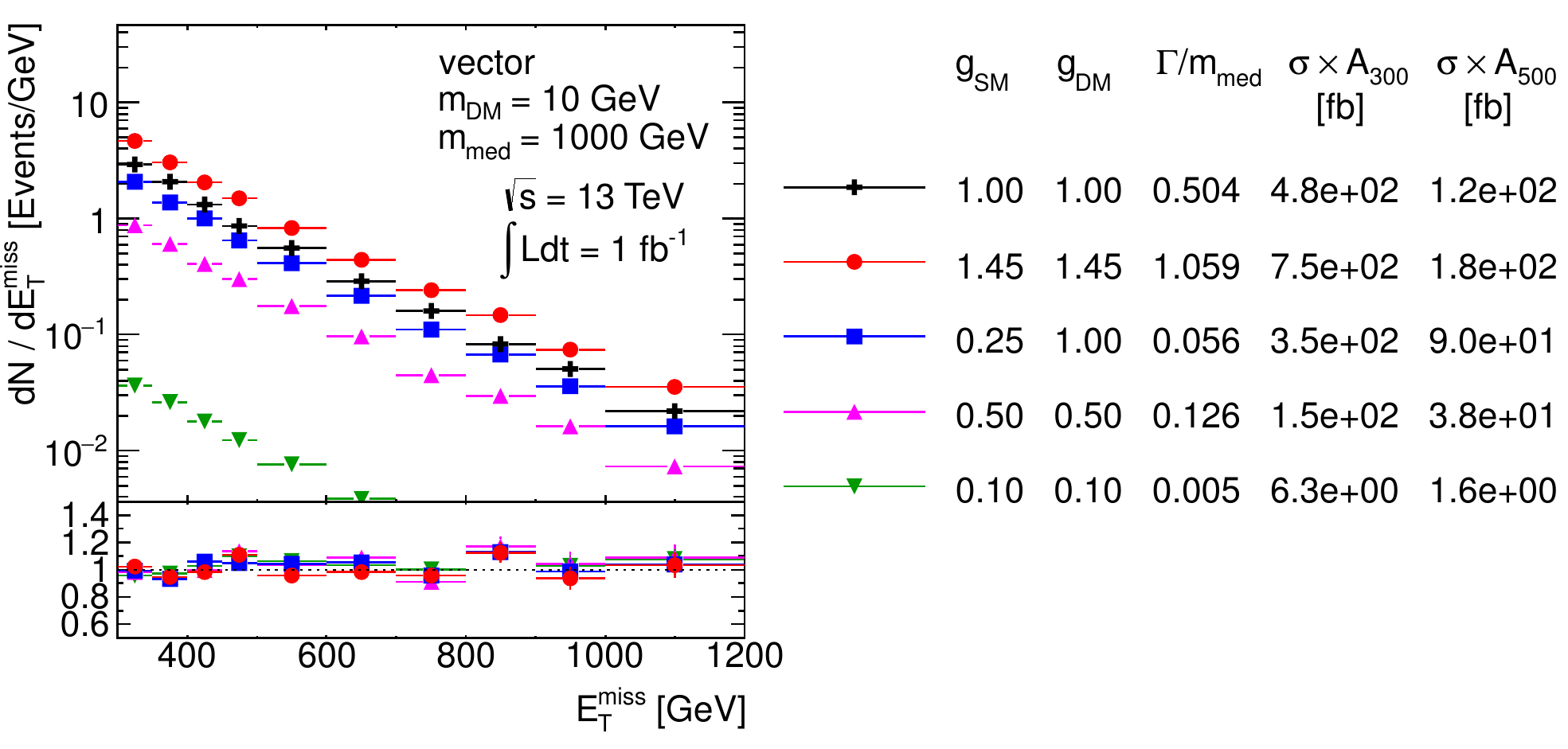}
	\caption[][-28pt]{Scan over couplings. The $\MET$ distribution is compared for the vector mediator models using the parameters as indicated. Ratios of the normalized distributions with respect to the first one are shown. $A_{300}$ and $A_{500}$ in the table denote the acceptance of the $\MET>300$~\gev and $\MET>500$~\gev cut, respectively. All figures in this Section have been obtained using the model implementation within the \powheg generator (\textsc{v3359}) \cite{Haisch:2013ata}, interfaced to \pythiaEight~\cite{Sjostrand:2007gs} for the parton shower. }
	\label{fig:monojet_scan_V_g}
\end{figure*}

To study the dependence of kinematic distributions on the coupling strength, samples were generated where a pair of $\mDM=10$~\gev Dark Matter particles is produced on-shell from the mediator of $\mMed=1$~\tev. 
Figure~\ref{fig:monojet_scan_V_g} compares the shapes of the \MET distribution for the different choices of the coupling strength. This is a generator-level prediction with no kinematic selections or detector simulation. Coupling values in the scan range 0.1--1.45, fixing $\gq=\gDM$, correspond to a rough estimate of the lower sensitivity of mono-jet analyses and a maximum coupling value such that $\Gamma_{\rm{min}} < \mMed$. We observe that the shapes of the \MET or jet \pT distributions do not depend on the couplings (and consequently the width) in the ranges considered. A large width of the mediator implies a broad integral over the contributing parton distributions, which might not be well approximated by the midpoint of this integral.  This study shows that the effect, in the \pT distribution of the observed gluon, is not important.

Based on similar findings for different choices of
\mMed and \mDM, we conclude that the shapes of
kinematic distributions are not altered
by coupling variations, neither for the on-shell mediator case where $\mMed>2\mDM$,
nor for the off-shell case where $\mMed<2\mDM$. Only the production cross sections change.
Differences in kinematic distributions are expected only close to the transition region between on-shell and off-shell mediators.

\begin{figure*}[!h]
	\centering
	\includegraphics[width=0.95\textwidth]{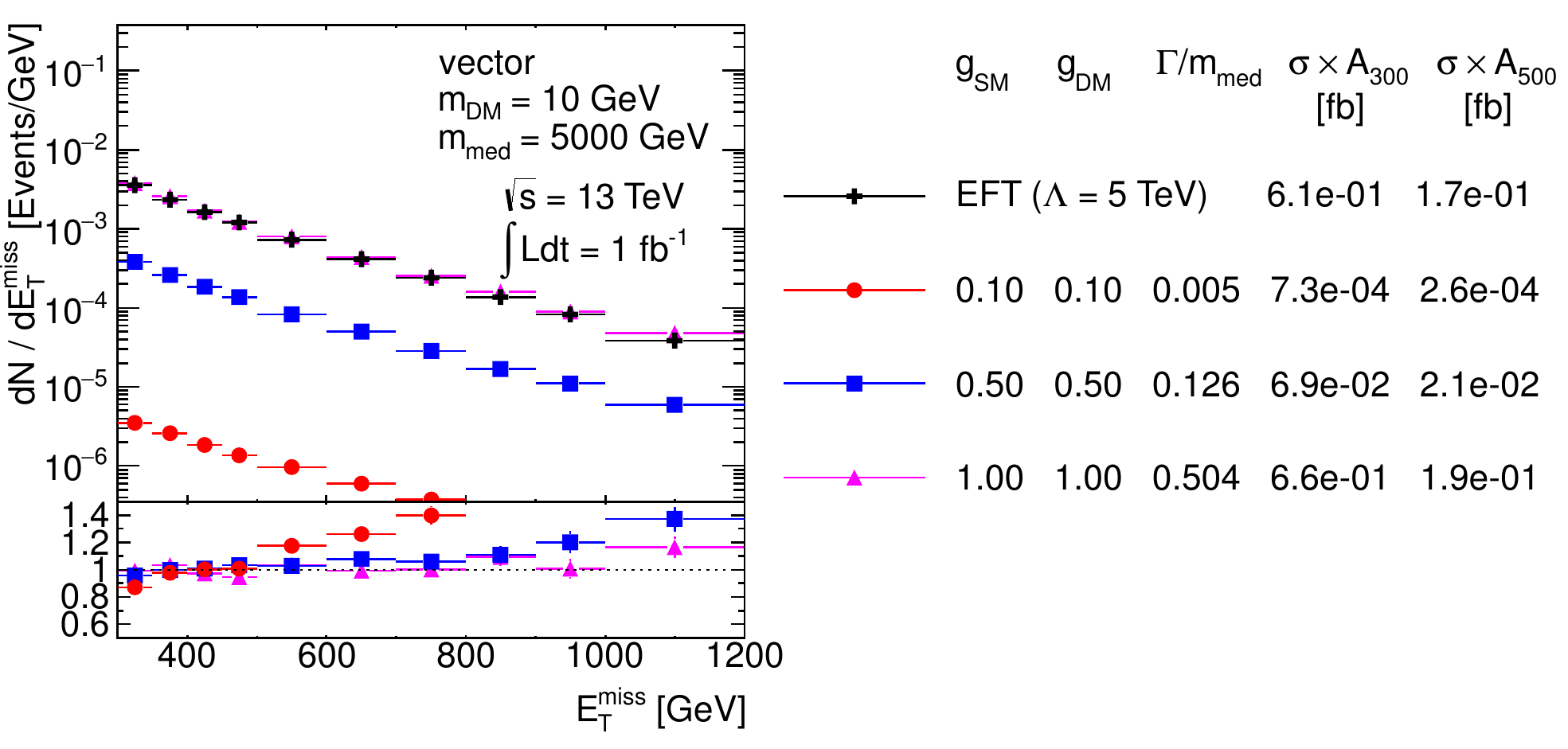}
	\caption[][-28pt]{Comparison of the $\MET$ distributions from the D5 EFT sample and the vector models with 5~\tev heavy mediator of various widths. Ratios of the normalized distributions with respect to the first one are shown. $A_{300}$ and $A_{500}$ in the table denote the acceptance of the $\MET>300$~\gev and $\MET>500$~\gev cut, respectively.}
	\label{fig:monojet_narrow}
\end{figure*}

Special care needs to be taken when coupling strengths are combined with extremely heavy mediators.
Figure\,\ref{fig:monojet_narrow} suggests a change in the shape of the
\MET distribution for a $\mMed=5$~\tev mediator
once $\Gamma_{\rm{min}}/\mMed$ is of the order of a percent or lower.

\begin{figure*}[!p]
\centering
\subfloat[]{
\includegraphics[width=0.7\textwidth]{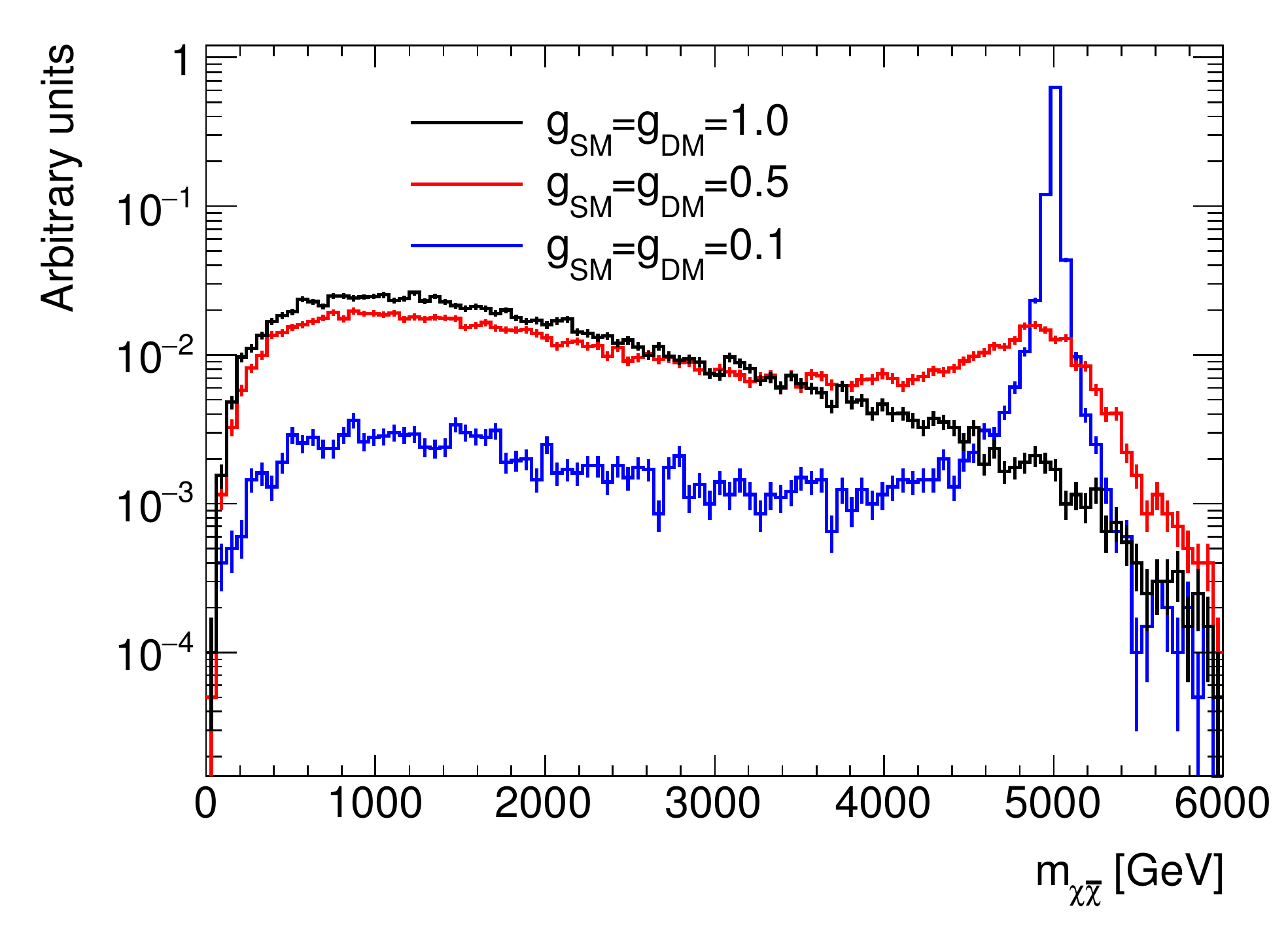}
\label{fig:monojet_mchichi_a}}
\hfill
\subfloat[]{
\includegraphics[width=0.7\textwidth]{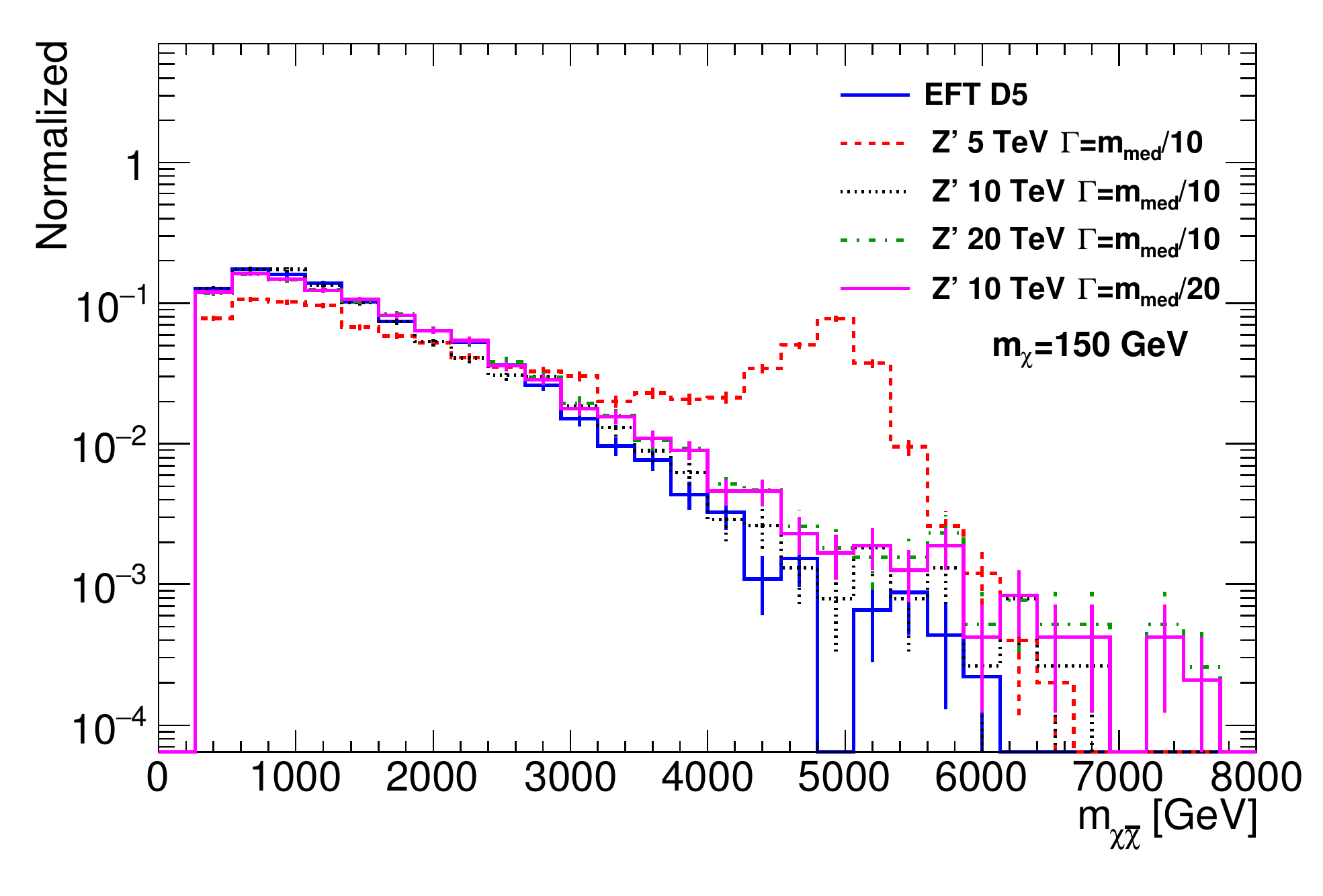}
\label{fig:monojet_mchichi_b}}
\caption[][14pt]{Invariant mass of the Dark Matter pair in the vector mediator samples with $\mDM=10$~\gev, $\mMed=5$~\tev and different coupling strengths (a).
A similar comparison is shown for the samples with different mediator masses considering $\Gamma_{\rm{min}}/\mMed=0.05$ and $0.1$ (b).
An EFT sample is also displayed in the latter case. The distributions are normalised to unit area.}
\label{fig:monojet_mchichi}
\end{figure*}

Such heavy mediators, although inaccessible with early LHC data, are interesting since they provide a good approximation for benchmark EFT models.
The observed difference among the simplified models in the plot arises from the fact that the region of low invariant masses of the Dark Matter pair, $m_{\bar{\chiDM}\chiDM}$, is suppressed due to narrow Breit-Wigner peak that only probes a narrow window of parton distribution functions. For wider mediators, the low mass region is significantly enhanced by parton distribution functions at low Bjorken $x$, as illustrated in Fig.\,\ref{fig:monojet_mchichi}(a).
This explains why the sample with the narrowest mediator in Fig.\,\ref{fig:monojet_narrow} is heavily suppressed in terms of production cross section and also gives different \MET shape.
Furthermore, Fig.\,\ref{fig:monojet_narrow} compares the vector model with 5~\tev mediator to the D5 EFT sample and reveals that the simplified models with larger mediator widths (e.g. for couplings of 1 where $\Gamma_{\rm{min}}/\mMed\sim0.5$) are the ones resembling the kinematics of contact interactions. This reflects the fact that in an EFT there is no enhancement due to on-shell mediators, leading to a closer resemblance to an off-shell regime where no peak in the $m_{\bar{\chiDM}\chiDM}$ distribution is present.
In case of narrow width mediators, e.g. $\Gamma_{\rm{min}}/\mMed\sim0.05$, even larger mediator masses need to be chosen in order to significantly suppress the peak in the $m_{\bar{\chiDM}\chiDM}$ distribution and reproduce the kinematic shapes of an EFT model. Figure\,\ref{fig:monojet_mchichi}(b) verifies that the choice of 10~\tev mediator mass is sufficient to achieve that.

Since kinematic distributions are robust to
changes in the specific values of coupling~\footnote{This applies
as long as heavy narrow mediators are generated without any truncation
of low-mass tails at the generator-level.},
the choice of $\gq=$0.25 and $\gDM=$1 is reasonable 
to reduce the parameter space to be scanned. 
There are no complications associated
with small couplings, but, also, the early part of Run~2 will not be
sensitive to them.  The range of couplings we recommend to generate limit the
calculated width of the mediator to be near or below \mMed.

For direct mediator searches, such as $q\bar q\to Z^\prime \to q\bar q$, different couplings ($\gq \ne \gDM$)
might also be considered. A scan in \gDM vs \gq can then be performed for a fixed mediator mass. Such searches
may restrict \gq to a greater degree than
\gDM.

\subsubsection{Scan over \mDM}

\begin{figure*}
\centering
\includegraphics[width=0.95\textwidth]{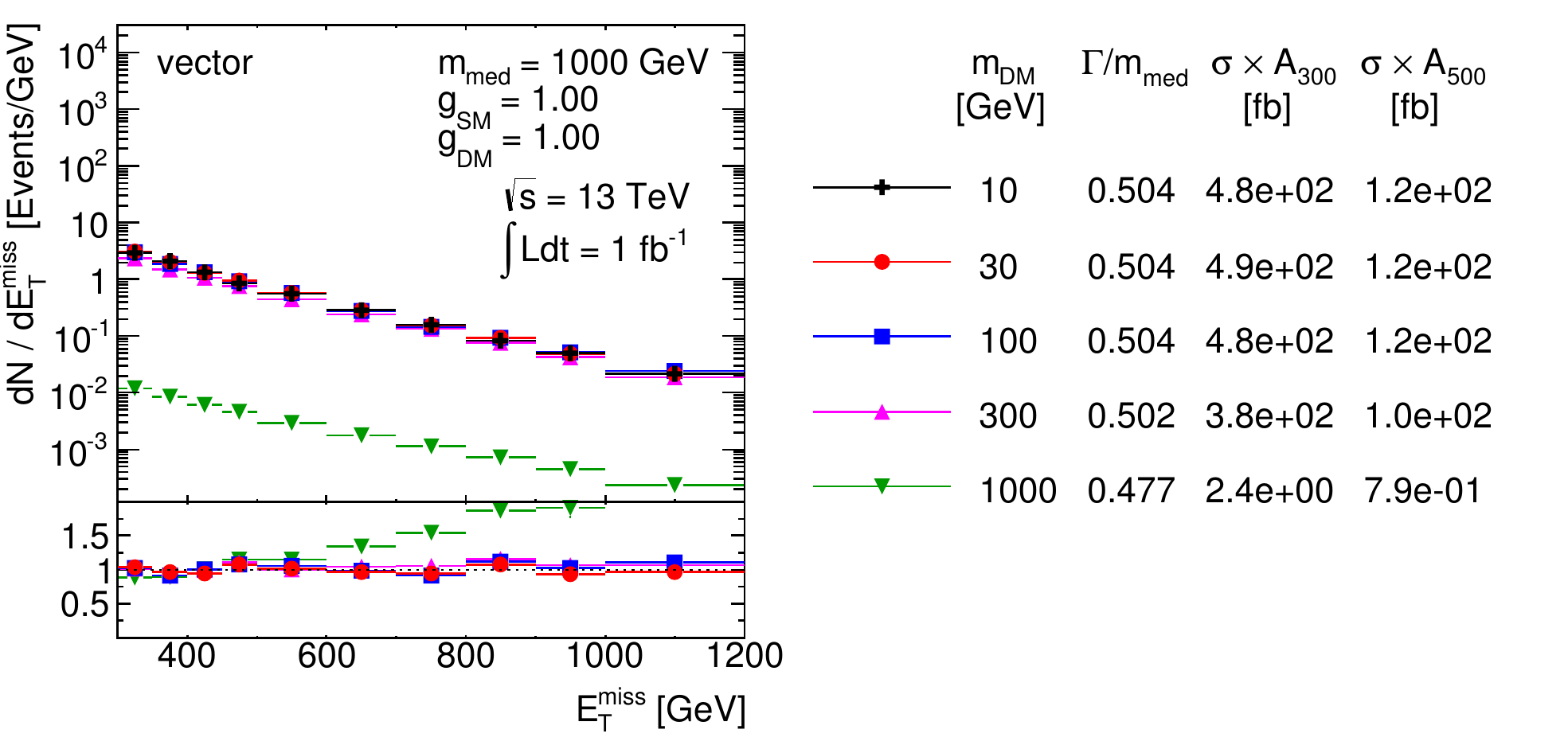}
\vspace{4\baselineskip}
\caption[][-84pt]{Scan over Dark Matter mass. The $\MET$ distribution is compared for the vector mediator models using the parameters as indicated. Ratios of the normalized distributions with respect to the first one are shown. $A_{300}$ and $A_{500}$ in the table denote the acceptance of the $\MET>300$~\gev and $\MET>500$~\gev cut, respectively.}
\label{fig:monojet_scan_V_mDM1000}
\end{figure*}

\begin{figure*}
\centering
\includegraphics[width=0.95\textwidth]{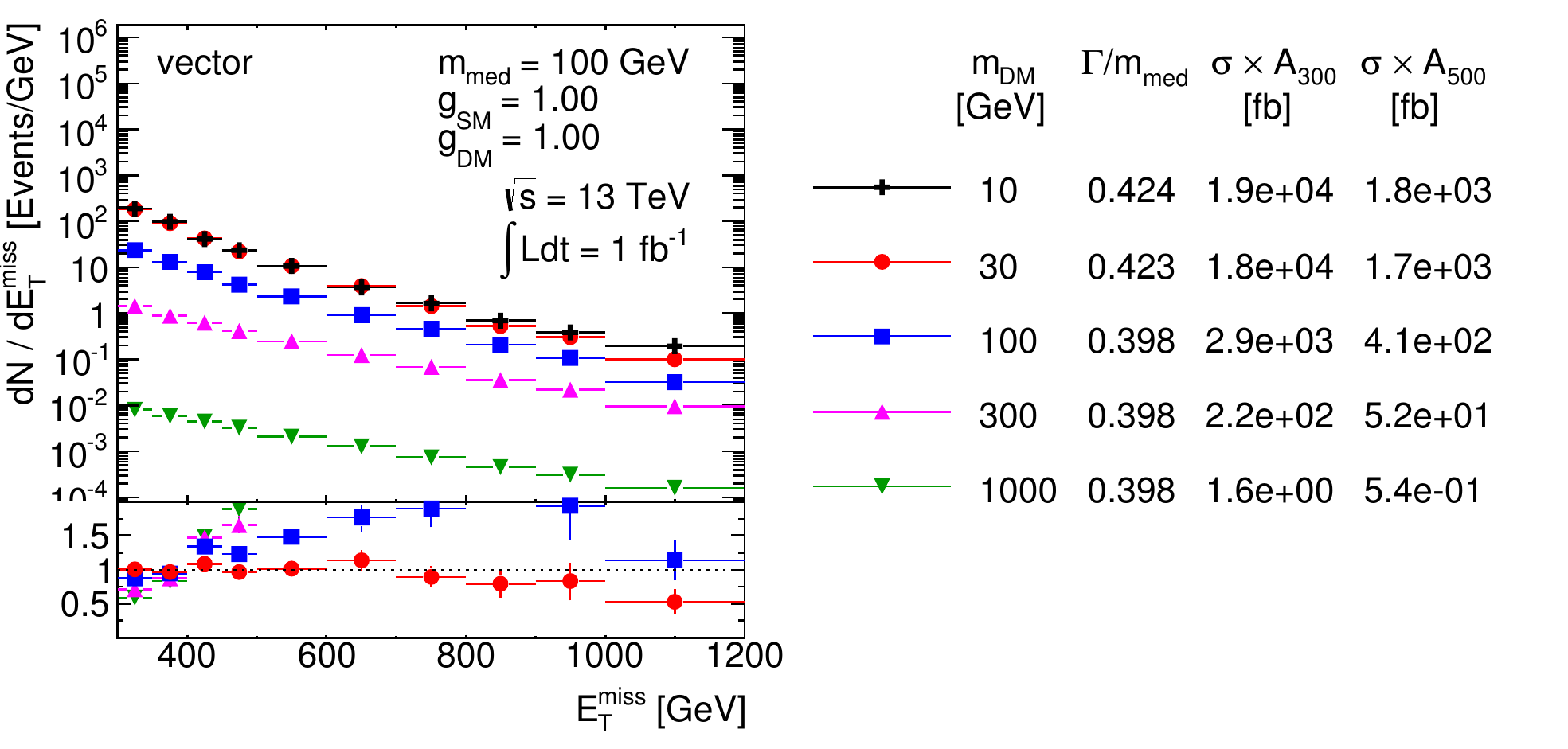}
\vspace{4\baselineskip}
\caption[][-84pt]{Scan over Dark Matter mass. The $\MET$ distribution is compared for the vector mediator models using the parameters as indicated. Ratios of the normalized distributions with respect to the first one are shown. $A_{300}$ and $A_{500}$ in the table denote the acceptance of the $\MET>300$~\gev and $\MET>500$~\gev cut, respectively.}
\label{fig:monojet_scan_V_mDM100}
\end{figure*}

For a fixed mediator mass \mMed and couplings, the Dark Matter mass falls into three regimes:
\begin{itemize}
\item[On-shell:] When $\Mmed \gg 2 \mDM$, most mediators are on-shell. The hardness of the ISR is set by \Mmed, and the kinematic distributions do not strongly depend on \mDM. This is illustrated in Fig.~\ref{fig:monojet_scan_V_mDM1000} for an example of \mMed=1~\tev 10~\gev $<\mDM<$ 300~\gev. The cross section decreases as the \mDM approaches $\mMed/2$. A coarse binning along $\mDM$ is sufficient.
\item[Threshold:] When $\Mmed \approx 2\mDM$, the production is resonantly enhanced, and both the cross section and kinematic distributions change more rapidly as a function of the two masses, and finer binning is needed in order to capture the changes.
\item[Off-shell:] When $\Mmed \ll 2 \mDM$, the Dark Matter pair is produced by an off-shell mediator. The mediator propagator gives an explicit suppression of $(\Mmed/Q)^2$ that suppresses hard ISR. The $\mDM=1$~\tev case, shown in Fig.~\ref{fig:monojet_scan_V_mDM1000}, and Figure\,\ref{fig:monojet_scan_V_mDM100} demonstrates that the \MET spectrum hardens with increasing \mDM, accompanied by the gradual decrease of the cross section. Due to the significant cross section suppression, it is not necessary to fully populate the parameter space. Imminent LHC searches are not expected to be sensitive to these signals.
\end{itemize}

\subsubsection{Scan over the mediator mass}

\begin{figure*}
\centering
\includegraphics[width=0.95\textwidth]{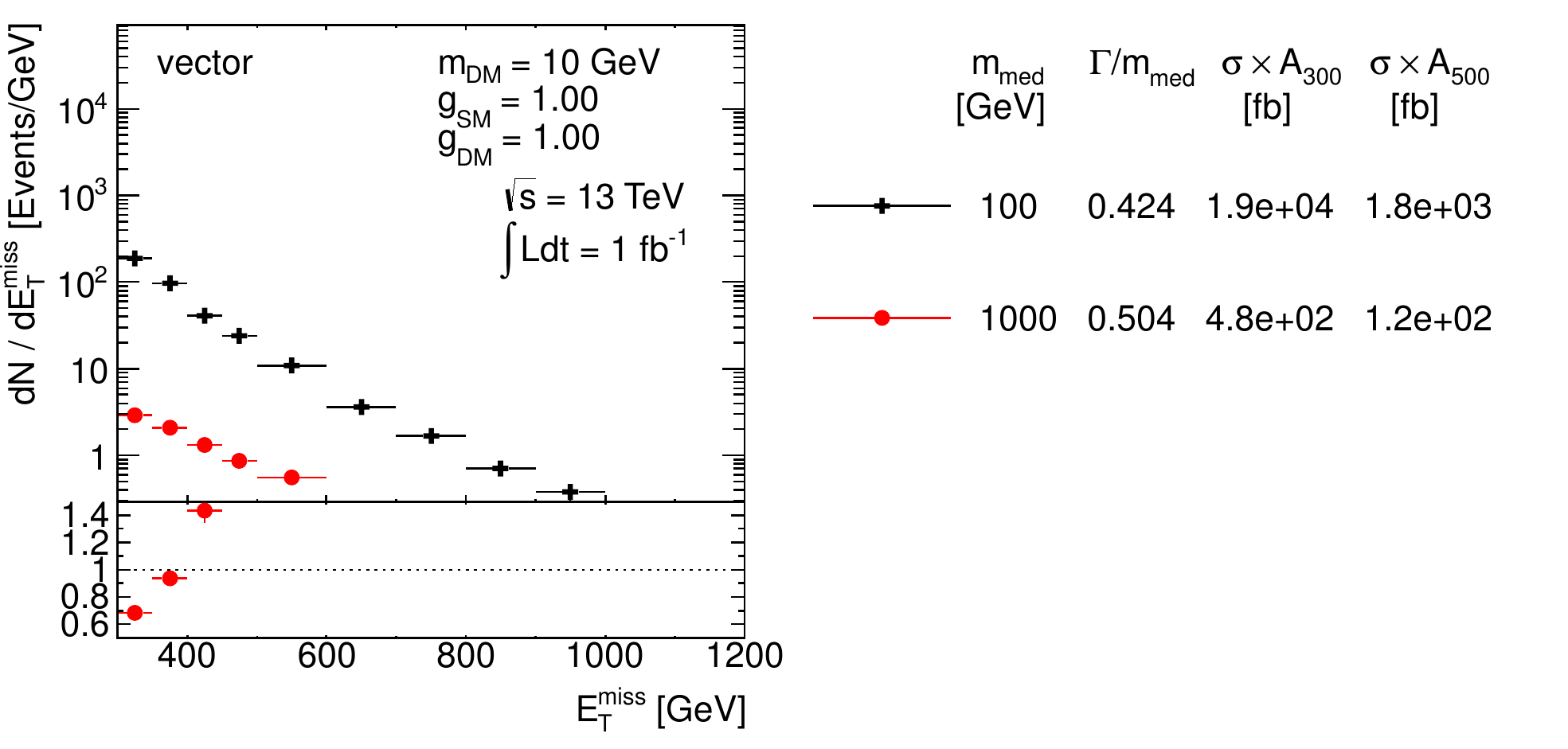}
\vspace{4\baselineskip}
\caption[][-84pt]{Scan over mediator mass. The $\MET$ distribution is compared for the vector mediator models using the parameters as indicated. Ratios of the normalized distributions with respect to the first one are shown. $A_{300}$ and $A_{500}$ in the table denote the acceptance of the $\MET>300$~\gev and $\MET>500$~\gev cut, respectively.}
\label{fig:monojet_scan_V_mMed10}
\end{figure*}

\begin{figure*}
\centering
\includegraphics[width=0.95\textwidth]{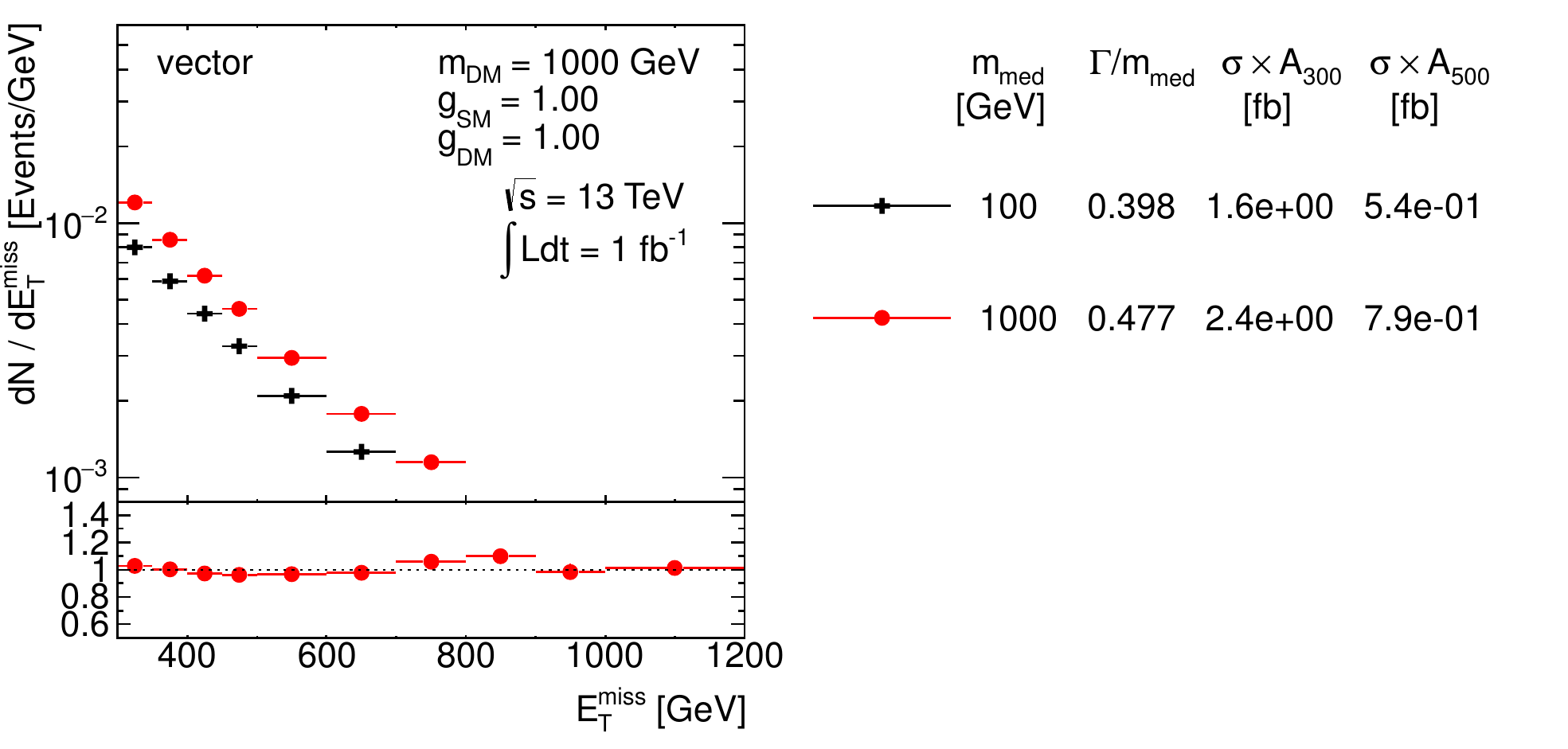}
\caption[][-28pt]{Scan over mediator mass. The $\MET$ distribution is compared for the vector mediator models using the parameters as indicated. Ratios of the normalized distributions with respect to the first one are shown. $A_{300}$ and $A_{500}$ in the table denote the acceptance of the $\MET>300$~\gev and $\MET>500$~\gev cut, respectively.}
\label{fig:monojet_scan_V_mMed1000}
\end{figure*}

Changing the mediator mass for fixed Dark Matter mass and couplings leads to significant differences in cross section and shapes of the kinematic variables for the on-shell regime, as shown in Fig.\,\ref{fig:monojet_scan_V_mMed10}. As expected, higher mediator masses lead to harder $\MET$ spectra.
On the other hand, the $\MET$ shapes are similar for off-shell mediators.  This
is illustrated in Fig.\,\ref{fig:monojet_scan_V_mMed1000}. Therefore, a coarse binning in \mMed is sufficient in the off-shell regime.

\subsubsection{Spin structure of the couplings}
\label{sec:monojet_spin}

\begin{figure*}
	\centering
	\includegraphics[width=0.95\textwidth]{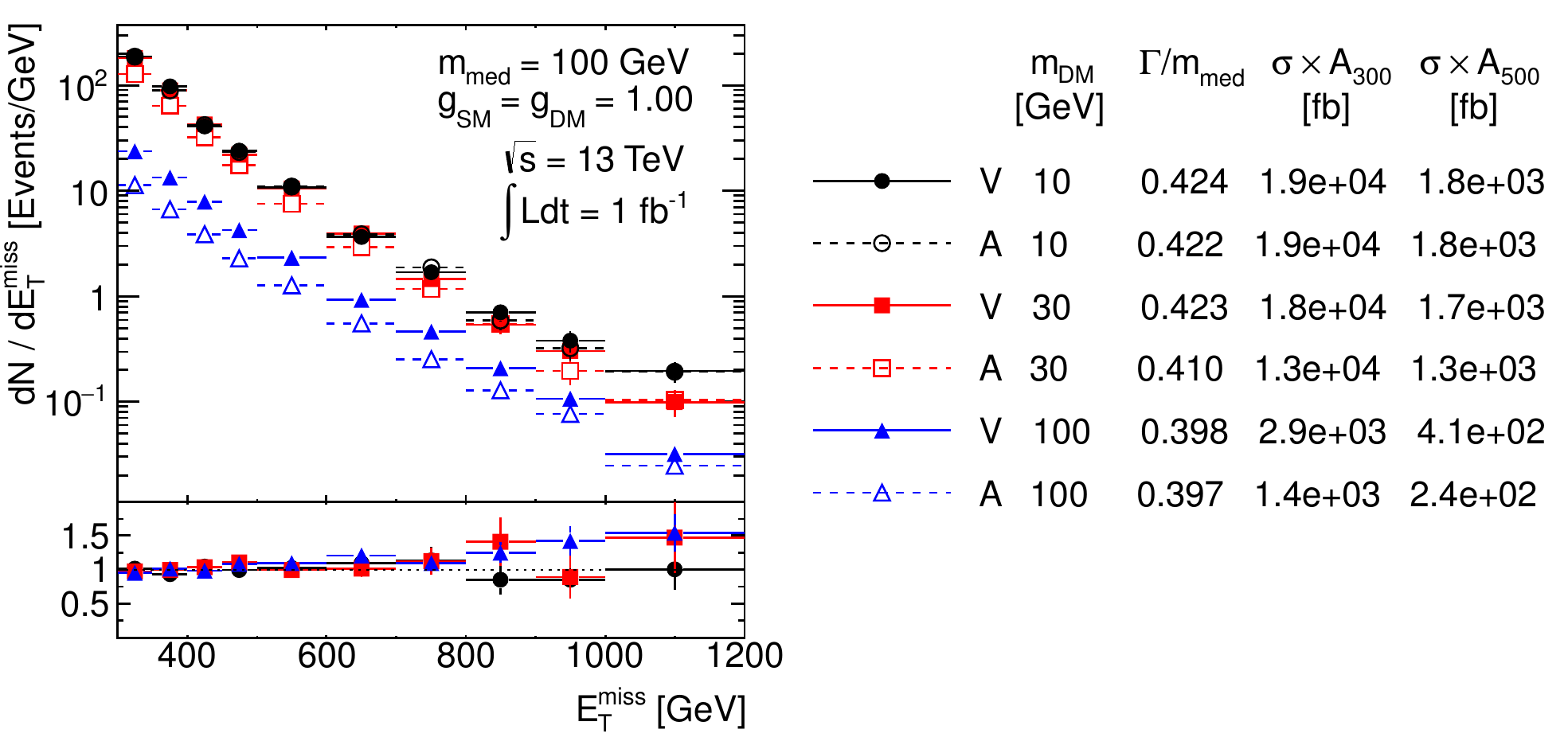}
	\caption[][-28pt]{Comparison of the pure vector and pure axial-vector couplings. The $\MET$ distribution is shown for the samples generated with $\mMed=100$~\gev and different Dark Matter masses. Ratios of the normalized distributions are shown for between the samples with coincident masses. $A_{300}$ and $A_{500}$ in the table denote the acceptance of the $\MET>300$~\gev and $\MET>500$~\gev cut, respectively.}
	\label{fig:monojet_VAmodels}
\end{figure*}

This section compares the kinematic properties of vector, axial-vector and mixed vector/axial-vector models.
The samples with pure vector and pure axial-vector couplings are compared for $\mMed=100$~\gev and different 
Dark Matter masses in Fig.\,\ref{fig:monojet_VAmodels}. 
No differences in the shape of the \MET distributions are observed between the samples with coincident masses. 
In the case of the on-shell mediators, where $2\mDM\ll\mMed$, the cross sections of the pure vector and pure axial-vector models are similar. With increasing Dark Matter mass towards the $2\mDM=\mMed$ transition and further into the off-shell regime, the relative difference between the cross sections of the two samples is increasing, with the vector ones having larger cross sections. 

\begin{figure*}
	\centering
	\includegraphics[width=0.95\textwidth]{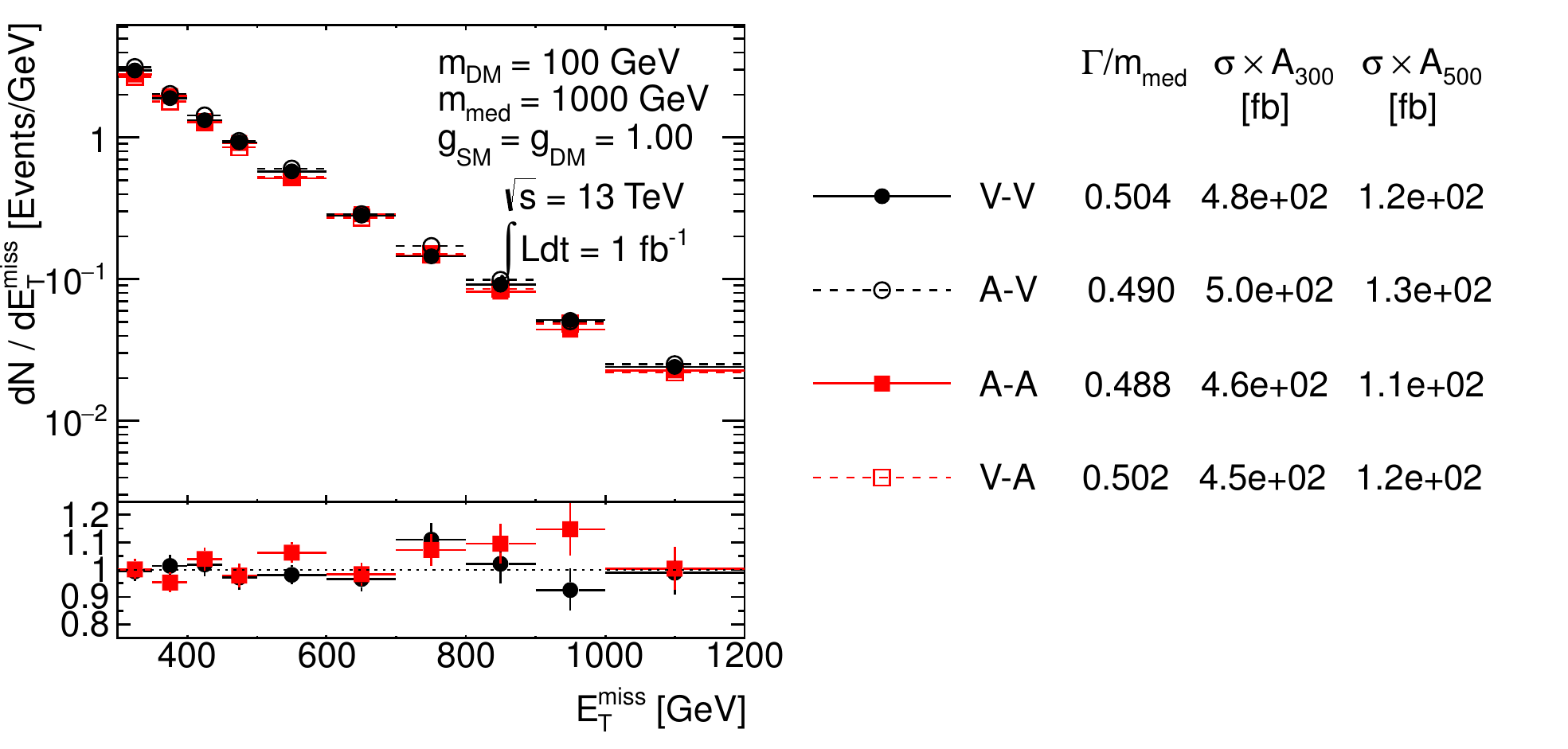}
	\caption[][-28pt]{Comparison of the pure vector, V-V, and pure axial-vector, A-A, couplings with mixed couplings, A-V and V-A where the first (second) letter indicates the Standard Model (dark sector) vertex. The $\MET$ distribution is shown for the samples generated with $\mDM=100$~\gev and $\mMed=1$~\tev. Ratios of the normalized distributions are shown for A-V over V-V and for V-A over A-A. $A_{300}$ and $A_{500}$ in the table denote the acceptance of the $\MET>300$~\gev and $\MET>500$~\gev cut, respectively.}
	\label{fig:monojet_scan_VA_mMed1000}
\end{figure*}

Figure\,\ref{fig:monojet_scan_VA_mMed1000} shows the samples generated with pure and mixed couplings for $\mDM=100$~\gev and $\mMed=1\,\tev$, i.e. where the mediator is on-shell. The mediator width between the pure vector and pure axial-vector couplings differ only by 2\% in this case, and $<10\%$ agreement between the cross sections is found. The mediator widths for the samples with the same type coupling to quarks agree at better than 1\% since the width is dominated by the quark contribution, as expected from
Eq.\,\ref{eq:monojet_min}.
No significant differences between the samples with same type Dark Matter coupling are seen, given the statistical precision of the generated samples. This is expected since the mediator is on-shell, and the details of the invisible decay are unimportant in cut-and-count searches.

\begin{figure*}
	\centering
	\includegraphics[width=0.95\textwidth]{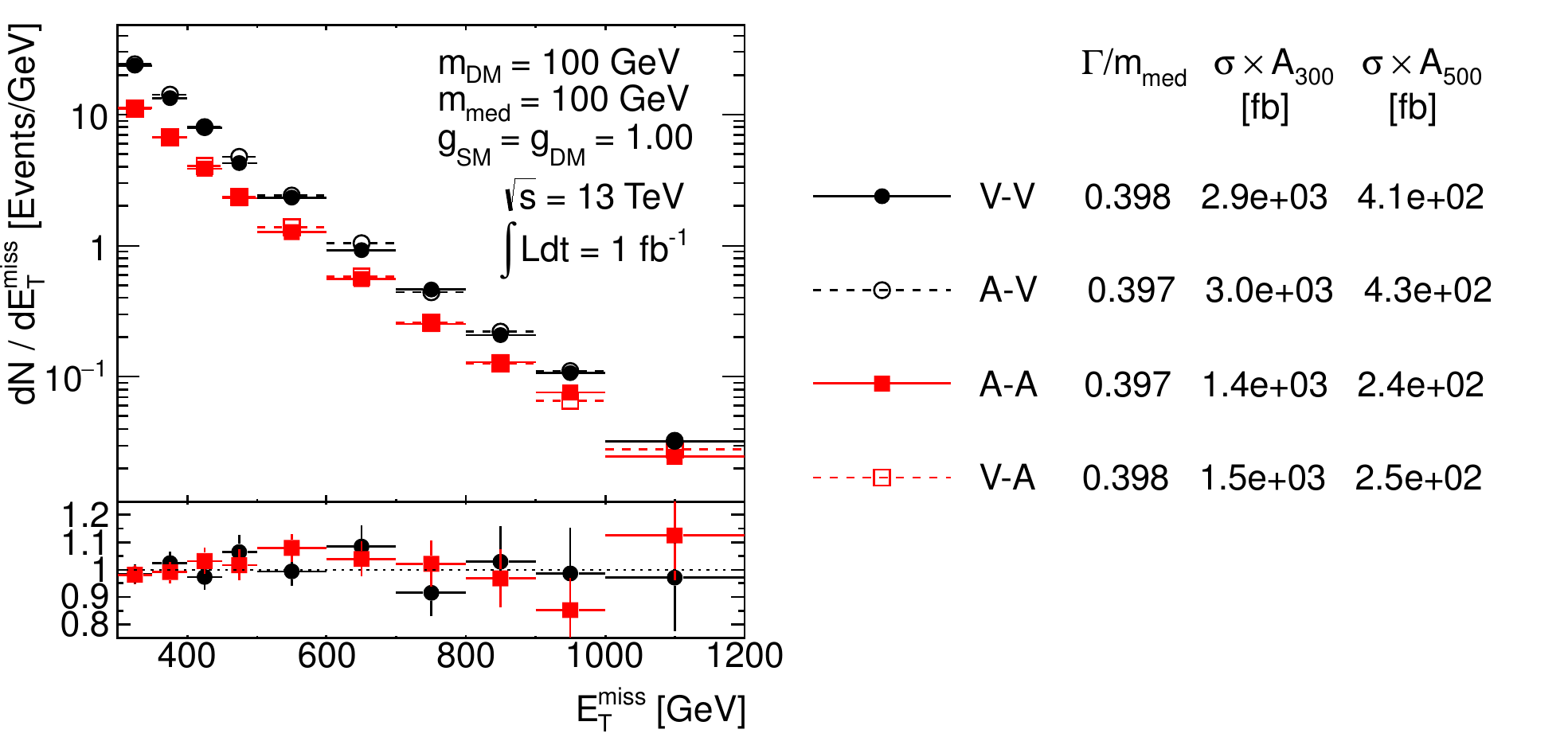}
	\caption[][-28pt]{Comparison of the pure vector, V-V, and pure axial-vector, A-A, couplings with mixed couplings, A-V and V-A where the first (second) letter indicates the Standard Model (Dark Sector) vertex. The $\MET$ distribution is shown for the samples generated with $\mDM=100$~\gev and $\mMed=100$~\gev. Ratios of the normalized distributions are shown for A-V over V-V and for V-A over A-A. $A_{300}$ and $A_{500}$ in the table denote the acceptance of the $\MET>300$~\gev and $\MET>500$~\gev cut, respectively. The suppression by $\beta^3$ for $\mDM\sim\mMed$ can be seen for the curves representing axial DM coupling.}
	\label{fig:monojet_scan_VA_mMed100}
\end{figure*}

For the off-shell case, shown in Fig.\,\ref{fig:monojet_scan_VA_mMed100} for $\mDM=100$~\gev and $\mMed=100$~\gev,
there is approximately a factor 2 difference
between the cross-sections of the samples with pure couplings is observed. As in the previous case, the samples with the same type coupling to Dark Matter are similar both in terms of cross sections and \MET shape. Since the contribution to the mediator width from Dark Matter is closed in this case, only the quark couplings define the width. Only couplings to light quarks are opened in the case of $\mMed=100$~\gev for which the differences between the partial widths of vector and axial-vector couplings are marginal. This explains the similar minimal widths for all four samples stated in Fig.\,\ref{fig:monojet_scan_VA_mMed100}.

In general, the coupling to quarks is not expected to play an important role in the kinematics as it is only needed to produce the mediator which is confirmed by the observations above. 
Based on this argument and on the observations above, we recommend to consider only the models with pure vector couplings or pure axial-vector couplings for simulation.

\subsubsection{Proposed parameter grid}

The final step in proposing a parameter grid is to evaluate the sensitivity
of Run-2 LHC data with respect to rate and/or kinematics.
The parameter scan focuses on two important regions, the light mediator region and  the heavy mediator limit to reproduce the EFT limit, 
and takes into account the projected sensitivities for the mono-jet analysis.

Considering simplified models also allows to discuss constraints from different search channels. In the case of the \schannel exchange, the results from the mono-jet final states, where the mediator decays to a DM pair, one can also take into account dijet constraints on the processes where the mediator decays back to Standard Model particles. The importance of the dijet results depend on the magnitude of the coupling $\gq$. We recommend to keep the two channels rather independent by choosing $\gq=0.25$ and $\gDM=1$, based on the findings given in Ref.\,\cite{Chala:2015ama}. Furthermore, it is also important to mention this choice leads to $\Gamma_{\rm{min}}/\mMed \lsim 0.06$. Note that the usual choice of $\gq=\gDM=1$ used in literature leads to $\Gamma_{\rm{min}}/\mMed \sim 0.5$, questioning the applicability of the narrow width approximation.

The expected upper limit at 95\% confidence level on the product of cross section, acceptance and efficiency, $\sigma\times A\times\epsilon$, in the final Run-1 ATLAS mono-jet analysis\,\cite{Aad:2015zva} is 51\,fb and 7.2\,fb  for $\MET>300$~\gev and $\MET>500$~\gev, respectively. Projected sensitivities for a 14~\tev\, mono-jet analysis are available from ATLAS~\cite{ATL-PHYS-PUB-2014-007}. These ATLAS studies estimate a factor of two increase in sensitivity with the 2015 data. 
The generator level cross section times efficiency times acceptance at $\MET>500$~\gev for the model with couplings $\gq=0.25$ and $\gDM=1$, a light Dark Matter particle of
\mDM=10~\gev and a \mMed=1~\tev vector mediator is at the order of 100\,fb, i.e. the early Run-2 mono-jet analysis is going to be sensitive to heavier mediators than this. The value of $\sigma\times \epsilon \times A$ at $\MET>500$~\gev for a 5~\tev vector mediator is at the order of 0.1\,fb, therefore this model lies beyond the reach of the LHC in the early Run-2. However, models with high enough mediators are still useful to reproduce the EFT result.

Following these arguments, \mMed grid points are chosen, roughly equidistant in a logarithmic scale: 10~\gev, 20~\gev, 50~\gev,  100~\gev, 200~\gev, 300~\gev, 500~\gev, 1000~\gev and 2000~\gev. In the threshold regime $\mMed=2\mDM$, the $\mDM$ grid points are taken at approximately $\mMed/2$, namely: 10~\gev, 50~\gev, 150~\gev, 500~\gev and 1000~\gev. Points on the on-shell diagonal are always chosen to be 5~\gev away from the threshold, to avoid numerical instabilities in the event generation. 
The detailed studies of the impact of the parameter changes on the cross section and kinematic distributions presented earlier in this section support removing some of the grid points and relying on interpolation. The optimized grids proposed for the vector and axial-vector mediators are given in Table.\,\ref{tab:mDMmMedScan_VA}.
One point at very high mediator mass (10~\tev) is added for each of the DM masses scanned, to aid the reinterpretation of results in terms of contact interaction operators (EFTs), as discussed in Section~\ref{sec:RecommendationEFTResults}. 


\begin{table}[!h]
\centering
\resizebox{\textwidth}{!}{
\begin{tabular}{| l |r r r r r r r r r r|}
\hline
\multicolumn{1}{|c|}{\mDM/\gev} & \multicolumn{10}{c|}{\mmed/\gev} \\
\hline
 1             &         10  & 20 & 50 & 100 & 200 & 300 & 500 &         1000  &                 2000   &         10000  \\
 10   	       &         10  & 15 & 50 & 100 &     &     &     &               &                        &     10000      \\
 50            &   10  & 
& 50 &  95 & 200 & 300 &     &               &                        &    10000       \\
 150           &         10  &    &    &     & 200 & 295 & 500 &    1000    &                  &     10000      \\
 500           &         10  &    &    &     &     &     & 500 &          995  &                 2000   &     10000      \\
 1000          &         10  &    &    &     &     &     &     &         1000  &                 1995   &         10000  \\
\hline
\end{tabular}
}
\caption{Simplified model benchmarks for \schannel simplified models (\spinone mediators 
decaying to Dirac DM fermions in the V and A case, taking the minimum width for \gq = 0.25 and \gDM = 1)}.

\label{tab:mDMmMedScan_VA}
\end{table}




Tables\,\ref{tab:widthV} and \ref{tab:widthA} give the $\Gamma_{\rm{min}}/\mMed$ ratio for the parameter grid proposed for vector and axial-vector \schannel models, respectively. The numbers range from $\sim0.02$ in the off-shell regime at $2\mDM>\mMed$ to $\sim0.06$ in the on-shell regime for heavy mediators where all coupling channels contribute.

\begin{table}
	\centering
	\resizebox{\textwidth}{!}{
		\begin{tabular}{| l |r r r r r r r r r r|}
			\hline
			\multicolumn{1}{|c|}{\mDM/\gev} & \multicolumn{10}{c|}{\mmed/\gev} \\
			&         10  & 20 & 50 & 100 & 200 & 300 & 500 &         1000  &                 2000   &         10000  \\
			\hline
			\hline
			1 & 0.049  & 0.051  & 0.051  & 0.051  & 0.051  & 0.051  & 0.056  & 0.056  & 0.056  & 0.056  \\
			10 & 0.022  & 0.024  & 0.054  & 0.052  &        &        &        &        &        & 0.056  \\
			50 & 0.022  &        & 0.025  & 0.025  & 0.055  & 0.053  &        &        &        & 0.056  \\
			150 & 0.022  &        &        &        & 0.025  & 0.025  & 0.061  & 0.058  &        & 0.056  \\
			500 & 0.022  &        &        &        &        &        & 0.029  & 0.030  & 0.060  & 0.057  \\
			1000 & 0.022  &        &        &        &        &        &        & 0.030  & 0.030  & 0.057  \\
			\hline
		\end{tabular}}
		\caption
		{Minimal width of the vector mediator exchanged in \schannel divided by its mass, assuming $\gq=0.25$ and $\gDM=1$. The numbers tabulated under $2\mDM=\mMed$ correspond to the width calculated for $\mMed-5$~\gev.}
		\label{tab:widthV}
	\end{table}
	
	\begin{table}
		\centering
		\resizebox{\textwidth}{!}{
			\begin{tabular}{| l |r r r r r r r r r r|}
				\hline
				\multicolumn{1}{|c|}{\mDM/\gev} & \multicolumn{10}{c|}{\mmed/\gev} \\
				&         10  & 20 & 50 & 100 & 200 & 300 & 500 &         1000  &                 2000   &         10000  \\
				\hline
				\hline
				1 & 0.045  & 0.049  & 0.051  & 0.051  & 0.051  & 0.051  & 0.053  & 0.055  & 0.056  & 0.056  \\
				10 & 0.020  & 0.022  & 0.047  & 0.050  &        &        &        &        &        & 0.056  \\
				50 & 0.020  &        & 0.025  & 0.025  & 0.045  & 0.048  &        &        &        & 0.056  \\
				150 & 0.020  &        &        &        & 0.025  & 0.025  & 0.044  & 0.053  &        & 0.056  \\
				500 & 0.020  &        &        &        &        &        & 0.027  & 0.029  & 0.050  & 0.056  \\
				1000 & 0.020  &        &        &        &        &        &        & 0.029  & 0.030  & 0.055  \\
				\hline
			\end{tabular}}
			\caption
			{Minimal width of the axial-vector mediator exchanged in \schannel divided by its mass, assuming $\gq=0.25$ and $\gDM=1$. The numbers tabulated under $2\mDM=\mMed$ correspond to the width calculated for $\mMed-5$~\gev.}
			\label{tab:widthA}
		\end{table}

\subsection{Additional considerations for $V$+\MET{} signatures}
\label{sec:bosonrad}

All models detailed in this Section are applicable to signatures where 
a photon, a W boson, a Z boson or a Higgs boson
is radiated from the initial state partons instead of a gluon. 
The experimental signature is identified as \textit{V+\MET} and it
has been sought by ATLAS and CMS in Refs.~\cite{Khachatryan:2014rwa,Aad:2014tda,Khachatryan:2014tva,ATLAS:2014wra,Aad:2013oja,Aad:2014vka}. 
This signature is also produced by the models described in 
Section~\ref{subsec:EWSpecificModels}. 

Monojet searches are generally more sensitive
with respect to final states including EW bosons, due to the much
larger rates of signal events featuring quark or gluon radiation with
respect to radiation of bosons~\cite{Zhou:2013fla},
in combination with the low branching ratios if leptons from
boson decays are required in the final state.
The rates for the Higgs boson radiation is too low for these models
to be considered a viable benchmark~\cite{Carpenter:2013xra}.
However, the presence of photons,
leptons from W and Z decays,
and W or Z bosons decaying hadronically
allow backgrounds to be rejected more effectively,
making Z/$\gamma$/W+\MET searches
still worth comparing with searches in the jet+\MET final state (see e.g. Ref.~\cite{Gershtein:2008bf}).

In the case of a \spinone mediator,
an example Feynman diagram for these processes can be constructed by taking
Fig.~\ref{fig:OP} and replacing the gluon with $\gamma,W$ or $Z$.

When the initial state radiation is a W boson, Run-1 searches have considered three benchmark cases, varying the relative coupling of the $W$ to $u$ and ${d}$ quarks.
The simplified model with a vector mediator mediator exchanged in the s-channel includes only the simplest of these cases, in which the $W$ coupling to $u$ and ${d}$ quarks is identical, as required naively by $\mathrm{SU}(2)$ gauge invariance.  With some more complex model building, other cases are possible.  The case in which the $u$ and ${d}$ couplings have opposite sign is particularly interesting, since this enhances the $W+\MET$ signal over the jet$+$\MET signal~\cite{Bell:2015sza,Bai:2012xg,Hamaguchi:2014pja}. An example of a model of this type is discussed in Appendix~\ref{app:monoWExtramodel}.



Simulations for the models in this Section have been done at the LO+PS level using  \madgraph 2.2.3 interfaced to \pythiaEight, 
and therefore no special runtime configuration is needed for pythia 8. Even though merging samples with different parton multiplicities is possible,  this has not been deemed necessary as the visible signal comes from the production of a heavy SM boson whose transverse momentum distribution is sufficiently well described at LO+PS level. 

In these $V$+\MET models, as in the case of the jet+\MET models, \pT of the boson or the \MET does not depend 
strongly on the width of the mediator. An example of the particle-level analysis acceptance using the
generator-level cuts from Ref.~\cite{Aad:2014tda}
for the photon+\MET analysis, but raising the photon $p_T$ cut
to 150~\gev, is shown in Figure~\ref{fig:DMV_EW_gamma_acceptance},
comparing a width that is set to $\Gamma=M_{med}/3$ to the
minimal width (the ratio between the two widths
ranges from 1.05 to 1.5 with increasing mediator masses).

\begin{table}[!h]
\begin{tabular}{| l |r r r r|}\hline
\multicolumn{5}{|c|}{Acceptance ratio for $\Gamma=\Gamma_{\rm min}$ vs
$\Gamma=\mMed/3$} \\ \hline 
\multicolumn{1}{|c|}{ } & \multicolumn{4}{c|}{\mdm/GeV}\\
\hline 
{\mMed/GeV}      & 10     & 50    & 200   & 400  \\ \hline
50   & 0.96   & 0.99  &       & 0.95 \\  
100  & 0.97   &       &       &      \\
300  & 1.00   & 1.02  &       &      \\
600  &        &       & 0.96  &      \\
1000 & 1.01   & 1.02  & 1.03  &      \\
3000 & 1.02   & 1.03  &       & 1.01 \\
\hline
\end{tabular}
    \caption{Analysis acceptance ratios for the photon+\MET analysis when varying the mediator width, in the
    case of a vector mediator exchanged in the \schannel. The figures shown in this Section
    have been obtained using a LO UFO model in \madgraph 2.2.3 interfaced to \pythiaEight
    for the parton shower.}
    \label{fig:DMV_EW_gamma_acceptance}
\end{table}

\begin{figure*}[h!]
\centering  
\subfloat[Leading photon transverse momentum distribution for the photon+\MET final state, for 
different mediator mass choices, for \mdm=10~\gev.\label{fig:DMV_EW_gamma_MET_SVMed}]{%
\includegraphics[width=0.49\textwidth]{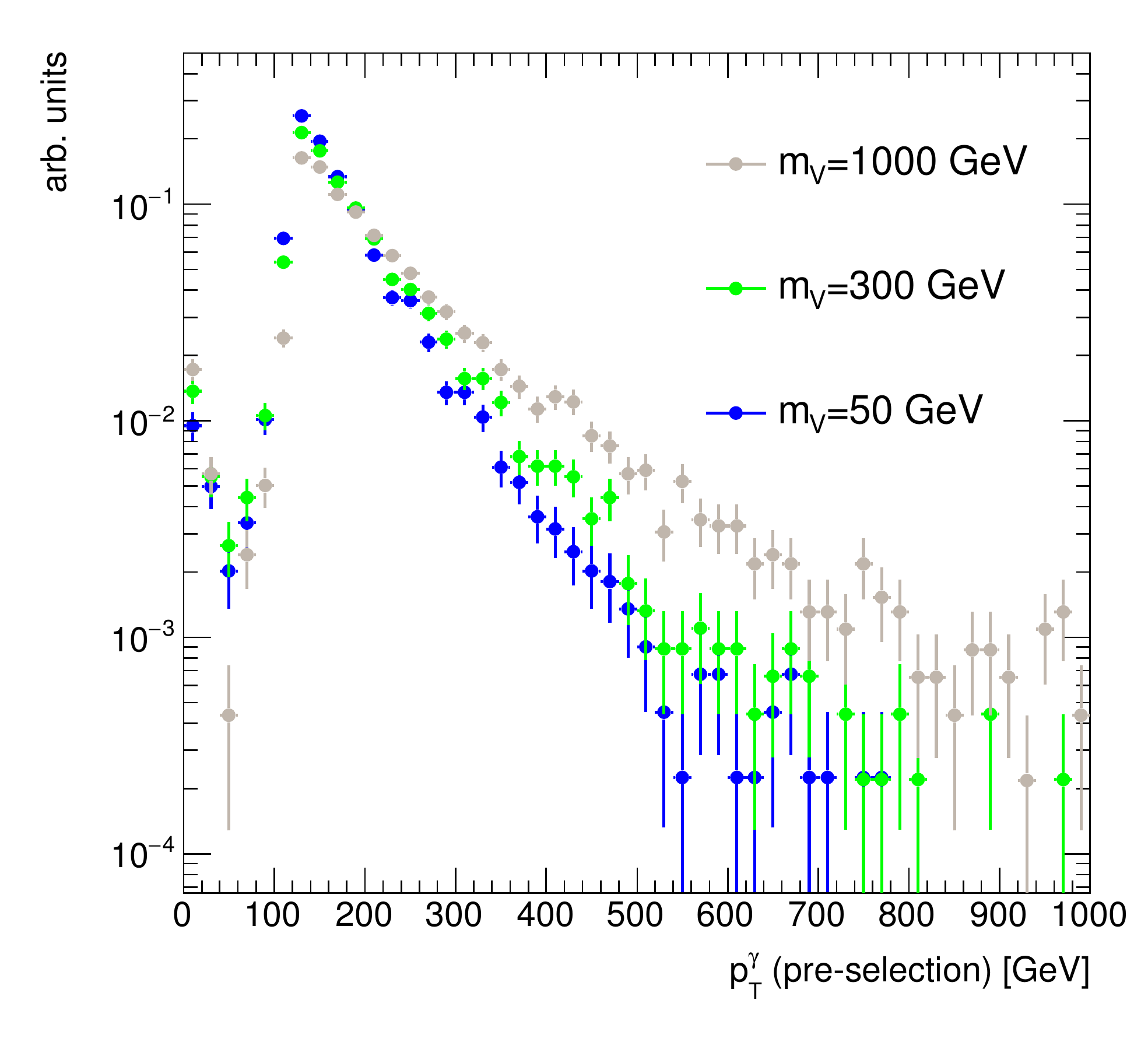}
}
\hfill
\subfloat[Leading photon transverse momentum distribution for the photon+\MET final state, 
for different DM mass choices, with \mMed=1~\tev.\label{fig:DMV_EW_gamma_pT_SVMed}]{%
		\includegraphics[width=0.49\textwidth]{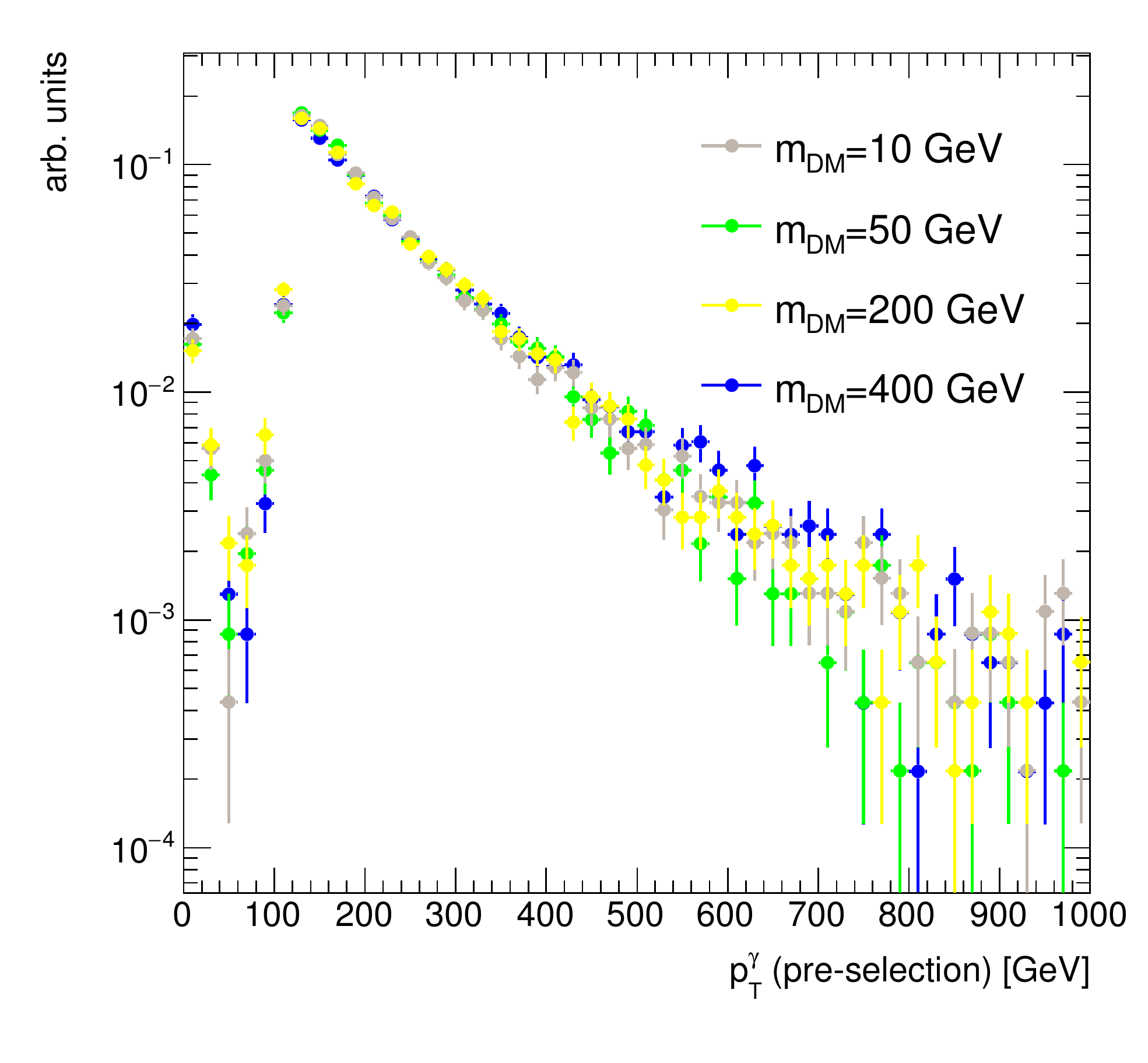}
}
\hfill
\subfloat[Missing transverse momentum distribution for the leptonic Z+\MET final state, 
for different mediator mass choices, for \mdm=15~\gev\label{fig:DMV_EW_Z_MET_SVMed}]{%
\includegraphics[width=0.49\textwidth]{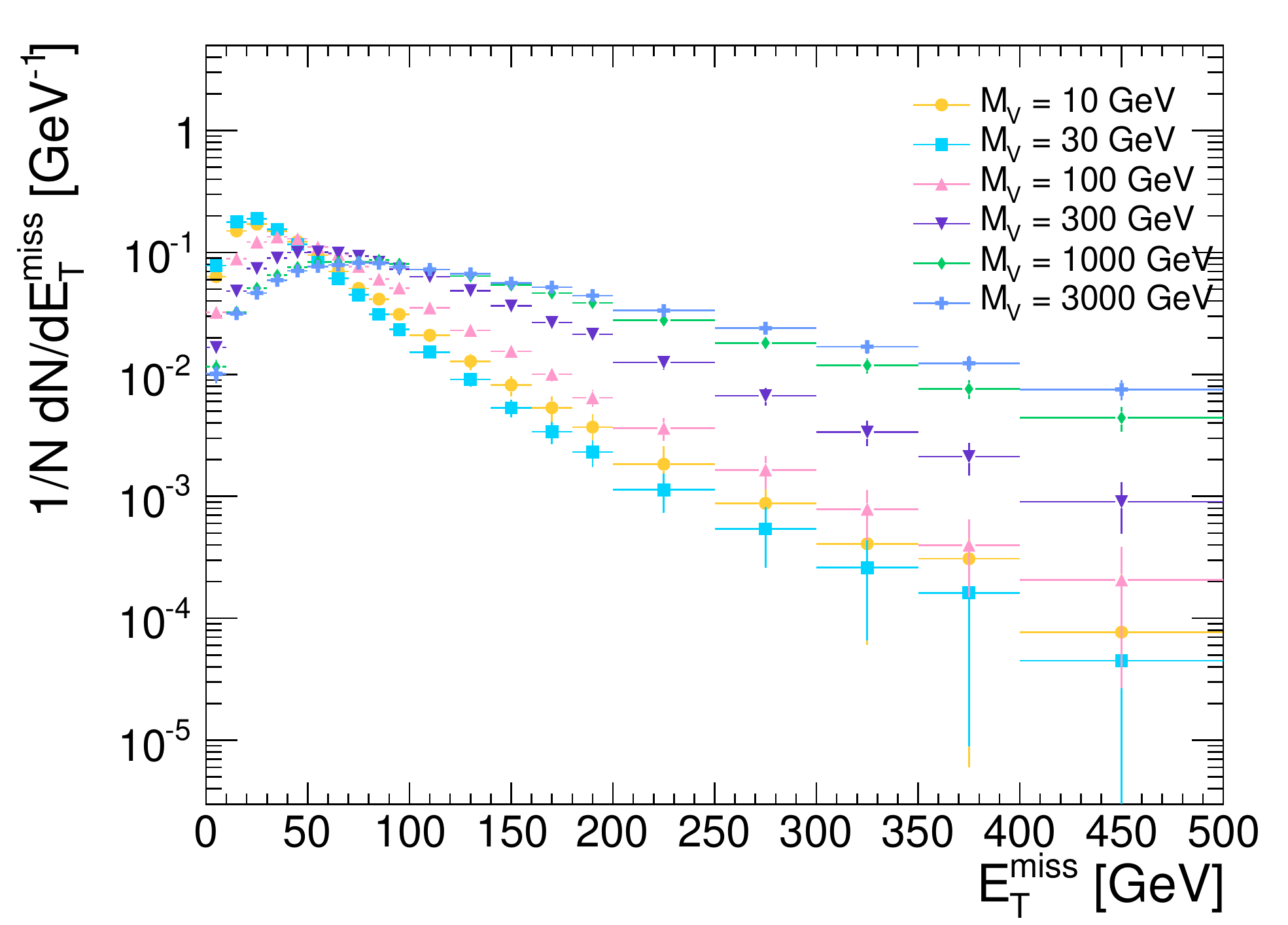}
}    
\hfill
\subfloat[Missing transverse momentum distribution for the hadronic W+\MET final state.\label{fig:DMV_EW_Whad_MET_SVMed}]{%
	\includegraphics[width=0.49\textwidth]{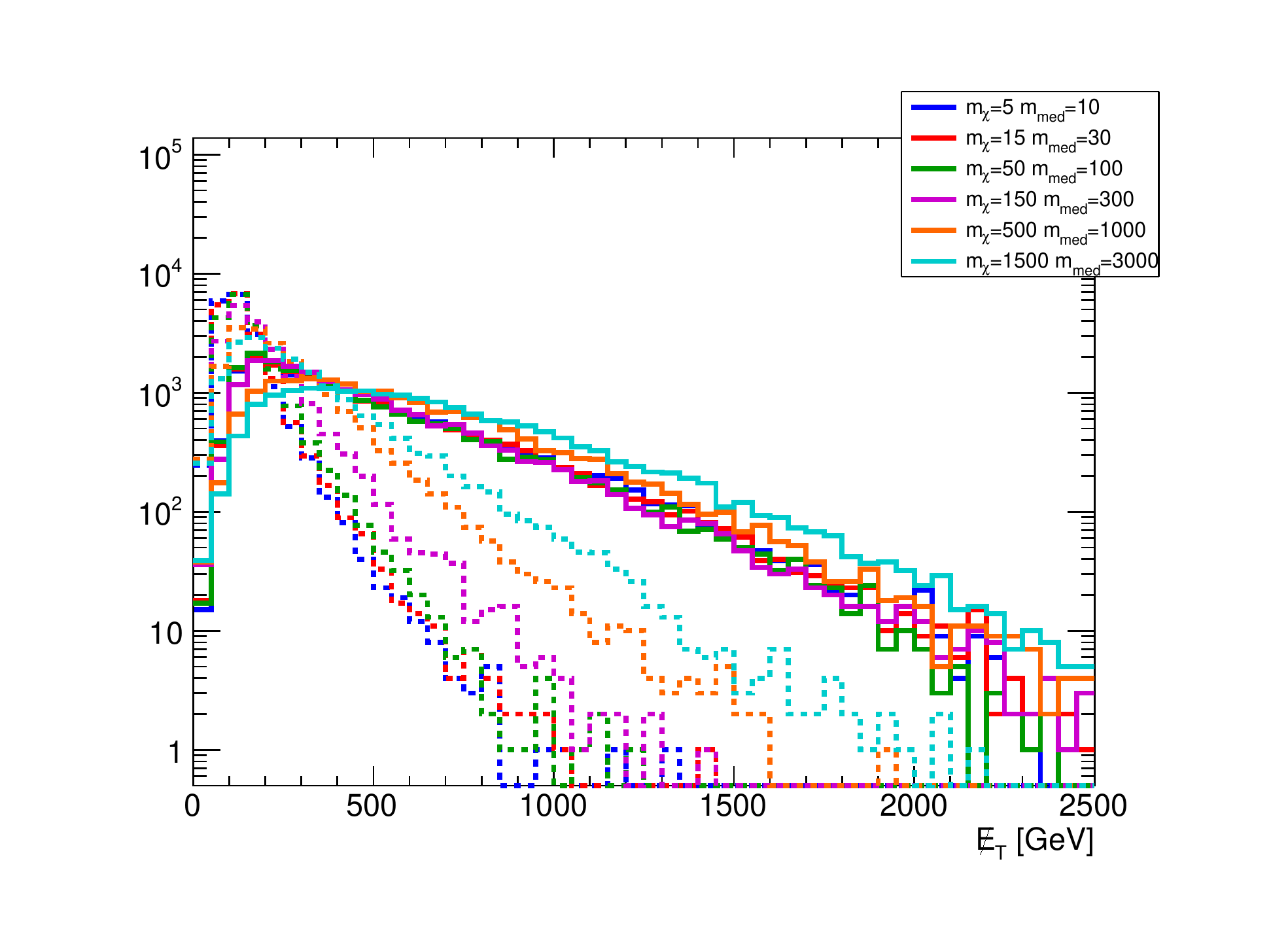}
}    
\caption{Kinematic distributions relevant for searches with W, Z and photons in the final state, 
for the simplified model
       with a vector mediator exchanged in the \schannel.}
\label{fig:DMV_EW_kinematics_SVMed}
\end{figure*}

Examples of relevant kinematic distributions for selected benchmark points are
shown in Fig.~\ref{fig:DMV_EW_kinematics_SVMed}. 

\section{Scalar and pseudoscalar mediator, \schannel exchange}
\label{sec:monojet_scalar}

\begin{figure}
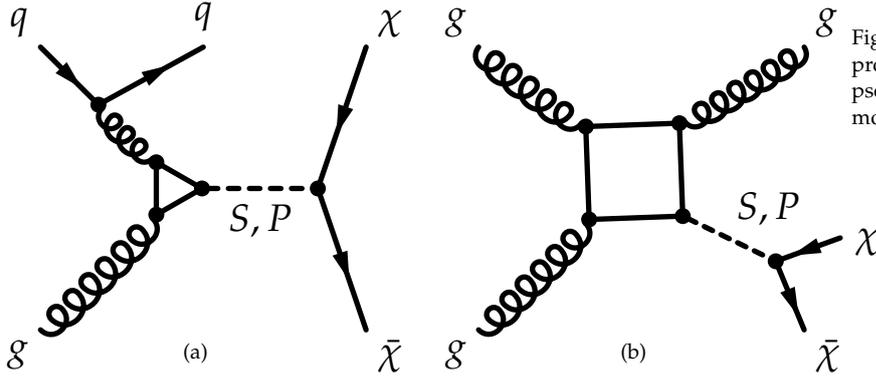

\centering
\unitlength=0.005\linewidth
	\subfloat[\label{subfig:appendixmodelSmonojetTopTriangle}]	{
	\begin{feynmandiagram}[appendixmodelSmonojetTopTriangle]
		\fmfleft{i1,i2}
		\fmfright{o1,o2}
		\fmftop{isr}
		\fmfbottom{pisr}
		\fmfpolyn{empty}{v}{3}
		\fmf{fermion}{i2,v6}
		\fmf{phantom}{i1,pv1,v1}
		\fmf{gluon,tension=0}{i1,v1}
		\fmf{gluon}{v6,v3}
		\fmf{dashes,label={\LARGE $S,,P$}}{v2,v5}
		\fmf{fermion,tension=1.2}{o2,v5,o1}
		\fmfdot{v1,v6,v2,v3,v5}
		\fmffreeze
		\fmf{phantom}{pv1,pisr}
		\fmf{fermion}{v6,isr}
		\fmflabel{\LARGE ${g}$}{i1}
		\fmflabel{\LARGE ${q}$}{i2}
		\fmflabel{\LARGE ${q}$}{isr}
		\fmflabel{\LARGE $\bar\chi$}{o1}
		\fmflabel{\LARGE $\chi$}{o2}
	\end{feynmandiagram}
}
	\subfloat[\label{subfig:appendixmodelSmonojetTopBox}]	{
	\begin{feynmandiagram}[appendixmodelSmonojetTopBox]
		\fmfleft{i1,i2}
		\fmfright{o1,o2,hisr,isr}
		\fmfpolyn{empty}{v}{4}
		\fmf{gluon}{i2,v4}
		\fmf{gluon}{i1,v1}
		\fmf{dashes,label={\LARGE $S,,P$}}{v2,vwimp}
		\fmflabel{\LARGE ${g}$}{i1}
		\fmflabel{\LARGE ${g}$}{i2}
		\fmflabel{\LARGE ${g}$}{isr}
		\fmflabel{\LARGE $\bar\chi$}{o1}
		\fmflabel{\LARGE $\chi$}{o2}
		\fmf{fermion}{o2,vwimp,o1}
		\fmfdot{v1,v2,v3,v4,vwimp}
		\fmf{gluon}{v3,isr}
	\end{feynmandiagram}
}
\setfloatalignment{t}
\vspace{0.5\baselineskip}
	\caption
	{
		One-loop diagrams of processes exchanging a scalar ($S$) or pseudoscalar ($P$) mediator, leading to a mono-jet signature. 
	}
	\label{fig:feyn_prod_S}
\end{figure}

In this section, we consider a parallel situation to the vector and axial-vector mediators in the previous sections: a real scalar or a pseudoscalar where the associated scalar is decoupled at higher energies\sidenote{This assumption does not hold in a UV-complete model where the two components of the complex scalar mediator would be approximately degenerate.  The complex scalar case could be studied separately in the case of heavy flavor final states given the sufficiently different kinematics.}. This section is largely based on Refs.~\cite{Buckley:2014fba,Harris:2014hga,Haisch:2015ioa} which contain a thorough discussion of these models. 

Assuming MFV, \spinzero resonances behave in a similar fashion as the SM Higgs boson. If the mediators are pure singlets of the SM, their interactions with quarks are not $SU(2)_L$ invariant. To restore this invariance, one could include the mixing of such mediators with the Higgs sector. This leads to extra interactions and a more complex phenomenology with respect to what considered in this Section (for a more complete discussion, see Refs.~\cite{Buckley:2014fba,Haisch:2015ioa}). In the interest of simplicity, we do not study models including those interactions in this report as early Run-2 benchmark models, but we give an example of a model of this kind in Appendix~\ref{sec:i2HDM}. 

Relative to the vector and axial-vector models discussed above, the scalar models are distinguished by the special consequences of the MFV 
assumption: the very narrow width of the mediator and its extreme sensitivity to which decays are kinematically available, and the loop-induced coupling to gluons. The interaction Lagrangians are

\begin{fullwidth}
  \begin{eqnarray} \mathcal{L}_{\phi} & = &
          \gdm \phi \bar{\chiDM}\chiDM+ \frac{\phi}{\sqrt{2}} \sum_i \left(g_u y_i^u \bar{u}_i u_i+g_d y_i^d \bar{d}_i d_i+g_\ell y_i^\ell \bar{\ell}_i \ell_i\right)\, , \label{eq:scalarlag} \\
    \mathcal{L}_{a} & = &
          i\gdm a \bar{\chiDM}\gamma_5\chiDM+ \frac{i a}{\sqrt{2}}\sum_i  \left(g_u y_i^u \bar{u}_i \gamma_5 u_i+g_d y_i^d \bar{d}_i \gamma_5 d_i+ \right. \nonumber \\
                                   & & \left. g_\ell y_i^\ell   \bar{\ell}_i \gamma_5 \ell_i\right) \,. \label{eq:pseudoscalarlag}
  \end{eqnarray}
\end{fullwidth}
where $\phi$ and $a$ are respectively the scalar and pseudoscalar mediators, and the Yukawa couplings $y_i^f$ are normalized to the Higgs vev as $y_i^f = \sqrt{2}m_i^f/v$.

The couplings to fermions are proportional to the SM Higgs couplings, yet one is still allowed to adjust an overall strength of the coupling to charged leptons and the relative couplings of $u$- and $d$-type quarks. As in the preceding sections, for the sake of simplicity and straightforward comparison, we reduce the couplings to the SM fermions to a single universal parameter $\gq \equiv g_u = g_d = g_\ell$. Unlike the vector and axial-vector models, the scalar mediators are allowed to couple to leptons.\sidenote{This contribution plays no role for most of the parameter space considered. The choice to allow lepton couplings follows Refs.~\cite{Buckley:2014fba,Harris:2014hga}.}

The relative discovery and exclusion power of each search can be compared in this framework.
However, we again emphasize the importance of searching the
full set of allowed channels in case violations of these simplifying assumptions
lead to significant modifications of the decay rates that
unexpectedly favor different
channels than the mix obtained under our assumptions. The coupling $\gdm$ parametrizes the entire dependence on the structure between the mediator and the dark sector.


Given these simplifications, the minimal set of parameters under consideration is
 \bea
  \left\{ \mDM,~ m_{\phi/a} = \mMed,~ \gdm,~ \gq \right\} \,.
 \eea
Fig.~\ref{fig:feyn_prod_S} shows the one-loop diagrams producing a jet+X signature. 
The full calculation of the top loop is available at LO for DM pair production in association 
with one parton.

The minimal mediator width (neglecting the small contributions from quarks other than top in the loop) is given by 
\begin{fullwidth}
  \begin{equation} \label{eq:width}
    \begin{split}
      \Gamma_{\phi,a} = & \sum_f N_c \frac{y_f^2 \gq^2 m_{\phi,a}}{16
        \pi} \left(1-\frac{4 m_f^2}{m_{\phi,a}^2}\right)^{x/2}
      + \frac{\gdm^2 m_{\phi,a}}{8 \pi} \left(1-\frac{4 \mDM^2}{m_{\phi,a}^2}\right)^{x/2}\\
      & + \frac{\alpha_s^2 y_t^2 \gq^2 m_{\phi,a}^3}{32 \pi^3 v^2}
      \left| f_{\phi,a}\left(\tfrac{4m_t^2}{m_{\phi,a}^2}
        \right)\right|^2
    \end{split}
  \end{equation}
\end{fullwidth}
where $x=3$ for scalars and $x=1$ for pseudoscalars. The loop integrals, with $f$ as complex functions, are
\begin{fullwidth}
  \bea \label{eq:fphifa}
  f_\phi (\tau) &=& \tau \left [ 1+ (1-\tau) \arctan^2 \left ( \frac{1}{\sqrt{\tau-1}} \right ) \right ]  \,, \\
  f_a (\tau) &=& \tau \arctan^2 \left ( \frac{1}{\sqrt{\tau-1}}
  \right) \, 
  \eea
\end{fullwidth}
where $\tau = 4 m_{t}^2/m_{\phi,a}^2$. 

The minimal widths for scalar and pseudo-scalar mediators with $\gq=\gDM=1$ are shown in Fig.\,\ref{fig:monojet_width_S}, illustrating the effect of choosing
the SM Higgs-like Yukawa couplings for the SM fermions.
For the mediator mass above twice the top quark mass $m_t$, the minimal width receives the dominant contribution from the top quark. For lighter mediator masses, Dark Matter dominates as the couplings to lighter quarks are Yukawa suppressed.

As shown in the diagram of Fig.~\ref{fig:feyn_prod_S}, the lowest order process of these models
already involves a one-loop amplitude in QCD, and only LO predictions are currently available. 
The generator used for the studies for the jet+\MET{} signature is \powheg\cite{Haisch:2013ata,Haisch:2015ioa,Alioli:2010xd,Nason:2004rx,Frixione:2007vw},
with \pythiaEight~\cite{Sjostrand:2007gs} for the parton shower; 
within this implementation,
the scalar and pseudoscalar mediator benchmark models are known at LO+PS accuracy. 

\subsection{Parameter scan}

\begin{figure*}[!htbp]
\centering
\includegraphics[width=0.95\textwidth]{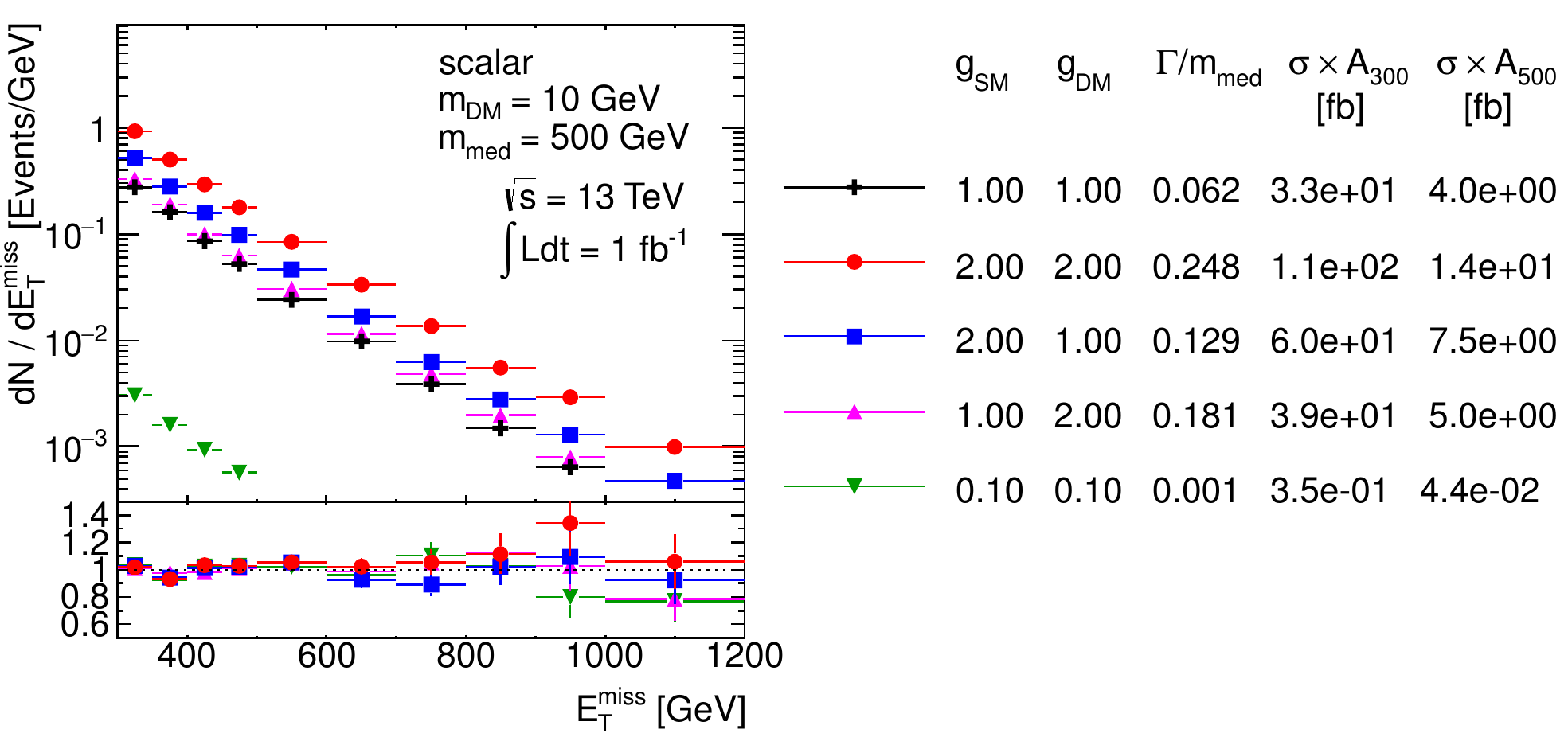}
\caption{Scan over couplings. The $\MET$ distribution is compared for the scalar mediator models using the parameters as indicated. Ratios of the normalized distributions with respect to the first one are shown. $A_{300}$ and $A_{500}$ in the table denote the acceptance of the $\MET>300$~\gev and $\MET>500$~\gev cut, respectively. Studies in all figures for the jet+\MET{} signature is \powheg,
	with \pythiaEight for the parton shower; }
\label{fig:monojet_scan_S_g}
\end{figure*}

\begin{figure}[!p]
\centering
\includegraphics[width=0.95\textwidth]{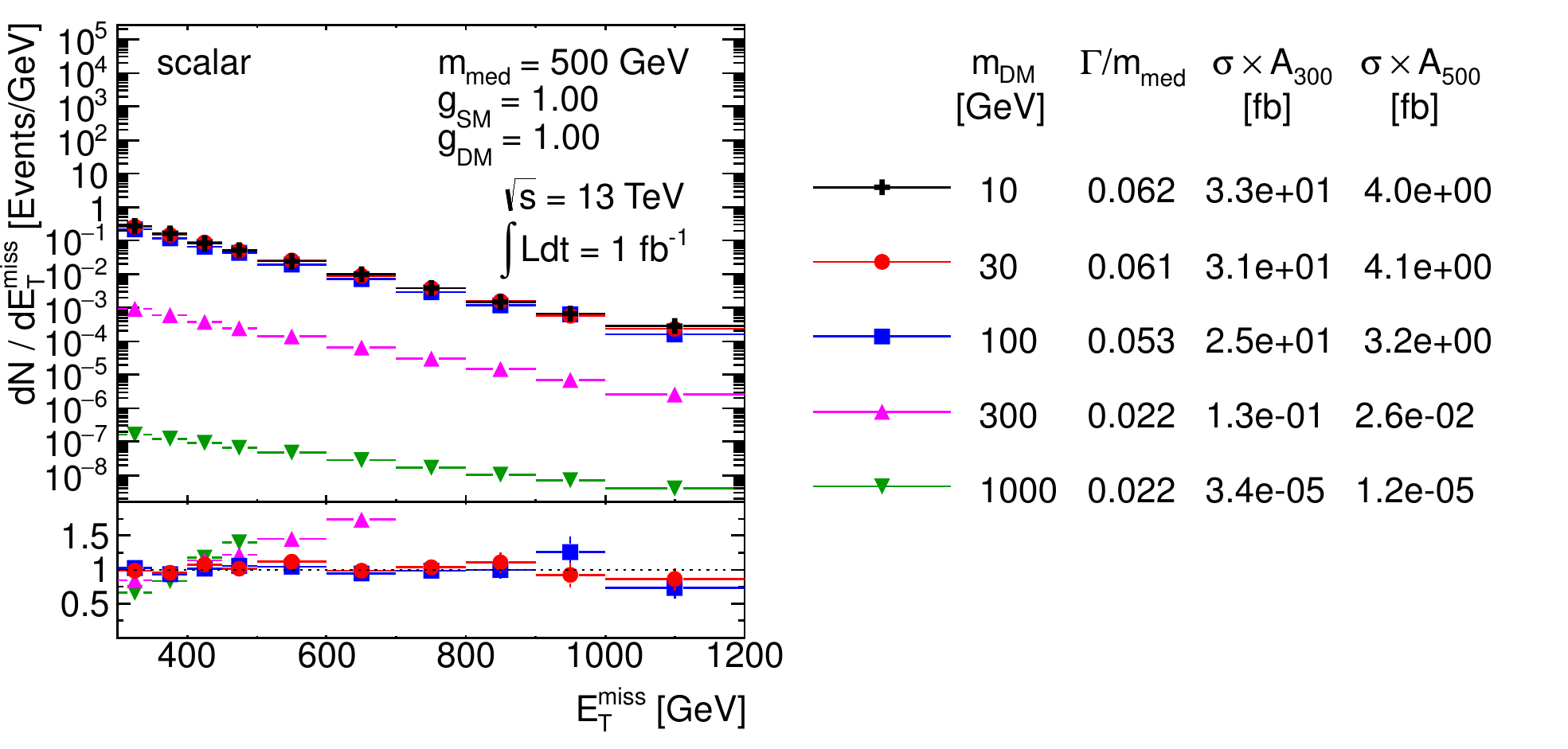}
\caption{Scan over Dark Matter mass. The $\MET$ distribution is compared for the scalar mediator models using the parameters as indicated. Ratios of the normalized distributions with respect to the first one are shown. $A_{300}$ and $A_{500}$ in the table denote the acceptance of the $\MET>300$~\gev and $\MET>500$~\gev cut, respectively.}
\label{fig:monojet_scan_S_mDM1000}
\end{figure}

\begin{figure}[!p]
\centering
\includegraphics[width=0.95\textwidth]{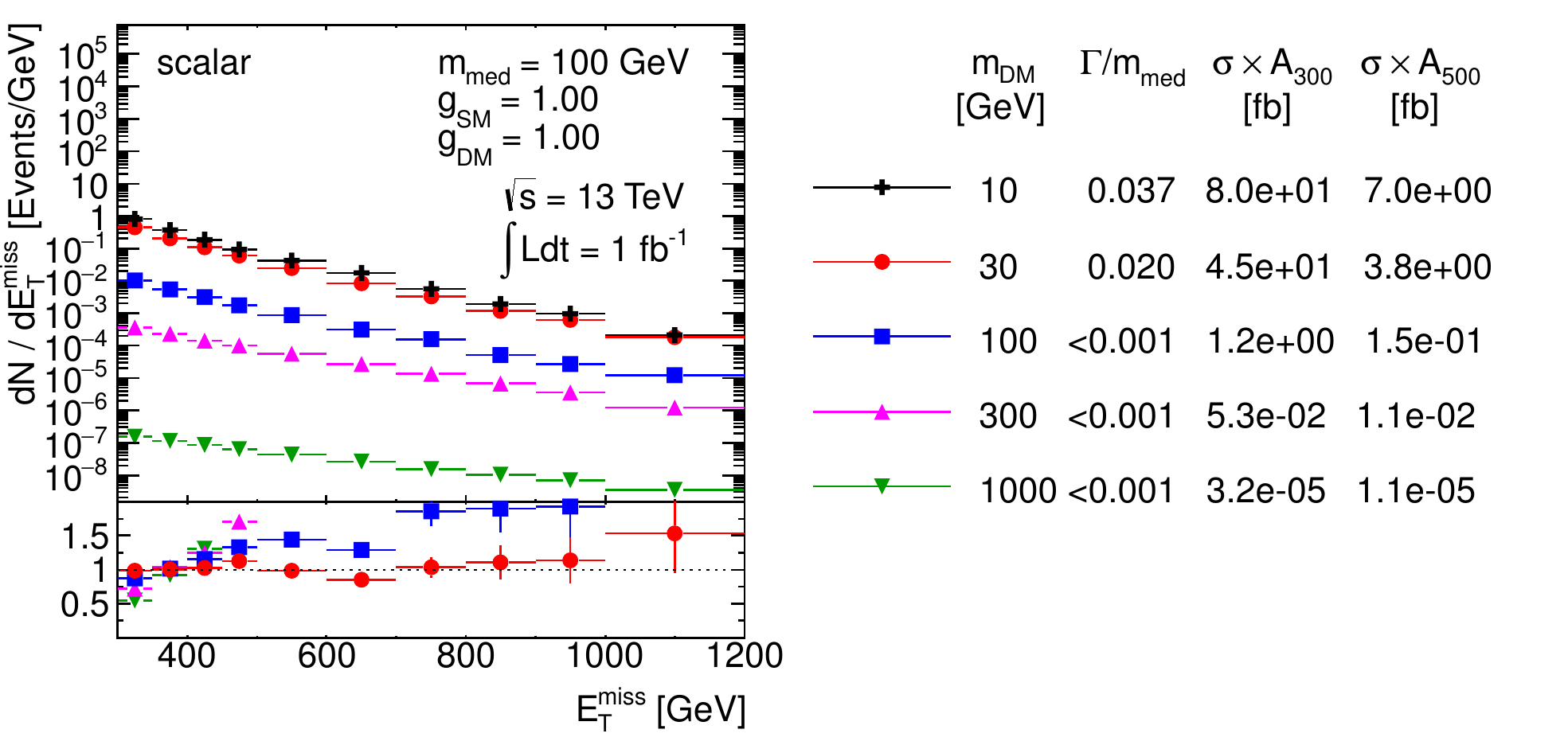}
\caption{Scan over Dark Matter mass. The $\MET$ distribution is compared for the scalar mediator models using the parameters as indicated. Ratios of the normalized distributions with respect to the first one are shown. $A_{300}$ and $A_{500}$ in the table denote the acceptance of the $\MET>300$~\gev and $\MET>500$~\gev cut, respectively.}
\label{fig:monojet_scan_S_mDM100}
\end{figure}

\begin{figure}[!p]
\centering
\includegraphics[width=0.95\textwidth]{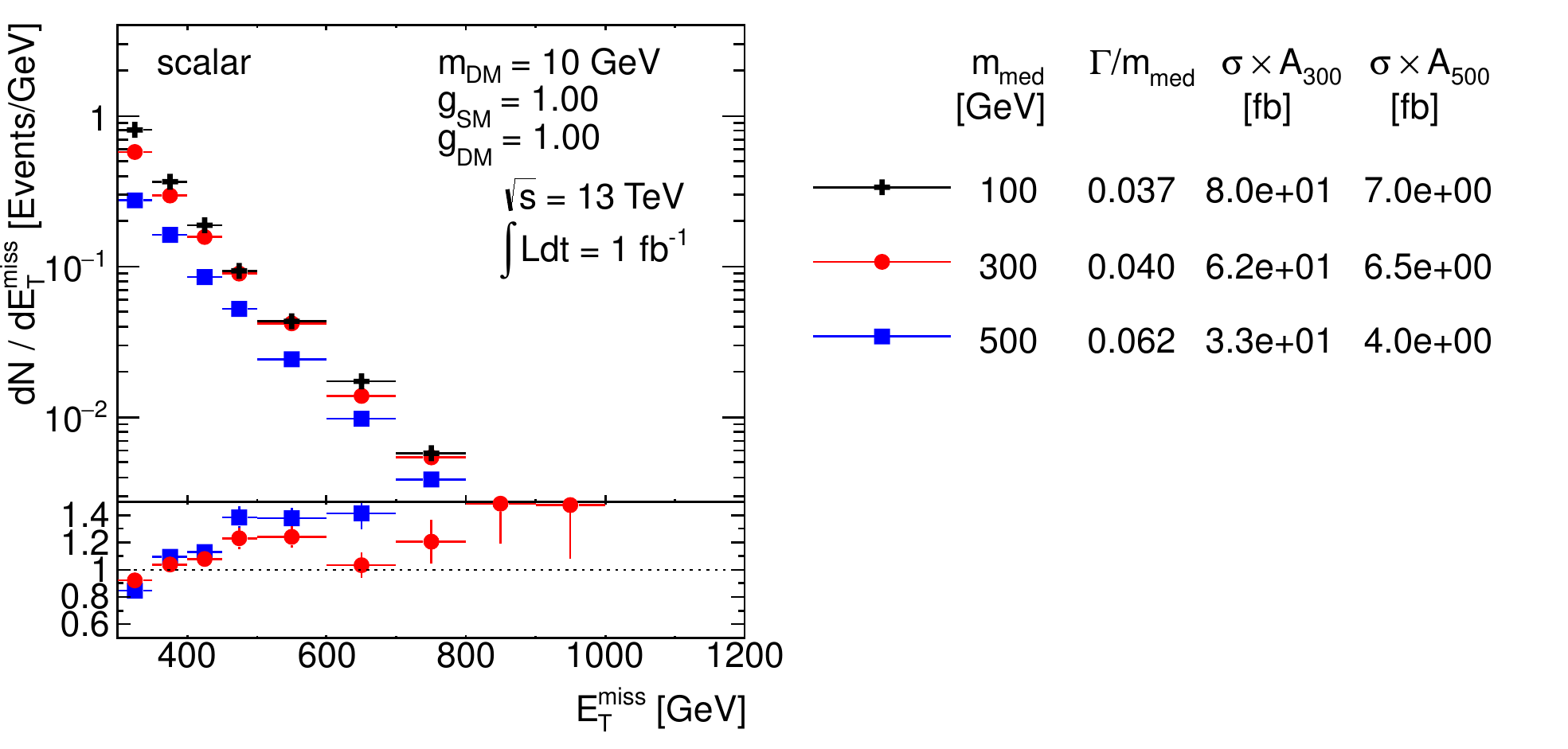}
\caption{Scan over mediator mass. The $\MET$ distribution is compared for the scalar mediator models using the parameters as indicated. Ratios of the normalized distributions with respect to the first one are shown. $A_{300}$ and $A_{500}$ in the table denote the acceptance of the $\MET>300$~\gev and $\MET>500$~\gev cut, respectively.}
\label{fig:monojet_scan_S_mMed10}
\end{figure}

\begin{figure}[!p]
\centering
\includegraphics[width=0.95\textwidth]{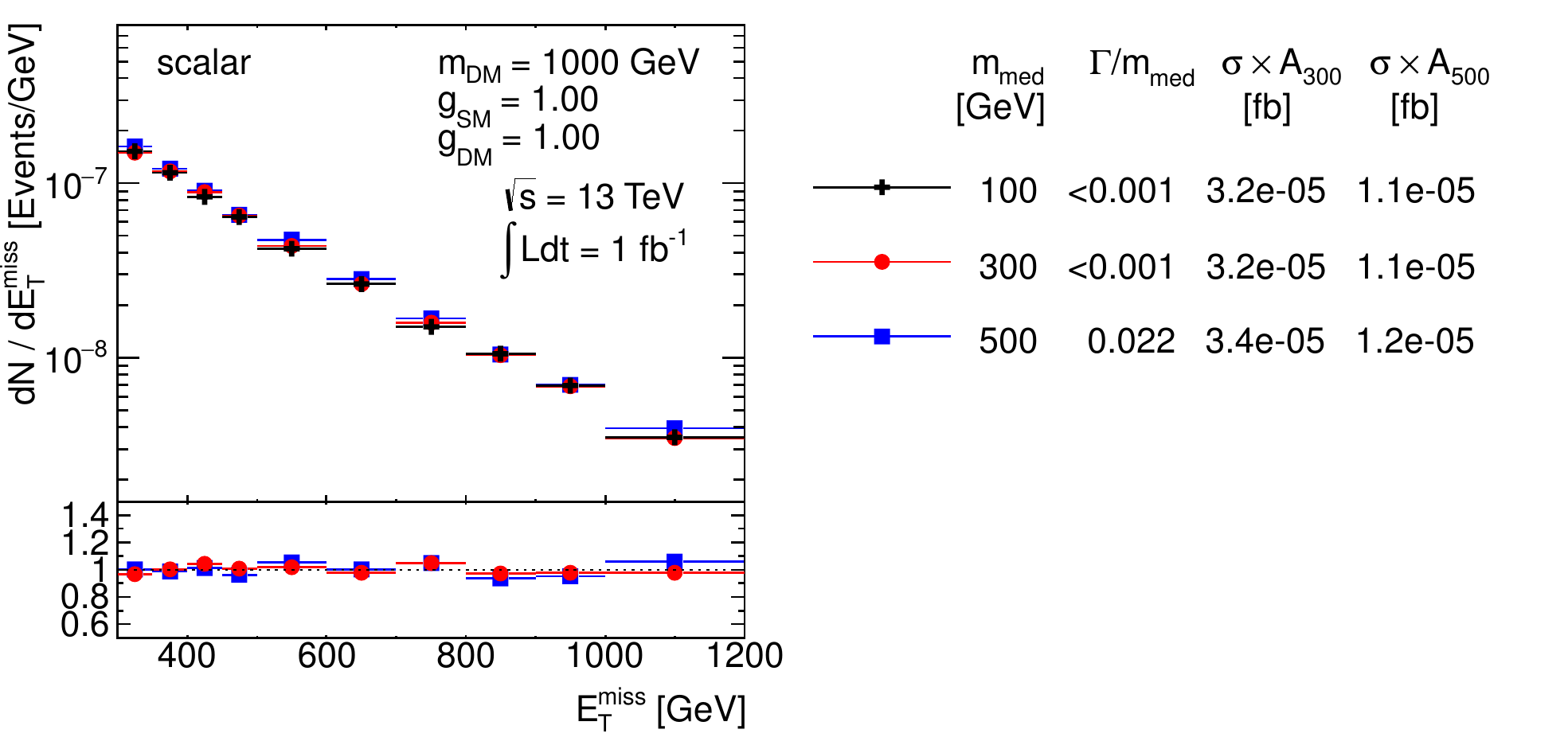}
\caption{Scan over mediator mass. The $\MET$ distribution is compared for the scalar mediator models using the parameters as indicated. Ratios of the normalized distributions with respect to the first one are shown. $A_{300}$ and $A_{500}$ in the table denote the acceptance of the $\MET>300$~\gev and $\MET>500$~\gev cut, respectively.}
\label{fig:monojet_scan_S_mMed1000}
\end{figure}

\begin{figure}
	\centering
	\includegraphics[width=0.95\textwidth]{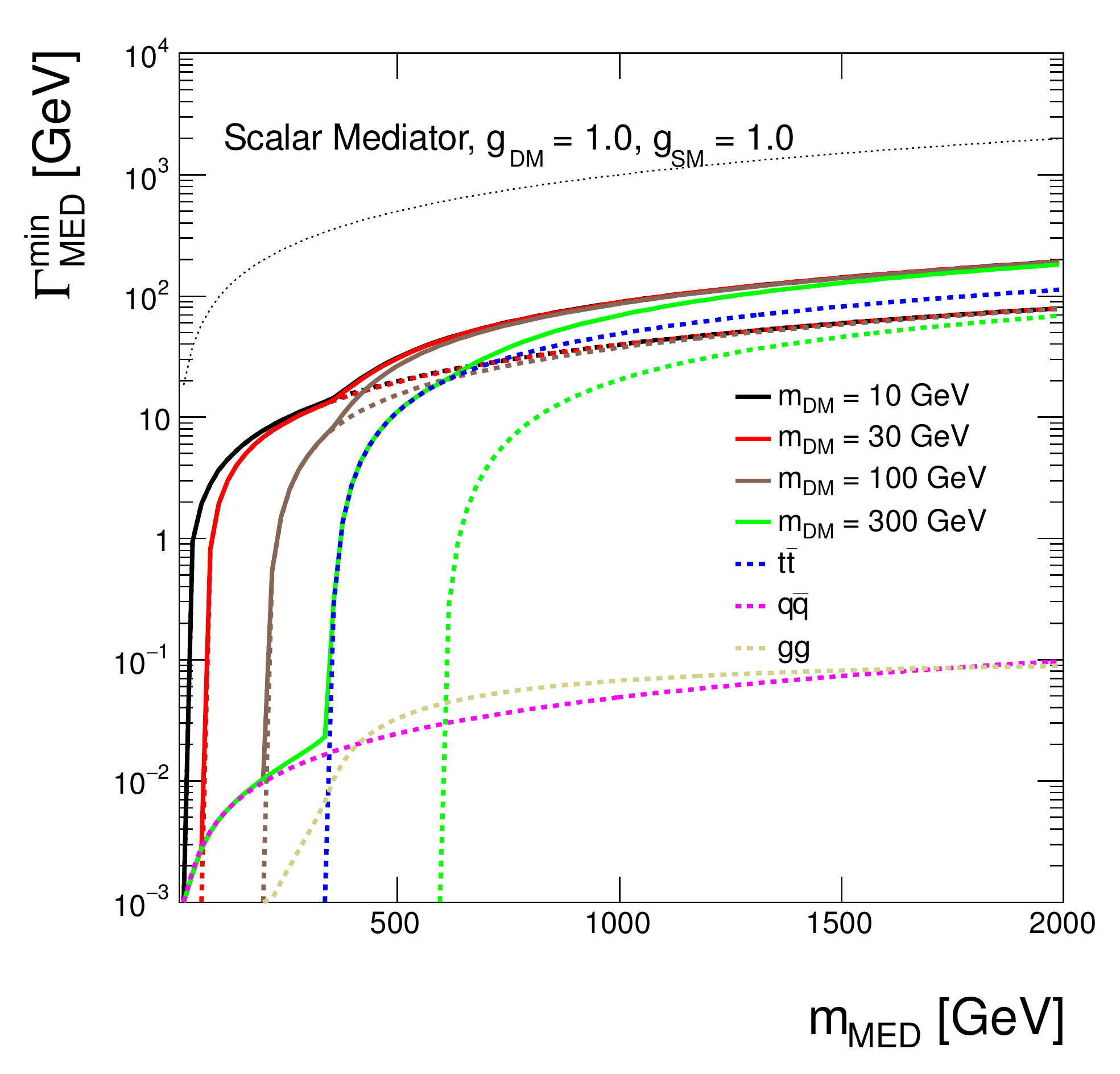}
	\includegraphics[width=0.95\textwidth]{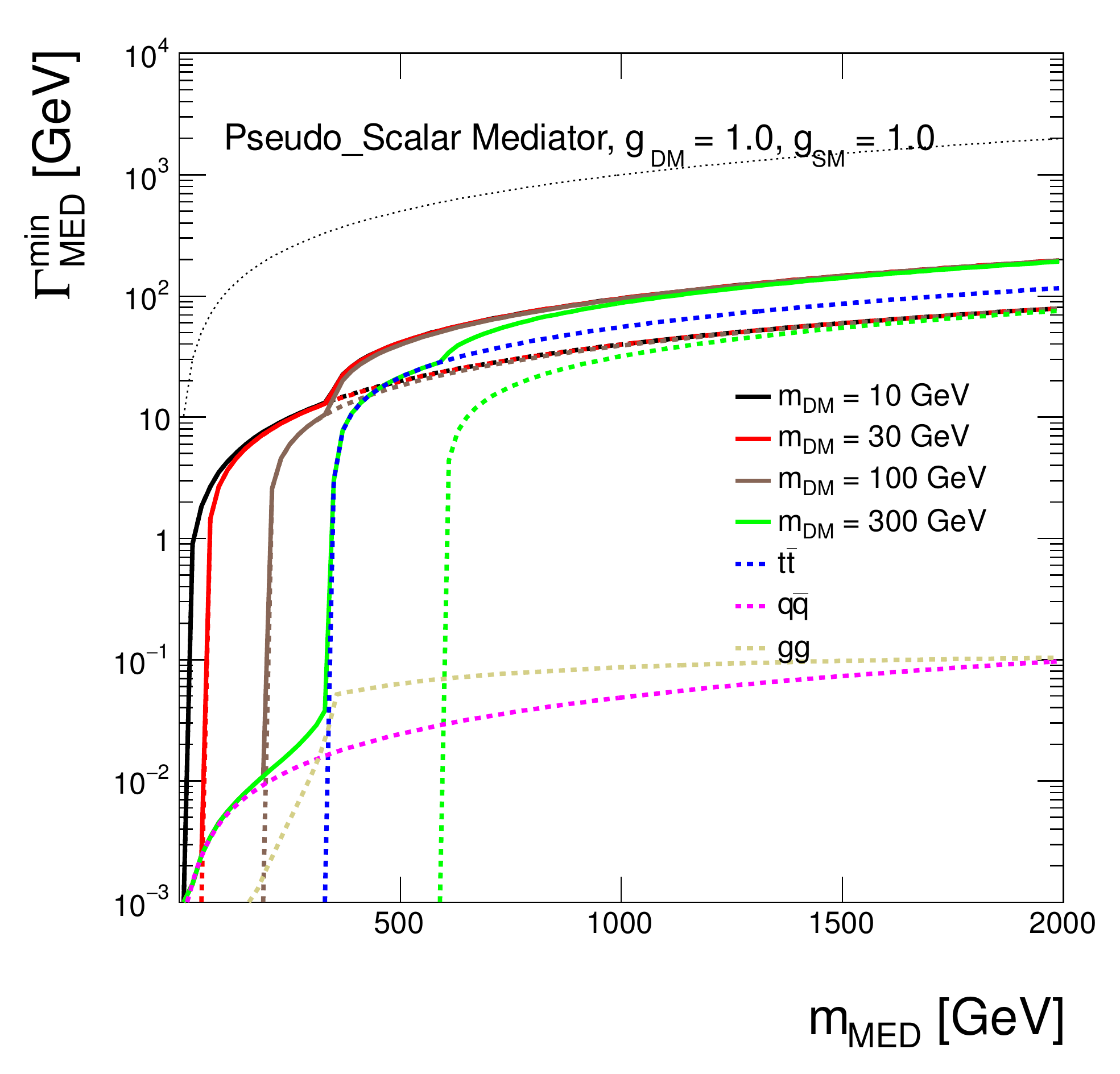}
	\caption{Minimal width as a function of mediator mass for scalar and pseudo-scalar mediator assuming couplings of 1. The total width is shown as solid lines for Dark Matter masses of \mDM=10~\gev, 30~\gev, 100~\gev and 300~\gev in black, red, brown and green, respectively. The individual contributions from Dark Matter are indicated by dotted lines with the same colors. The contribution from all quarks but top is shown as magenta dotted line and the contribution from top quarks only is illustrated by the dotted blue line. The dotted beige line shows the contribution from the coupling to gluons. The dotted black line shows the extreme case $\Gamma_{\rm{min}}=\mMed$.}
	\label{fig:monojet_width_S}
\end{figure}

Similarly as in the case of the vector and axial-vector couplings
of \spinone mediators, scans in the parameter space are performed also for the scalar and pseudo-scalar couplings of the \spinzero mediators
in order to decide on the optimized parameter grid for the presentation of Run-2 results. Figures\,\ref{fig:monojet_scan_S_g}-
\ref{fig:monojet_scan_S_mMed1000} show the scans over the couplings, Dark Matter mass and mediator mass and the same conclusions apply as in Section\,\ref{sec:monojet_V}.

A scan over the mediator mass is shown in Fig.\,\ref{fig:monojet_scan_S_mMed1000} where \mMed = 300~\gev and 500~\gev are chosen to be below and above $2m_t$. The off-shell case is assumed by taking an extreme limit ($\mDM=1$~\tev) in order to study solely the effects of the couplings to quarks. 
No differences in the kinematic distributions are observed and also the cross sections remain similar in this case. No significant changes appear for mediator masses around the $2m_t$ threshold.

\begin{figure*}
	\centering
	\includegraphics[width=0.95\textwidth]{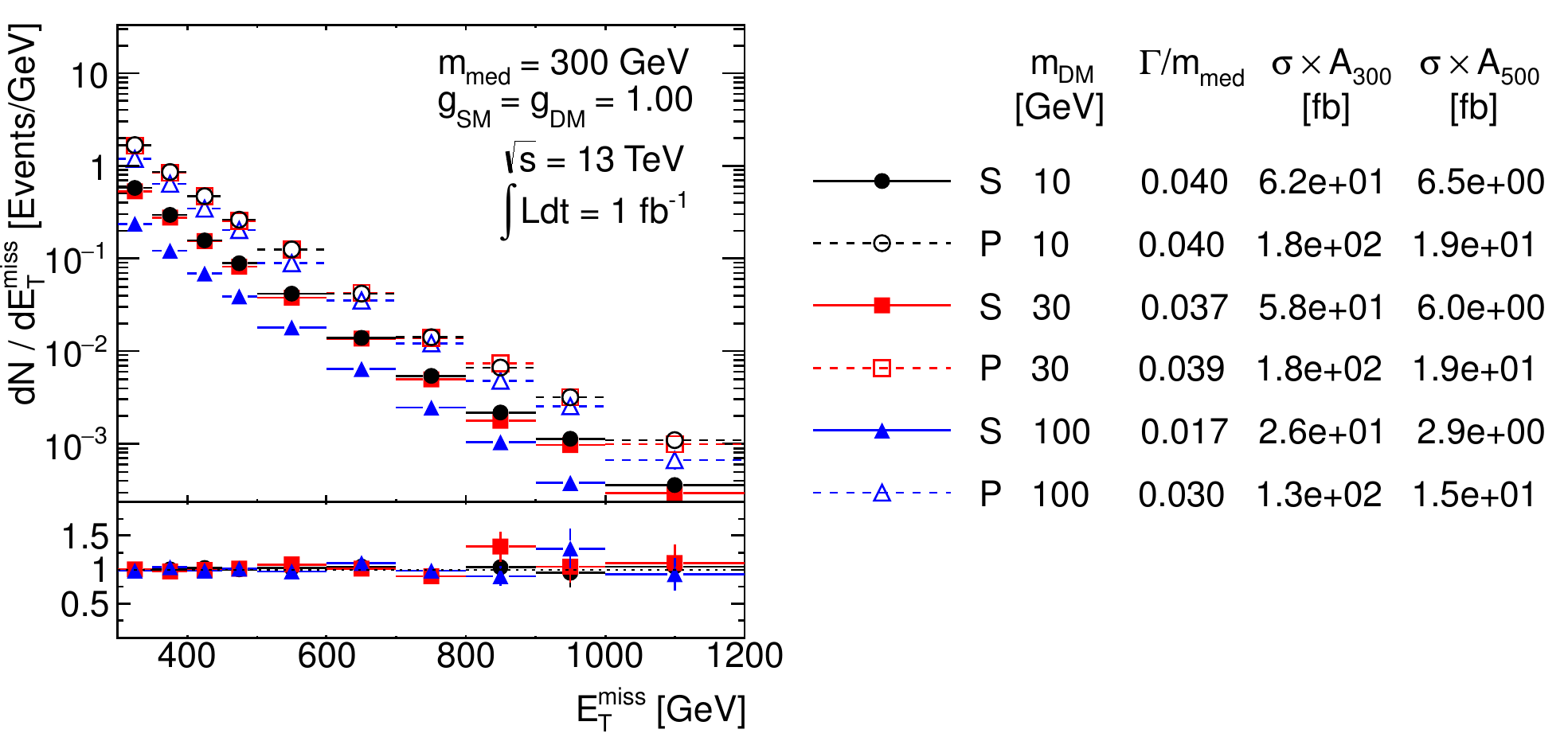}
	\caption{Comparison of the $\MET$ distributions for the scalar and pseudoscalar models for different $\mMed=300\,\gev$ and different Dark Matter masses. 
		Ratios of the normalized distributions with respect to the first one are shown. $A_{300}$ and $A_{500}$ in the table denote the acceptance of the $\MET>300$~\gev and $\MET>500$~\gev cut, respectively.}
	\label{fig:monojet_SPmodels}
\end{figure*}

It can be seen in Fig.~\ref{fig:monojet_SPmodels} that the kinematics for the scalar and pseudoscalar models coincides when considering the diagrams in Fig.~\ref{fig:feyn_prod_S}. 
For this reason, we recommend to fully simulate only one of the two models.
No preference is given between the two models as they have the
same kinematics, although it is worth noting that the pseudo-scalar model has been used for a Dark Matter interpretation of the DAMA signal 
and of the galactic center excess~\cite{Arina:2014yna}.
Like in the case of the vector and axial-vector models described in Section~\ref{sec:monojet_spin}, the differences between the cross sections for the scalar and pseudo-scalar samples with the same $\mDM$ and $\mMed$ are increasing with the Dark Matter mass for fixed mediator mass, with the pseudo-scalar model yielding larger cross sections. There is an increasing difference between the minimal widths close to the $2\mDM=\mMed$ threshold.

\subsubsection{Proposed parameter grid}

The optimized parameter grid in the $\mMed$--$\mDM$ plane for scalar and pseudo-scalar mediators is motivated by similar arguments as in the previous section. Therefore, a similar pattern is followed here, with the exception of taking $\gq=\gDM=1$. The choice of $\gq=0.25$ for the vector and axial-vector models is motivated by suppressing constraints from di-jets, which is not a concern in the scalar and pseudo-scalar mediator case. Here a di-jet signal emerges only at the 2-loop level through diagrams where the mediator is produced via gluon-gluon fusion and decays back into two gluons through a top loop. The strong loop suppression renders such signals unobservable at the LHC. Further constraints on the scalar and pseudo-scalar mediators may emerge from searches in $t\bar{t}$ final states. Studies of the electroweak effects to $t\bar{t}$ production suggest that one can only expect percent level contributions for $\gq \sim O(1)$ \cite{Haisch:2013fla}. Therefore, keeping $\gq=\gDM=1$ is a reasonable choice in the case of the scalar and pseudo-scalar mediators. Contrary to the vector and axial-vector models, note that couplings of 1 lead to $\Gamma_{\rm{min}}/\mMed \lsim 0.1$, ensuring the narrow width approximation is applicable. Furthermore, the sensitivity to the highest mediator masses has to be re-evaluated. The generator level cross section times the acceptance at $\MET>500$~\gev for the model with couplings $\gq=\gDM=1$, light Dark Matter of \mDM=10~\gev and a \mMed=500~\gev scalar mediator is at the order of 10\,fb, i.e. just at the edge of the early Run-2 sensitivity. Increasing the mediator mass to 1~\tev pushes the product $\sigma\times A$ down to approximately 0.1\,fb, below the LHC sensitivity. Therefore, we choose to remove the 2~\tev mediator mass from the grid and present the final grid with 33 mass points only, as shown in Tab.\,\ref{tab:mDMmMedScan_SP}. One point at very high mediator mass (10~\tev) is added for each of the DM masses scanned, to aid the reinterpretation of results in terms of contact interaction operators (EFTs).

\begin{table}[!h]
\centering
\begin{tabular}{| l |r r r r r r r r r|}
\hline
\multicolumn{1}{|c|}{\mDM (\gev)} & \multicolumn{9}{c|}{\mmed (\gev)} \\
\hline
 1             &         10  & 20 & 50 & 100 & 200 & 300 & 500 &         1000  &         10000  \\
 10   	       &         10  & 15 & 50 & 100 &     &     &     &               &         10000  \\
 50            &         10  &    & 50 &  95 & 200 & 300 &     &               &         10000  \\
 150           &         10  &    &    &     & 200 & 295 & 500 &         1000  &         10000  \\
 500           &         10  &    &    &     &     &     & 500 &          995  &         10000  \\
 1000          &         10  &    &    &     &     &     &     &         1000  &         10000  \\
\hline
\end{tabular}

\caption{Simplified model benchmarks for \schannel simplified models (\spinzero mediators 
decaying to Dirac DM fermions in the scalar and pseudoscalar case, taking the minimum width for \gq = 1 and \gDM = 1)}.

\label{tab:mDMmMedScan_SP}
\end{table}


For the parameter grid for scalar and pseudo-scalar mediator \schannel exchange, the $\Gamma_{\rm{min}}/\mMed$ ratio is given in Tables\,\ref{tab:widthS} and \ref{tab:widthP}, respectively. In the on-shell regime, the ratio is between 0.04 and 0.1. Very narrow resonances with $\Gamma_{\rm{min}}/\mMed<0.001$ correspond to the mass points where the mediator is off-shell. Note that the loop-induced contribution from gluons is ignored in the width calculation.

\begin{table}
	\centering
	\resizebox{\textwidth}{!}{
		\begin{tabular}{| l |r r r r r r r r r|}
			\hline
			\multicolumn{1}{|c|}{\mDM/\gev} & \multicolumn{9}{c|}{\mmed/\gev} \\
			&         10  & 20 & 50 & 100 & 200 & 300 & 500 &         1000  & 10000  \\
			\hline
			\hline
   1 & 0.040  & 0.040  & 0.040  & 0.040  & 0.040  & 0.040  & 0.062  & 0.089  & 0.099  \\
  10 &$<$0.001&$<$0.001& 0.040  & 0.040  &        &        &        &        & 0.099  \\
  50 &$<$0.001&        &$<$0.001&$<$0.001& 0.040  & 0.040  &        &        & 0.099  \\
 150 &$<$0.001&        &        &        &$<$0.001&$<$0.001& 0.062  & 0.089  & 0.099  \\
 500 &$<$0.001&        &        &        &        &        & 0.022  & 0.049  & 0.099  \\
1000 &$<$0.001&        &        &        &        &        &        & 0.049  & 0.099  \\
			\hline
		\end{tabular}}
		\caption
		{Minimal width of the scalar mediator exchanged in \schannel divided by its mass, assuming $\gq=\gDM=1$. The loop-induced gluon contribution is ignored. The numbers tabulated under $2\mDM=\mMed$ correspond to the width calculated for $\mMed-5$~\gev.}
		\label{tab:widthS}
	\end{table}

	\begin{table}
		\centering
		\resizebox{\textwidth}{!}{
			\begin{tabular}{| l |r r r r r r r r r|}
				\hline
				\multicolumn{1}{|c|}{\mDM/\gev} & \multicolumn{9}{c|}{\mmed/\gev} \\
				&         10  & 20 & 50 & 100 & 200 & 300 & 500 &         1000  & 10000  \\
				\hline
				\hline
   1 & 0.040  & 0.040  & 0.040  & 0.040  & 0.040  & 0.040  & 0.083  & 0.095  & 0.099  \\
  10 &$<$0.001&$<$0.001& 0.040  & 0.040  &        &        &        &        & 0.099  \\
  50 &$<$0.001&        &$<$0.001&$<$0.001& 0.040  & 0.040  &        &        & 0.099  \\
 150 &$<$0.001&        &        &        &$<$0.001&$<$0.001& 0.083  & 0.095  & 0.099  \\
 500 &$<$0.001&        &        &        &        &        & 0.043  & 0.056  & 0.099  \\
1000 &$<$0.001&        &        &        &        &        &        & 0.056  & 0.099  \\
				\hline
			\end{tabular}}
			\caption
			{Minimal width of the pseudo-scalar mediator exchanged in \schannel divided by its mass, assuming $\gq=\gDM=1$. The loop-induced gluon contribution is ignored. The numbers tabulated under $2\mDM=\mMed$ correspond to the width calculated for $\mMed-5$~\gev.}
			\label{tab:widthP}
		\end{table}

\subsection{Additional considerations for $V+\MET$ signatures}
\label{sub:EW_Scalar}

The discussion of parameters for the model with a color-singlet, \spinzero mediator
parallels that in Section~\ref{subsec:MonojetLikeModels}. 

Even though the sensitivity of mono-boson searches to this model is low and it may not
be in reach of early LHC searches, this model can be generated for W, Z and photon searches 
in order to reproduce the kinematics of contact interaction operators that are further 
described in Section~\ref{sub:EW_EFT_Dim5}, to aid later reinterpretation.  

Other models of dark matter that couple dominantly to electroweak gauge bosons through either
pseudo-scalar or vector mediators can be found in Ref.~\cite{Lee:2012ph}.

%

\subsection{\texorpdfstring{Additional considerations for $t \bar{t}$ and $b \bar{b}$+\MET signatures}{Additional considerations for ttbar/bbbar+\MET signatures}}
\label{subsec:DMHFModels}

\begin{figure}[h!]
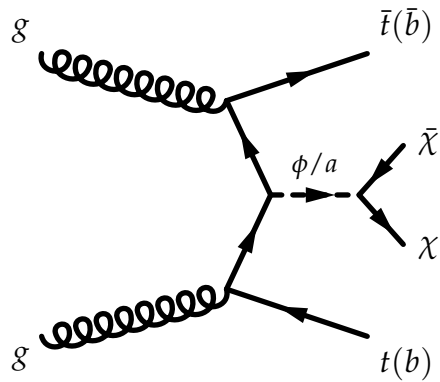

\centering
  \unitlength=0.005\textwidth
  \begin{feynmandiagram}[modelTTbarMET]
    \fmfleft{i1,i2}
    \fmfright{o1,o4,o5,o3}
    \fmf{gluon}{i1,v1}
    \fmf{gluon}{i2,v2}
    \fmf{fermion}{o1,v1,v3,v2,o3}
    \fmf{scalar}{v3,v4}
    \fmf{fermion}{o5,v4,o4}
    \fmf{phantom,tension=0,label={$\phi/a$}}{v3,v4}    
    \fmflabel{\Large $g$}{i1}
    \fmflabel{\Large $g$}{i2}
    \fmflabel{\Large $t (b)$}{o1}
    \fmflabel{\Large $\bar t (\bar b)$}{o3}
    \fmflabel{\Large $\chi$}{o4}
    \fmflabel{\Large $\bar \chi$}{o5}    
  \end{feynmandiagram}
\caption{Representative Feynman
diagram showing the pair production of Dark Matter particles in
association with $t\bar t$ (or $b\bar b$).}
\label{fig:TTbarPhi}
\setfloatalignment{t}
\end{figure}

With the MFV assumption, the top and bottom
quark can play an important  role in the phenomenology.
The scalar and pseudoscalar mediator models predict not only
the monojet process described in Section~\ref{sec:monojet_scalar}, but also production of Dark Matter
in association with top (or bottom) pairs, as illustrated in Fig.~\ref{fig:TTbarPhi}. 
Dedicated searches including jets from heavy flavor quarks in the final state
can be designed for this signature. Another class of simplified models,  
which includes a Dark Matter interpretation among many others, and yields a single
top quark in the final state, is detailed in Appendix~\ref{sec:singletop}. 

In addition to the $t\bar t$+DM models illustrated in Fig.~\ref{fig:TTbarPhi}, 
some theoretically motivated scenario (e.g. for high $tan\beta$ in 2HDM in the pMSSM) 
privilege the coupling of \spinzero mediators to down generation quarks.
This assumption motivates the study of final states involving $b$-quarks 
as a complementary search to the $t\bar
t$+DM models, to directly probe the $b$-quark coupling. 
An example of such a model can be found in Ref.~\cite{Buckley:2014fba}
and can be obtained by replacing top quarks with $b$ quarks in Fig.~\ref{fig:TTbarPhi}.
Note that, because of the kinematics features of $b$ quark production relative
to heavy $t$ quark production, a $b\bar b$+DM final state may only yield one
experimentally visible $b$ quark, leading to a mono-$b$ signature in a model that conserves $b$ flavor.

Dedicated implementations of these models for the work of this Forum are available at LO+PS accuracy,
even though the state of the art is set to improve on a timescale beyond that for early Run-2 DM searches
as detailed in Section~\ref{sec:TTBar_implementation}.
The studies in this Section have been produced using a leading order UFO model within \madgraph 2.2.2
~\cite{Alwall:2014hca,Alloul:2013bka,Degrande:2011ua} 
using \pythiaEight for the parton shower. 

\subsubsection{Parameter scan}


The parameter scan for the dedicated $t\bar{t}$+\MET{} searches has been studied in detail to target the production 
mechanism of DM associated with heavy flavor quarks, and shares many details of the scan for the scalar model with a gluon radiation.
The benchmark points scanning the model parameters have been selected to ensure that the kinematic features of the 
parameter space are sufficiently represented. Detailed studies were performed to identify points in the \mdm, 
$m_{\phi,a}$, $\gDM$, $\gq$ (and $\Gamma_{\phi,a}$) parameter space that differ significantly from each other 
in terms of expected detector acceptance. Because missing transverse momentum is the key observable for searches, the 
mediator $p_{T}$ spectra is taken to represent the main kinematics of a model. Another consideration in determining the set 
of benchmarks is to focus on the parameter space where we expect the searches to be sensitive during the 2015 LHC run. 
Based on a projected integrated luminosity of $30\,{\rm fb}^{-1}$ expected for 2015, we disregard model points with a 
cross section times branching ratio smaller than $0.1\,{\rm fb}$, corresponding to a minimum of one expected event 
assuming a 0.1\% efficiency times acceptance. 

\begin{figure}[!ht]
  \begin{center}
    \includegraphics[width=0.95\textwidth]{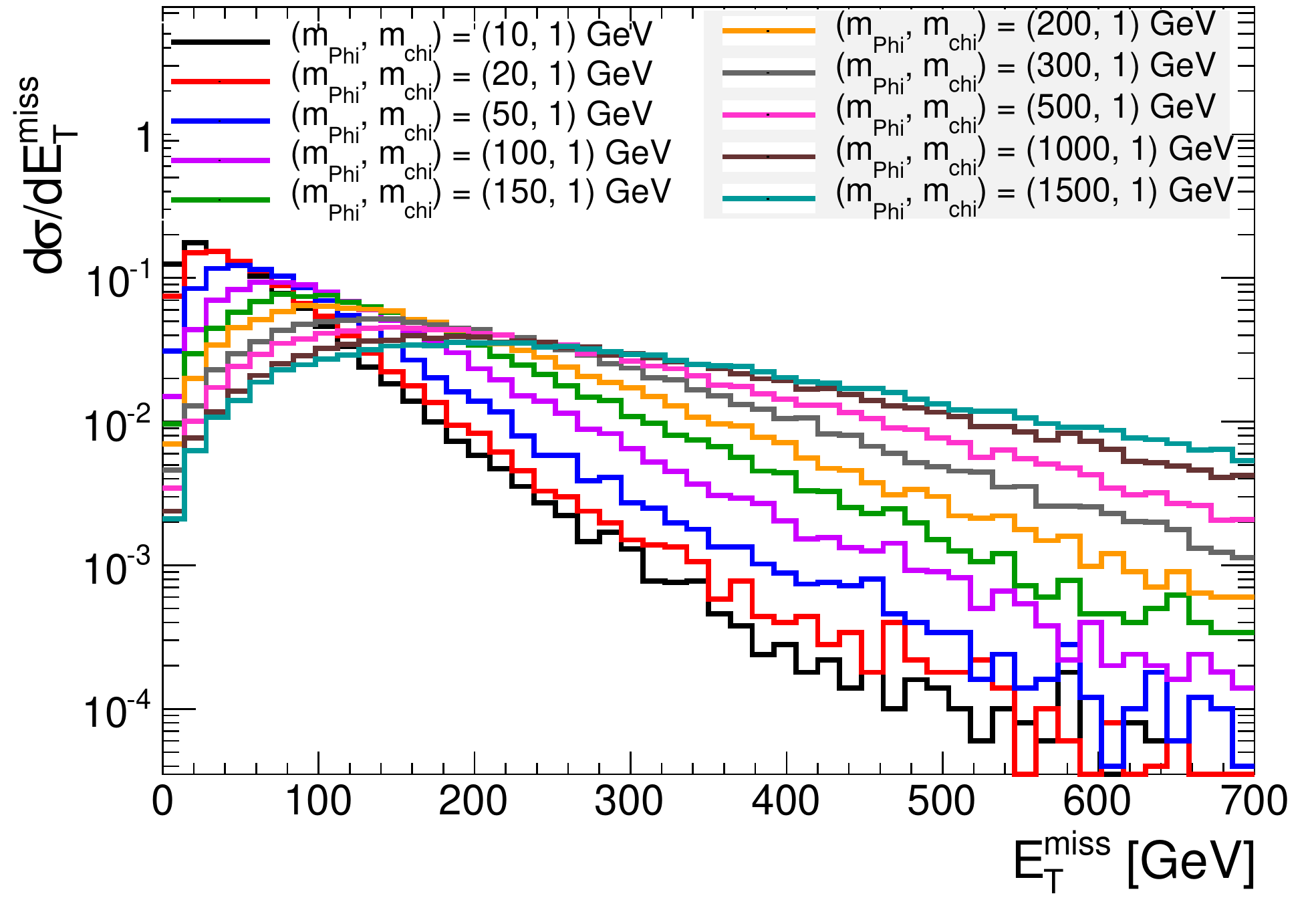}
    \caption{\label{fig:scanPhi} Example of the dependence of the kinematics on the scalar mediator mass in the $t\bar{t}$+\MET{} signature. The Dark Matter mass is fixed to be \mdm=$1 {\rm GeV}$.}
\end{center}
\end{figure}

\begin{figure}[!ht]
  \begin{center}
    \includegraphics[width=0.95\textwidth]{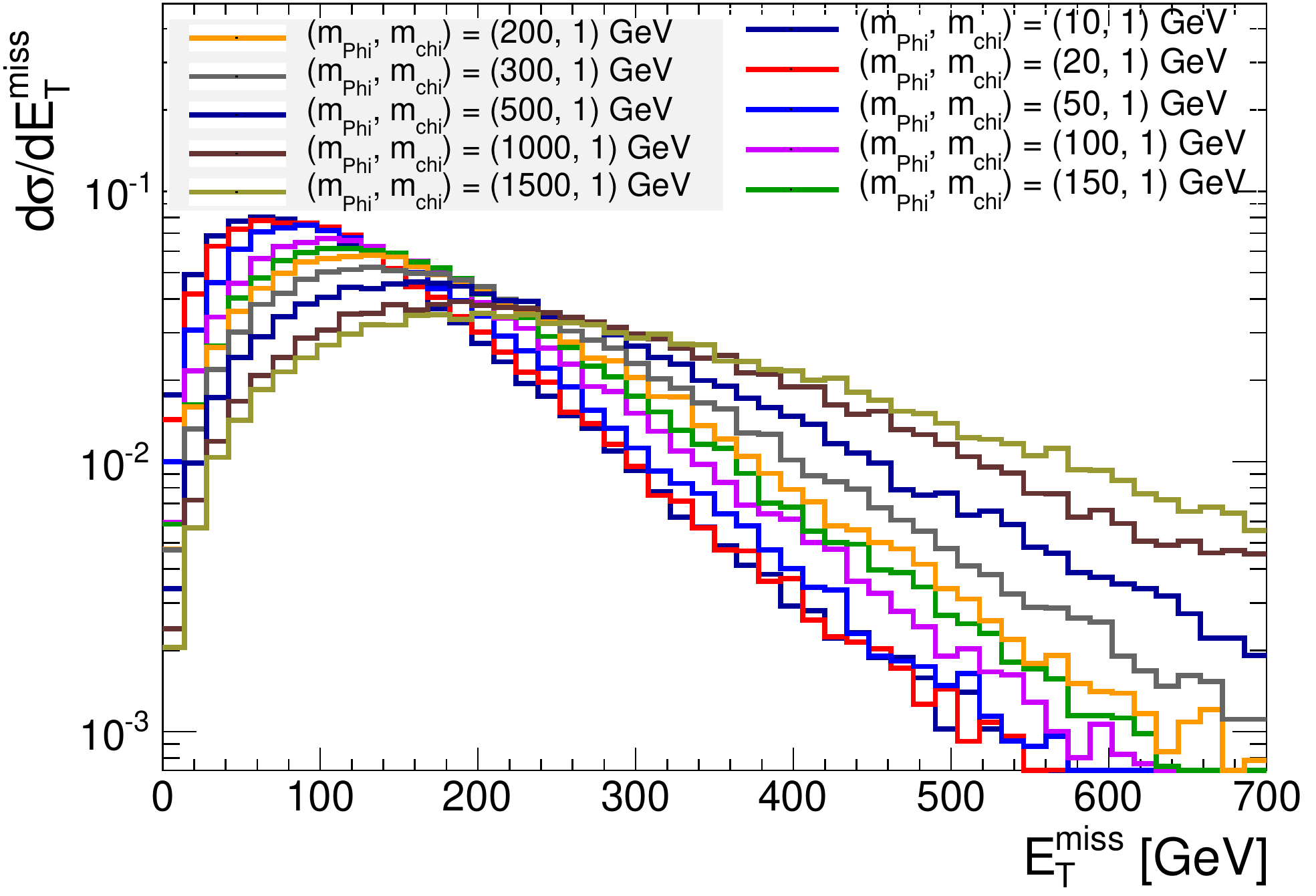}
    \caption{\label{fig:scanPhiPseudo} Example of the dependence of the kinematics on the pseudoscalar mediator mass in the $t\bar{t}$+\MET{}. The Dark Matter mass is fixed to be \mdm=$1 {\rm GeV}$. All figures concerning the $t\bar{t}$+\MET{}  signature have been produced using a leading order model within \madgraph 2.2.2, using \pythiaEight for the parton shower.}
\end{center}
\end{figure}

The kinematics is most dependent on the masses \mdm and $m_{\phi,a}$. Figure~\ref{fig:scanPhi} 
and~\ref{fig:scanPhiPseudo} show typical dependencies for scalar and pseudoscalar couplings respectively.
Typically, the mediator $p_T$ spectrum broadens with larger $m_{\phi,a}$. 
The kinematics are also different between on-shell ($\mMed>2\mdm$) and off-shell ($\mMed<2\mdm$) mediators as discussed in Section~\ref{sec:monojet_scalar}. 
Furthermore, the kinematic differences in the \MET{} spectrum between scalar and pseudoscalar are larger for light mediator 
masses with respect to heavier mediators. It is therefore important to  
choose benchmark points covering on-shell and off-shell mediators with sufficient granularity, including the
transition region between on-shell and off-shell mediators. 

\begin{figure}[!ht]
  \begin{center}
    \includegraphics[width=0.95\textwidth]{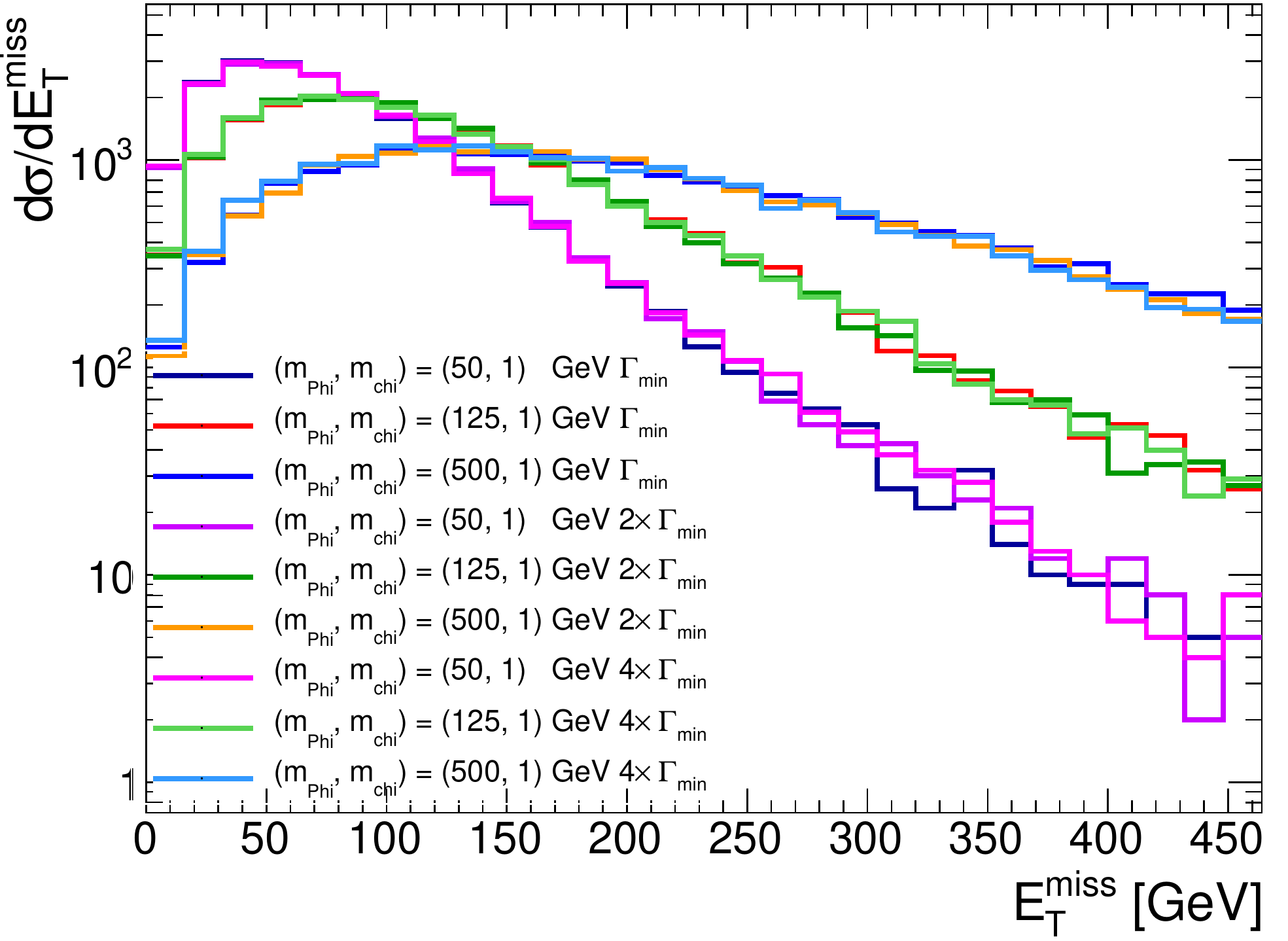}
    \caption{\label{fig:widthsmallscan} Study of the dependence of kinematics on the width of a scalar mediator $t\bar{t}$+\MET{}. The width is increased up to four times the minimal width for each mediator and Dark Matter mass combination. 
    }
\end{center}
\end{figure}

Typically only weak dependencies on couplings are observed (see Fig~\ref{fig:widthsmallscan}) where the variation with width of the integral over parton distributions is unimportant. As shown in Section~\ref{sub:parameter_scan_monojet}, for couplings $\sim O(1)$ the width is large enough that the $p_T$ of the mediator is determined mainly by the PDF. 

At large mediator masses ($\sim 1.5\,{\rm TeV}$) or very small couplings ($\sim 10^{-2}$), width effects are significant, but these regimes have production cross sections that are too small to be relevant for $30\,{\rm fb}^{-1}$ and are not studied here. However, with the full Run~2 dataset, such models may be within reach. 

\begin{figure}[!ht]
  \begin{center}
    \includegraphics[width=0.95\textwidth]{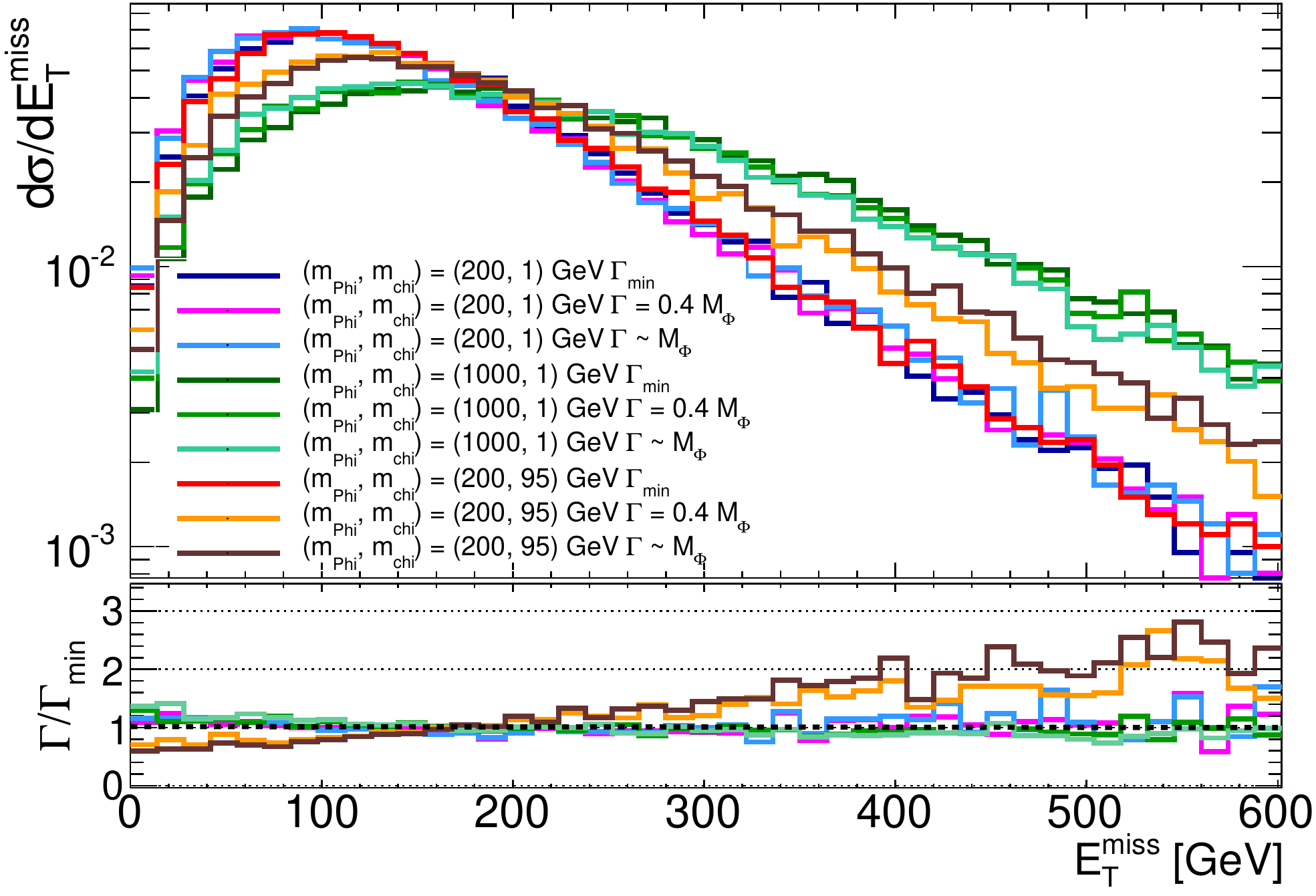}
    \vspace{2mm}
    \caption{\label{fig:widthlargescan} Dependence of the kinematics on the width of a scalar mediator $t\bar{t}$+\MET{}. The width is increased up to the mediator mass. Choices of mediator and Dark Matter masses such that $m_{\phi,a}$ is slightly larger than $2m_\chi$ is the only case that shows a sizeable variation of the kinematics as a function of the width.  
    }
\end{center}
\end{figure}

Another case where the width can impact the kinematics is when $m_{\phi,a}$ is slightly larger than $2m_\chi$. Here, the width determines the relative contribution between on-shell and off-shell mediators. An example is given in Fig.~\ref{fig:widthlargescan}. As the minimal width choice pursued in this document is the most conservative one, this effect can be neglected in order to reduce the number of benchmark points to be generated. 


The points for the parameter scan chosen for this model are listed in Table~\ref{tab:mDMmMedScan_SP}, chosen
to be harmonized with those for other analyses employing the same scalar model as benchmark. 
Based on the sensitivity considerations above, DM masses are only simulated up to 500 GeV (but the 5 TeV mediator point is retained)
leading to a total of 24 benchmark points. However for these searches we recommend to generate and simulate scalar and pseudoscalar
models separately, as the kinematics differs due to the different coupling of the mediator to the final state top quarks in the two cases,
as shown in Figs.~\ref{fig:scanPhi} and ~\ref{fig:scanPhiPseudo}.

Similar studies were performed in the $b \bar b$ case. It was found that they 
show the same weak dependence of the kinematics of the event on the mediator width.
The same benchmark parameters of the $t\bar t$ case could then be chosen.

\section{Colored scalar mediator, \tchannel exchange}
\label{sec:monojet_t_channel}


The preceding sections address models with a Dirac fermion coupled to
the SM through exchange of a neutral \spinzero or \spinone particle in an
\schannel process.  A \tchannel process may couple the SM and DM
directly, leading to a different phenomenology.
For completeness, we examine a
model where $\chiDM$ is a Standard Model (SM) singlet, a Dirac
fermion
; the
mediating particle, labeled $\phi$, is a charged scalar color triplet and the
SM particle is a quark. Such models have been studied in
Refs.~\cite{An:2013xka,Papucci:2014iwa,Bai:2013iqa,Tait:2013,Chang:2013oia,Bell:2012rg}. 
However, these models have not been studied as extensively as others in this Forum.

Following the example of Ref.~\cite{Papucci:2014iwa}, the interaction Lagrangian is written as

\begin{equation}
\mathcal{L}_{\mathrm{int}} = g \sum_{i=1,2} (\phi_{(i),L} \bar{Q}_{(i),L} + \phi_{(i),u,R} \bar{u}_{(i),R} + \phi_{(i),d,R} \bar{d}_{(i),R}) \chiDM
\end{equation}
where $Q_{(i),L}$, $u_{(i),R}$ and $d_{(i),R}$ are the SM quarks of the $i$-th generation and $\phi_{(i),L}$, $\phi_{(i),u,R}$ and $\phi_{(i),d,R}$ are the corresponding mediators, which 
(unlike the \schannel mediators) must be heavier than $\chiDM$. 
These mediators have SM gauge representations under $(SU(3), SU(2))_Y$ of $(3,2)_{-1/6}$, $(3,1)_{2/3}$ and $(3,1)_{-1/3}$ respectively. Variations of the model previously studied in the literature include coupling to the left-handed quarks only~\cite{Chang:2013oia, Busoni:2014haa}, to the $\phi_{(i),u,R}$ \cite{Tait:2013} or $\phi_{(i),d,R}$ \cite{Papucci:2014iwa, Yavin:14092893}, or some combination~\cite{Bai:2013iqa, An:2013xka}.

The minimal width of each mediator is expressed, using the example of decay to an up quark, as

\begin{equation}
\begin{split}
\Gamma (\phi_{(i)} \rightarrow \bar{u}_{(i)} \chiDM) &= \frac{g_{(i)}^2}{16 \pi M_{\phi_{(i)}}^3}(M_{\phi_{(i)}}^2 - m_{u_{(i)}}^2 - \mDM^2) 		\\
& \times
\sqrt{(M_{\phi_{(i)}}^2 - (m_{u_{(i)}} + \mDM)^2)(M_{\phi_{(i)}}^2 - (m_{u_{(i)}}-\mDM)^2)} \, ,
\end{split}
\end{equation}
which reduces to 

\begin{equation}
\frac{g_{(i)}^2 M_{\phi_{(i)}}}{16 \pi} \left(1 - \frac{\mDM^2}{M_{\phi_{(i)}}^2} \right)^2
\end{equation}
in the limit $M_{\phi_{(i)}}, \mDM \gg m_{u_{(i)}}$.


The generation index $i$ for $\phi_{(i)}$ is linked to the incoming
fermion(s), and it runs on all three quark generations due to 
the MFV assumption. 
Ref.~\cite{Papucci:2014iwa} considers two extreme cases for this model in terms of cross-sections: 
the case in which all mediator flavors are present, leading to the maximal cross-section, and
the case in which only right-handed down-type mediators are present. 
Neither of the models in this reference include couplings to the third quark generation, leading to a violation of the MFV assumption. In the case of purely down-type right-handed squarks this is still safe from flavor constraints. Furthermore, reintroducing the third generation squarks would lead to models that produce qualitatively similar signals in the mono-jet and SUSY squark searches, the main difference being the production cross-section. At the same time the presence of third generation squarks will lead to further constraints from other searches such as those for mono-bjets, for stops and for sbottoms, as discussed in Sec.~\ref{sec:singleb}. 
The studies in this Section
are performed using a model with a mediator coupling to all three generation, following Ref.~\cite{Bell:2012rg}. 
Further differences between the two models (hypercharge, chirality) only lead to a change in the cross-section. 
The LO UFO model is interfaced to \madgraph{} v2.2.3, but it was not possible to go beyond parton-level studies
and interface those models to a parton shower in time for the conclusion of this Forum. 
The state of the art for calculating these models is LO+PS, and the implementation of
multi-parton merging has been studied in detail~\cite{Maltoni:2015twa,deAquino:2012ru,Alwall:2008qv,Papucci:2014iwa},
and further studies should be undertaken prior to generating signal samples for early Run-2 LHC searches. 

\begin{figure}
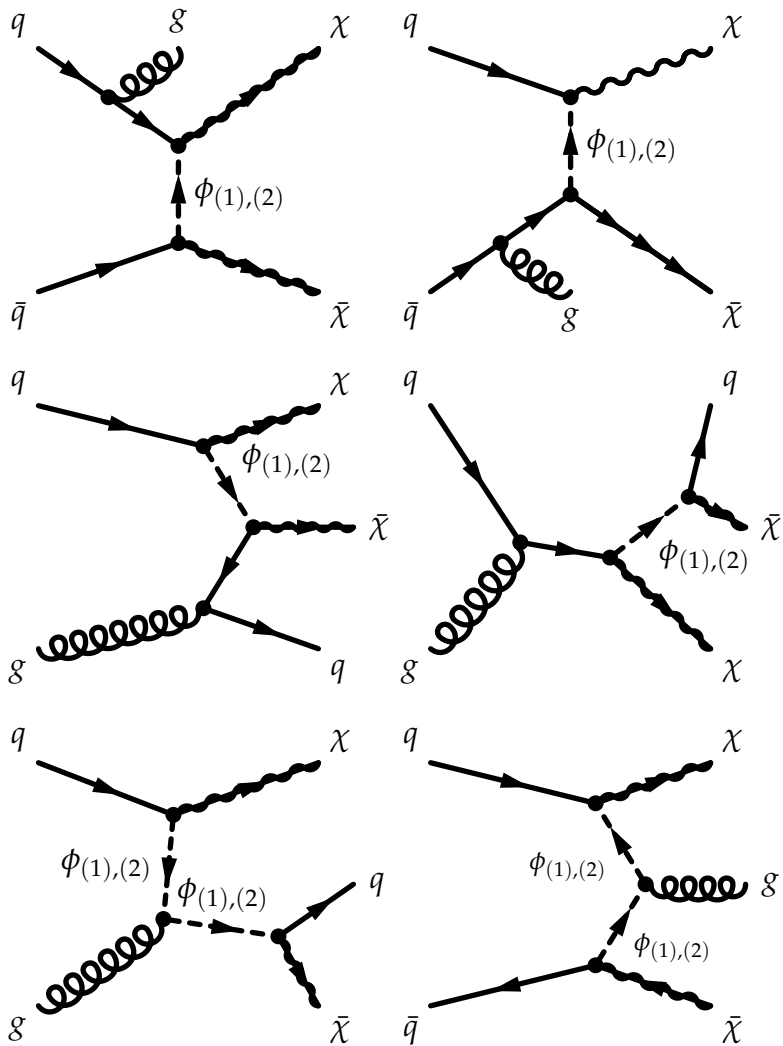

  \unitlength=0.0043\linewidth
  \vspace{3\baselineskip}
  \begin{feynmandiagram}[modelTmonojetA]
    \fmfleft{i1,i2}
    \fmfright{o1,o2}
    \fmftop{isr}
    \fmflabel{\Large ${\bar{q}}$}{i1}
    \fmflabel{\Large ${q}$}{i2}
    \fmf{fermion}{i2,visr,v2}
    \fmf{phantom}{v2,pisr,o2}
    \fmf{phantom,tension=0}{pisr,isr}
    \fmf{fermion,tension=0}{v2,o2}
    \fmf{wiggly,tension=0}{v2,o2}
    \fmf{fermion}{i1,v1,o1}
    \fmf{wiggly,tension=0}{v1,o1}
    \fmf{scalar,label={\Large $\phi_{(1),,(2)}$}}{v1,v2}
    \fmflabel{\Large ${\bar{\chiDM}}$}{o1}
    \fmflabel{\Large ${\chiDM}$}{o2}
    \fmfdot{v1,v2,visr}
    \fmf{gluon,tension=0}{visr,isr}
    \fmflabel{\Large ${g}$}{isr}
    \fmfdot{v1,v2,visr}
  \end{feynmandiagram}\quad
  \begin{feynmandiagram}[modelTmonojetAbar]
    \fmfleft{i1,i2}
    \fmfright{o1,o2}
    \fmfbottom{isr}
    \fmflabel{\Large ${\bar{q}}$}{i1}
    \fmflabel{\Large ${q}$}{i2}
    \fmf{fermion}{i2,v2}
    \fmf{phantom}{v2,o2}
    \fmf{phantom,tension=0}{pisr,isr}
    \fmf{fermion,tension=0}{v1,o1}
    \fmf{wiggly,tension=0}{v2,o2}
    \fmf{fermion}{i1,visr,v1}
    \fmf{fermion}{v1,pisr,o1}
    \fmf{wiggly,tension=0}{v2,o2}
    \fmf{scalar,label={\Large $\phi_{(1),,(2)}$}}{v1,v2}
    \fmflabel{\Large ${\bar{\chiDM}}$}{o1}
    \fmflabel{\Large ${\chiDM}$}{o2}
    \fmfdot{v1,v2,visr}
    \fmf{gluon,tension=0}{visr,isr}
    \fmflabel{\Large ${g}$}{isr}
    \fmfdot{v1,v2,visr}
  \end{feynmandiagram}\\\vspace{3\baselineskip}
  \begin{feynmandiagram}[modelTmonojetC]
    \fmfleft{i1,i2}
    \fmfright{o1,o2,o3}
    \fmf{fermion}{i2,v3}
    \fmf{fermion,tension=0}{v3,o3}
    \fmf{wiggly}{v3,o3}
    \fmf{gluon}{i1,v1}
    \fmf{fermion}{v2,v1,o1}
    \fmf{scalar,label={\Large $\phi_{(1),,(2)}$}}{v3,v2}
    \fmf{fermion,tension=0}{v2,o2}
    \fmf{wiggly}{v2,o2}
    \fmflabel{\Large ${q}$}{i2}
    \fmflabel{\Large ${g}$}{i1}
    \fmflabel{\Large $\chiDM$}{o3}
    \fmflabel{\Large $q$}{o1}
    \fmflabel{\Large $\bar{\chiDM}$}{o2}
    \fmfdot{v1,v2,v3}
  \end{feynmandiagram}\quad
  \begin{feynmandiagram}[modelTmonojetB]
    \fmfleft{i1,i2}
    \fmfright{o1,o2,o3}
    \fmf{gluon}{i1,v1}
    \fmf{fermion}{i2,v1}
    \fmf{fermion,tension=2}{v1,v2}
    \fmf{scalar,label={\Large $\phi_{(1),,(2)}$}}{v2,v3}
    \fmf{fermion}{v3,o3}
    \fmf{fermion,tension=0}{v2,o1}
    \fmf{wiggly}{v2,o1}
    \fmf{fermion,tension=0}{v3,o2}
    \fmf{wiggly}{v3,o2}
    \fmflabel{\Large ${q}$}{i2}
    \fmflabel{\Large ${g}$}{i1}
    \fmflabel{\Large $\chiDM$}{o1}
    \fmflabel{\Large $q$}{o3}
    \fmflabel{\Large $\bar{\chiDM}$}{o2}
    \fmfdot{v1,v2,v3}
  \end{feynmandiagram}\\\vspace{3\baselineskip}
  \begin{feynmandiagram}[modelTmonojetD]
    \fmfleft{i1,i2}
    \fmfright{o1,o2,o3}
    \fmf{gluon}{i1,v1}
    \fmf{fermion}{i2,v2}
    \fmf{scalar,label={\Large $\phi_{(1),,(2)}$}}{v2,v1}
    \fmf{scalar,label={\Large $\phi_{(1),,(2)}$}}{v1,v3}
    \fmf{fermion}{v3,o2}
    \fmf{fermion,tension=0}{v3,o1}
    \fmf{wiggly}{v3,o1}
    \fmf{fermion,tension=0}{v2,o3}
    \fmf{wiggly}{v2,o3}
    \fmflabel{\Large ${q}$}{i2}
    \fmflabel{\Large ${g}$}{i1}
    \fmflabel{\Large $\chiDM$}{o3}
    \fmflabel{\Large $q$}{o2}
    \fmflabel{\Large $\bar{\chiDM}$}{o1}
    \fmfdot{v1,v2,v3}
  \end{feynmandiagram}\quad
  \begin{feynmandiagram}[modelTmonojetE]
    \fmfleft{i1,i2}
    \fmfright{o1,o2,o3}
    \fmf{fermion}{i2,v3,o3}
    \fmf{fermion}{o1,v1,i1}
    \fmf{scalar,label={$\phi_{(1),,(2)}$}}{v1,v2}
    \fmf{scalar,label={$\phi_{(1),,(2)}$}}{v2,v3}
    \fmf{gluon}{v2,o2}
    \fmf{wiggly,tension=0}{v3,o3}
    \fmf{wiggly,tension=0}{v1,o1}
    \fmflabel{\Large ${q}$}{i2}
    \fmflabel{\Large ${\bar{q}}$}{i1}
    \fmflabel{\Large $\chiDM$}{o3}
    \fmflabel{\Large $g$}{o2}
    \fmflabel{\Large $\bar{\chiDM}$}{o1}
    \fmfdot{v1,v2,v3}
  \end{feynmandiagram}
\caption{Leading order mono-jet \tchannel processes, adapted from \cite{Papucci:2014iwa}.}\label{fig:tchannelMonojet}
\end{figure}
\begin{figure}
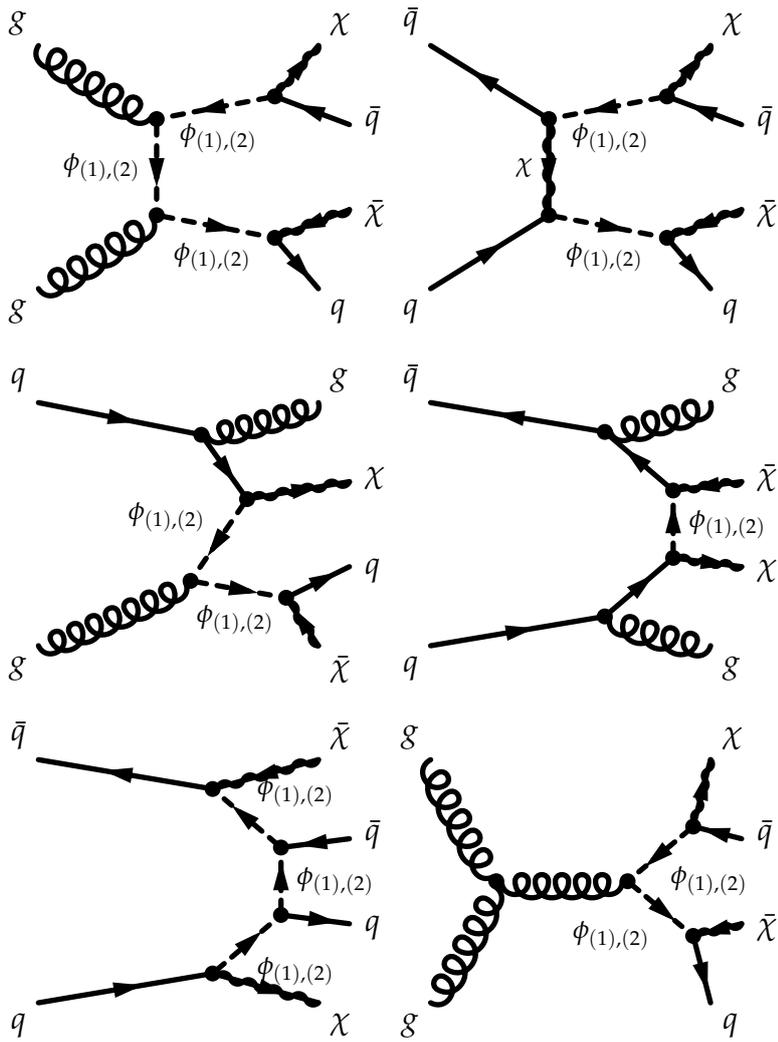

  \unitlength=0.0043\linewidth
  \begin{feynmandiagram}[modelTDijetA]
    \fmfleft{i1,i2}
    \fmfright{o1,o2,o3,o4}
    \fmf{gluon}{i2,v2}
    \fmf{gluon}{i1,v1}
    \fmf{scalar}{v4,v2,v1,v3}
    \fmf{phantom,tension=0,label={$\phi_{(1),,(2)}$}}{v2,v1}
    \fmf{phantom,tension=0,label={$\phi_{(1),,(2)}$}}{v4,v2}
    \fmf{phantom,tension=0,label={$\phi_{(1),,(2)}$}}{v1,v3}
    \fmf{fermion}{o3,v4,o4}
    \fmf{fermion}{o2,v3,o1}
    \fmf{wiggly,tension=0}{v4,o4}
    \fmf{wiggly,tension=0}{v3,o2}
    \fmflabel{\Large ${g}$}{i2}
    \fmflabel{\Large ${g}$}{i1}
    \fmflabel{\Large $\chiDM$}{o4}
    \fmflabel{\Large $\bar{\chiDM}$}{o2}
    \fmflabel{\Large $\bar{q}$}{o3}
    \fmflabel{\Large $q$}{o1}
    \fmfdot{v1,v2,v3,v4}
  \end{feynmandiagram}\quad
  \begin{feynmandiagram}[modelTDijetB]
    \fmfleft{i1,i2}
    \fmfright{o1,o2,o3,o4}
    \fmf{fermion}{v2,i2}
    \fmf{fermion}{i1,v1}
    \fmf{scalar}{v4,v2}
    \fmf{scalar}{v1,v3}
    \fmf{phantom,tension=0,label={$\phi_{(1),,(2)}$}}{v4,v2}
    \fmf{phantom,tension=0,label={$\phi_{(1),,(2)}$}}{v1,v3}
    \fmf{phantom,tension=0,label={$\chiDM$}}{v2,v1}
    \fmf{fermion}{v2,v1}
    \fmf{wiggly,tension=0}{v2,v1}
    \fmf{fermion}{o3,v4,o4}
    \fmf{fermion}{o2,v3,o1}
    \fmf{wiggly,tension=0}{v4,o4}
    \fmf{wiggly,tension=0}{v3,o2}
    \fmflabel{\Large ${\bar{q}}$}{i2}
    \fmflabel{\Large ${q}$}{i1}
    \fmflabel{\Large $\chiDM$}{o4}
    \fmflabel{\Large $\bar{\chiDM}$}{o2}
    \fmflabel{\Large $\bar{q}$}{o3}
    \fmflabel{\Large $q$}{o1}
    \fmfdot{v1,v2,v3,v4}
  \end{feynmandiagram}\\\vspace{3\baselineskip}
  \begin{feynmandiagram}[modelTDijetC]
    \fmfleft{i1,i2}
    \fmfright{o1,o2,o3,o4}
    \fmf{fermion}{i2,v4,v3}
    \fmf{gluon}{v4,o4}
    \fmf{fermion}{v3,o3}
    \fmf{wiggly,tension=0}{v3,o3}
    \fmf{scalar,label={$\phi_{(1),,(2)}$}}{v3,v2,v1}
    \fmf{gluon}{i1,v2}
    \fmf{fermion}{v1,o2}
    \fmf{fermion}{o1,v1}
    \fmf{wiggly,tension=0}{o1,v1}
    \fmflabel{\Large ${q}$}{i2}
    \fmflabel{\Large ${g}$}{i1}
    \fmflabel{\Large $g$}{o4}
    \fmflabel{\Large $\chiDM$}{o3}
    \fmflabel{\Large $q$}{o2}
    \fmflabel{\Large $\bar{\chiDM}$}{o1}
    \fmfdot{v1,v2,v3,v4}
  \end{feynmandiagram}\quad
  \begin{feynmandiagram}[modelTDijetD]
    \fmfleft{i1,i2}
    \fmfright{o1,o2,o3,o4}
    \fmf{fermion}{i1,v1,v2}
    \fmf{fermion}{v3,v4,i2}
    \fmf{gluon}{v4,o4}
    \fmf{gluon}{v1,o1}
    \fmf{scalar,label={$\phi_{(1),,(2)}$}}{v2,v3}
    \fmf{fermion}{o3,v3}
    \fmf{wiggly,tension=0}{o3,v3}
    \fmf{fermion}{v2,o2}
    \fmf{wiggly,tension=0}{v2,o2}
    \fmflabel{\Large ${\bar{q}}$}{i2}
    \fmflabel{\Large $q$}{i1}
    \fmflabel{\Large ${g}$}{o4}
    \fmflabel{\Large ${g}$}{o1}
    \fmflabel{\Large $\chiDM$}{o2}
    \fmflabel{\Large $\bar{\chiDM}$}{o3}
    \fmfdot{v1,v2,v3,v4}
  \end{feynmandiagram}\\\vspace{3\baselineskip}
  \begin{feynmandiagram}[modelTDijetE]
    \fmfleft{i1,i2}
    \fmfright{o1,o2,o3,o4}
    \fmf{fermion}{i1,v1,o1}
    \fmf{fermion}{o4,v4,i2}
    \fmf{scalar,label={$\phi_{(1),,(2)}$}}{v1,v2,v3,v4}
    \fmf{fermion}{v2,o2}
    \fmf{fermion}{o3,v3}
    \fmf{wiggly,tension=0}{v4,o4}
    \fmf{wiggly,tension=0}{v1,o1}
    \fmflabel{\Large ${\bar{q}}$}{i2}
    \fmflabel{\Large $q$}{i1}
    \fmflabel{\Large ${\bar{q}}$}{o3}
    \fmflabel{\Large ${q}$}{o2}
    \fmflabel{\Large $\chiDM$}{o1}
    \fmflabel{\Large $\bar{\chiDM}$}{o4}
    \fmfdot{v1,v2,v3,v4}
  \end{feynmandiagram}\quad
  \begin{feynmandiagram}[modelTDijetGGG]
    \fmfleft{i1,i2}
    \fmfright{o1,o2,o3,o4}
    \fmf{gluon}{i2,v1}
    \fmf{gluon}{i1,v1}
    \fmf{gluon}{v1,v2}
    \fmf{scalar}{v4,v2,v3}
    \fmf{phantom,tension=0,label={$\phi_{(1),,(2)}$}}{v4,v2}
    \fmf{phantom,tension=0,label={$\phi_{(1),,(2)}$}}{v2,v3}
    \fmf{fermion}{o3,v4,o4}
    \fmf{fermion}{o2,v3,o1}
    \fmf{wiggly,tension=0}{v4,o4}
    \fmf{wiggly,tension=0}{v3,o2}
    \fmflabel{\Large ${g}$}{i2}
    \fmflabel{\Large ${g}$}{i1}
    \fmflabel{\Large $\chiDM$}{o4}
    \fmflabel{\Large $\bar{\chiDM}$}{o2}
    \fmflabel{\Large $\bar{q}$}{o3}
    \fmflabel{\Large $q$}{o1}
    \fmfdot{v1,v2,v3,v4}
  \end{feynmandiagram}
\caption{Leading order two-jet \tchannel processes, adapted from \cite{Papucci:2014iwa}.}\label{fig:tchannelDijet}
\end{figure}

The leading-order processes involved in \MET{}+jet production are shown
in Fig. \ref{fig:tchannelMonojet}. 
This model can also give a signal in the \MET + di-jet
channel when, for example, the \chiDM is exchanged in the
\tchannel and the resulting $\phi$ pair each decay to a jet +
\chiDM. Fig.~\ref{fig:tchannelDijet} shows the leading order diagrams.
Except for the $gg$ induced process, di-jet production
through the third-generation mediator $\phi_{(3),u}$ is not possible, 
and production through $\phi_{(3),d}$ is suppressed. 
However, if the coupling $g$ includes a Yukawa coupling proportional to the quark mass, 
and $g$ is sufficiently large, LHC searches will still be sensitive to this model, 
as explained in Section~\ref{sec:singleb}.

The diagram involving the \tchannel exchange
of $\chiDM$ is strongly dependent upon the Dirac fermion assumption.
For a Majorana fermion, $q\bar q,\bar q\bar q,$ and $qq$ production
would be possible with the latter having a pronounced enhancement
at the LHC.

This model is similar to the simplified model considered in SUSY searches, 
implemented as the MSSM with only light squarks and
a neutralino, except for two distinct points:  the $\chiDM$ is
a Dirac fermion and the coupling $g$ is not limited to be
weak scale ($g\ll 1$).
In the MSSM, most of these processes are sub-dominant, even
if resonantly enhanced, because the production is proportional
to weak couplings.
In the more general theories
considered here, $g$ is free to take on large values of order 1 or
more, and thus diagrams neglected in MSSM simulation can occur at a
much higher rate here. While constraints from SUSY jets+\MET analyses
on MSSM models can be recast to apply to the specific model in this report, 
DM searches should also directly test their sensitivity to the MSSM benchmark models.

The state of the art calculation for these models is LO and they can be interfaced with a parton shower program.
The studies in this Section use a LO model implementation within \madgraph v2.2.3, but no parton shower could be employed
in the time-frame of the conclusions of this Forum. Further implementation details can be found in Section~\ref{sec:tchannel_implementation}.

\subsection{Parameter scan}

As for the \schannel models, we adopt the simplifying assumption that the mediator masses and 
couplings are equal for each flavor and handedness. 
The free parameters are then

\begin{equation}
\{ \mDM,\, \Mphi,\, g\}.
\end{equation}

Ref.~\cite{Papucci:2014iwa} studies the parameter space and obtains
bounds on this model from LHC Run-1 mono-jet and dijets+\MET data. 
The Forum did not exhaustively compare the kinematic distributions of the \tchannel models 
as done in the \schannel case. 
In particular, the absence of a parton shower simulation can affect some of the conclusions on
the points and sensitivity chosen. 
While this means the conclusions on the parameter scan below should be taken with more caution, 
the model is plausible and distinctive, and it should be included in the design of early Run-2 LHC searches.

\begin{figure}
	\centering
	\subfloat[\MET{} distribution for a 200 GeV \tchannel mediator, when varying the couplings.]{
		\includegraphics[width=0.8\textwidth]{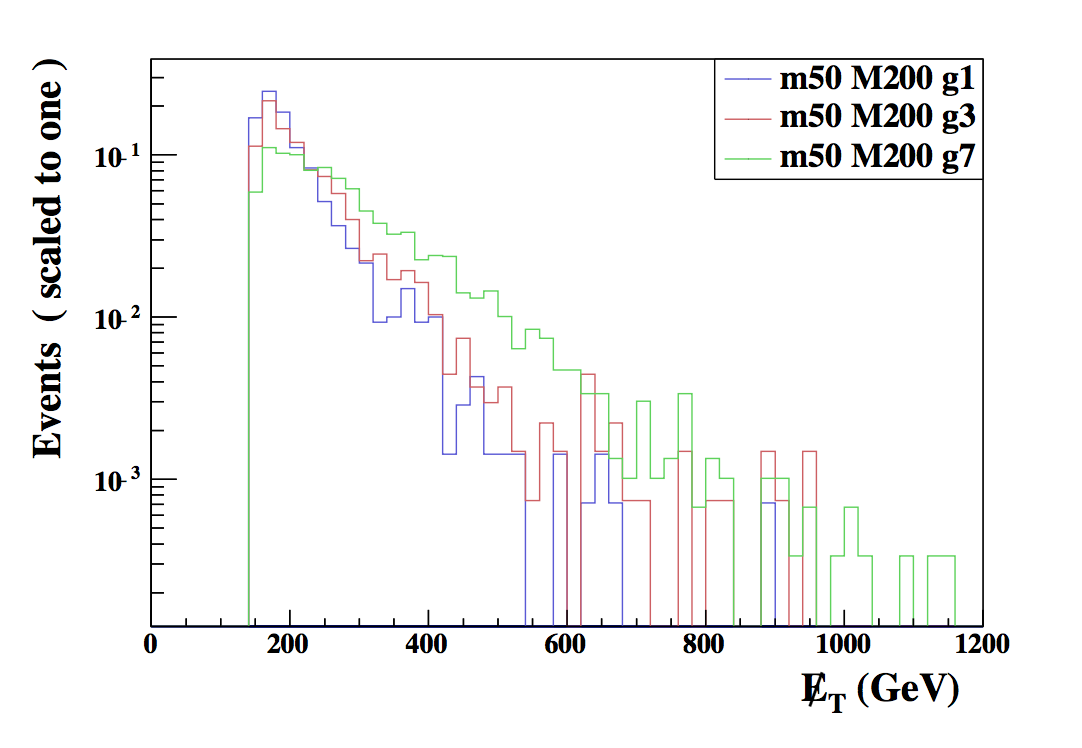}
		\label{fig:monojet_tchannel_coupling}}
	\hfill
	\subfloat[Leading jet \pT distribution for a 2 TeV \tchannel mediator with small ($g$=0.5) to large ($g$=7) couplings with a DM mass of 1 GeV]{
		\includegraphics[width=0.8\textwidth]{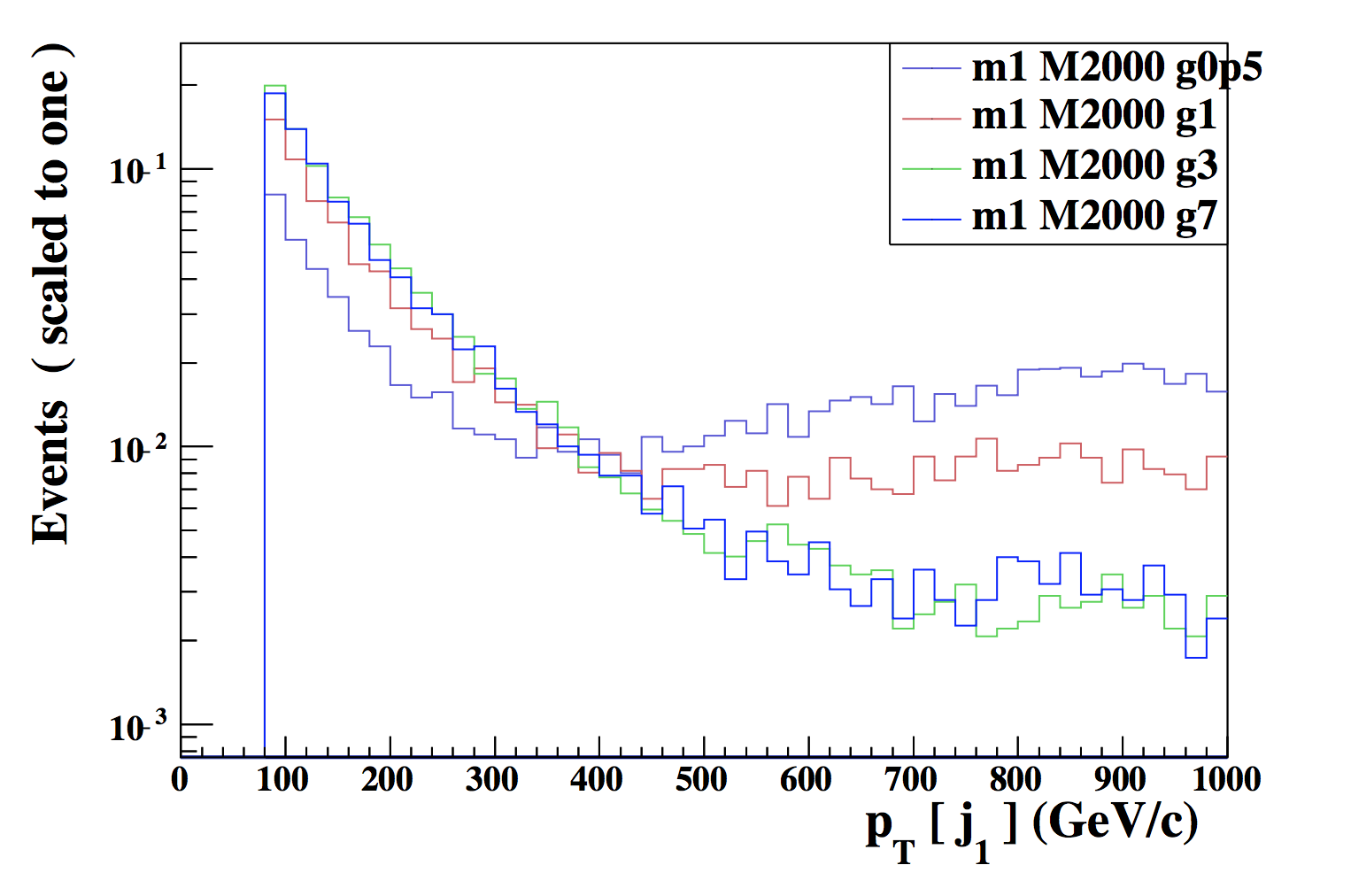}
		\label{fig:monojet_tchannel_edge}}
	\caption[][14pt]{Kinematic distributions normalized to unit area from the \tchannel model from Ref.~\cite{Bell:2012rg}, using MadAnalysis \cite{Conte:2012fm,Conte:2014zja} and simplified
		analysis cuts on the leading jet \pT$>$150 GeV and $\eta <$ 2.8, \MET{}$>$150 GeV. For these models, a LO UFO model is interfaced to \madgraph{} v2.2.3, and studies are at parton-level only. }
	\label{fig:monojet_tchannel}
\end{figure}

As in the \schannel models, scans should be performed over
$\mDM$ and $\Mphi$. The viable ranges of both parameters nearly
coincide with the scan proposed for the \schannel. For the early Run-2 searches, we
recommend to generate and fully simulate a subset of the \schannel mono-jet grid that accounts
for the on-shell and off-shell regions.
In contrast to the \schannel case, the
bounds one obtains from \MET{}+X searches depend strongly on the width
of the mediator, as is visible in Figs.~5 and 6 of
Ref.~\cite{Papucci:2014iwa} and in Fig.~\ref{fig:monojet_tchannel} (a), 
except in the heavy mediator limit ($\Mphi \approx 2$\,TeV). This figure has been obtained applying
a simplified analysis selection (cuts on the leading jet \pT$>$150 GeV and $\eta <$ 2.8, \MET{}$>$150 GeV.) using MadAnalysis~\cite{Conte:2014zja,Dumont:2014tja}.
Figure~\ref{fig:monojet_tchannel} (b) also shows that, if the DM mass is low and the mediator is produced on-shell and its width is narrow,
the cross-section is dominated by $qg \rightarrow q\chi\chi$ diagram. The mediator energy is then
split evenly between the light DM particles and the quark, leading to a broad enhancement at \mMed/2. 

Points with distinct kinematic distributions for a preliminary scan in \{ \mDM,\, \Mphi,\, g\} are selected 
taking into account the expected sensitivity of Run-2 searches, and requiring at least 100 events
to pass the kinematic cuts outlined for Fig.~\ref{fig:monojet_tchannel} in 25 \invfb of collected data, and respect $\Gamma/\mMed < 1$.
They are outlined in Table~\ref{tab:tchannel_parameter_scan}. The conclusions in this table may change
when a parton shower is employed together with multiparton matching.

\begin{table}[!h]
	\centering
	\resizebox{\textwidth}{!}{
		\begin{tabular}{| l |r r r r r r r |c|}
			\hline
			\multicolumn{1}{|c|}{\mDM/\gev} & \multicolumn{7}{c|}{\mmed/\gev} & \multicolumn{1}{c|} {couplings} \\
			\hline
			1             &         10  &  50 & 100 & 300 &     &        &          & 0.1,  1, 3, 7 \\
			1             &             &     &     &     & 500 & 1000   &          & 0.25, 1, 3, 7 \\
			1             &             &     &     &     &     &        &  2000    &    1, 3, 7\\
			\hline
			50            &             &  55 &     &     &     &        &          & 0.1, 1, 3, 4$\pi$ \\
			50            &             &     & 200 & 300 &     &        &          & 0.1, 1, 3, 7      \\
			\hline
			500           &             &     &     &     & 550 &        &          & 1, 3 \\
			500           &             &     &     &     &     &  1000  &          & 0.25, 1, 3 \\
			500          &              &     &     &     &     &        &  2000    & 3           \\
			\hline
			1000           &             &     &     &    &     &  1100  &          & 3, 4$\pi$ \\
			1000           &             &     &     &    &     &        & 2000     & 4$\pi$ \\
			\hline
		\end{tabular}
	}
	\caption{Simplified model benchmark points for \tchannel simplified model (\spinzero mediators 
		coupling to Dirac DM fermions, taking the minimum width.)}.	
	\label{tab:tchannel_parameter_scan}
\end{table}

\subsection{Additional considerations for $V+\MET$ signatures}
\label{sub:EW_TChannel}

The models and parameters with emission of an EW boson 
generally follow those in Section~\ref{sec:monojet_t_channel}.
even though different diagrams are involved.  
A representative Feynman diagram can be
constructed by replacing a final-state gluon in Fig.~\ref{fig:tchannelMonojet}
with a $\gamma,W,Z$ boson, but radiation of electroweak bosons directly from the mediator 
also leads to a mono-boson signature. 

The models considered in Section~\ref{sec:monojet_t_channel}
present a relevant difference concerning final states with an electroweak boson.
In the model in~\cite{Bell:2012rg}, both right- and left-handed mediators can radiate a Z boson,
while only the left-handed mediator in~\cite{Bell:2012rg} allows for W and Z radiation.

\begin{figure}[h!]
	\centering  
	\includegraphics[width=0.95\textwidth]{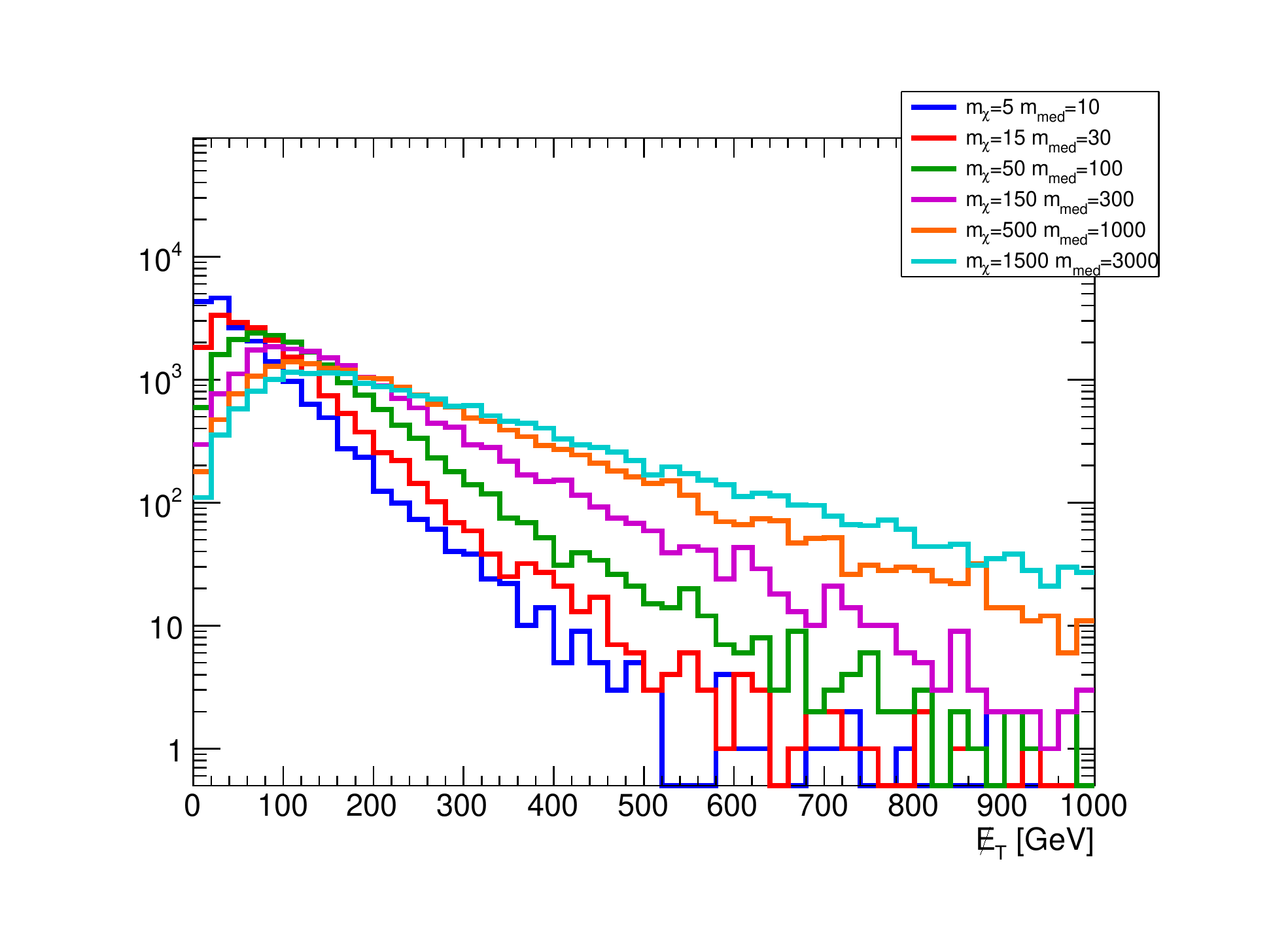}
	\caption{Missing transverse momentum distribution for the hadronic Z+\MET final state,
		for the simplified model with a colored scalar mediator exchanged in the \tchannel.}
	\label{fig:TChan_EW_Zhad_MET}
\end{figure}

\begin{figure}[h!]
	\centering  
	\includegraphics[width=0.95\textwidth]{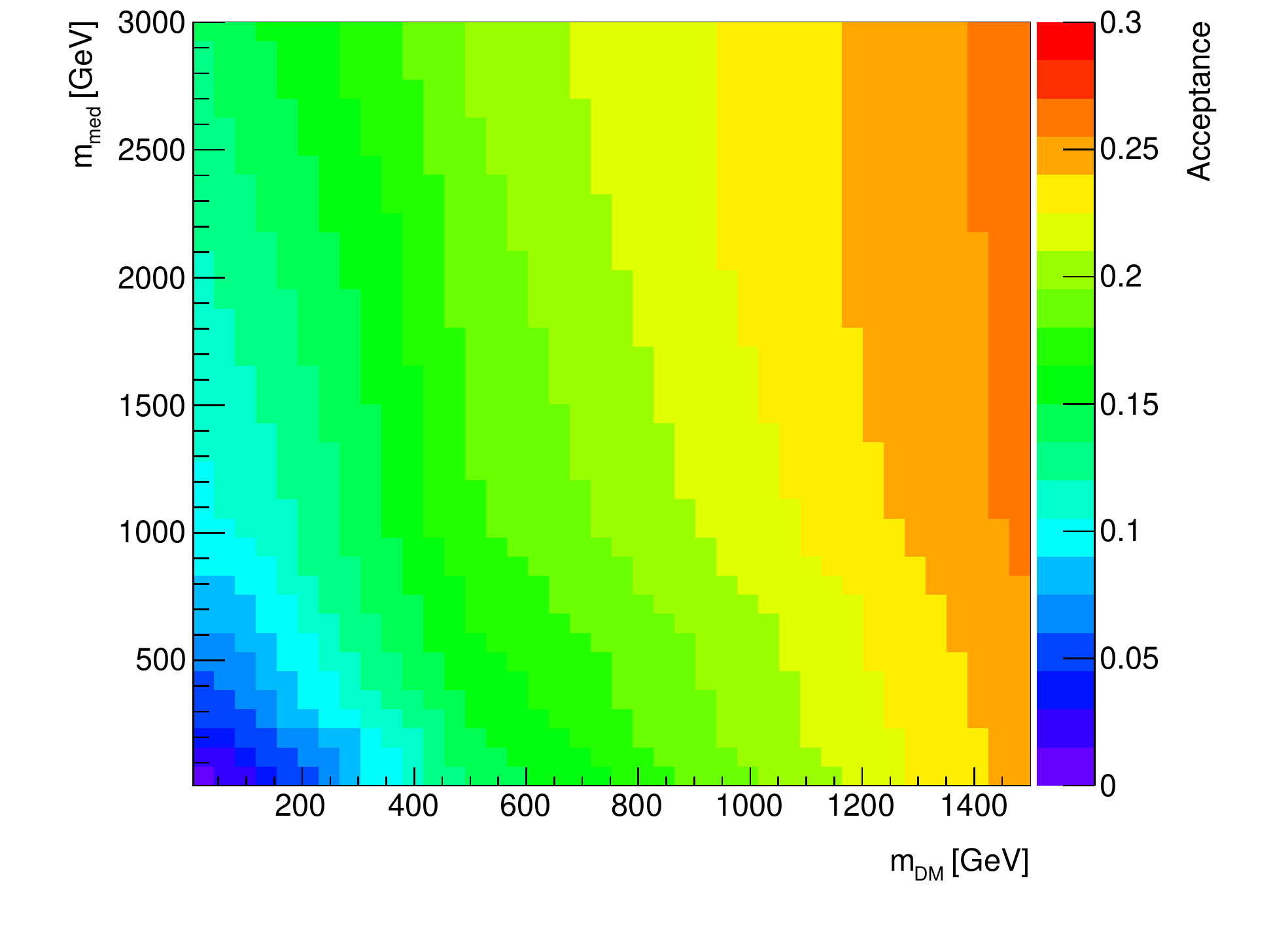}
	\caption{Acceptance for the hadronic Z+\MET final state,
		for the simplified model with a colored scalar mediator exchanged in the \tchannel.}
	\label{fig:TChan_EW_Zhad_acc}
\end{figure}

The studies in this Section use the LO+PS UFO model from~\cite{Bell:2012rg} in \madgraph v2.2.3, 
using \pythia 8 for the parton shower.
Figure~\ref{fig:TChan_EW_Zhad_MET} shows the \MET distribution for the hadronic Z+\MET final state, 
with varying DM and mediator mass, before any selection. 
The acceptance for a series of basic analysis selections
(\MET$>$350~\gev, leading jet $p_T >$ 40~\gev, minimum azimuthal angle between jet and \MET > 0.4) 
applied at the generator level is shown in Figure~\ref{fig:TChan_EW_Zhad_acc}. 

The discussion of the parameter scan for the \tchannel model
in the case of signatures including EW bosons
parallels that of the monojet case for mediator and DM masses,
but no kinematic dependence on the width is observed, so a coupling scan is not needed. 


%
%

 \subsection{\texorpdfstring{Additional considerations for signatures with $b-$quarks + \MET}{Additional considerations for signatures  with b-quarks + \MET}}
 \label{sec:singleb}
 Models of bottom-flavored Dark Matter that are closely related to the \tchannel mediated model from this 
Section have been proposed in Refs.~\cite{Lin:2013sca,Agrawal:2014una}. 
We describe the $b$-FDM model of Ref.~\cite{Agrawal:2014una}, created to explain the Galactic Center (GC) 
gamma-ray excess observed in data collected by the Fermi-LAT collaboration~\cite{Daylan:2014rsa,Calore:2014xka}. 
This model favors couplings to third-generation quarks via Yukawa couplings, 
therefore respecting the MFV assumption. 


The model contains a Dirac fermion transforming as a flavor triplet, exclusively coupling
to right-handed down-type quarks. The third component of the triplet $\chi_b$ comprises the 
cosmological DM. Within the MFV framework, the other fermions in the flavor triplet can be 
made sufficiently heavy and weakly-coupled that they can be neglected in the analysis.
A flavor singlet, color triplet scalar field $\Phi$ mediates the interactions between the DM 
and the Standard Model quarks.  The model is similar to the MSSM with a light bottom squark and neutralino, 
and is thus a flavor-specific example of a \tchannel model. 
Similar top-flavored models can exist, as e.g. in Refs.~\cite{Kumar:2013hfa,Batell:2013zwa}. 
In the case where the top coupling is the main DM coupling, 
the signal is very similar to a signal from a stop quark, since unlike the other t-channel cases there is no top
in the initial state parton distribution functions (PDFs). This is the reason why it wasn't considered as an additional model. More recent
literature shows that other flavor states could also contribute to LHC signals, as shown in Ref.~\cite{Kilic:2015vka},
but such models will have to be investigate on a longer timescale with respect to that of this Forum. 

The Lagrangian considered is given by
\begin{equation}
  - {\mathcal{L}} \supset g \Phi^* \bar\chi_b b_R  + {\rm h.c.}
\end{equation}


This model is known at LO+PS accuracy, and the studies in this Section use a LO model implementation within \madgraph v2.2.3
interfaced to \pythiaEight for the parton shower. Further implementation details can be found in Section~\ref{sec:TTBar_implementation}.
 
\subsubsection{Parameter scan}

In this model, the interference of diagrams with QCD production of the mediator (which scale as $g^2_s$) with diagrams that are proportional to the coupling $g$ in the $b+$\MET{} and $b\bar{b}$+\MET{} final states. In the case of large couplings, this is not conducive to a simple scaling behavior that would allow us to reduce the number of points to be simulated. This can be seen in Fig.~\ref{fig:g_comp}.

A full study of the parameter scan for this model was not available for this report; thus for early Run-2 searches we recommend scanning a range of possible widths as discussed in a more limited way than for the \tchannel mono-jet, spanning from the minimal width to a value approaching the particle limit, e.g. $g=0.5,1,2,3$. A coupling benchmark such as $g=1$ should be considered for each mass point since this would be a distinctive feature of this benchmark from SUSY models with sbottom squarks (see Section~\ref{sec:monojet_t_channel} for further discussion).

\begin{figure*}[h!]
  \includegraphics[width=0.49\textwidth]{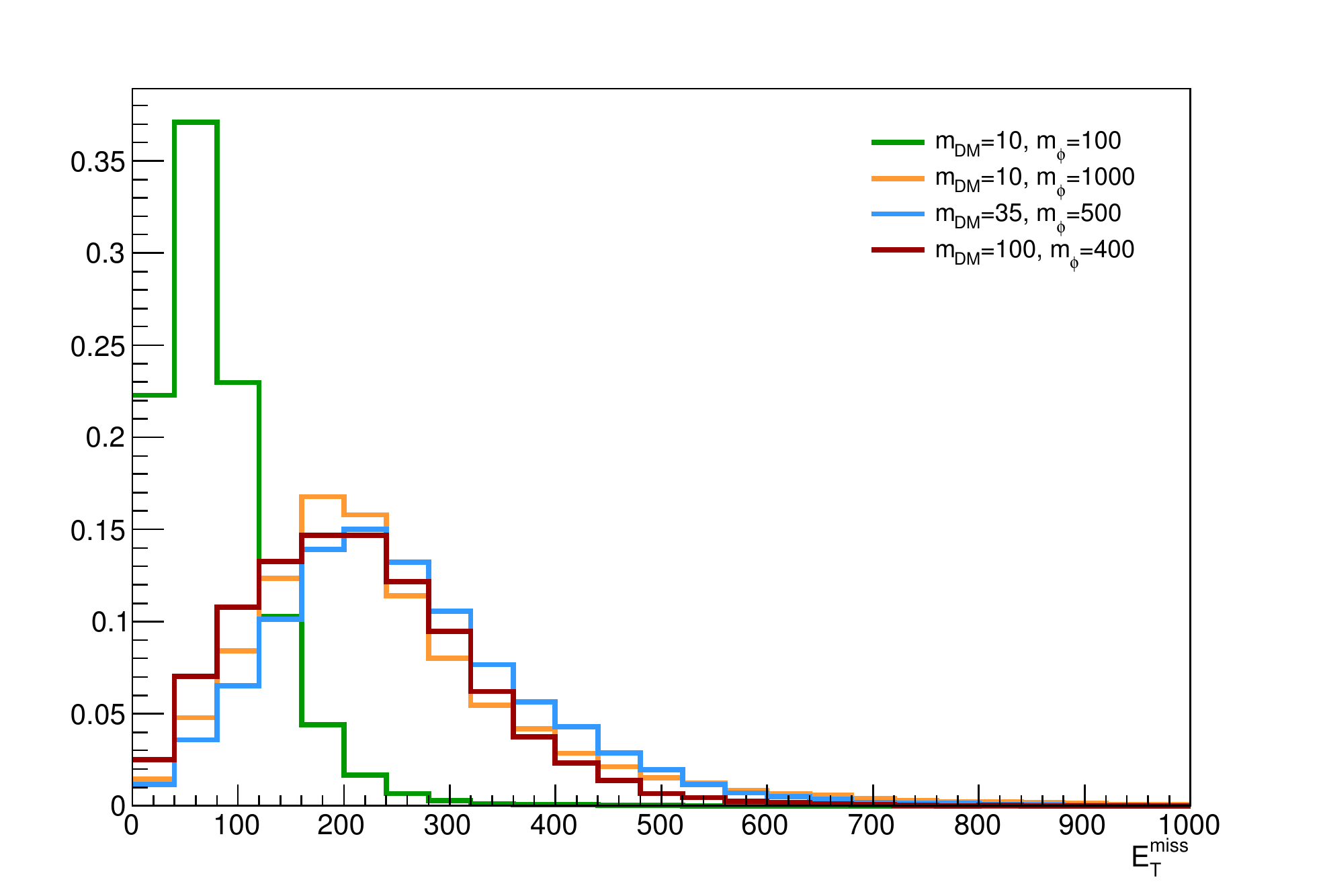}\quad
  \includegraphics[width=0.49\textwidth]{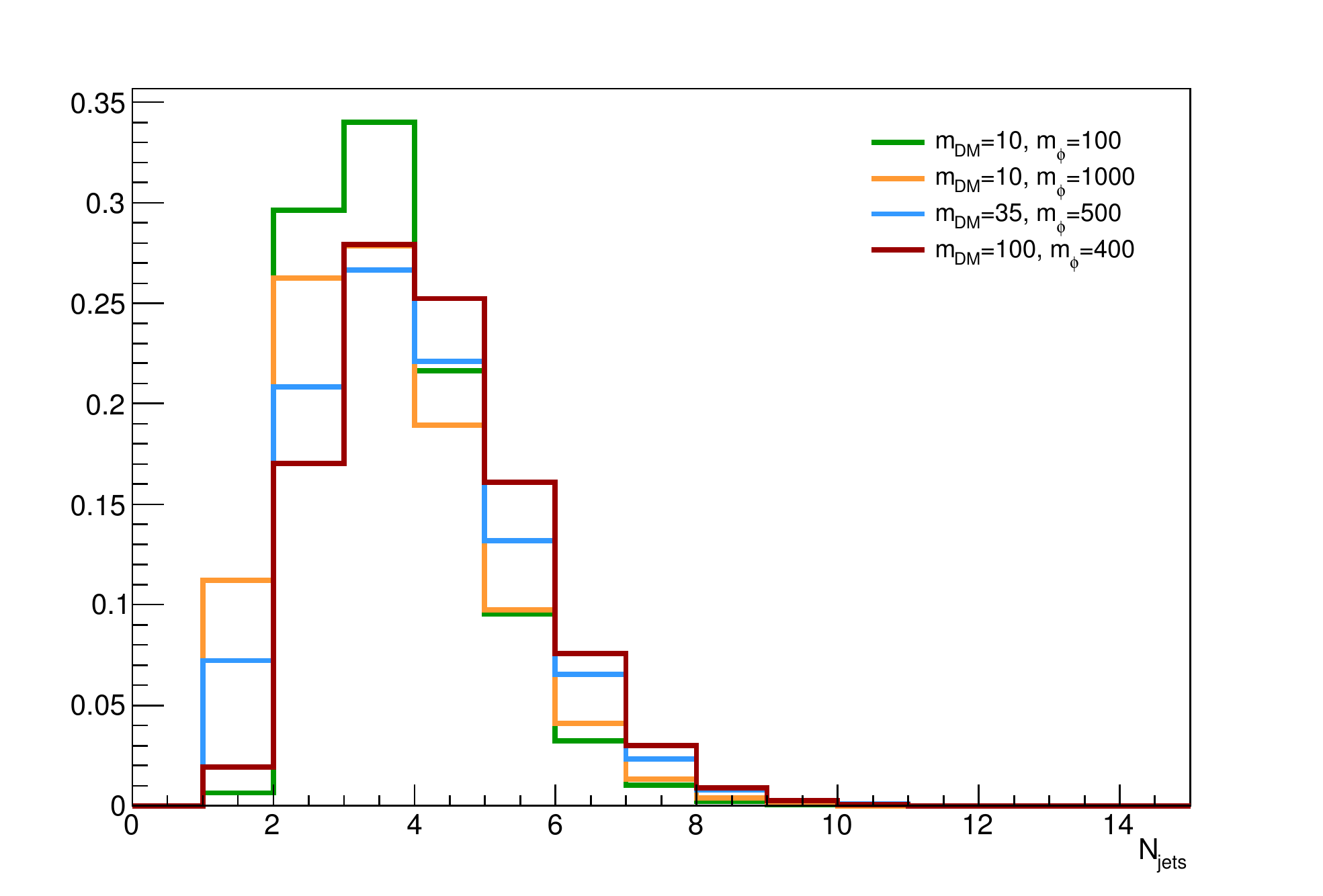}
  \caption{\MET (left) and jet multiplicity (right) for various DM and mediator masses and couplings normalized to the relic density observed in the early universe. Studies in this section use a LO UFO model implementation within \madgraph v2.2.3
  	interfaced to \pythiaEight for the parton shower.}\label{fig:relic}
\end{figure*}

\begin{figure*}[h!]
  \includegraphics[width=0.49\textwidth]{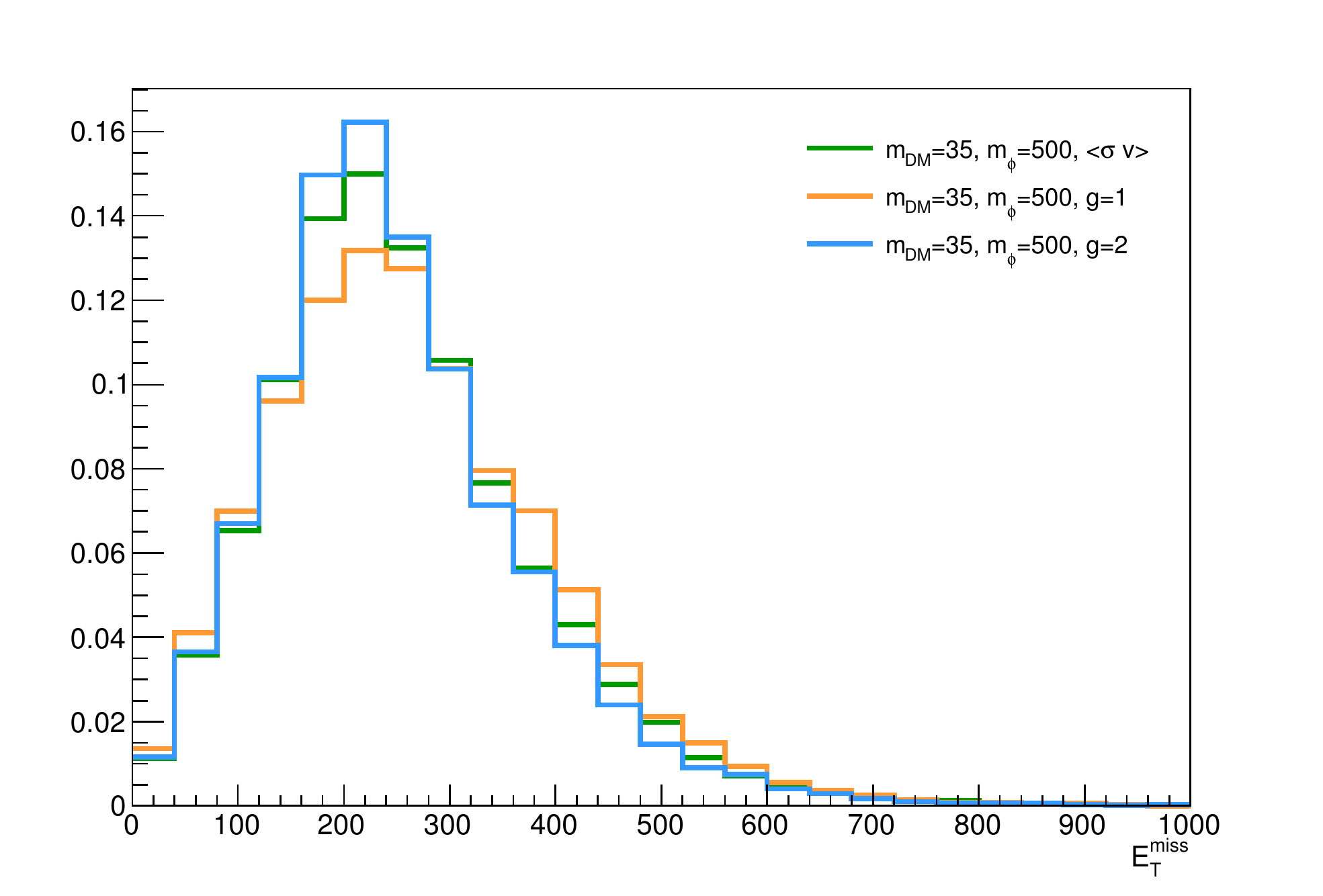}\quad
  \includegraphics[width=0.49\textwidth]{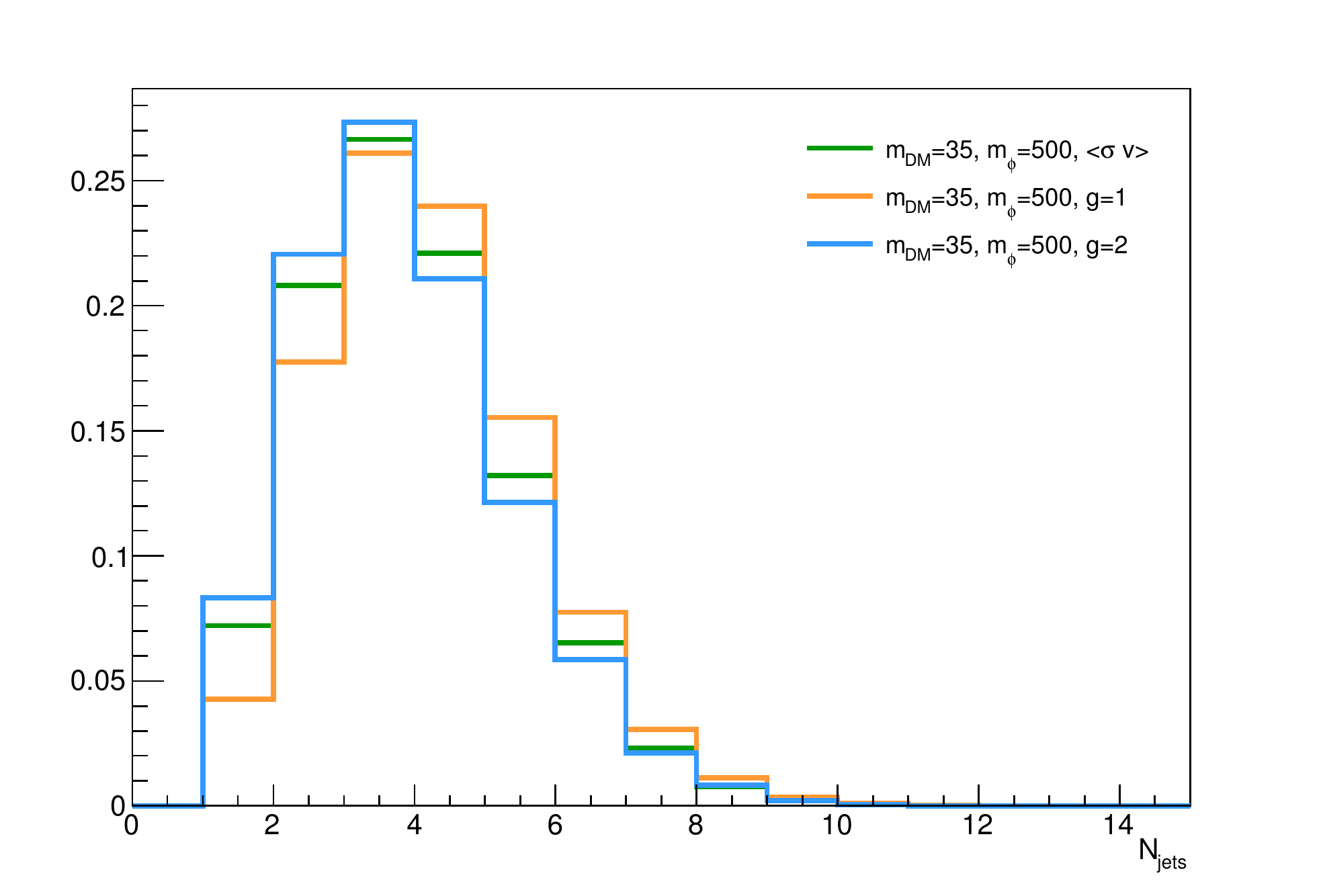}
  \caption{\MET (left) and jet multiplicity (right) for $\mDM=35$~GeV and $\MPhi=500$~GeV for varying couplings of $g=1,2$} \label{fig:g_comp}
\end{figure*}

A scan of Dark Matter and mediator masses should be done in the on-shell region $\MPhi > \mDM + m_b$, since the cross-sections in the off-shell region are too small to be probed with early LHC data, spanning from 10 to 500 GeV in \mDM and from 10 to 1300 GeV in \MPhi. Examples of the kinematic distributions produced by this model are shown in Fig.~\ref{fig:relic}~\footnote{Following the grounding assumptions in this report, the normalization to the relic density is considered only in these example plots rather than as a necessary ingredient for the parameter scan of this model.}.



\section{ \Spintwo mediator}
\label{sec:spintwo}

In models with extra dimensions, the Kaluza-Klein excitations of the graviton could also serve as a mediator between the Standard Model and dark sector physics. This kind of model was not studied in the forum and is not included in the recommendations, but models such as Ref.~\cite{Lee:2013bua,Lee:2014caa} may warrant further study on a longer timescale. 




\section{Presentation of results for reinterpretation of \schannel mediator models}
\label{sec:monojet_scaling}

The aim of the parameter grid optimization done for the \schannel models in the previous sections is to reduce the parameter space that must be simulated.  We then need a procedure for populating the full parameter space by using the simulated grid points.  We recommend doing this as follows:

\begin{itemize}
\item When the dependences on parameters are known, the cross
  sections and efficiencies at general points can be calculated from
  the grid data.
\item In other cases, this information can be obtained by interpolation
  between the grid points.  We have chosen the grid points so that the
  dependence is sufficiently smooth that this will be possible.
\end{itemize}

The results of the scan over the couplings presented in the previous sections indicate that there are no changes in kinematic distributions for different choices of the coupling strengths. This means that the acceptance remains the same in the whole $\gq$--$\gDM$ plane and it is sufficient to perform the detector simulation only for one single choice of $\gq, \gDM$. The resulting truth-level selection acceptance and the detector reconstruction efficiency can then be applied to all remaining grid points in the $\gq$--$\gDM$ plane where only the generator-level cross section needs to be known. This significantly reduces the computing time as the detector response is by far the most CPU-intensive part of the Monte Carlo sample production.
However, the number of generated samples can be reduced even further
if a parameterization of the cross section dependence from one grid point to another exists.
In this section, we describe the details of a cross section scaling procedure that
can be used to reinterpret results for a fixed coupling for \schannel mediator models.
The studies in this section employ the \powheg~\cite{Haisch:2015ioa} generator. 

The propagator for the \schannel exchange is written in a Breit-Wigner
form as $\displaystyle \frac{1}{q^2-\mMed^2 + i\mMed\Gamma}$, where $q$ is the momentum transfer calculated from the two partons entering the hard process after the initial state radiation, which is equivalent to the momentum of the Dark Matter pair~\footnote{Using a running width and replacing the denominator of the propagator with $q^2 - \mMed^2 + \complexi\,Q^2\,\frac{\Gamma}{\mMed}$ should be considered in the case of wide mediators~\cite{Bardin:1989qr}.}. 
The size of the momentum transfer with respect to the mediator mass allows us to identify three cases:
\begin{itemize}
	\item off-shell mediator, when $q^2 \gg \mMed^2$ leading to suppressed cross sections,
	\item on-shell mediator, when $q^2 \sim \mMed^2$ leading to enhanced cross sections,
	\item effective field theory (EFT) limit when $q^2 \ll \mMed^2$.
\end{itemize}
In the case of the off-shell mediator and the EFT limit, the first and second term in the propagator dominate, respectively, which reduces the dependence on the mediator width. Therefore, in these cases one can approximate the cross section as

\begin{equation}
\sigma \propto \gq^2\gDM^2.
\end{equation}
The on-shell regime is the most interesting one as it gives the best chances for a discovery at the LHC given the cross section enhancement. The propagator term with the width cannot be neglected in this case and, in the narrow width approximation which requires $\Gamma \ll \mMed$ (this is not necessarily the case in the benchmarks considered in the scans), one can integrate

\begin{equation}
\int \frac{ds}{(s-\mMed^2)^2 + \mMed^2\Gamma^2} = \frac{\pi}{\mMed\Gamma}
\label{eq:monojet_int}
\end{equation}
which further implies the cross section scaling

\begin{equation}
\sigma \propto \frac{\gq^2\gDM^2}{\Gamma}.
\label{eq:monojet_scaling}
\end{equation}
The narrow width approximation is important here as it ensures an integration over parton distribution functions (PDFs) can be neglected. In other words, it is assumed the integrand in Eq.\,\ref{eq:monojet_int} is non-zero only for a small region of $s$, such that the PDFs can be taken to be constant in this range.
By simplifying the dependence of the minimal width on the couplings as $\Gamma \sim \gq^2+\gDM^2$, one can approximate this scaling rule in the extreme cases as follows

\begin{eqnarray}
\sigma &\propto& \frac{\gq^2\gDM^2}{\gq^2+\gDM^2} \xrightarrow{\gq \ll \gDM} \gq^2 \label{eq:monojet_gSM} \\
\sigma &\propto& \frac{\gq^2\gDM^2}{\gq^2+\gDM^2} \xrightarrow{\gq \gg \gDM} \gDM^2 \label{eq:monojet_gDM} \;.
\end{eqnarray}
However, it is important to keep in mind that this formula omits color and multiplicity factors as well as possible Yukawa suppression, and there is no simple scaling rule for how the cross section changes with the Dark Matter mass and the mediator mass, or for mediators with a large width, because PDFs matter in such cases as well.
Therefore, the scaling procedure outlined above is expected to work only for fixed masses and fixed mediator width, assuming the narrow width approximation applies.

\begin{figure*}
	\centering
	\includegraphics[width=0.49\textwidth]{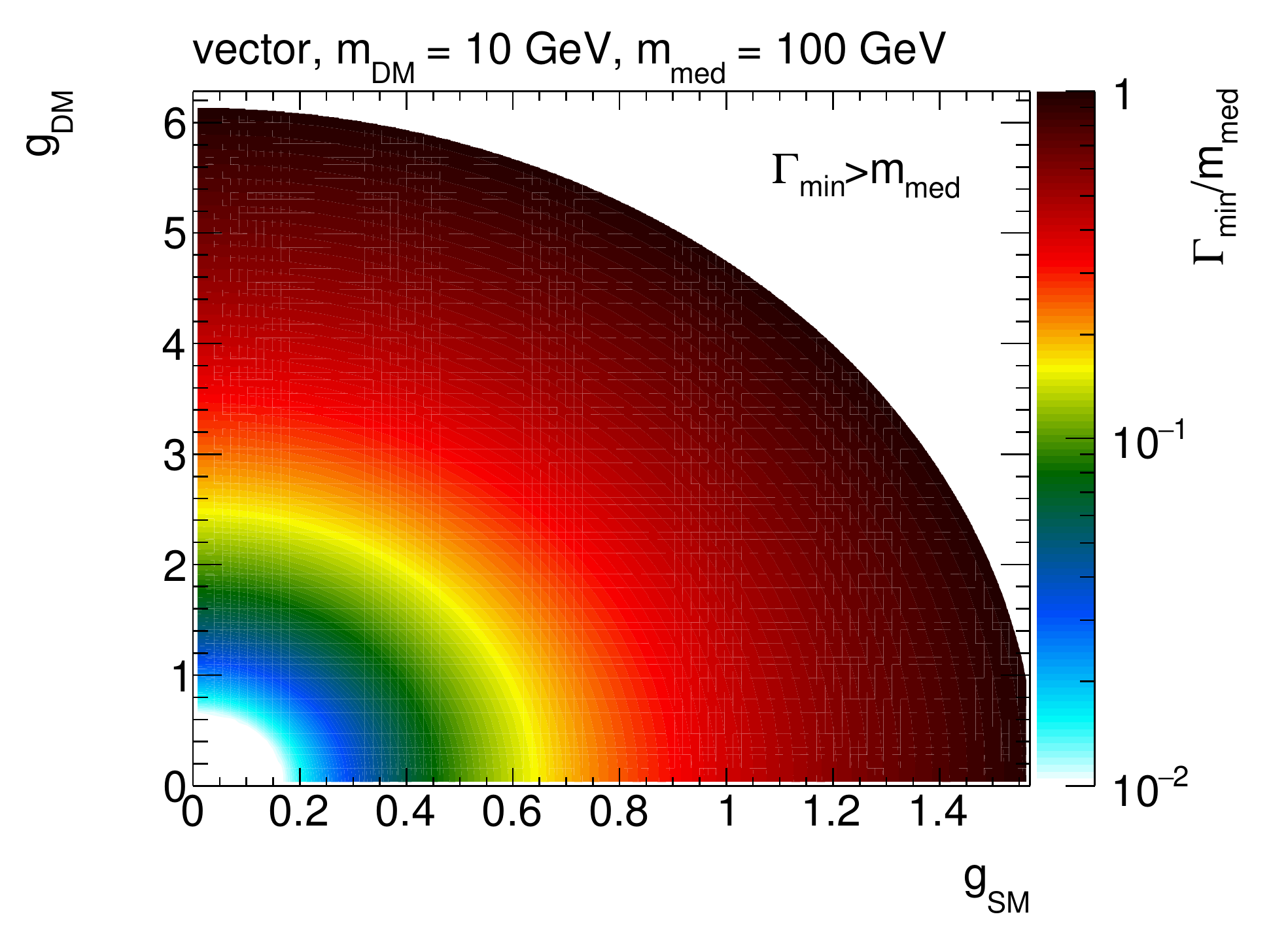}
	\includegraphics[width=0.49\textwidth]{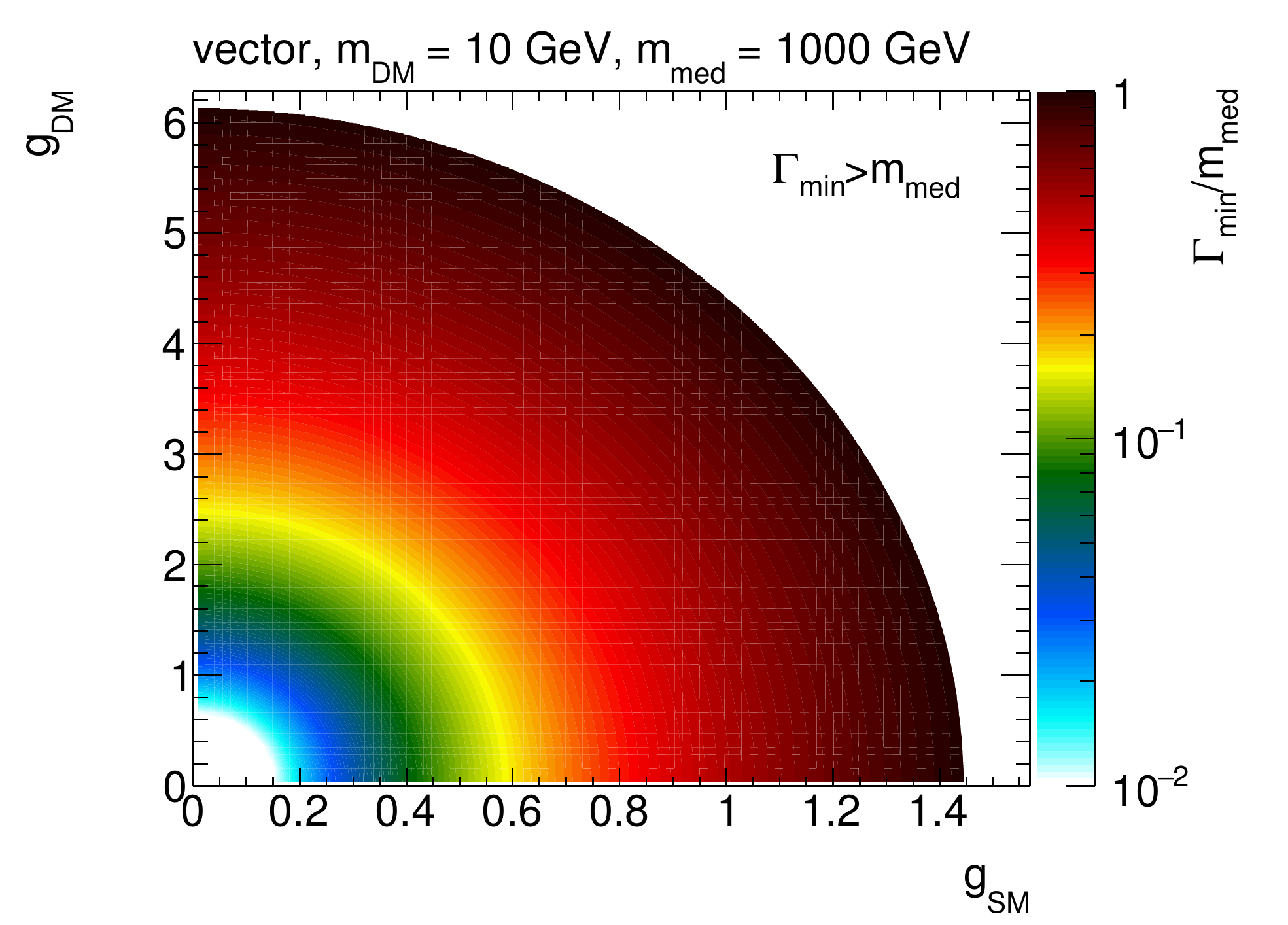}\\
	\includegraphics[width=0.49\textwidth]{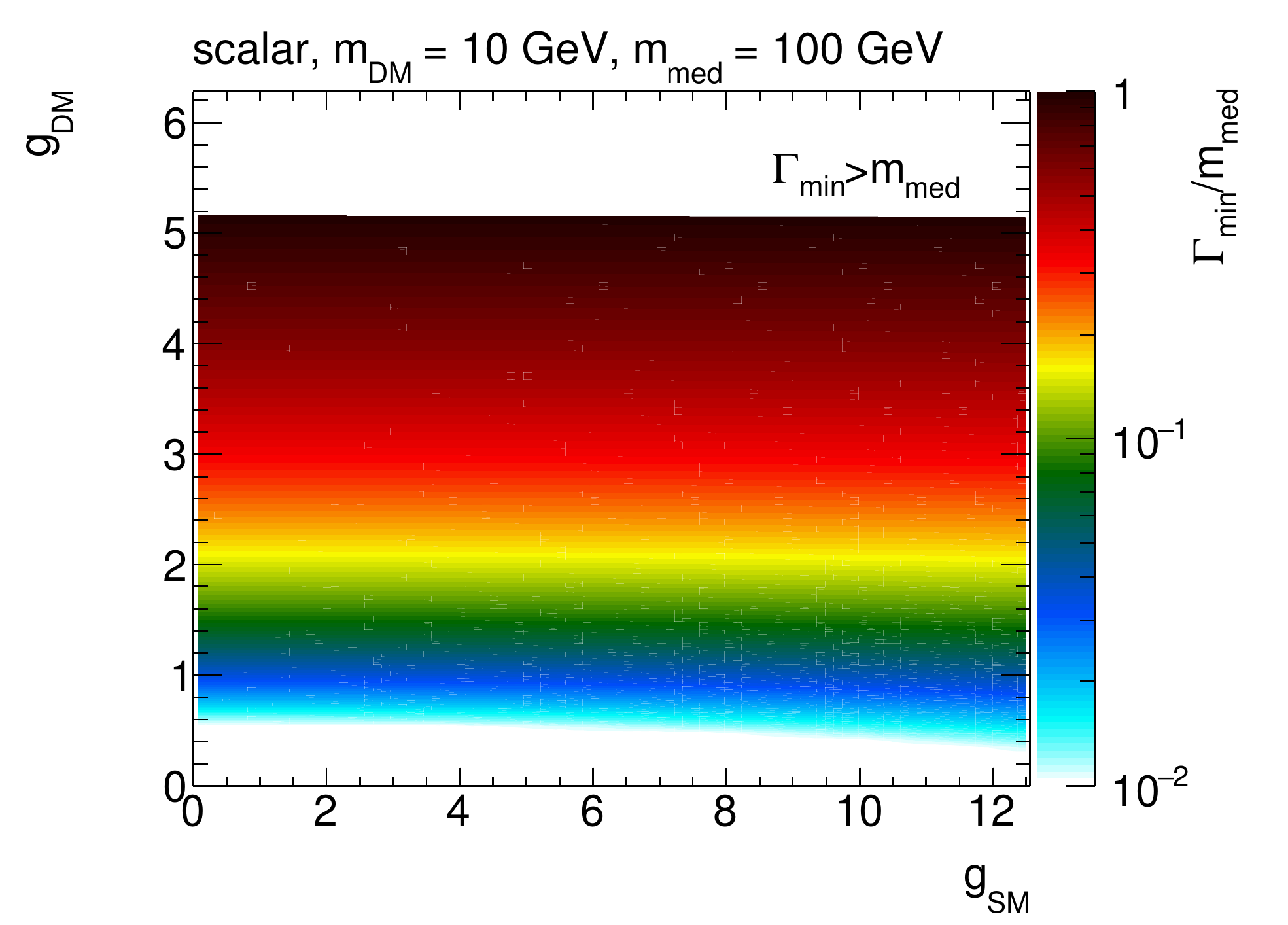}
	\includegraphics[width=0.49\textwidth]{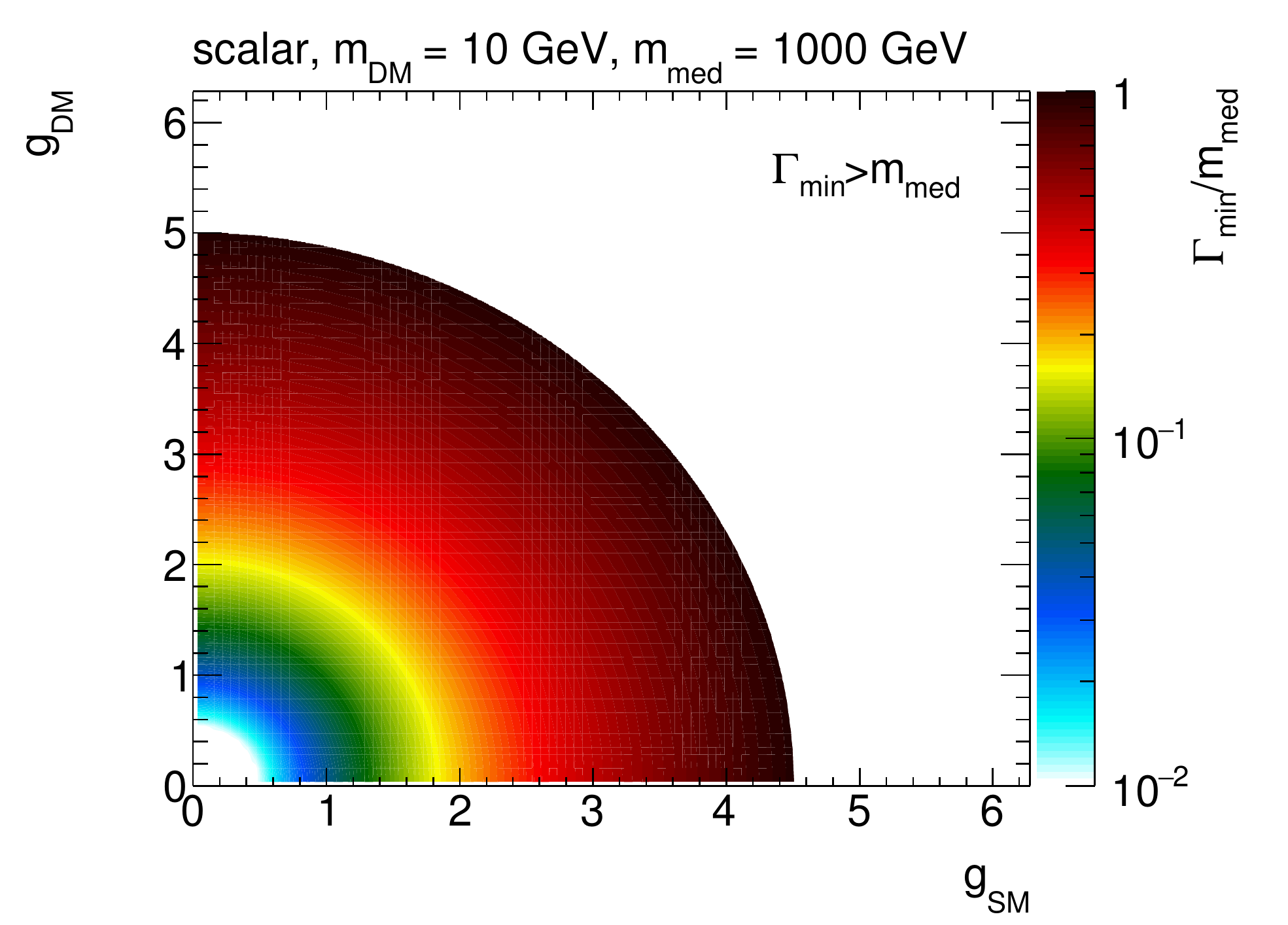}
	\caption{Minimal width over the mediator mass for vector (top) and scalar (bottom) mediators as a function of the individual couplings $\gq$ and $\gDM$, assuming $\mMed=100$~\gev (left) and $\mMed=1$~\tev (right). $\mDM=10$~\gev is considered in all cases.
		Only the cases with $\Gamma_{\rm{min}}<\mMed$ are shown.}
	\label{fig:monojet_width}
\end{figure*}

Figure\,\ref{fig:monojet_width} shows the minimal width over the mediator mass in the $\gq$--$\gDM$ plane for vector and scalar mediators for $\mMed=100$~\gev and 1000~\gev, taking $\mDM=10$~\gev.
The individual colors indicate the lines of constant width, along which the cross section scaling may work for narrow mediators.
The limiting case $\Gamma_{\rm{min}}=\mMed$ defines the upper values of the couplings below which the narrow width approximation can be considered and provides more stringent constraint than the perturbative limit $\gq=\gDM=4\pi$.
For vector and axial-vector mediators, the minimal width is predominantly defined by $\gq$ due to the number of quark flavors and the color factor. 
On the contrary, both the Standard Model and Dark Matter partial width have comparable contributions in case of scalar and pseudo-scalar mediators if the top quark channel is open ($\mMed>2m_t$). However, mostly $\gDM$ defines the minimal width for $\mMed<2m_t$ due to the Yukawa-suppressed light quark couplings.

%

\begin{figure}
	\centering
	\includegraphics[width=0.95\textwidth]{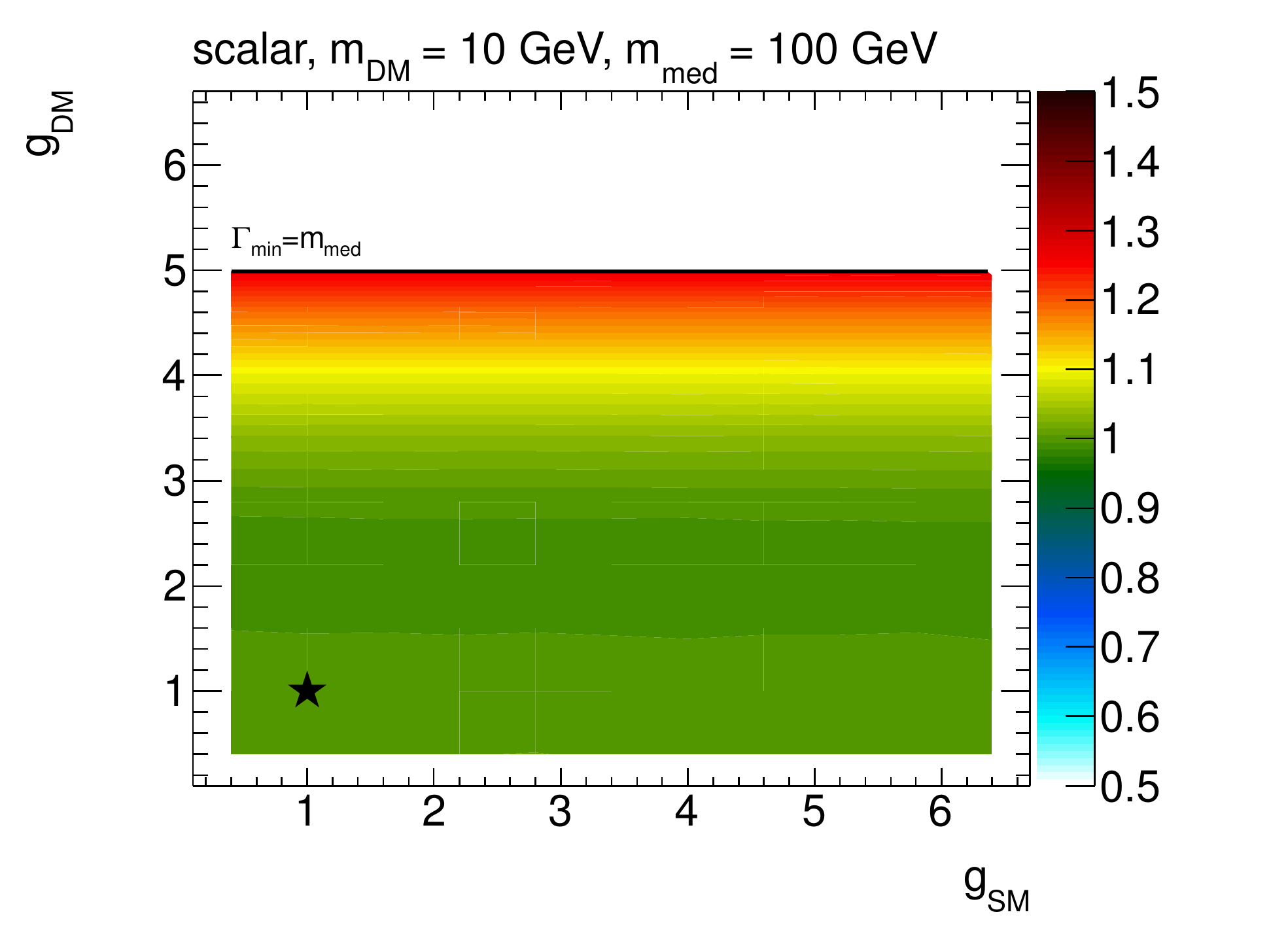}
	\includegraphics[width=0.95\textwidth]{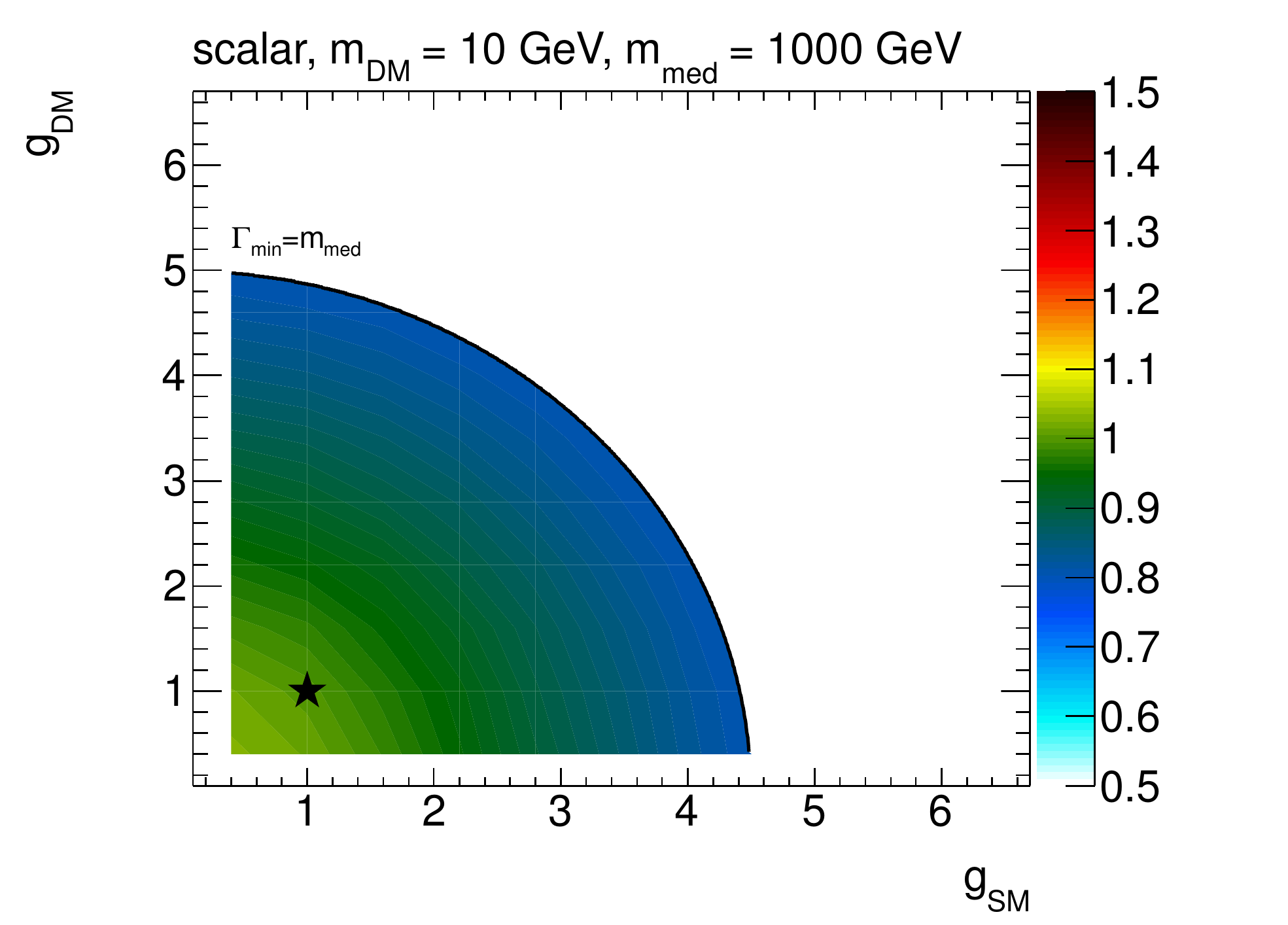}\\
	\caption{Ratio of the rescaled and generated cross sections in the $\gq$--$\gDM$ plane. The point at $\gq=\gDM=1$, taken as a reference for the rescaling, is denoted by a star symbol.
		Scalar model with $\mMed=100$~\gev (left) and 1~\tev (right) is plotted for $\mDM=10$~\gev.
		The limiting case $\Gamma_{\rm{min}}=\mMed$ is indicated by a black line and no results are shown beyond.}
	\label{fig:monojet_scaling}
\end{figure}
\begin{figure*}
	\centering
	\includegraphics[page=1, trim=310 200 310 200, clip, width=0.49\textwidth]{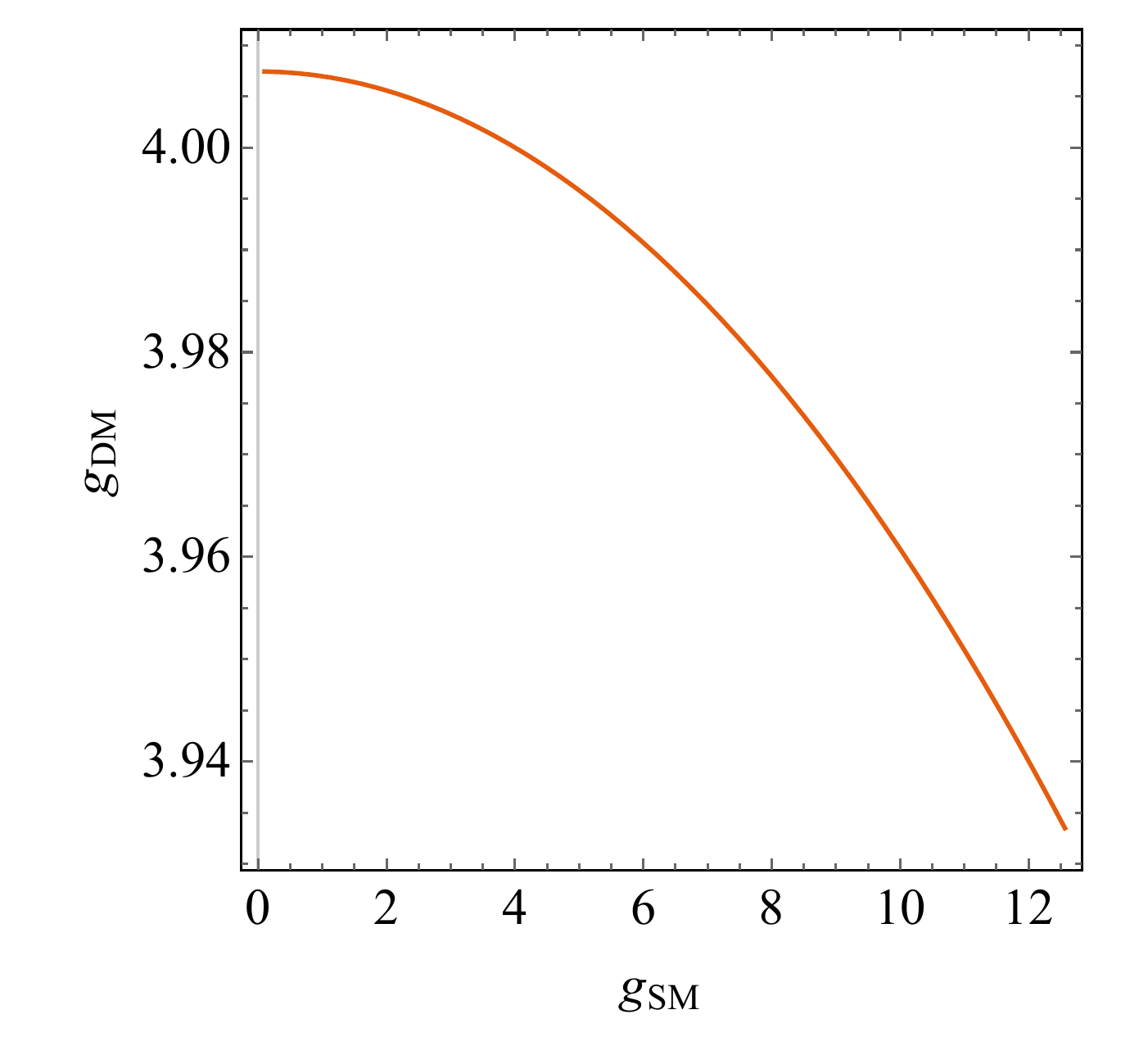}
	\includegraphics[page=2, trim=305 195 305 195, clip, width=0.49\textwidth]{figures/monojet/rescalingexercise.pdf}\\
	\includegraphics[page=3, trim=300 190 300 190, clip, width=0.49\textwidth]{figures/monojet/rescalingexercise.pdf}
	\caption{Scaling along the lines of constant width. The line of constant width for $\mMed=300$~\gev and $\mDM=100$~\gev, intercepting $\gq=\gDM=4$ is shown on left. The generated and rescaled cross sections are compared in the middle, the corresponding ratio is shown on right.}
	\label{fig:monojet_scaling_constwidth}
\end{figure*}

The performance of the cross section scaling is demonstrated in Fig.\,\ref{fig:monojet_scaling} 
where two mass points $\mMed=100$~\gev and 1~\tev with $\mDM=10$~\gev are chosen
and rescaled from the starting point $\gq=\gDM=1$ according to Eq.\,\ref{eq:monojet_scaling} to populate the whole $\gq$--$\gDM$ plane. This means the width is not kept constant in this test and this is done in purpose in order to point out deviations from the scaling when the width is altered. For each mass point, the rescaled cross section is compared to the generator cross section and the ratio of the two is plotted.
For the given choice of the mass points, the scaling seems to work approximately within the precision of $\sim20\%$ in the region where $\Gamma_{\rm{min}}<\mMed$.
Constant colors indicate the lines along which the cross section scaling works precisely and there is a remarkable resemblance of the patterns shown in the plots of the mediator width. To prove the scaling along the lines of constant width works, one such line is chosen in Fig.\,\ref{fig:monojet_scaling_constwidth} for a scalar mediator, defined by $\mMed=300$~\gev, $\mDM=100$~\gev, $\gq=\gDM=1$, and the rescaled and generated cross sections are found to agree within 3\%.

\subsection{Proposed parameter grid for cross-section scaling}

We propose to deliver collider results in the $\gq$--$\gDM$ plane using the following prescription, to ease reinterpretation through cross-section scaling:
\begin{itemize}
	\item Since the shapes of kinematic quantities do not change for different couplings, use the acceptance and efficiency for the available $\mDM=50$~\gev, $\mMed=300$~\gev grid point from the $\mMed$--$\mDM$ plane for the scalar and pseudo-scalar mediator. In case of the vector and axial-vector mediator, use the grid point $\mDM=150$~\gev, $\mMed=1$~\tev.
	\item Generate additional samples in order to get generator cross sections only. For scalar and pseudo-scalar mediator, choose $\mDM=50$~\gev, $\mMed=300$~\gev with the following values for $\gq=\gDM$: 0.1, 1, 2, 3. For vector and axial vector mediator, choose $\mDM=150$~\gev, $\mMed=1$~\tev with the following values for $\gq=\gDM$: 0.1, 0.25, 0.5, 0.75, 1, 1.25, 1.5. The upper values are defined by the minimal width reaching the mediator mass. 
	\item Rescale the generator cross sections for on-shell resonance production along the lines of constant width in order to populate the whole $\gq$--$\gDM$ plane in the region $\Gamma_{\rm{min}}<\mMed$.
The scaling follows from Eq.\,\ref{eq:monojet_scaling} which for the constant width implies:
\begin{equation}
\sigma' = \sigma \times \frac{\gq'^2\gDM'^2}{\gq^2\gDM^2} \;.
\end{equation}
\end{itemize}



\subsection{Rescaling to different mediator width}
\label{paragraph:nonminimalwidth}

In general it is also important to consider a larger mediator width than $\Gamma_{\rm{min}}$ in order to accommodate additional interactions of the mediator with the visible and hidden sector particles~\cite{Buckley:2014fba,Harris:2014hga}. If the narrow width approximation applies, the cross section scaling method described above can be used to reinterpret the results presented for the minimal width, since multiplying the width by factor $n$ is equivalent to changing the coupling strength by factor $\sqrt{n}$, i.e.

\begin{equation}
\sigma(\gq,\gDM, n\Gamma_{\rm{min}}(\gq,\gDM)) \propto \frac{\gq^2 \gDM^2}{\Gamma_{\rm{min}}(\sqrt{n}\gq,\sqrt{n}\gDM)} \;.
\end{equation}
The cross section for the sample with couplings $\gq$ and $\gDM$ and modified mediator width $\Gamma = n\Gamma_{\rm{min}}$ can therefore be rescaled from a sample generated with the minimal width corresponding to the couplings scaled by $\sqrt{n}$ as described in the following formula.

\begin{equation}
\sigma(\gq,\gDM, n\Gamma_{\rm{min}}(\gq,\gDM)) = \frac{1}{n^2} \sigma(\sqrt{n}\gq,\sqrt{n}\gDM,\Gamma_{\rm{min}}(\sqrt{n}\gq,\sqrt{n}\gDM))
\end{equation}
The advantage of doing this is in the fact that no event selection and detector response needs to be simulated since the changes in couplings do not have an effect on the shapes of kinematic distributions.

It should be noted again that this procedure is only useful when the narrow width approximation applies. Care must be taken to ensure that is the case. For example, in the vector and axial-vector cases, one quickly breaks this approximation even for small $n$.

\subsection{\texorpdfstring{Additional considerations for $t \bar{t}$ and $b \bar{b}$+\MET signatures}{Additional considerations for ttbar/bbbar+\MET signatures}}

The cross-section scaling considerations shown in Sec.~\ref{sec:monojet_scaling} still apply for the reactions in the scalar and psuedoscalar models with explicit $b$ and $t$ quarks. Here 
we detail the specific studies done for the  $t\bar t$ model. 

Given that the kinematics are similar for all couplings $g \simeq 1$, we recommend to generate only samples with $\gDM = \gq = 1$. It follows from this that these benchmark points should be a good
approximation for non-unity couplings and for $\gDM \neq \gq$, provided
that the sample is rescaled to the appropriate cross section times
branching ratio.

\begin{figure}[!ht]
	\begin{center}
		\includegraphics[width=0.65\textwidth]{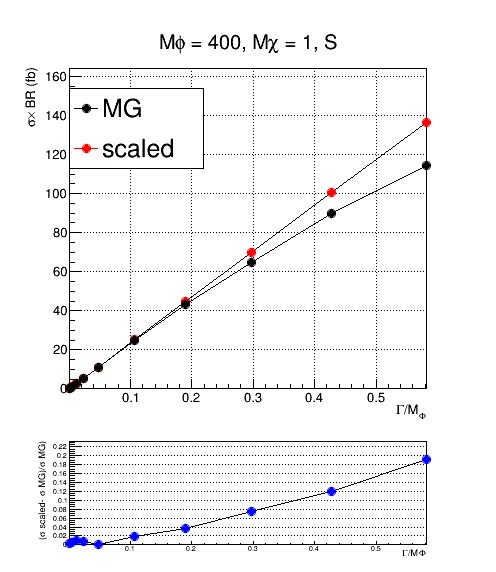}
		\vspace{2mm}
		\caption{\label{fig:xsec_scaling} An example comparing a simple cross section scaling versus the computation from the \madgraph generator, for a scalar $t \bar{t}$+\MET model with $m_{\phi}=400\,{\rm GeV}$, $\mdm=1\,{\rm GeV}$ and all couplings set to unity. In this example, the scaling relationship holds for $\Gamma_{\phi}/m_{\phi}$ below $0.2$, beyond which finite width effects become important and the simple scaling breaks down.}
	\end{center}
\end{figure}

While the simple scaling function
\begin{equation}
\sigma'\times BR' = [\sigma \times BR]
\times \left( {\gq' \over \gq} \right)^2
\times \left( {\gDM'\over \gDM} \right)^2
\times {\Gamma\over\Gamma'}
\end{equation}
is sufficient for a limited range of coupling values (see Fig.~\ref{fig:xsec_scaling} for example), 
this scaling is only approximate (up to 20\%)  and relies on the narrow width approximation, ignoring PDFs effects. 

\chapter{Specific models for signatures with EW bosons}
\label{subsec:EWSpecificModels}

In this Section, we consider specific models with a photon, a W boson, a Z boson or a Higgs boson in the final state (\textit{V+\MET} signature), accompanied by Dark Matter particles that either couple directly to the boson or are mediated by a new particle. The common feature of those models is that they provide different kinematic distributions with respect to the models described in Section~\ref{subsec:MonojetLikeModels}.


\begin{figure}[h!]
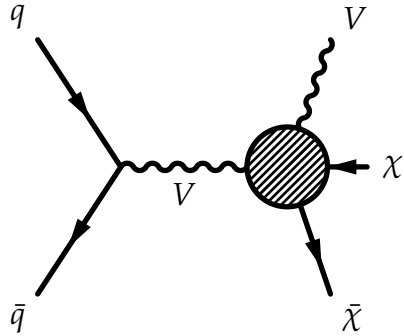

	\centering
	\vspace{\baselineskip}
	\unitlength=0.0045\textwidth
	\begin{feynmandiagram}[modelVeft5pt]
		\fmfleft{i1,i2}
		\fmfright{o1,o2,o3}
		\fmf{photon,label={\Large $V$}}{v1,v2}
		\fmfv{decor.shape=circle,decor.filled=shaded, decor.size=30}{v2}
		\fmf{fermion}{i2,v1,i1}
		\fmf{fermion}{o2,v2,o1}
		\fmflabel{\Large ${\bar{q}}$}{i1}
		\fmflabel{\Large ${q}$}{i2}
		\fmflabel{\Large ${\bar{\chiDM}}$}{o1}
		\fmflabel{\Large ${\chiDM}$}{o2}
		\fmf{photon}{o3,v2}
		\fmflabel{\Large ${V}$}{o3}
	\end{feynmandiagram}
	
	\vspace{\baselineskip}
	\caption{Sketch of benchmark models including a contact interaction
		for V+MET searches, adapted from~\cite{Nelson:2013pqa}. \label{fig:VPlusMET_EFT}}
\end{figure}

The models considered in this Section can be divided into two categories:
\begin{description}
	\item[V-specific simplified models] These models postulate direct couplings of new mediators
	to bosons, e.g. they couple the Higgs boson to a new vector or to a new scalar~\cite{Carpenter:2013xra,Berlin:2014cfa}. 
	\item[Models involving a SM singlet operator including a boson pair that couples to Dark Matter through a contact interaction]
	Shown on the right-hand side of Figure~\ref{fig:VPlusMET_EFT},
	these models allow for a contact interaction vertex that directly couples the boson to Dark Matter~\cite{Cotta:2012nj, Carpenter:2012rg, Crivellin:2015wva,Berlin:2014cfa}.
	These models are included in this report devoted to simplified models since 
	UV completions for most of these operators proceed through loops and are not available to date. 
	These models provide a benchmark to motivate signal regions that are unique to searches with
	EW final states and would otherwise not be studied. However, we recommend to use these models
	as placeholders and emphasize model-independent results especially in signal regions tailored to these models. 
	Wherever results are interpreted in terms of these operators, a truncation procedure
	to ensure the validity of the EFT should be employed, as detailed in the next Section (Sec. ~\ref{sec:EFTValidity}). 
\end{description}

The following Sections describe the models within these categories,
the parameters for each of the benchmark models chosen,
the studies towards the choices of the parameters to be scanned.

\section{Specific simplified models including EW bosons, tailored to Higgs+MET searches}
\label{sec:monoHiggs}

\begin{figure}[!htpb!tpd]
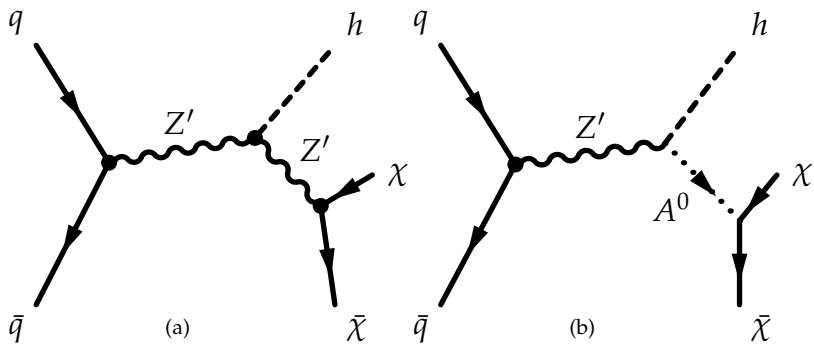

	\centering
	\unitlength=0.0046\textwidth
	\subfloat[\label{subfig:modelMonoHZprimeqq}]{
		\begin{feynmandiagram}[modelMonoHZprimeqq]
		\fmfleft{i1,i2}
		\fmfright{o1,o2,o3}
		\fmf{fermion}{i2,v1,i1}
		\fmflabel{\Large $q$}{i2}
		\fmflabel{\Large $\bar{q}$}{i1}
		\fmf{photon,label={\Large \Zprime}}{v1,v2}
		\fmf{photon,label={\Large \Zprime},label.angle=-100,label.distance=5}{v2,v3}
		\fmf{dashes}{v2,o3}
		\fmflabel{\Large $h$}{o3}
		\fmf{fermion}{o2,v3,o1}
		\fmflabel{\Large ${\bar{\chiDM}}$}{o1}
		\fmflabel{\Large ${\chiDM}$}{o2}
		\fmfdot{v1,v2,v3}
	\end{feynmandiagram}
	}
	\subfloat[\label{subfig:modelMonoHSimplifiedA0}]{
	\begin{feynmandiagram}[modelMonoHSimplifiedA0]
		\fmfleft{i1,i2}
		\fmfright{o1,o2,o3}
		\fmf{fermion}{i2,v1,i1}
		\fmflabel{\Large $q$}{i2}
		\fmflabel{\Large $\bar{q}$}{i1}
		\fmf{photon,label={\Large \Zprime}}{v1,v2}
		\fmf{dashes}{v2,o3}
		\fmflabel{\Large $h$}{o3}
		\fmf{dots_arrow,label={\Large $A^0$},label.side=right}{v2,v3}
		\fmf{fermion,tension=2}{o2,v3,o1}
		\fmflabel{\Large ${\bar{\chiDM}}$}{o1}
		\fmflabel{\Large ${\chiDM}$}{o2}
		\fmfdot{v1}
	\end{feynmandiagram}
	}
	\caption
	{
		Examples of Feynman diagrams leading to Higgs+\MET events: 
                (a) 
                a model with a vector mediator (\Zprime) 
		coupling with DM and with the Higgs boson $h$,
and
                (b) 
		a 2HDM model with a new invisibly decaying pseudoscalar $A^0$ 
		from the decay of an on-shell resonance \Zprime giving rise to a Higgs+\MET signature
.
	}
	\label{fig:feyn_prod_monoH}
\end{figure}

\begin{figure}[!htpb!tpd]
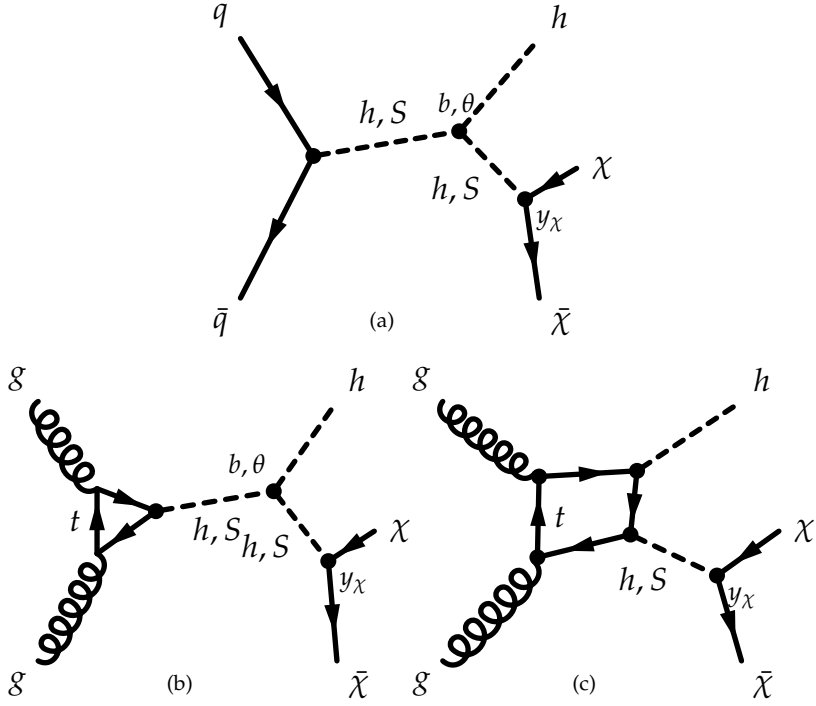

	\centering
	\unitlength=0.0046\textwidth
	\subfloat[\label{subfig:modelMonoHbaryonicqq}]{
		\begin{feynmandiagram}[modelMonoHbaryonicqq]
			\fmfleft{i1,i2}
			\fmfright{o1,o2,o3}
			\fmf{fermion}{i2,v1,i1}
			\fmflabel{\Large $q$}{i2}
			\fmflabel{\Large $\bar{q}$}{i1}
			\fmf{dashes,label={\Large $h,,S$}}{v1,v2}
			\fmf{dashes,label={\Large $h,,S$},label.side=right}{v2,v3}
			\fmf{dashes}{v2,o3}
			\fmflabel{\Large $h$}{o3}
			\fmfv{label={$b,,\theta$},label.a=100,label.distance=3w}{v2}
			\fmfv{label={$y_{\chiDM}$},label.a=-45,label.distance=5w}{v3}
			\fmf{fermion}{o2,v3,o1}
			\fmflabel{\Large ${\bar{\chiDM}}$}{o1}
			\fmflabel{\Large ${\chiDM}$}{o2}
			\fmfdot{v1,v2,v3}
		\end{feynmandiagram}
	}\\\vspace{\baselineskip}
	\subfloat[\label{subfig:modelMonoHbaryonicgg}]{
		\begin{feynmandiagram}[modelMonoHbaryonicgg]
			\fmfleft{i1,i2}
			\fmfright{o1,o2,o3}
			\fmf{gluon}{i1,vt1}
			\fmf{gluon}{i2,vt2}
			\fmflabel{\Large $g$}{i2}
			\fmflabel{\Large $g$}{i1}
			\fmf{fermion,label={\Large $t$}}{vt1,vt2}
			\fmf{fermion}{vt2,v1,vt1}
			\fmf{dashes,label={\Large $h,,S$},label.side=right}{v1,v2}
			\fmf{dashes,label={\Large $h,,S$},label.side=right,label.d=5}{v2,v3}
			\fmf{dashes}{v2,o3}
			\fmflabel{\Large $h$}{o3}
			\fmfv{label={$b,,\theta$},label.a=120,label.distance=3w}{v2}
			\fmfv{label={$y_{\chiDM}$},label.a=-45,label.distance=5w}{v3}
			\fmf{fermion}{o2,v3,o1}
			\fmflabel{\Large ${\bar{\chiDM}}$}{o1}
			\fmflabel{\Large ${\chiDM}$}{o2}
			\fmfdot{v1,v2,v3}
		\end{feynmandiagram}
	}
	\subfloat[\label{subfig:modelMonoHbaryonicggS}]{
	\begin{feynmandiagram}[modelMonoHbaryonicggS]
		\fmfleft{i1,i2}
		\fmfright{o1,o2,o3}
		\fmf{gluon}{i1,vt1}
		\fmf{gluon}{i2,vt2}
		\fmflabel{\Large $g$}{i2}
		\fmflabel{\Large $g$}{i1}
		\fmf{fermion,label={\Large $t$}}{vt1,vt2}
		\fmf{fermion}{vt2,v2,v1,vt1}
		\fmf{dashes}{v2,o3}
		\fmflabel{\Large $h$}{o3}
		\fmf{dashes,label={\Large $h,,S$},label.side=right,label.d=5}{v1,v3}
		\fmfv{label={$y_{\chiDM}$},label.a=-45,label.distance=5w}{v3}
		\fmf{fermion}{o2,v3,o1}
		\fmflabel{\Large ${\bar{\chiDM}}$}{o1}
		\fmflabel{\Large ${\chiDM}$}{o2}
		\fmfdot{v1,v2,v3,vt1,vt2}
	\end{feynmandiagram}

	}
	\caption
	{
		Examples of Feynman diagrams leading to Higgs+\MET events for a model with a scalar mediator ($S$) 
		coupling with DM and with the Higgs boson $h$. 
	}
	\label{fig:feyn_prod_monoH_S}
\end{figure}

Three benchmark simplified models \cite{Carpenter:2013xra,Berlin:2014cfa} 
are recommended for Higgs+\MET searches:
\begin{itemize}
	\item A model where a vector mediator ($Z_B^\prime$) is exchanged in the \schannel, 
	radiates a Higgs boson, and decays into two DM particles (Fig.~\ref{fig:feyn_prod_monoH} (a)). As in Section \ref{sec:monojet_V}, we conservatively omit couplings of the $\Zprime_B$ to leptons.
    \item A model where a scalar mediator $S$ is emitted from the Higgs boson and decays to a pair of DM particles (Fig.~\ref{fig:feyn_prod_monoH_S}).
	\item A model where a vector \Zprime is produced resonantly and decays into a Higgs boson
	plus an intermediate heavy pseudoscalar particle $A^0$, in turn decaying into two DM particles (Fig. \ref{fig:feyn_prod_monoH} (b)). 
\end{itemize}

\begin{figure}[hbpt!]
	\centering
	\subfloat[High mediator mass]{
		\includegraphics[width=0.75\linewidth]{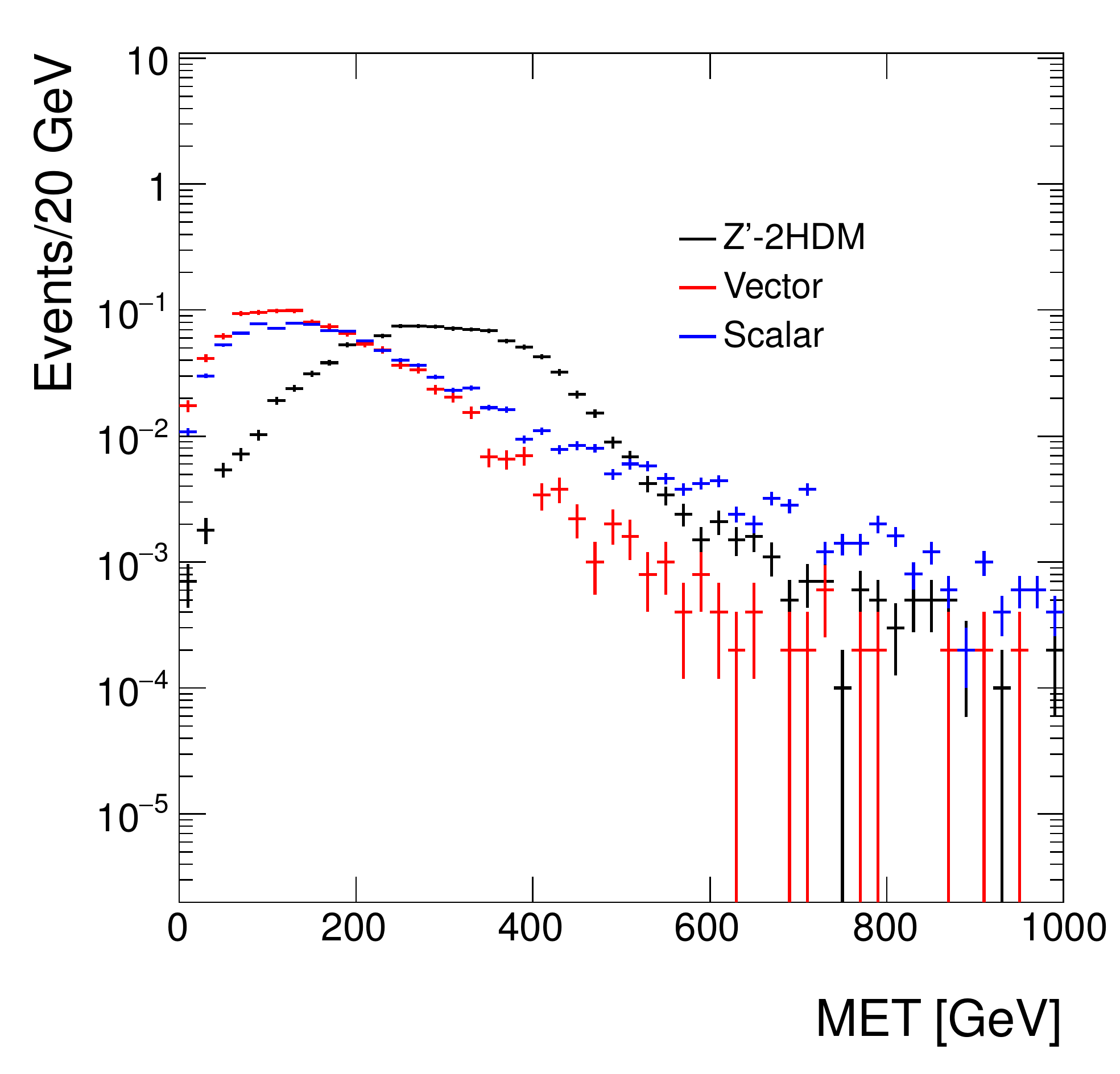} \label{fig:met_cmp_high}
	}\\
	\subfloat[Low mediator mass]{
		\includegraphics[width=0.75\linewidth]{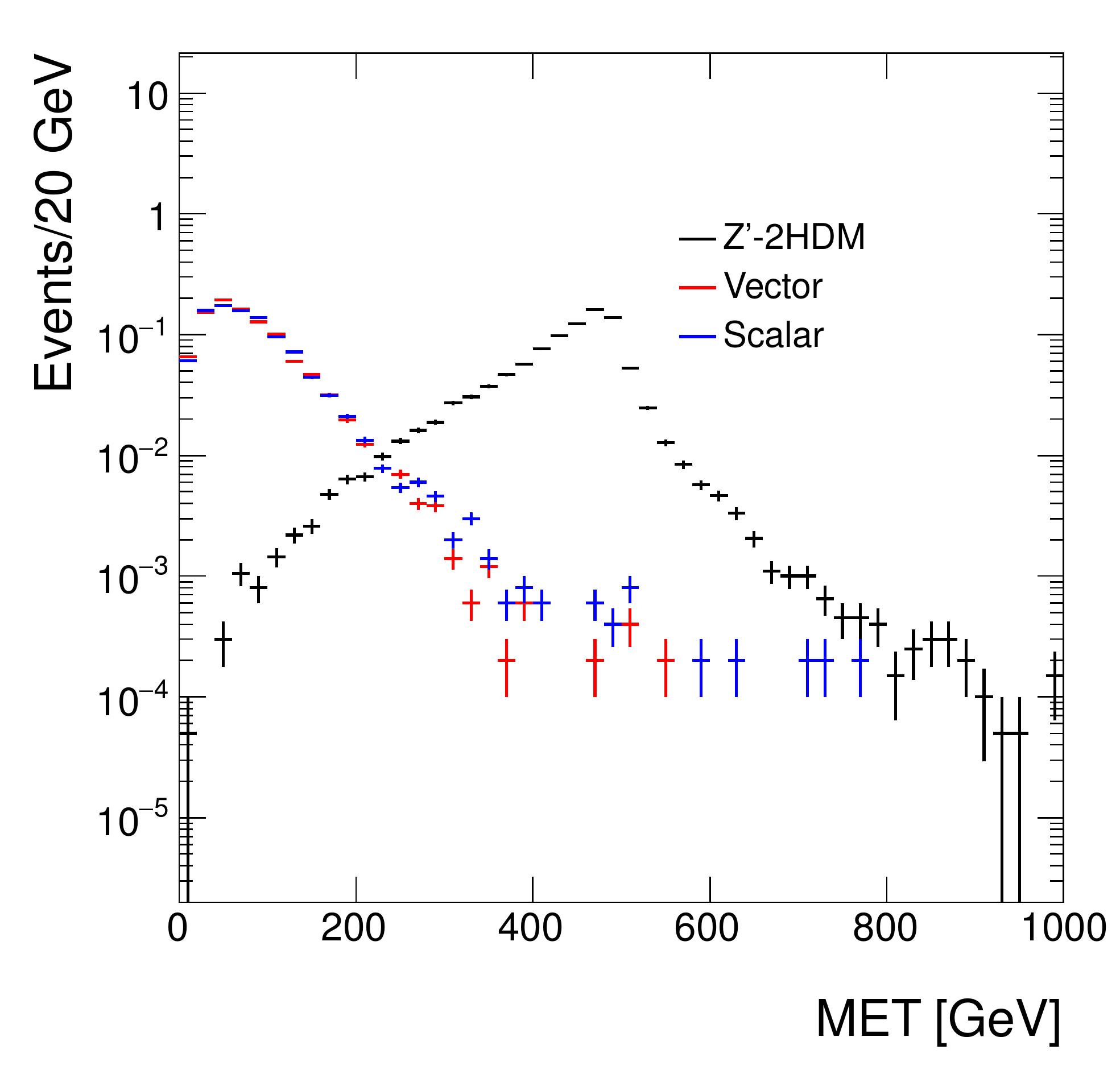} \label{fig:met_cmp_low}
	}
	\caption{Comparison of the missing transverse momentum distributions at generator level in different 
		simplified models leading to a Higgs+\MET signature. The model parameter settings are detailed in the text. The figures in this Section have been obtained using LO UFO models within \madgraph v2.2.3, interfaced to \pythia 8 for the parton shower.  		
		\label{fig:METSimpMonoHiggs}}
\end{figure}

These models are kinematically distinct from one another, as shown in the comparison of the 
\MET spectra in Fig.~\ref{fig:METSimpMonoHiggs} for high and low masses of the pseudoscalar mediator. 
Figure~\ref{fig:METSimpMonoHiggs} (a) shows the \MET distribution 
for models with high mediator masses ($m_{S} = 1$~\tev, $m_{\Zprime} = 1$~\tev, $m_{A^0} = 1$~\tev)
and DM mass of either 50 ($Z_B'$ and $A^0$ models) or 65~\gev (scalar mediator model).
Figure~\ref{fig:METSimpMonoHiggs} (b)  shows the \MET distribution 
for models with low pseudoscalar mediator masses ($m_{Z_B'} = 100$~\gev, $m_{\Zprime} = 1$~\tev, $m_{A^0} = 100$~\gev)
and DM mass of 1~\tev for all models. 

Predictions for this class of models have been so far considered at LO+PS, even though they could be extended to NLO+PS in the near future. The studies in this Section
have been performed using a model within \madgraph v2.2.3, interfaced to \pythia 8 for the parton shower.  
The implementation details for these models are discussed in Section~\ref{sec:monoHImplementation}.

\subsection{\MET+Higgs from a baryonic \Zprime}

The model shown in Fig.~\ref{fig:feyn_prod_monoH} (a)
postulates a new gauge boson \Zprime corresponding to a new $U(1)_B$ baryon 
number symmetry. The stable baryonic states included in this model are the DM candidate particles.
The mass of the \Zprime boson is acquired through a baryonic Higgs $h_B$, which mixes with the 
SM Higgs boson. 

The interactions between the \Zprime, the quarks and the DM are described by 
the following Lagrangian:   

\be \label{ZprimeDM}
	\mathcal{L} =  \gq  \bar q \gamma^\mu q  Z_\mu' +
%
	 \gDM  \bar\chiDM \gamma^\mu \chiDM Z_\mu' .
\ee

The quark couplings \gq are fixed to be equal to one third of the gauge coupling $g_B$, 
while the DM coupling to the \Zprime are proportional to the baryon number and to the gauge coupling 
($g_{\chiDM} = B g_B$). No leptonic couplings of the \Zprime are allowed, thus evading dilepton constraints. 
After incorporating the mixing of the baryonic and SM Higgs bosons, this model is 
is described by the following Lagrangian term at energies below $m_{\Zprime}$~\footnote{The operator 
	in Eqn.~\ref{U1Beft} is an effective one, to highlight the two main terms. The full dimension-4 simplified
	model is used in the model for event generation.}: 

\be \label{U1Beft}
 \mathcal{L}_{\rm eff} = - \frac{\gq \gDM }{m_{\Zprime}^2} \bar{q} \gamma^\mu q \bar\chiDM \gamma_\mu \chiDM \Big( 1 + \frac{g_{h \Zprime \Zprime} }{m_{\Zprime}^2} h \Big) \, ,
\ee

The first term of this equation
is the standard \modelDMV model in the large $M_{Z^\prime}$ limit.  This term can lead
to a monojet signature, which can be also used to constrain this model.
The second term describes the interaction between the \Zprime and the SM Higgs boson,
via the coupling $g_{h \Zprime \Zprime} = \frac{m_{\Zprime}2 \sin\theta}{v_B}$, where
$\sin\theta$ is the mixing angle between the SM Higgs and the baryonic Higgs $h_B$, and $v_B$ is the
Baryonic Higgs vacuum expectation value. 


In its most general form, this model can contribute to mono-Z signals due to the \Zprime mixing with the Z or photon. Note that EWSB and $ U(1)_B $ breaking do not lead to this mixing at tree-level. Instead, kinetic mixing occurs between the $ U(1)_Y $ and $ U(1)_B $ gauge bosons due to the gauge invariant term $ F^{\mu\nu}_Y F_{B\mu\nu} $. This mixing is a free parameter which we assume to be small in order to focus on the mono-Higgs signature. Mixing may also occur due to radiative corrections, however this is model dependent so we choose to ignore this here.


The predictions of the model depend upon the two additional
parameters beyond an \schannel simplified model, namely the
mixing angle between baryonic Higgs $h_B$ and the SM-like Higgs boson $\sin\theta$ and the coupling of the mediator to SM-like Higgs boson, $g_{h\Zprime \Zprime}$.
Thus, a full model is specified by:

\be
\left\{\mMed ,\, \mDM ,\, \gDM ,\, \gq ,\, \sin\theta ,\, g_{h\Zprime \Zprime}\right\}.
\ee

\subsubsection{Parameter scan} 

The width of the \Zprime mediator is calculated using all possible decays to SM particles (quarks) and to pairs of DM particles if kinematically allowed
as in the \modelDMV model.

\begin{figure}[htpb!]
	\includegraphics[width=0.75\linewidth]{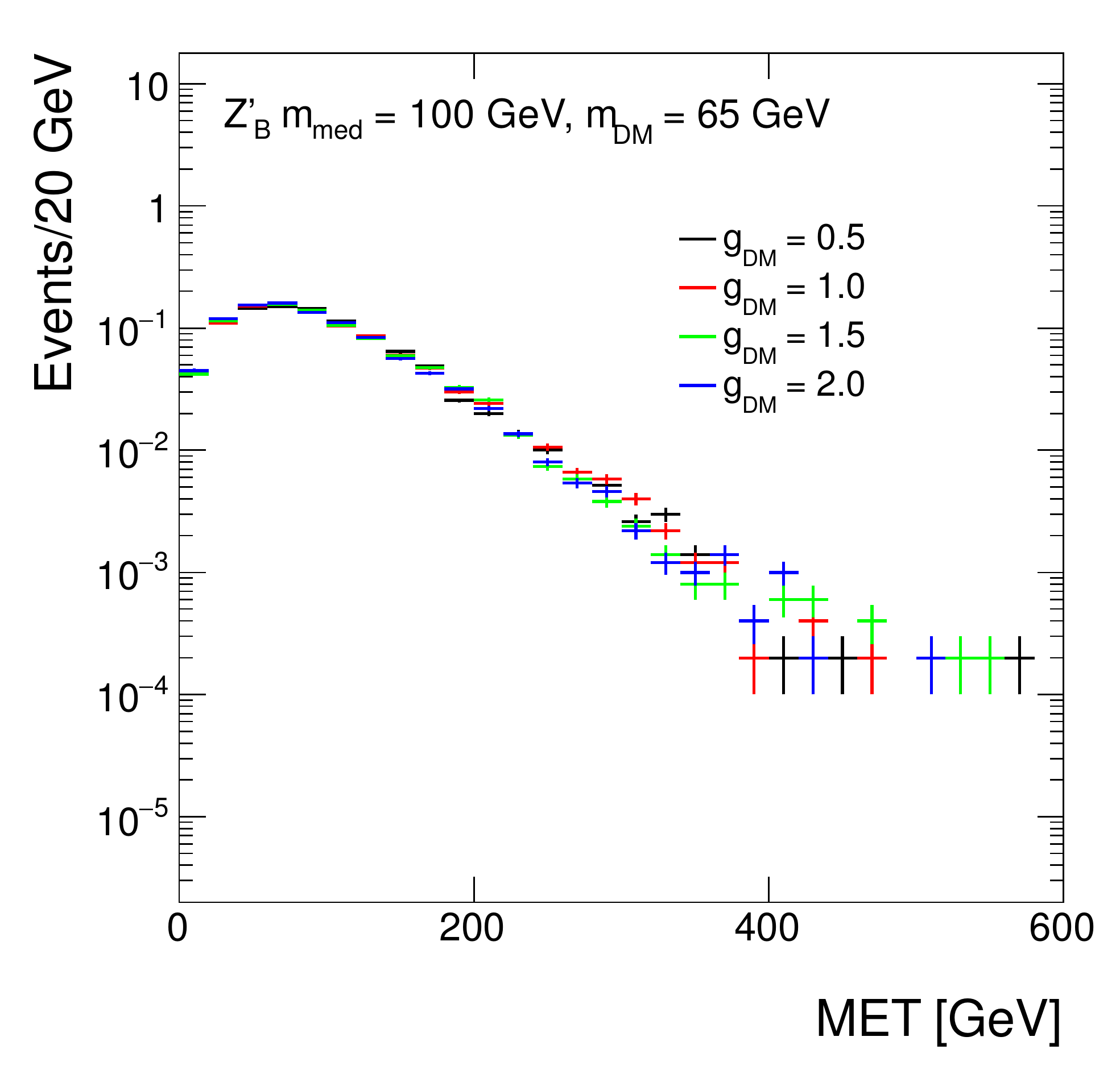}\\
	\includegraphics[width=0.75\linewidth]{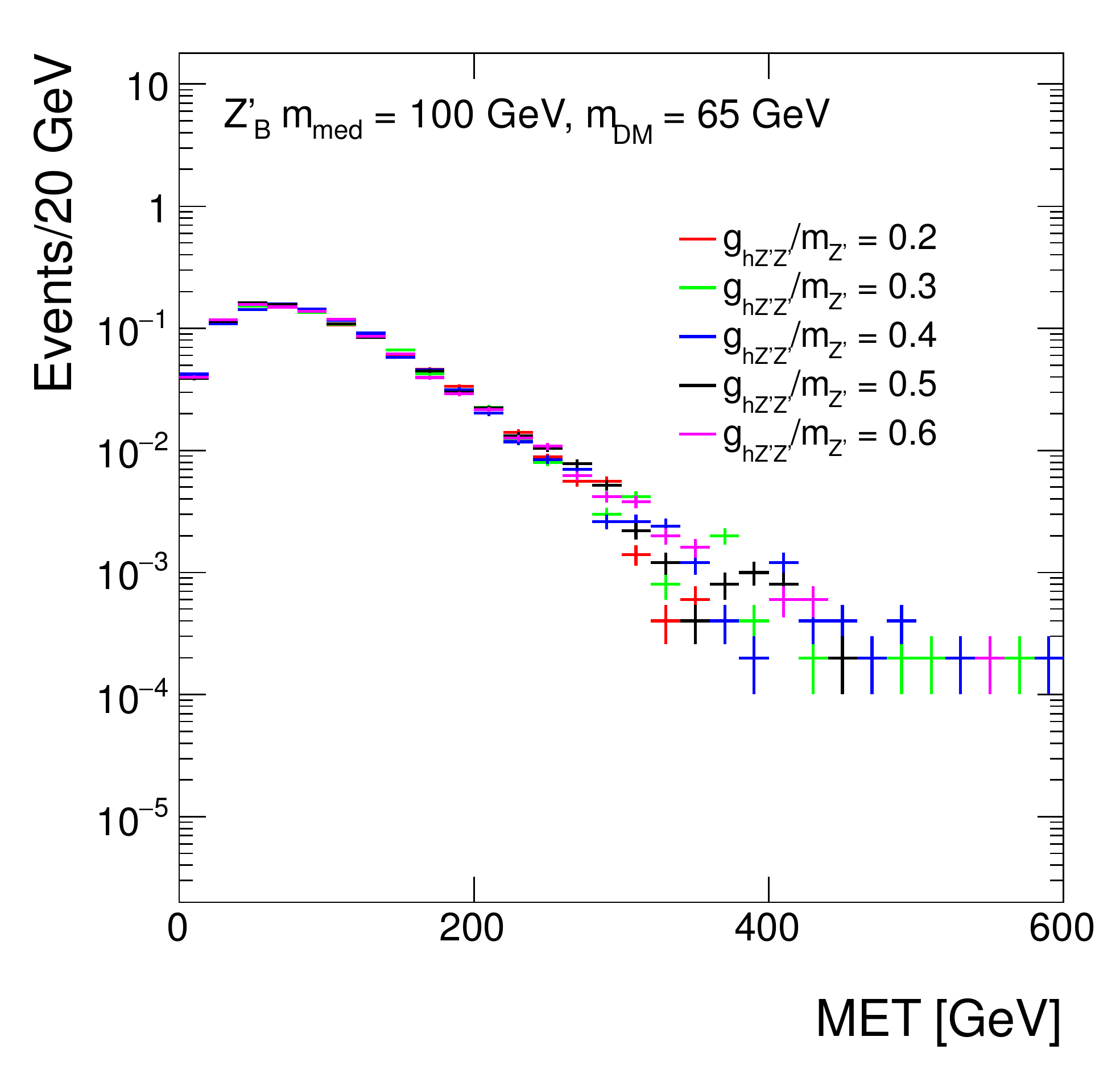}
	\caption{Missing transverse momentum distributions at generator level in the vector 
		mediator scenario for different values of: the mediator-dark matter coupling \gDM (left),
		and the coupling between the mediator and the SM-like Higgs boson, scaled by the mediator mass, 
		$g_{h \Zprime \Zprime}/m_{\Zprime}$ (right).
		\label{fig:metVectorCoupling}}
\end{figure}

\begin{figure}[htpb!]
	\includegraphics[width=0.75\linewidth]{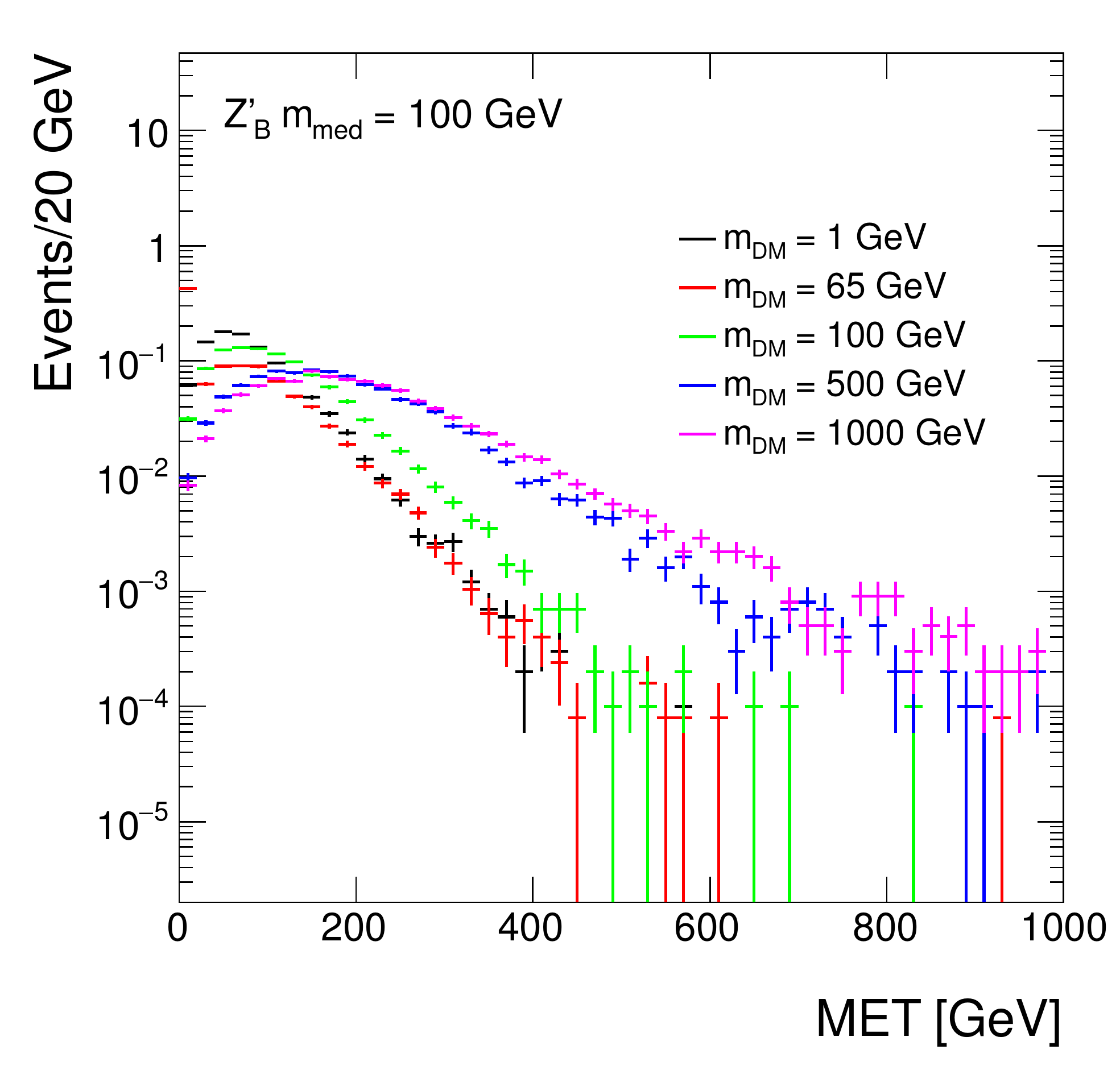}\\
	\includegraphics[width=0.75\linewidth]{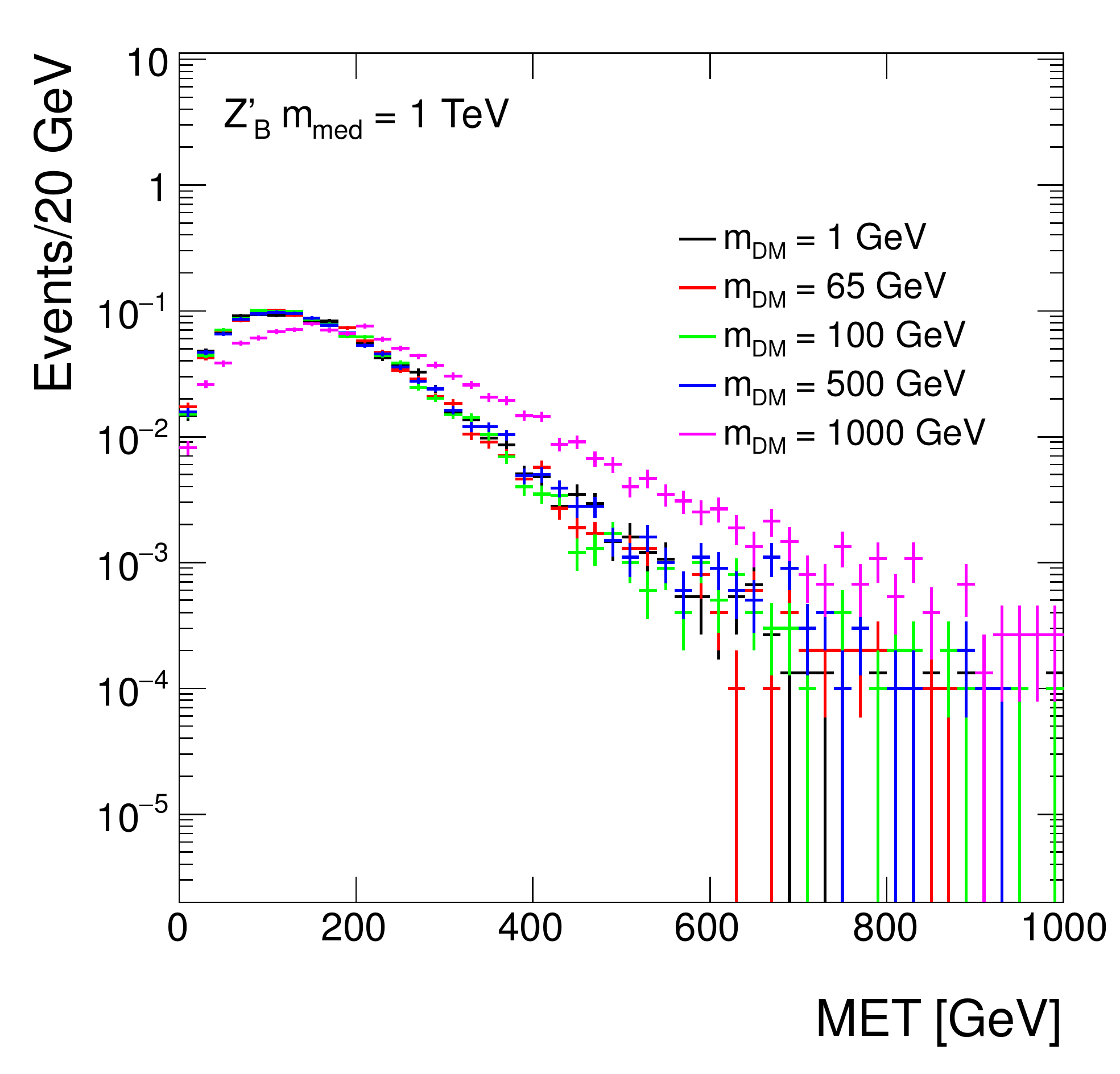}
	\caption{Missing transverse momentum distributions at generator level in the vector 
		mediator scenario: for different values of the dark matter mass \mDM 
		and a mediator mass of \mmed = 100~\gev (left) and \mmed = 1~\tev (right).
		\label{fig:metVectorMass} }
\end{figure}

The dependence of the missing transverse momentum (\MET) on the model parameters 
is studied by varying the parameters one at a time. The variation of parameters 
other than \mMed and \mDM does not result in significant 
variations of the \MET spectrum, as shown in Figures~\ref{fig:metVectorCoupling}. 
Figure~\ref{fig:metVectorMass} shows that for an on-shell mediator, 
varying \mDM with the other parameters fixed does not affect the \MET distribution, while 
the distribution broadens significantly in the case of an off-shell mediator. 
For this reason, the same grid in \mmed, \mdm as for the vector mediator
of the jet+\MET search (Table~\ref{tab:mDMmMedScan_VA}) is chosen as a starting point. 
The coupling $g_{h\Zprime \Zprime}$, along with \gq and \gDM, are subject to perturbativity bounds:

$$\gq, \gDM < 4\pi $$
and

$$  g_{h \Zprime \Zprime} < \sqrt{4\pi}m_{\Zprime}\sin\theta$$ 
The value $g_{h \Zprime \Zprime}/m_{\Zprime} = 1$ is chosen as a benchmark value for the generation 
of Monte Carlo samples since it maximizes the cross section (as shown in the following paragraph)
without violating the bounds. The mediator-DM coupling \gDM is fixed to 1, and  
the mediator-quark $g_{q}$ coupling is fixed to 1/3. 
The kinematic distributions do not change as a function of these parameters, so 
results for other values of  $g_{h \Zprime \Zprime}/m_{\Zprime}$, \gDM and \gq can be 
obtained through rescaling by the appropriate cross sections. 

\begin{figure}[htpb!]
	\centering
	\subfloat[Leading $b-$jet transverse momentum]{
		\includegraphics[width=0.75\linewidth]{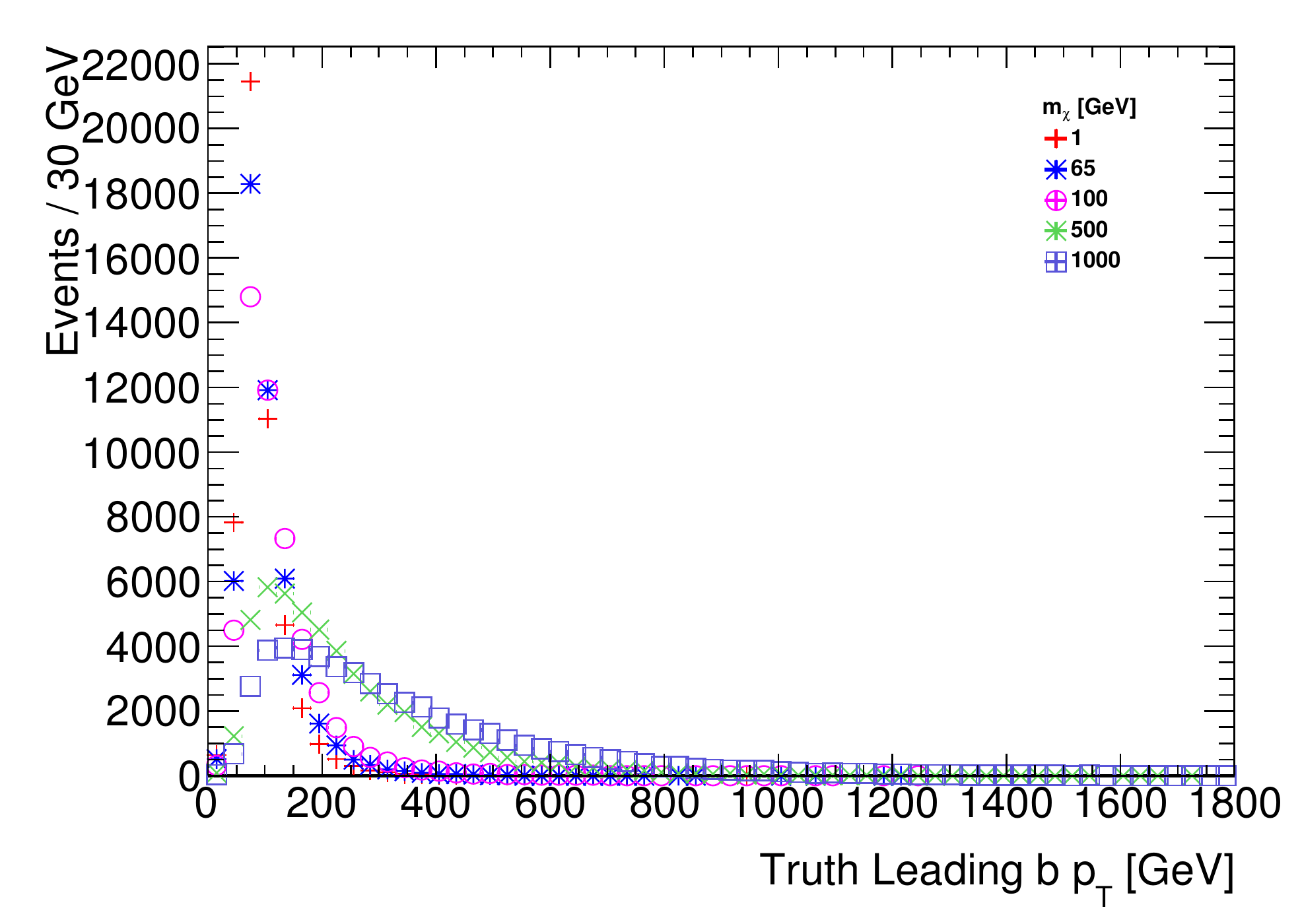} 
	}\hfill
	\subfloat[Leading $b-$jet pseudorapidity]{
		\includegraphics[width=0.75\linewidth]{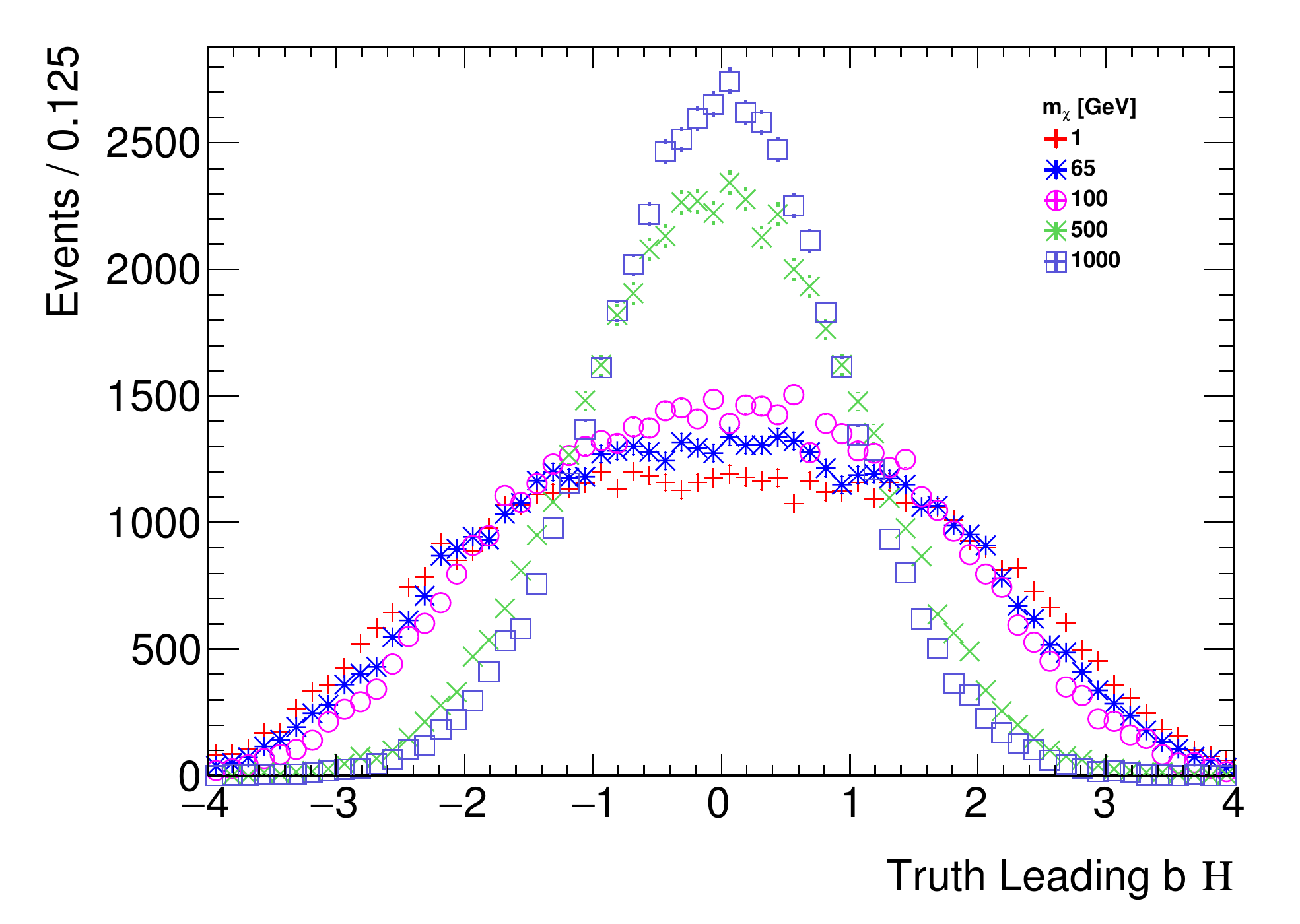} 
	}\hfill
	\subfloat[Angular distance between the two leading $b-$jets]{
		\includegraphics[width=0.75\linewidth]{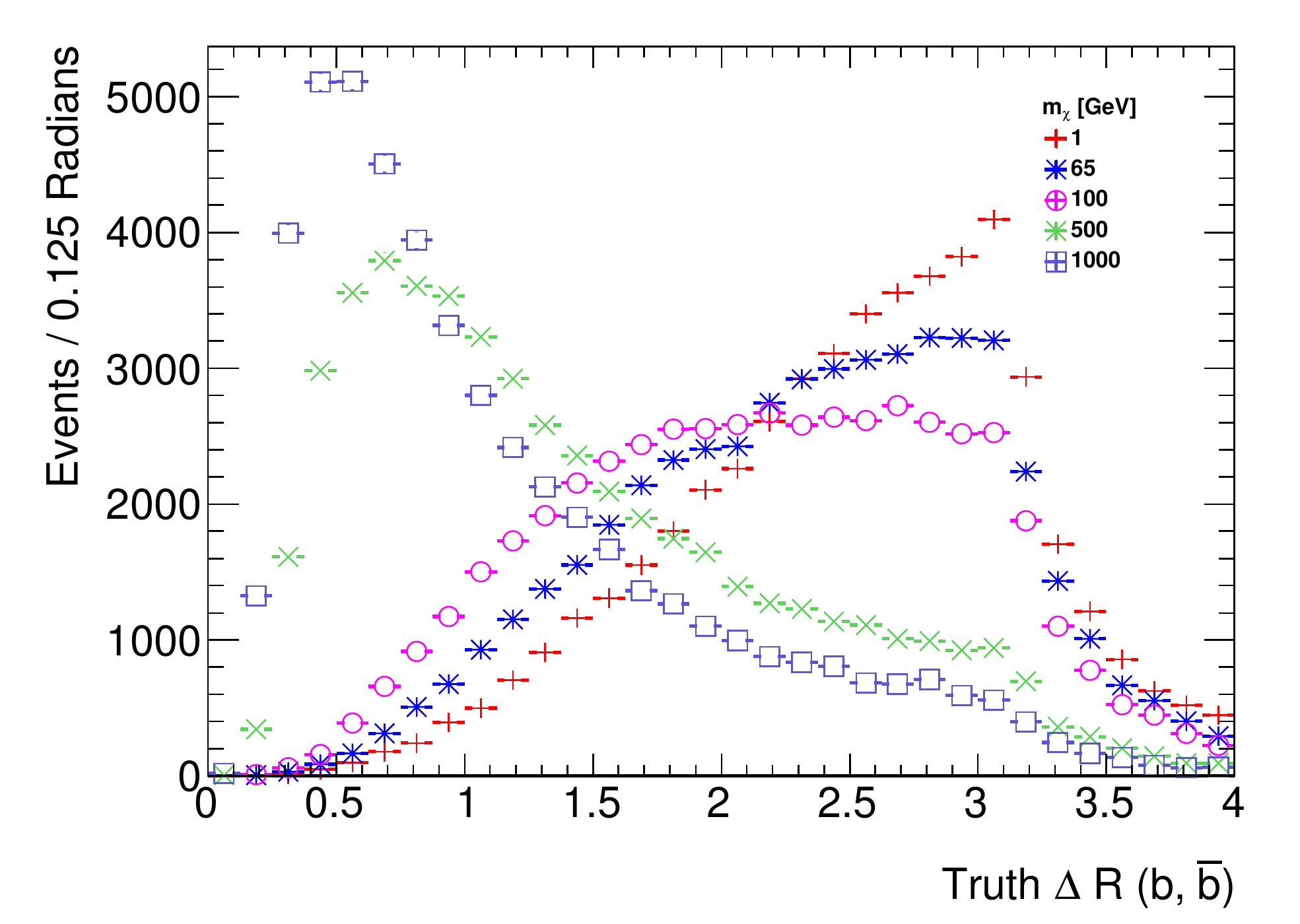} 
	}
	\caption{Comparison of the kinematic distributions for the two leading $b-$jets (from the Higgs decay) in the vector \Zprime simplified model, 
		when fixing the \Zprime mass to 100~\gev and varying the DM mass. 
		\label{fig:VectorHbb_100}}
\end{figure}

\begin{figure}[htpb!]
	\centering
	\subfloat[Leading $b-$jet transverse momentum]{
		\includegraphics[width=0.75\linewidth]{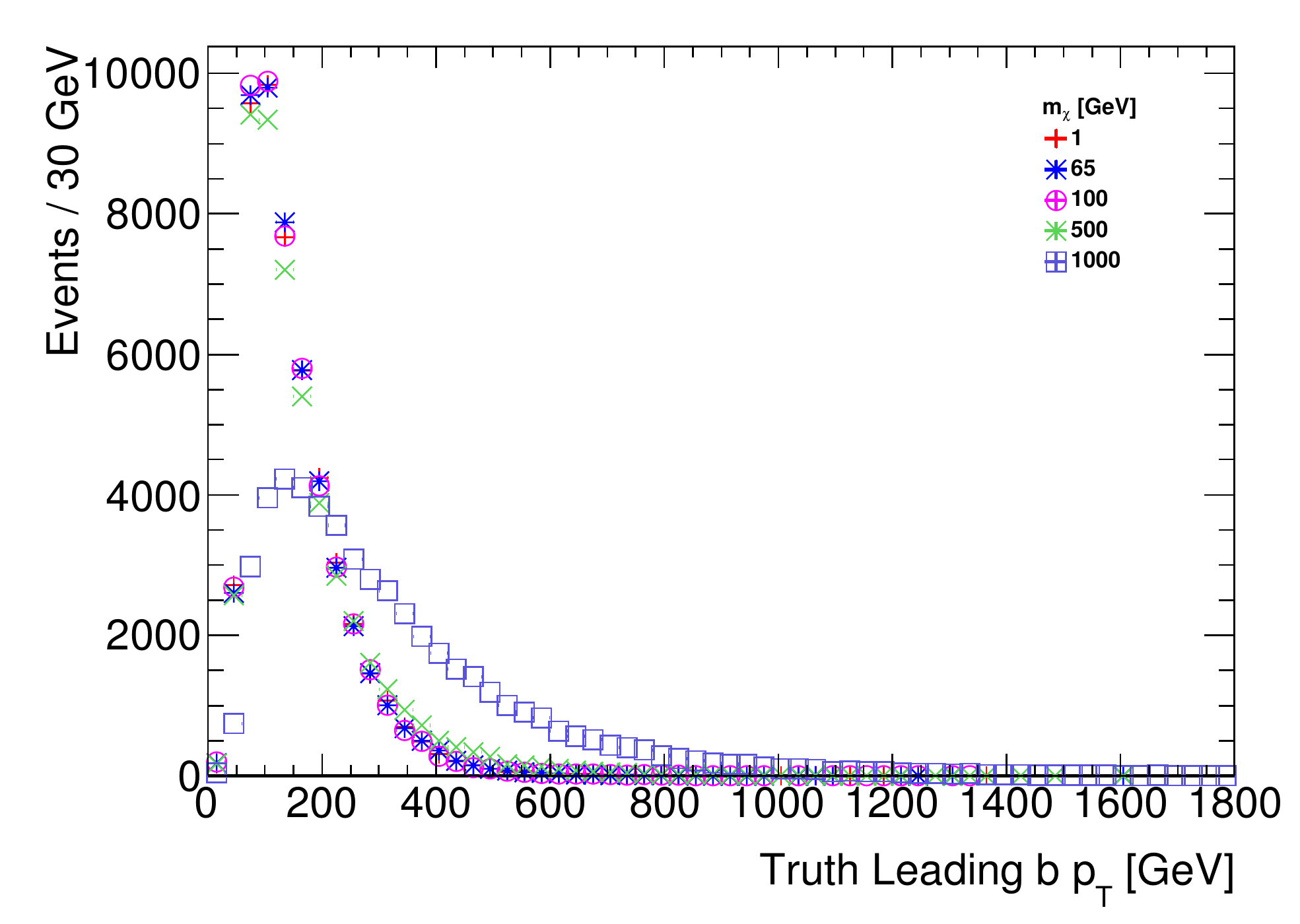} 
	}\hfill
	\subfloat[Leading $b-$jet pseudorapidity]{
		\includegraphics[width=0.75\linewidth]{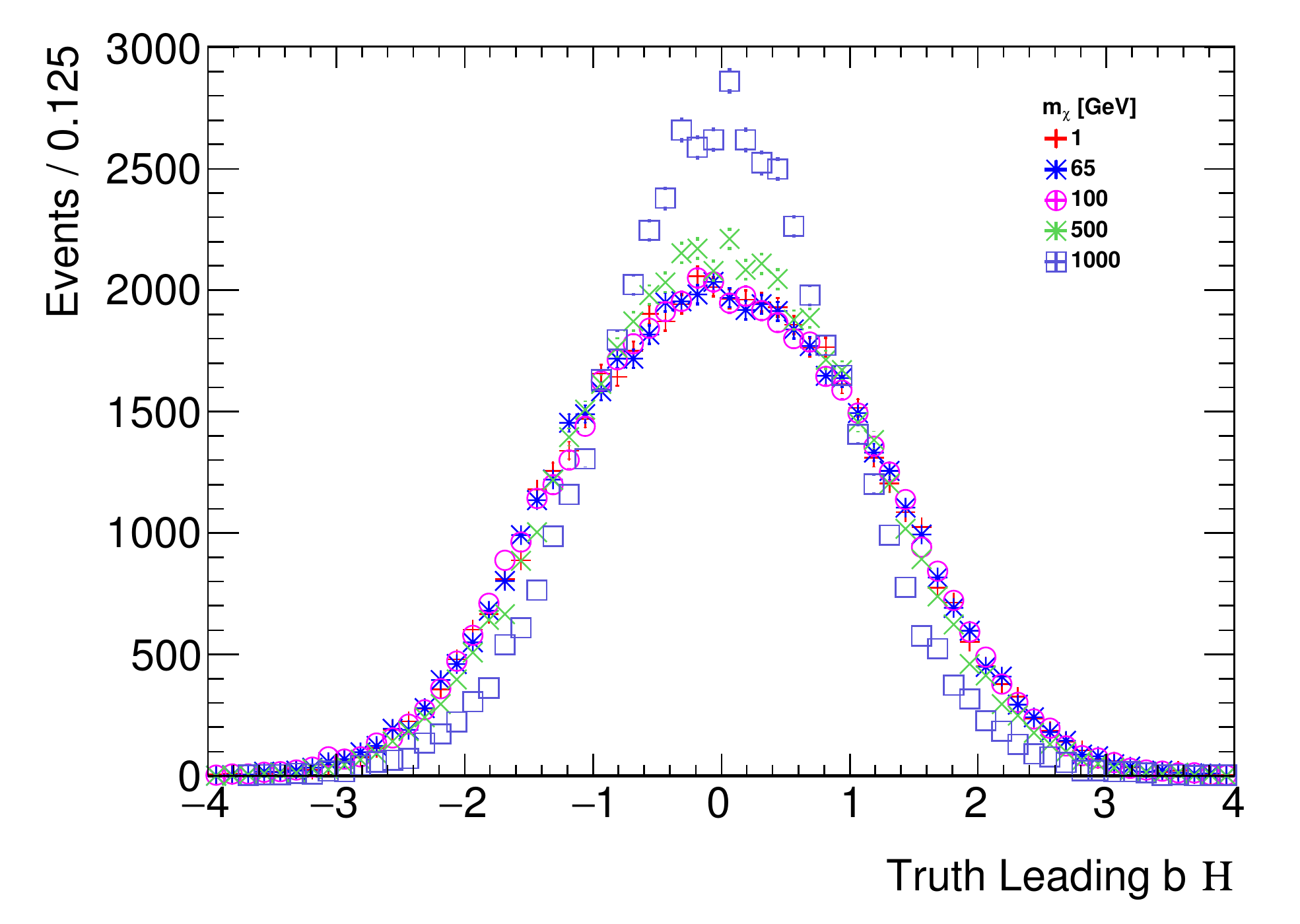} 
	}\hfill
	\subfloat[\text{Angular separation of the two leading $b$-jets}]{
		\includegraphics[width=0.75\linewidth]{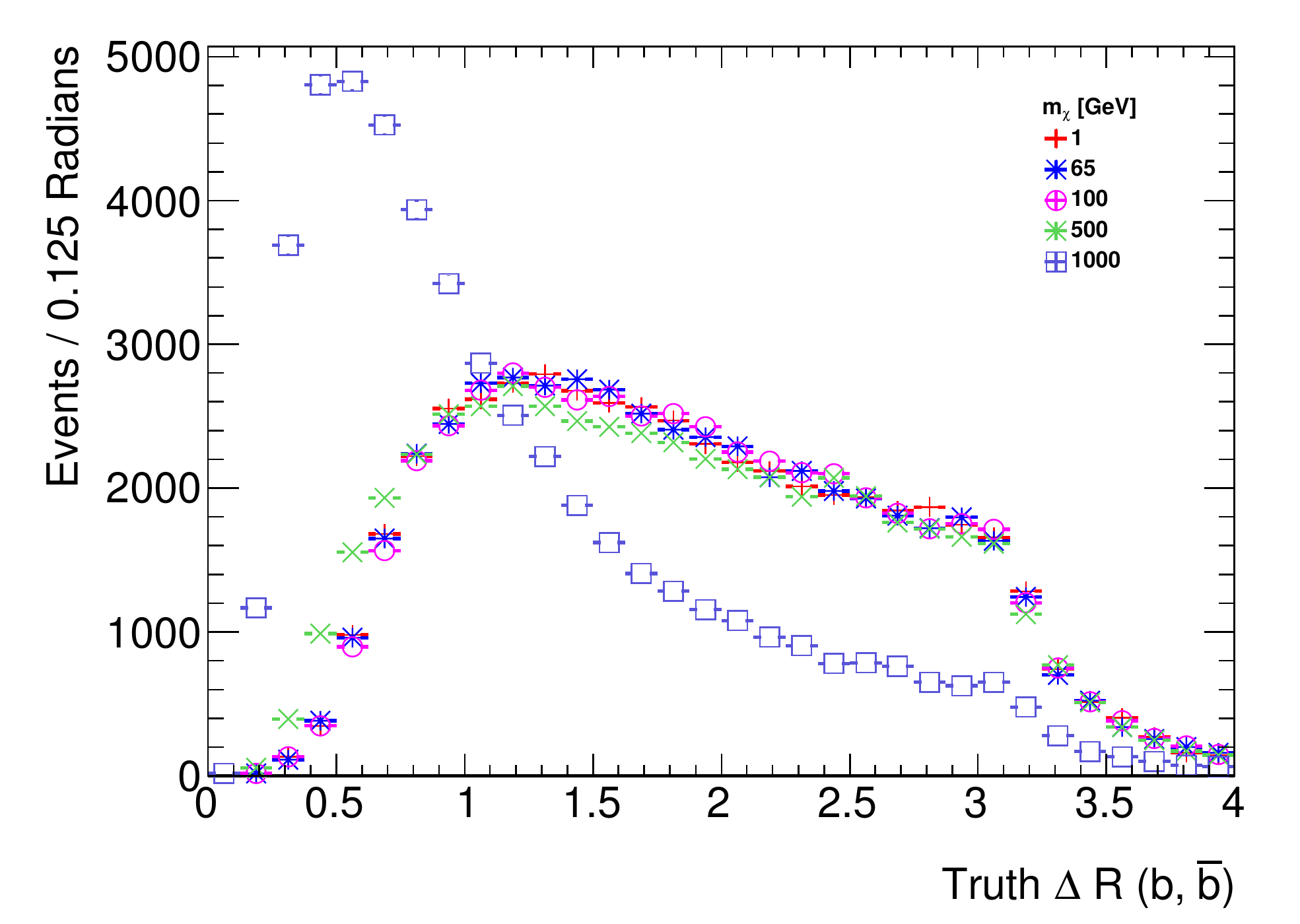} 
	}
	\caption{Comparison of the kinematic distributions for the two leading jets from the Higgs decay in the vector \Zprime simplified model, 
		when fixing the \Zprime mass to 1000~\gev and varying the DM mass. 
		\label{fig:VectorHbb_1000}}
\end{figure}

Figs~\ref{fig:VectorHbb_100} and ~\ref{fig:VectorHbb_1000} show the kinematic distributions for the two leading jets
in the $H \to \bar b b$ decay channel, for two values of the mediator mass and varying the DM mass.  

Analyses should perform further studies, beyond those studies performed for the forum, 
to estimate the reach of the analysis with respect to all points in the grid and therefore decide 
on a smaller set of grid points to be generated.

\subsection{\MET+Higgs from a scalar mediator}

A real scalar singlet $S$ coupling to DM can be introduced as a portal between SM and the dark sector 
through the Higgs field. The most general scalar potential is detailed in Ref.~\cite{O'Connell:2006wi}, 
including terms that break \Ztwo. 
The \Ztwo symmetry, which causes the new scalar to also be a DM candidate, is not covered in this report, but follows Ref.~\cite{Carpenter:2013xra}
introducing an additional coupling to DM that breaks \Ztwo and leads to a new invisible decay of $S$. 
For this reason, no symmetry is broken and no new interactions arise, so there is no dependence on the vacuum
expectation value of $S$: a shift in the field leads to a redefinition of the model couplings. 
The new scalar $S$ mixes with the SM Higgs boson, and couples to DM through a Yukawa term $y_\chiDM$. 
The relevant terms in the scalar potential are:

\begin{align}
&V \supset a |H|^2 S + b |H|^2 S^2 + \lambda_h |H|^4 \notag \\
& \;\;\longrightarrow \tfrac{1}{2} a (h +  v)^2 S + \tfrac{1}{2}b (h +  v)^2 S^2 + \frac{\lambda_h}{4} (h +  v)^4 ,
\label{singlethiggsmix}
\end{align}
where $a,b$ are new physics couplings and $\lambda_h$ is the Higgs quartic coupling.  

The additional Lagrangian terms for this model are: 

\be \label{LintScalar2}
\mathcal{L} \supset - y_\chiDM \bar\chiDM \chiDM (  \cos\theta\:S - \sin\theta\: h ) - \frac{m_q}{v} \bar q q (\cos\theta\: h + \sin\theta\: S )  \,
\ee
where $\theta$ is the mixing angle between the Higgs boson and the new scalar.

Mono-Higgs signals in this second model arise through processes shown in Fig.~\ref{fig:feyn_prod_monoH_S} (a,b), or through 
the radiation of a Higgs boson from the  $t$ quark in the production loop, in Fig.~\ref{fig:feyn_prod_monoH_S} (c). 
The first two processes depend on the $h^2 S$ and $h S^2$ cubic terms in Eq.~\eqref{singlethiggsmix}.  
At leading order in $\sin\theta$, these terms are:

\be
V_{\rm cubic} \approx \frac{\sin\theta}{v} ( 2 m_h^2 + m_S^2) h^2 S  + b \, v \, h \, S^2 + ...
\ee
with $a$ and $\lambda_h$ expressed in terms of $\sin\theta$ and $m_h^2$, respectively.  
At leading order of $\sin\theta$, the $h^2 S$ term is fixed once the mass eigenvalues $m_h, m_S$ 
and mixing angle are specified.  The $h\,S^2$ term is not fixed and remains a free parameter of the model, depending on 
the new physics coupling $b$. 

This model also has mono-X signatures through $h/S$ mixing. This model is related to the scalar model discussed in Sec.~\ref{sec:monojet_scalar}
in the case of $m_S \gg m_h$ or $m_h \gg m_S$ and  \mMed equal to the lighter of the two masses, albeit with different mono-Higgs signatures
due to the $h S^2$ vertex. 

\subsubsection{Parameter scan}

The model is described by five parameters: 

\begin{enumerate}
	\item the Yukawa coupling of heavy scalar to dark matter, \gDM (also referred to as $y_\chiDM$) 
	\item the mixing angle between heavy scalar and SM-like Higgs boson, $\sin\theta$;
	\item the new physics coupling, $b$;
	\item mass of heavy scalar, $m_{S}$, also termed \mmed;
	\item mass of dark matter. \mDM;
\end{enumerate}

The mixing angle is constrained from current Higgs data
to satisfy $\cos\theta = 1$ within 10\% and therefore $\sin\theta \lesssim 0.4$. This provides a starting point 
for the parameter scan in this model: we recommend to set $\sin\theta = 0.3$. 

\begin{figure}[hbpt!]
	\begin{center}
		\includegraphics[width=0.75\linewidth]{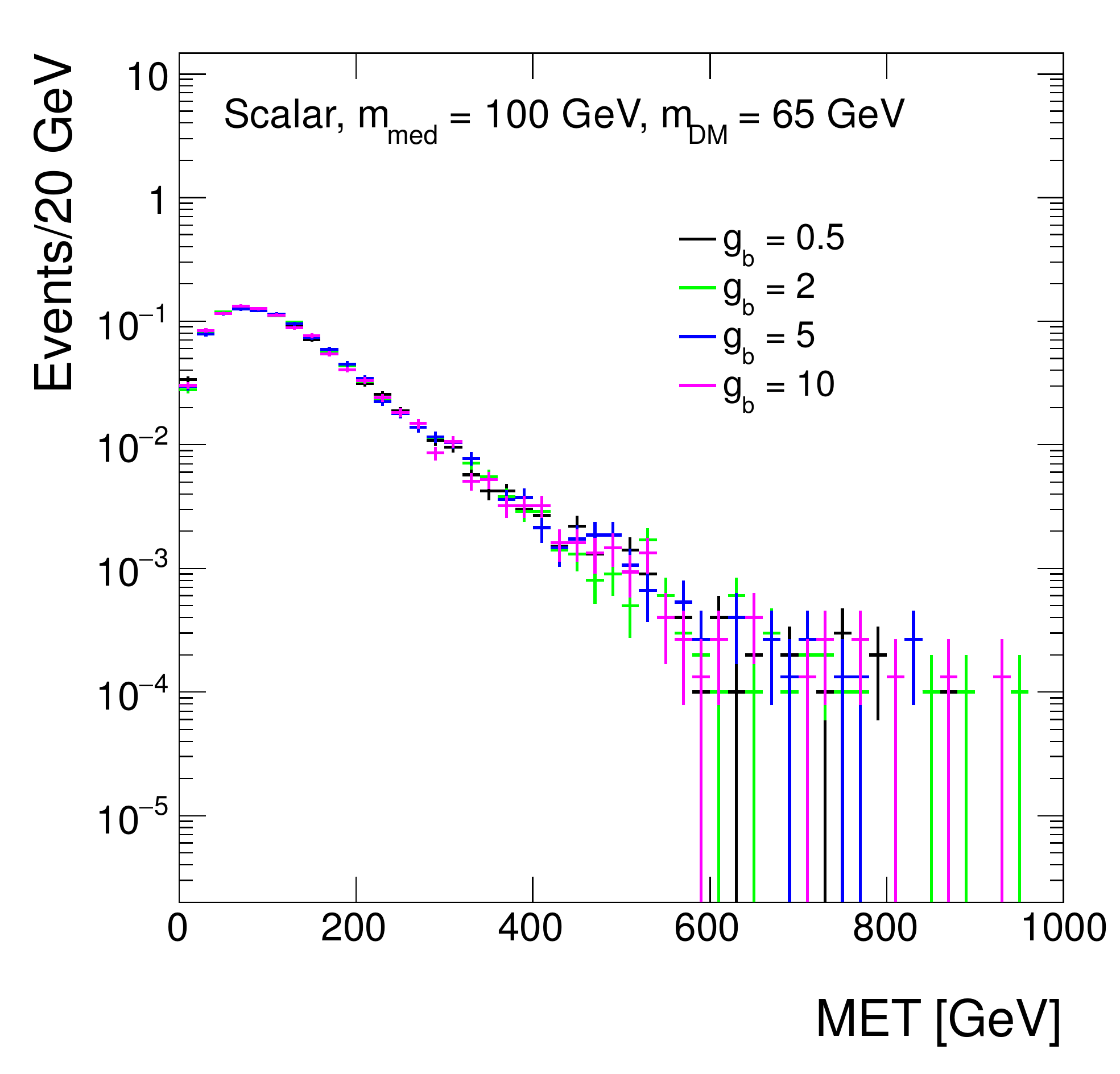}\\
		\includegraphics[width=0.75\linewidth]{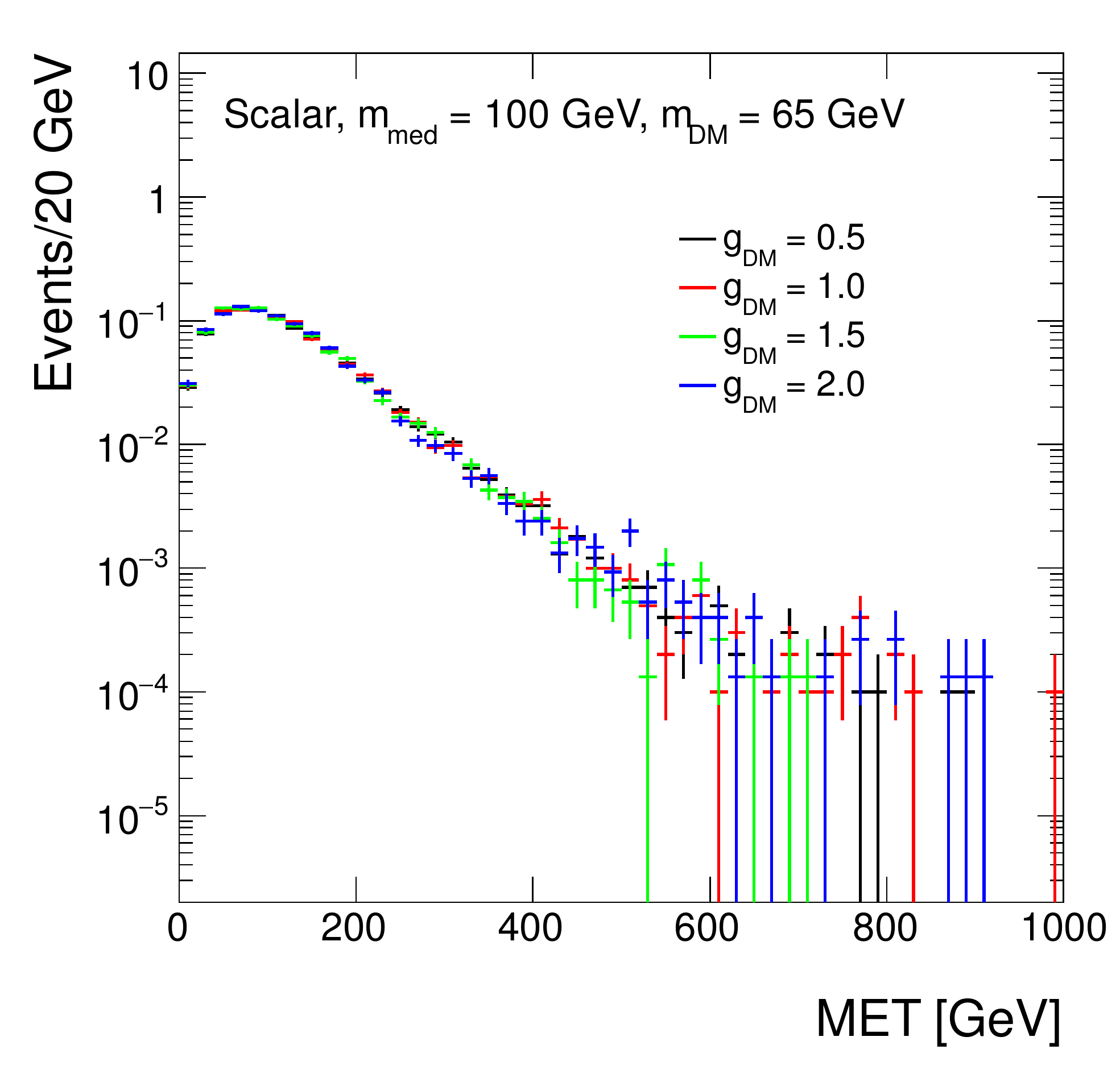}
		\caption{Missing transverse momentum distributions at generator level in the scalar 
			mediator scenario, for different values of: the new physics coupling $g_b$ (left),
			and the mediator-dark matter coupling \gDM (right).
			\label{fig:metScalarCoupling}}
	\end{center}
\end{figure}

\begin{figure}[hbpt!]
	\begin{center}
		\includegraphics[width=0.75\linewidth]{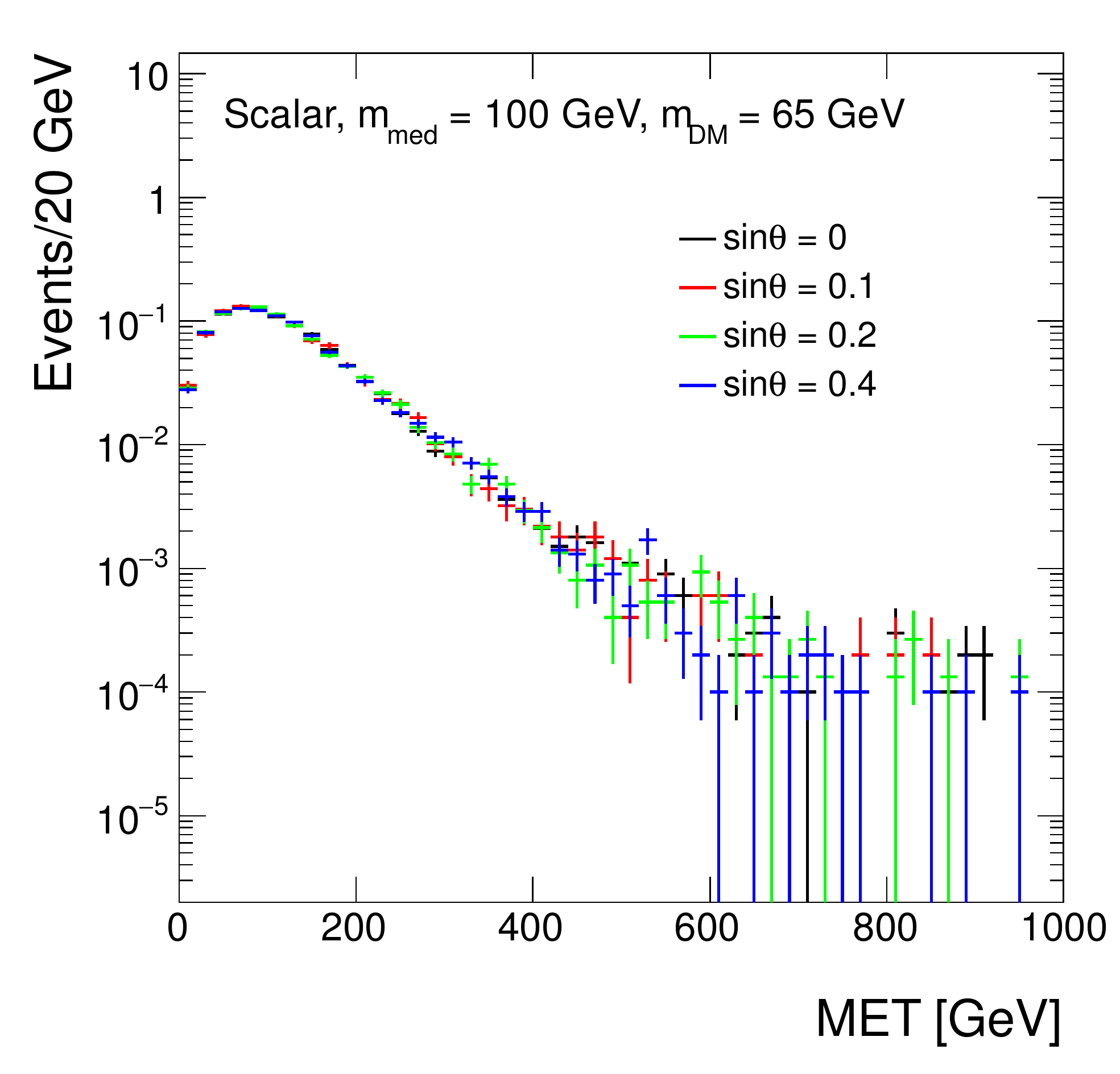}
		\caption{Missing transverse momentum distributions at generator level in the scalar 
			mediator scenario: for different values of the mixing angle $\sin\theta$.
			\label{fig:metScalarCoupling2} }
	\end{center}
\end{figure}

\begin{figure}[hbpt!]
	\begin{center}
		\includegraphics[width=0.75\linewidth]{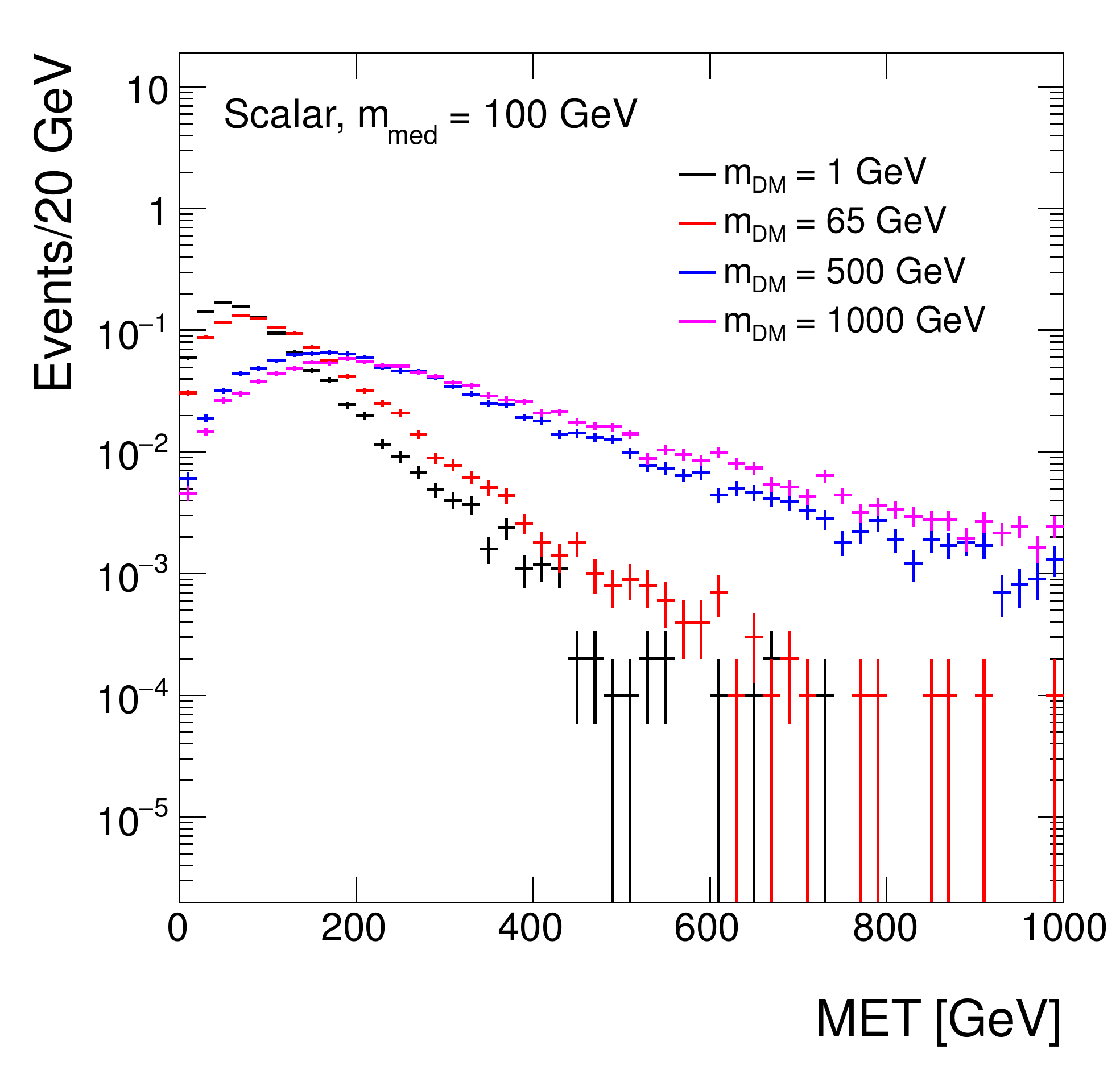}\\
		\includegraphics[width=0.75\linewidth]{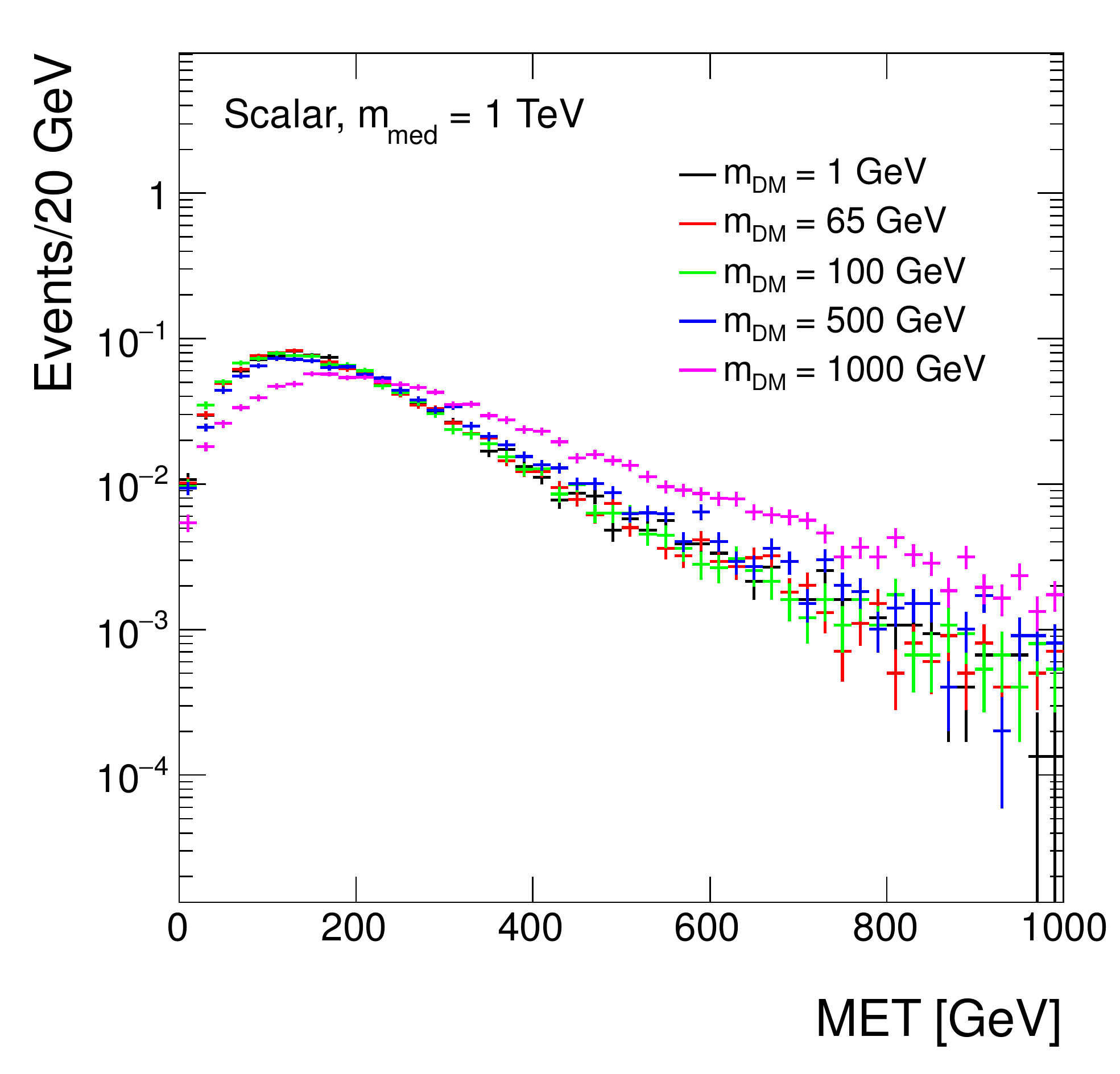}
		\caption{Missing transverse momentum distributions at generator level in the scalar 
			mediator scenario: for different values of the dark matter mass \mDM 
			and a mediator mass of $\mMed = 100~\gev$ (left) and $\mMed = 1~\tev$ (right).
			\label{fig:metScalarMass}}
	\end{center}
\end{figure}

Figure~\ref{fig:metScalarCoupling2} shows that
there is no dependence of the kinematics from the value of this angle, and different values can be obtained via rescaling
the results for this mixing angle according to the relevant cross-section. It can also be observed from Figures~\ref{fig:metScalarMass} and~\ref{fig:metScalarCoupling} 
that the kinematics of this model follows that of the equivalent jet+\MET model: only small changes are observed
in the on-shell region, while the relevant distributions diverge when the mediator is off-shell. 
For this reason, the same grid in \mmed, \mdm as for the scalar mediator
of the jet+\MET search (Table~\ref{tab:mDMmMedScan_SP}) is chosen as a starting point. 
The Yukawa coupling to DM $y_{DM}$ is set to 1, the 
new physics coupling between scalar and SM Higgs $b$ = 3. Results for other values can be obtained via a 
rescaling of the results for these parameters. 

\begin{figure}[hbpt!]
	\centering
	\subfloat[Leading $b-$jet transverse momentum]{
		\includegraphics[width=0.75\linewidth]{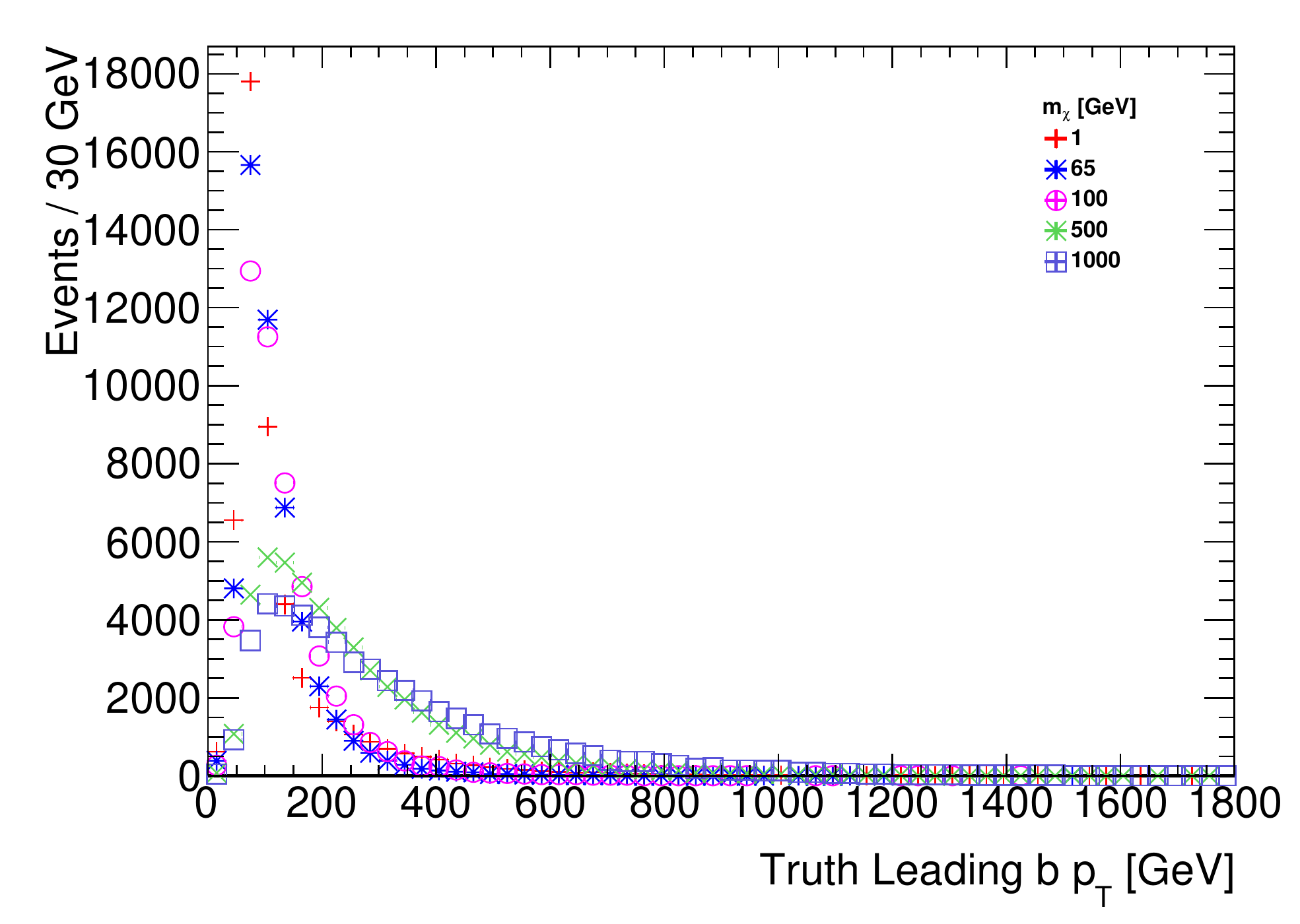} 
	}
	\hfill
	\subfloat[Leading $b-$jet pseudorapidity]{
		\includegraphics[width=0.75\linewidth]{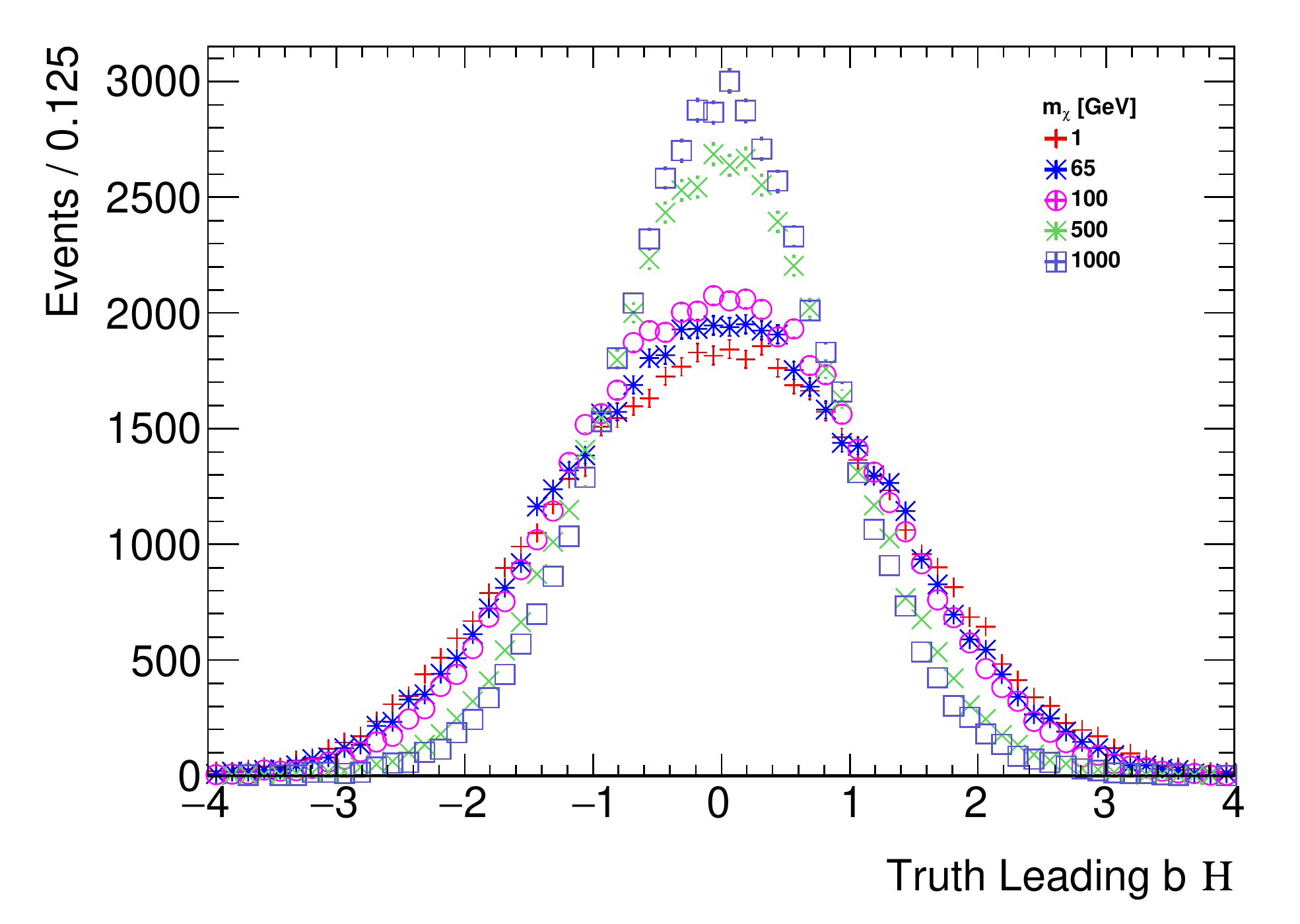} 
	}\hfill
	\subfloat[Angular distance between the two leading $b-$jets]{
		\includegraphics[width=0.75\linewidth]{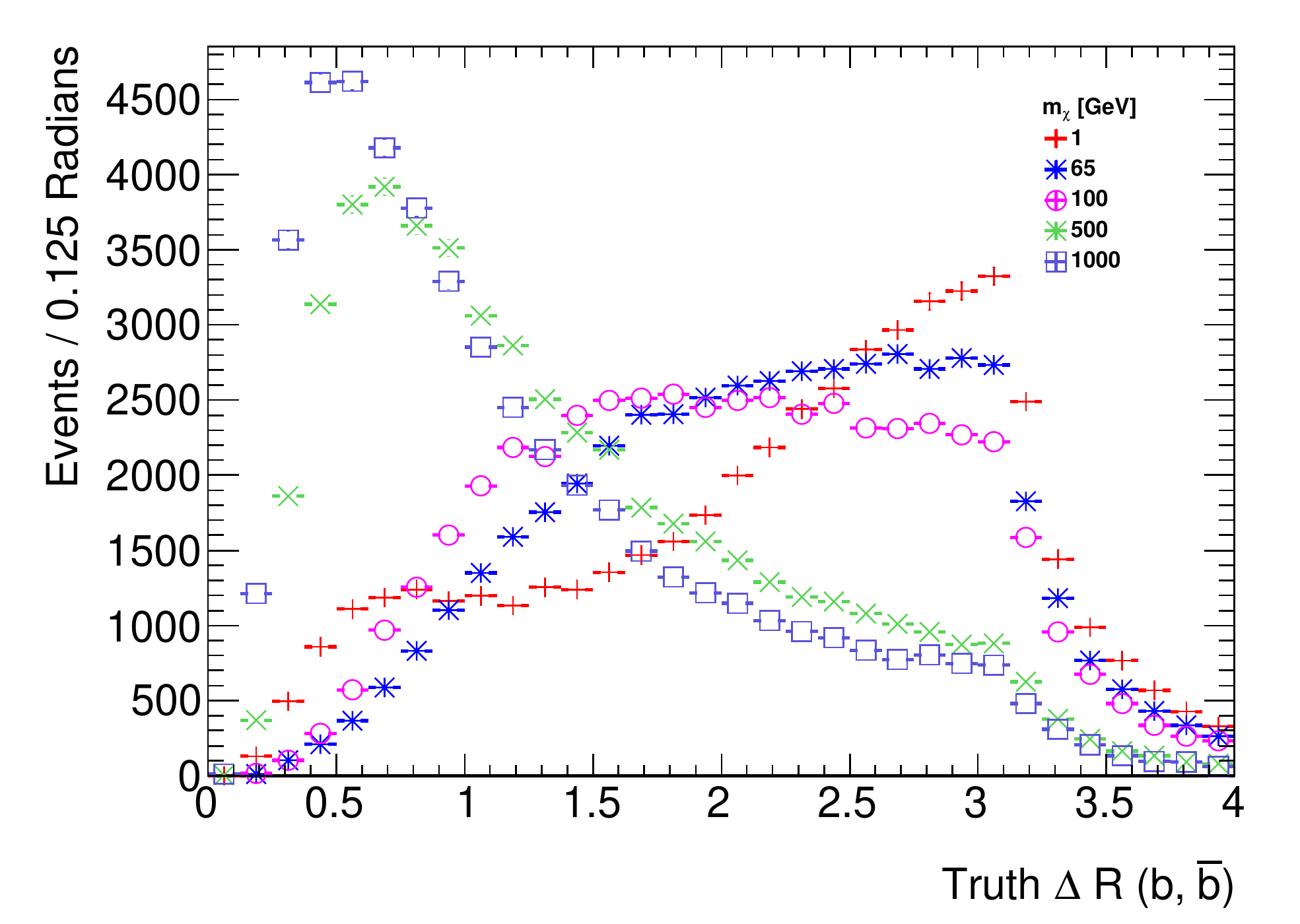} 
	}

	\caption{Comparison of the kinematic distributions for the two leading jets from the Higgs decay in the scalar simplified model, 
		when fixing the new scalar mass to 100~\gev and varying the DM mass. 
		\label{fig:ScalarHbb_100}}
\end{figure}

\begin{figure}[hbpt!]
	\centering
	\subfloat[Leading $b-$jet transverse momentum]{
		\includegraphics[width=0.75\linewidth]{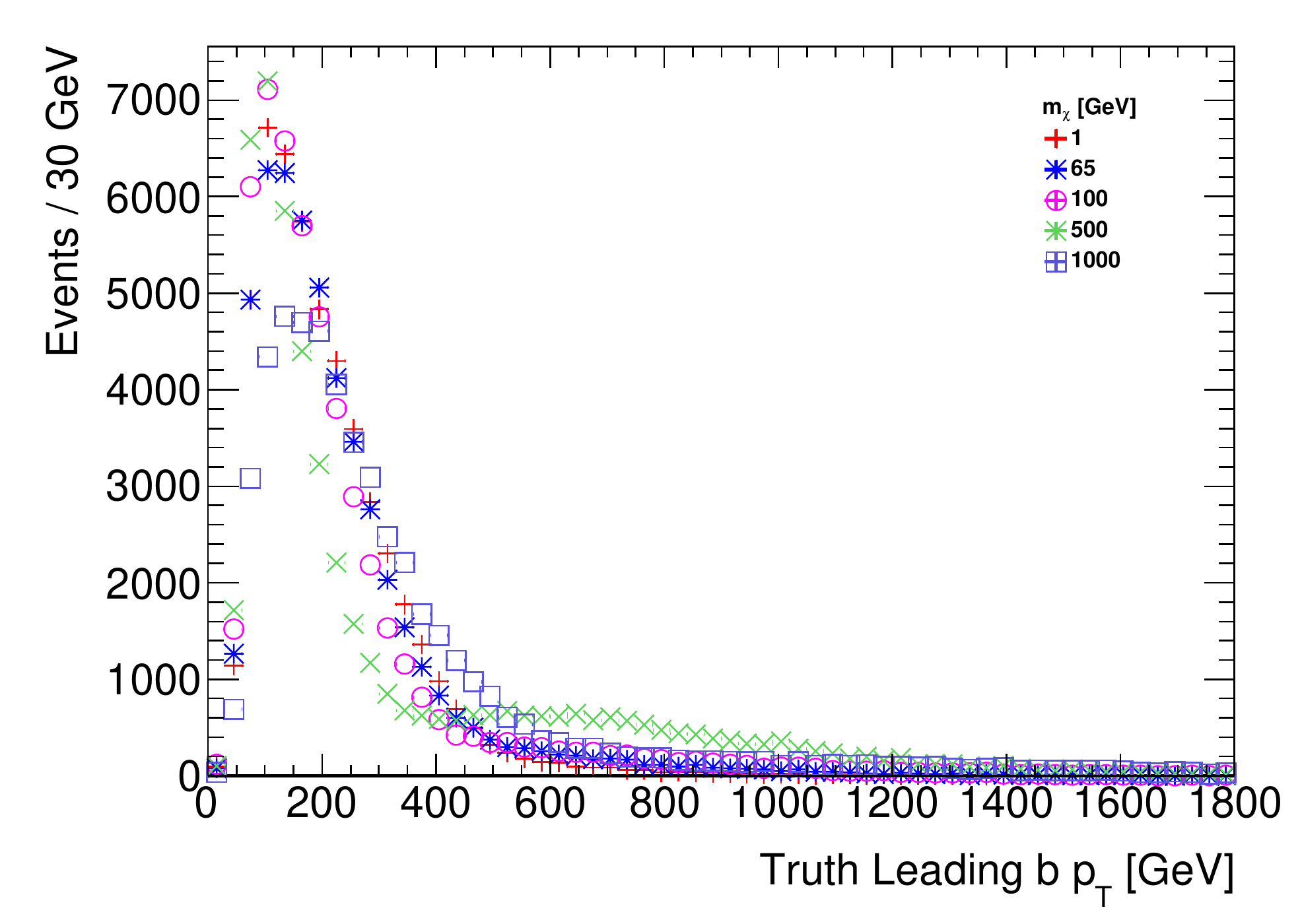} 
	}
	\hfill
	\subfloat[Leading $b-$jet pseudorapidity]{
		\includegraphics[width=0.75\linewidth]{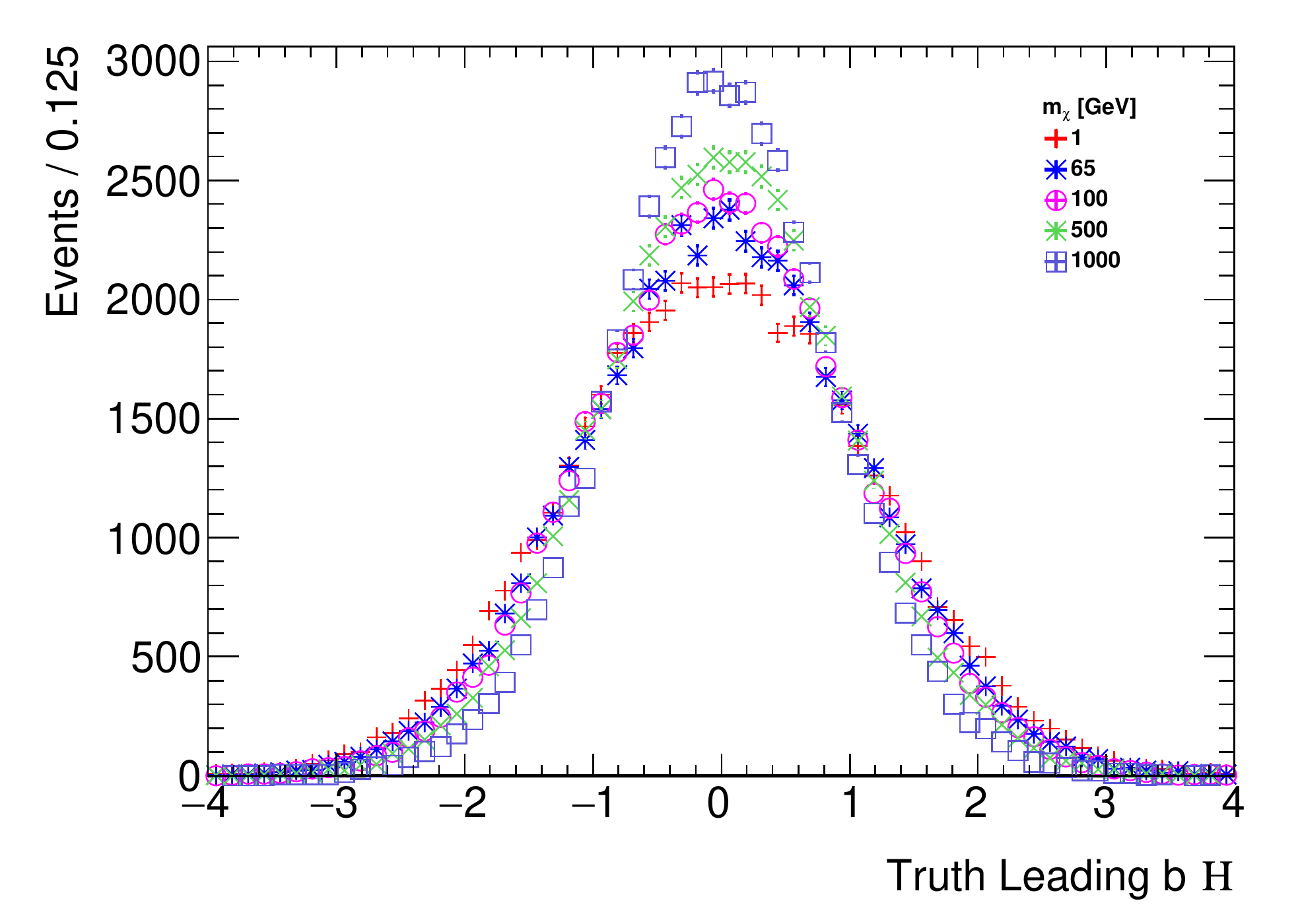} 
	}
	\hfill
	\subfloat[Angular distance between the two leading $b-$jets]{
		\includegraphics[width=0.75\linewidth]{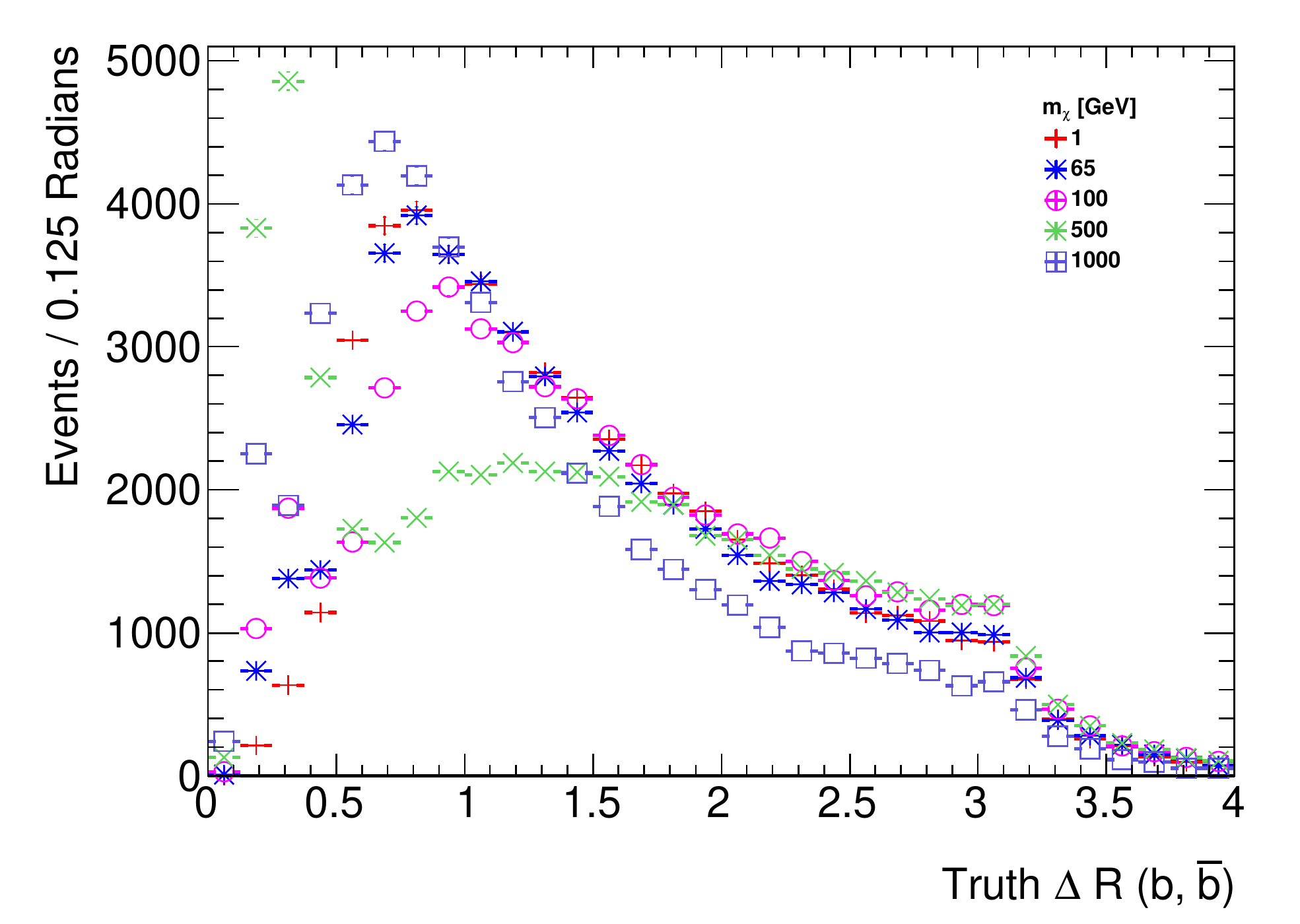} 
	}
	\caption{Comparison of the kinematic distributions for the two leading jets from the Higgs decay in the scalar simplified model, 
		when fixing the new scalar mass to 1000~\gev and varying the DM mass. 
		\label{fig:ScalarHbb_1000}}
\end{figure}

Figs. ~\ref{fig:ScalarHbb_100} and ~\ref{fig:ScalarHbb_1000} show the kinematic distributions for the two leading jets
in the $H \to \bar b b$ decay channel, for two values of the mediator mass and varying the DM mass.  

\subsection{Higgs+\MET signal from 2HDM model with a \Zprime and a new pseudoscalar}

In this simplified model~\cite{Berlin:2014cfa}, a new \Zprime resonance decays to a Higgs boson $h$ 
plus a heavy pseudoscalar state 
$A^0$ in the 2HDM framework, which in turn decays to a DM pair. This model is 
represented in the diagram in Fig. \ref{fig:feyn_prod_monoH} (b).

The motivation for coupling the dark matter to the pseudoscalar is that dark matter coupling to a Higgs or \Zprime boson is generically 
constrained by other signal channels and direct detection.
A reason to consider this model
is that it has different kinematics  due to the on-shell \Zprime production, 
where for heavy \Zprime masses the \MET and $p_T$ spectra are much harder.
This model can satisfy electroweak precision tests and constraints from dijet resonance searches, 
and still give a potentially observable Higgs+\MET signal.
 
 This model comprises two doublets, where $\Phi_u$ couples to up-type quarks and $\Phi_d$ couples to down-type
 quarks and leptons:

 \begin{equation}
 -{\mathcal{L}} \supset  y_u Q \tilde \Phi_u \bar u + y_d Q \Phi_d \bar d + y_e L \Phi_d \bar e  + {\rm h.c.}
 \end{equation}
 
 After electroweak symmetry breaking, the Higgs doublets attain vacuum expectation values $v_u$ and $v_d$, and in unitary gauge the doublets are parametrized as

 \begin{align}
 \Phi_d &= \frac{1}{\sqrt{2}}
 \begin{pmatrix}
 -\sin{\beta} \ H^+ \\ v_d - \sin{\alpha} \ h + \cos{\alpha} \ H - i \sin{\beta} \ A^0
 \end{pmatrix} 
 \quad , \nonumber \\
 \Phi_u &= \frac{1}{\sqrt{2}}
 \begin{pmatrix}
 \cos{\beta} \ H^+ \\ v_u + \cos{\alpha} \ h + \sin{\alpha} \ H + i \cos{\beta} \ A^0
 \end{pmatrix}
 \end{align}
 where $h,H$ are neutral CP-even scalars,
 $H^\pm$ is a charged scalar, and $A^0$ is a neutral CP-odd scalar. 
 In this framework, $\tan{\beta} \equiv v_u/v_d$, and $\alpha$ is the mixing angle that diagonalizes 
 the $h - H$ mass squared matrix. This model also contains an additional scalar singlet $\phi$
 that leads to spontaneous symmetry breaking. 
We take $\alpha = \beta - \pi/2$, in the 
limit where $h$ has SM-like couplings to fermions and 
gauge bosons as per Ref.~\cite{Craig:2013hca}, and $\tan{\beta} \ge 0.3$ 
as implied from the perturbativity of the top Yukawa coupling. 
The Higgs vacuum expectation values lead to $Z-\Zprime$ mass mixing, with a small mixing parameter given by 
 \begin{align}
 \epsilon & \equiv \frac{1}{M_{\Zprime}^2 - M_Z^2} \frac{g g_z}{2 \cos{\theta_w}} ( z_d v_d^2 + z_u v_u^2) \nonumber \\
 & =  \frac{(M_Z^0)^2}{M_{\Zprime}^2 - M_Z^2} \frac{2 g_z \cos \theta_w}{g}  z_u \sin^2 \beta, 
 \label{eq:epsilon}
 \quad
 \end{align}
 where $z_i$ are the $\Zprime$ charges of the two Higgs doublets, and  $g$ and $g_z$ related to the mass-squared
 values in absence of mixing  $(M_Z^0)^2 = g^2(v_d^2+ v_u^2)/(4\cos^2{\theta_w}) $ and
 $(M_{Z'}^0)^2 = g_z^2 ( z_d^2 v_d^2 + z_u^2 v_u^2 + z_\Phi^2
 v_\Phi^2)$. 
    
The production cross section for this model scales as $(g_z)^2$, as the decay width for this process
to leading order in $\epsilon$ (Eq.~\ref{eq:epsilon}) is
\begin{equation}
\Gamma_{\Zprime \to h A^0} =  (g_z \sin \beta \cos \beta)^2 \frac{|p|}{24 \pi} \frac{|p|^2}{M_{\Zprime}^2}.
\end{equation}
where the center of mass momentum for the decay products
$\displaystyle |p| = \frac{1}{2 M_{\Zprime}} \sqrt{ (M_{\Zprime}^2 - (m_h + m_{A^0})^2)
(M_{\Zprime}^2 - (m_h - m_{A^0})^2)}$.
The $\Zprime$ can also decay to $Zh$, leading to the same signature if the $Z$ decays invisibly. The partial width for this decay is:
\begin{equation}
\Gamma_{Z' \to hZ}  = (g_z \sin \beta^2)^2 \frac{|p|}{24 \pi} \left( \frac{ |p|^2 }{M_{Z'}^2} + 3 \frac{M_Z^2}{M_{Z'}^2} \right),
\end{equation}. We recommend to generate these two decays separately and combine them at a later stage. 

   
 
\subsubsection{Parameter scan}
 
 The model is described by five parameters:
 \begin{itemize}
 	\item the pseudoscalar mass $M_{A^0}$,
 	\item the DM mass \mDM,
 	\item the \Zprime mass, $M_{\Zprime}$,
        \item $\tan{\beta} (\equiv v_u/v_d)$,
 	\item the \Zprime coupling strength $g_z$. 
 \end{itemize}
 
 To study the signal production and kinematic dependencies on these parameters, 
 we produced signal samples varying each of the five parameters through 
 \madgraph for the matrix element, \pythiaEight for the parton shower, and DELPHES\cite{deFavereau:2013fsa}  for a parameterized detector-level simulation.
 
\begin{figure}[htpb!]
\centering
\subfloat[\MET distribution]{
	\includegraphics[width=0.75\linewidth]{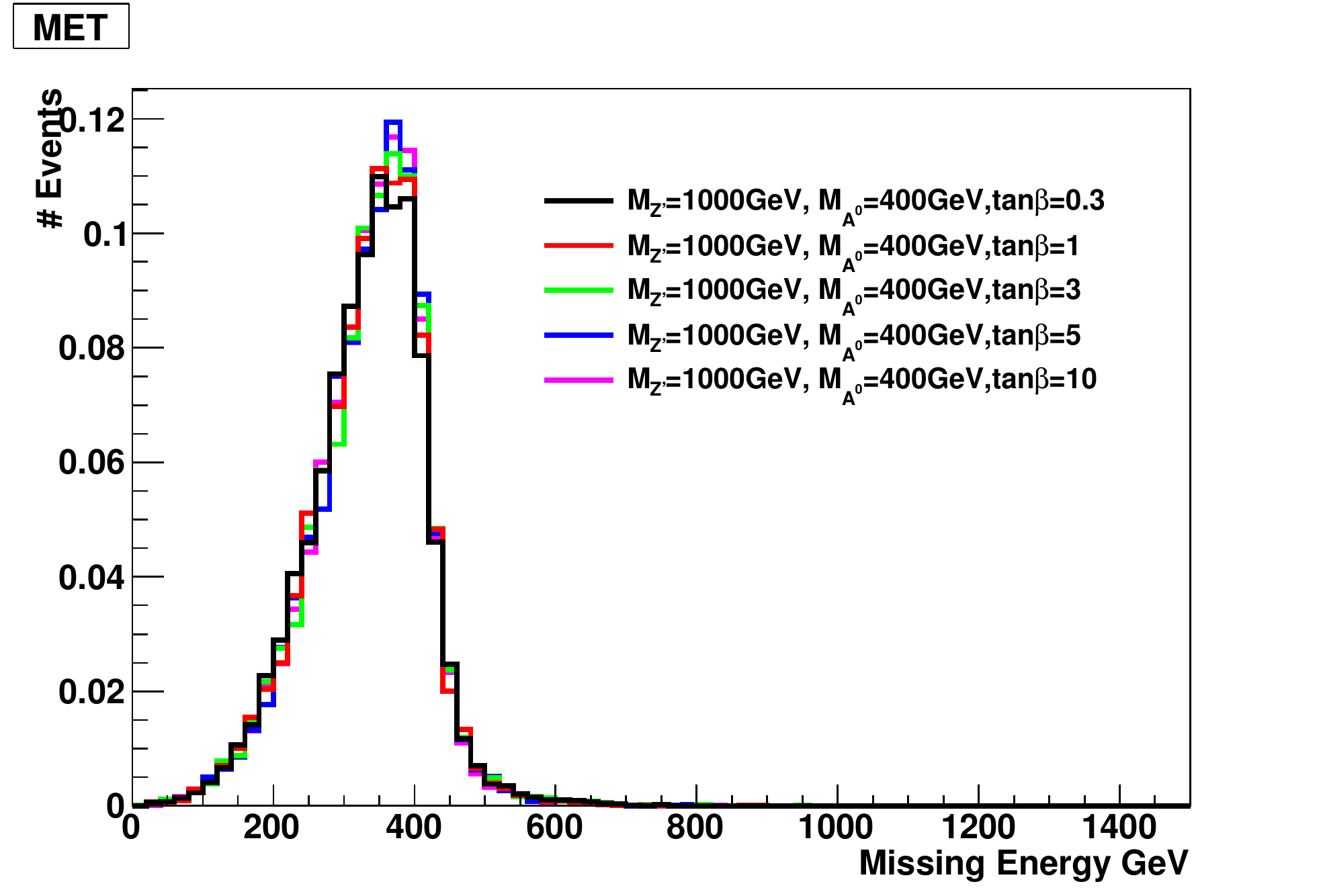}
}
\hfill
\subfloat[$\Delta\phi$ distance between the two $b-$ jets]{
	\includegraphics[width=0.75\linewidth]{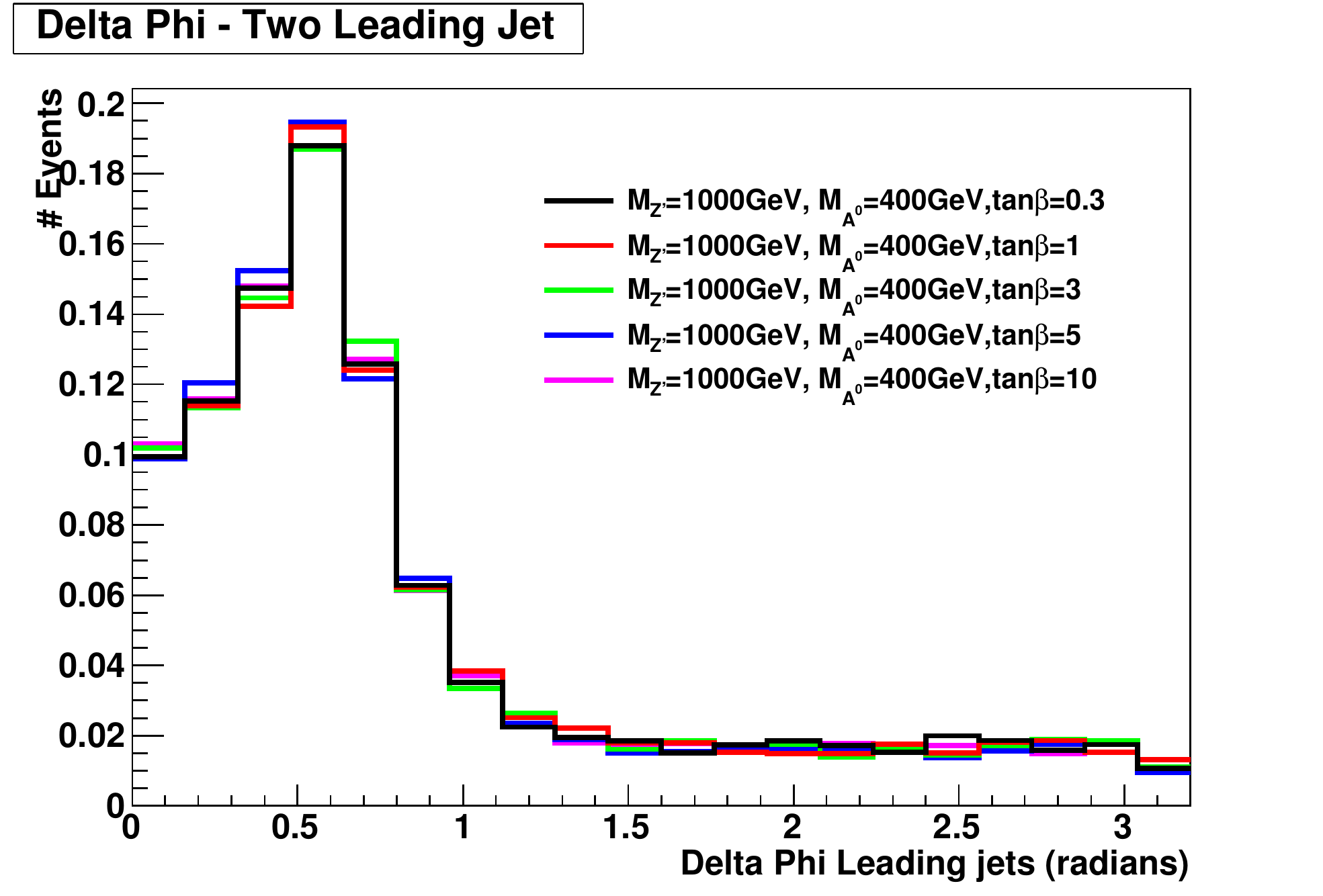}
}
\caption{Kinematic distributions of the signal process varying $\tan{\beta}$, in the case of a Higgs boson decaying into two $b$ quarks,
	after parameterized detector simulation: no kinematic dependence is observed}
\label{fig:DMH_tanbeta}
\end{figure}

 As seen in Fig.~\ref{fig:DMH_tanbeta}, variations of $\tan{\beta}$ does not lead to any kinematic 
 difference and the production cross section simply scales as a function of $\tan{\beta}$. Hence 
we recommend to fix $\tan{\beta}$ to unity in the signal generation.

Similarly, variations of $g_z$ do not lead to any kinematic changes. 
The value of $g_z$ for a given $M_{\Zprime}$ and $\tan \beta$ can be set according to the maximum value allowed by electroweak global 
fits and dijet constraints, as described in~\cite{Berlin:2014cfa}. Since this parameter does not influence the kinematics, 
we leave it up to individual analyses on whether they generate benchmark points only according to these external constraints.

\begin{figure}[htpb!]
	\centering
	\subfloat[\MET distribution]{
		\includegraphics[width=0.75\linewidth]{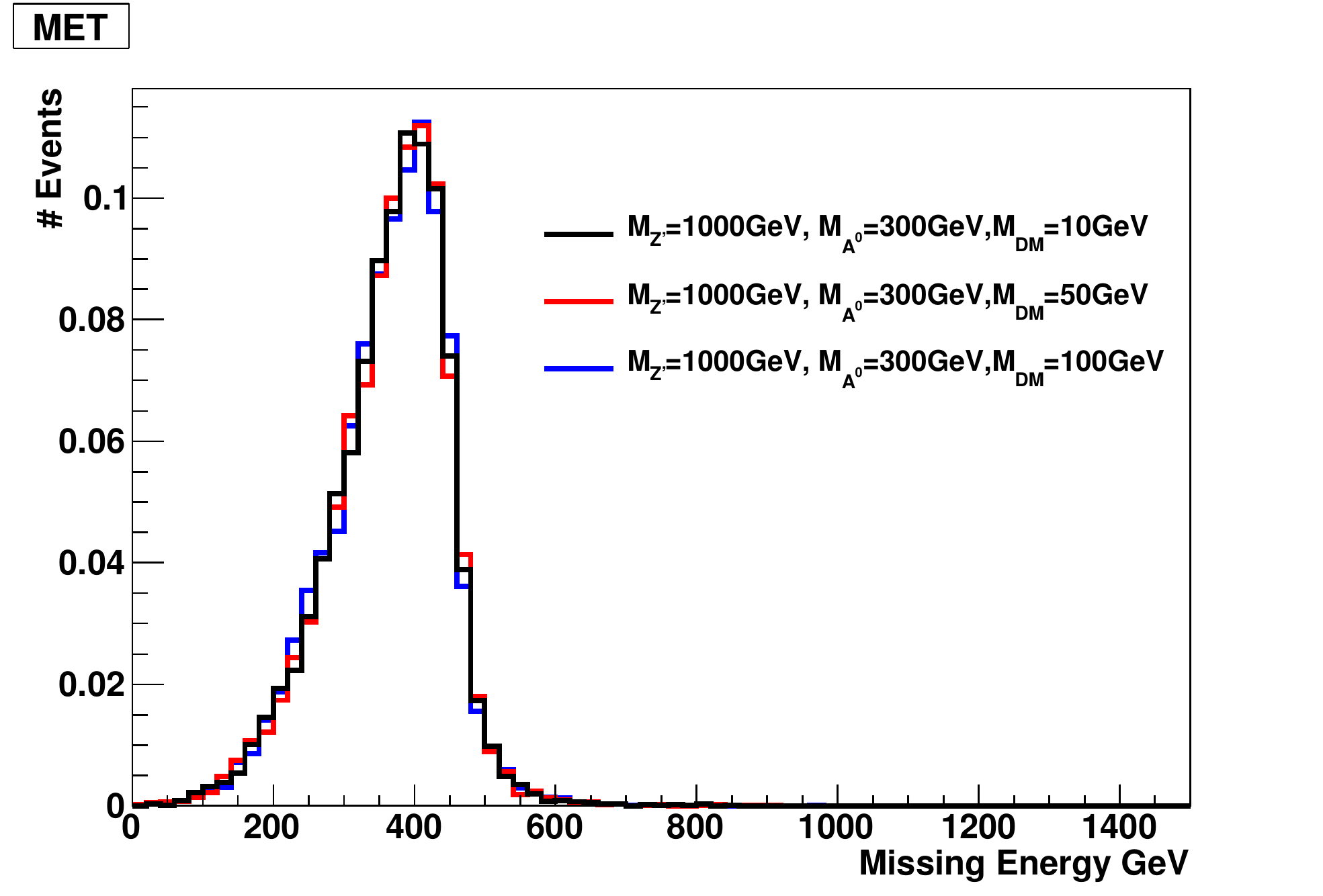}
	}
	\hfill
	\subfloat[$\Delta\phi$ distance between the two $b-$ jets]{
		\includegraphics[width=0.75\linewidth]{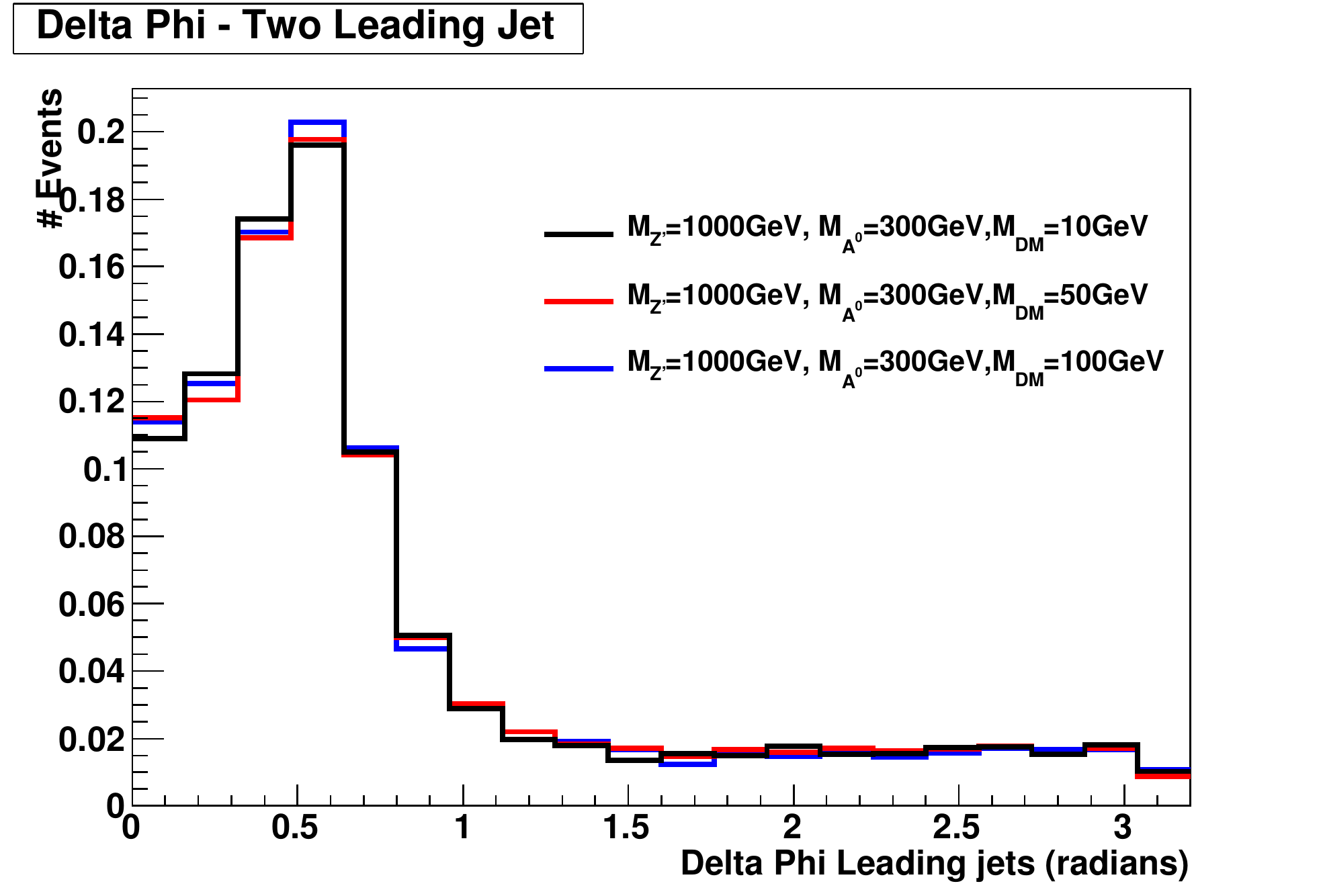}
	}
	\caption{Kinematic distributions of the signal process varying \mDM: minimal kinematic dependency on \mDM as expected when $A^0$ is produced on-shell. Plots shown for $M_{\Zprime}=1000$~\gev, $M_{A^0}=300$~\gev.}
	\label{fig:DMH_mdm}
\end{figure}

\begin{figure}[hbpt!]
  	\centering
  		\includegraphics[width=0.75\linewidth]{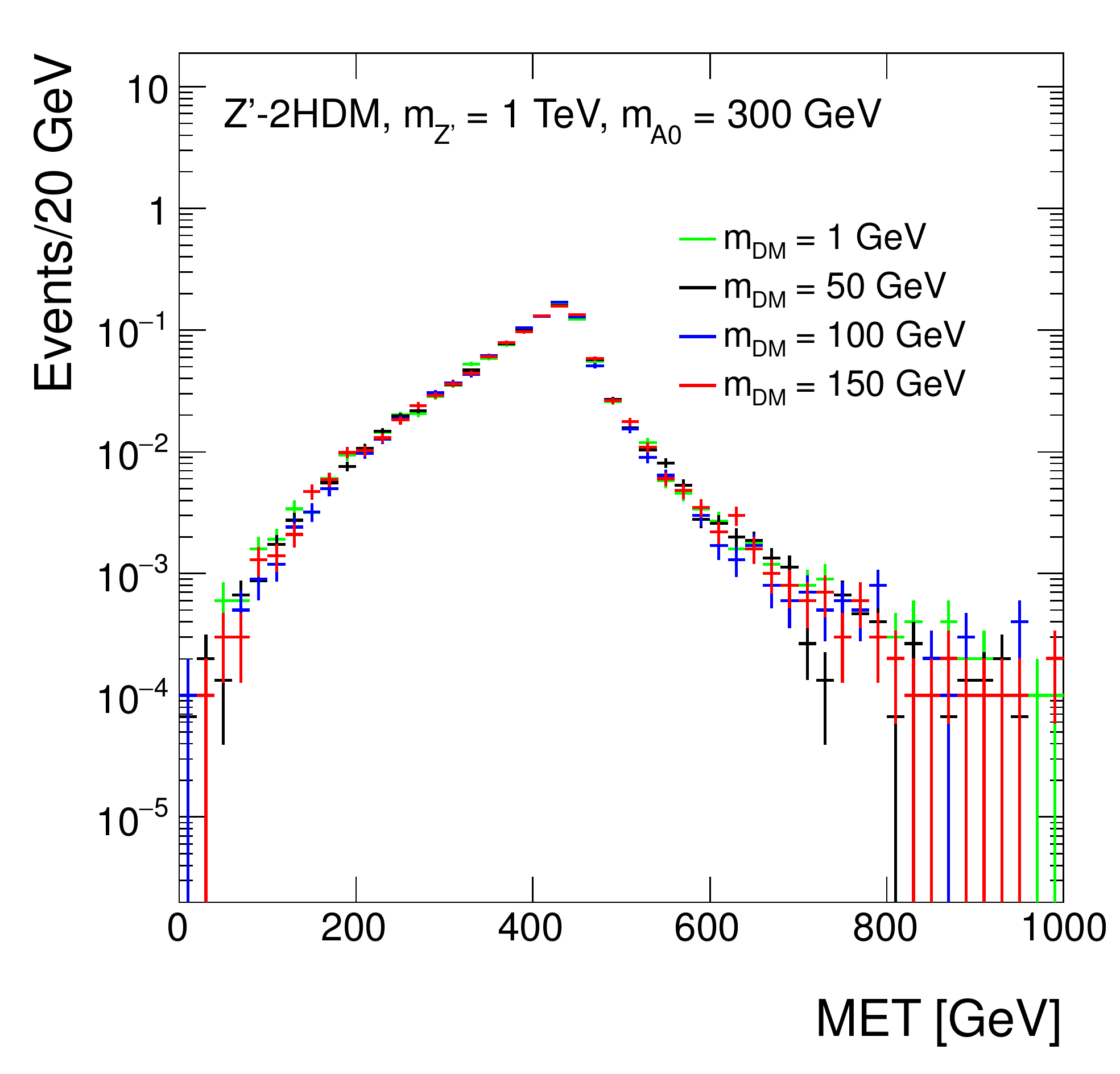}
  		\includegraphics[width=0.75\linewidth]{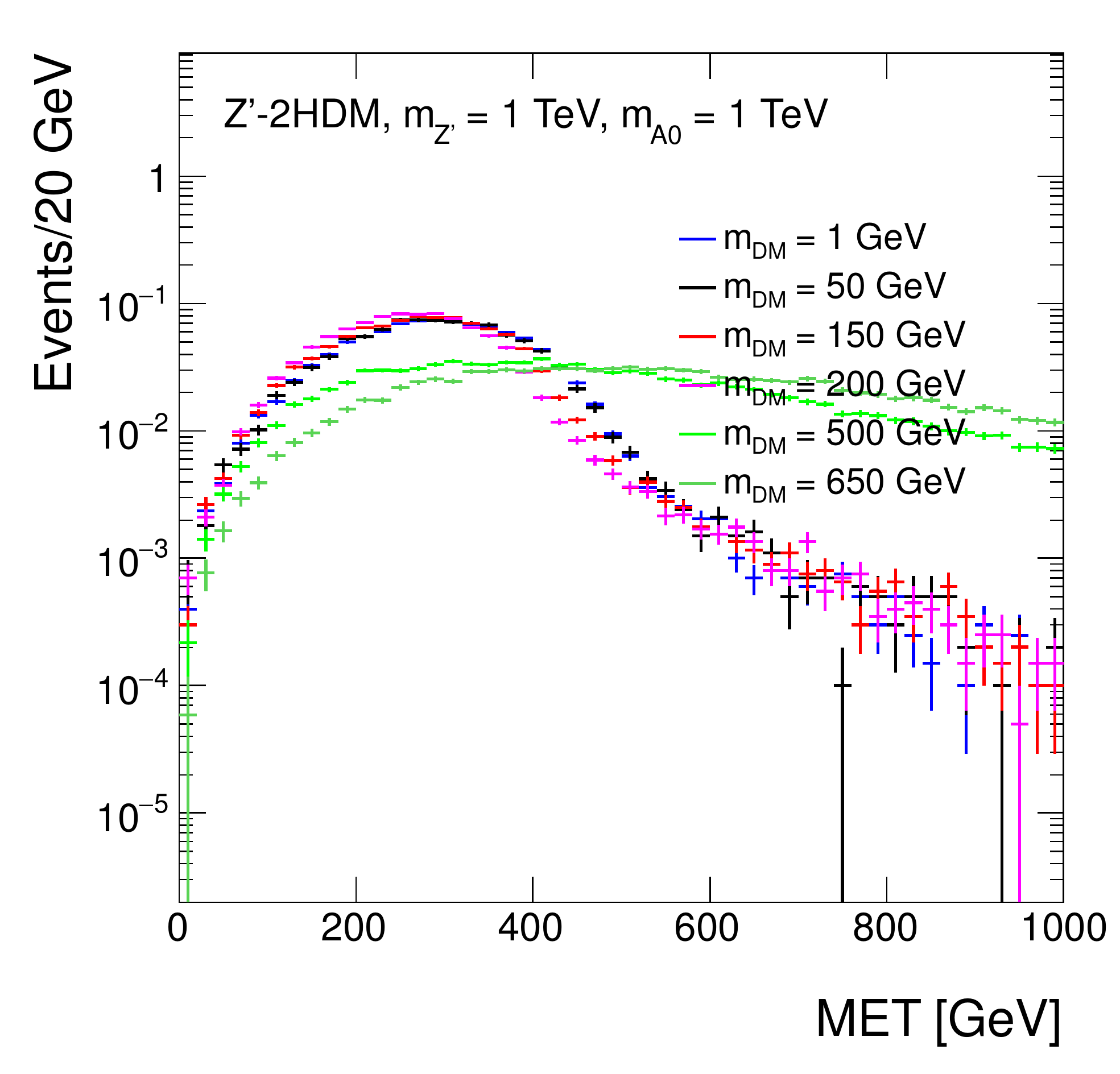}
  		\caption{Missing transverse momentum distributions at generator level in the \Zprime+2HDM 
  			scenario for different values of the dark matter mass \mDM, with 
  			$m_{\Zprime}$ = 1~\tev and $m_{A^0}$ = 300~\gev (left) and $m_{A^0}$ = 1~\tev (right).
  			\label{fig:zprimeDecay}}
  \end{figure}

Since the DM pair are produced as a result of the decay of $A^0$, there are minimal kinematic changes when varying \mDM
as long as $\mDM<M_{A^0}/2$ so that $A^0$ production is on-shell, as shown in Fig.~\ref{fig:DMH_mdm} and~
\ref{fig:zprimeDecay} (before detector simulation). 

We recommend to produce signal events for a fixed $g_z=0.8$, $\tan{\beta}=1$ and $\mDM=100$~\gev. For these values, we scan the 2-D parameter space of ${M_{\Zprime}, M_{A^0}}$ with $M_{\Zprime}=600, 800, 1000, 1200, 1400$~\gev, and $M_{A^0}=300, 400, 500, 600, 700, 800$~\gev with $M_{A^0} < M_{\Zprime}-m_h$, for a total of 24 points. The choice of scan is justified by the sensitivity study in~\cite{Berlin:2014cfa}: the expected LHC sensitivity for Run-2 is up to $M_{\Zprime} \sim 1.5$~\tev.
For the parameter scan, the DM mass is fixed to 100~\gev. For two $M_{\Zprime}$, $M_{A^0}$ value sets, we vary the DM mass to obtain sample cross section for rescaling results. 
All LO cross sections for the various parameter scan points are reported in Appendix~\ref{app:EWSpecificModels_Appendix}.
The parameter scan excludes the off-shell region, as the cross-sections are suppressed and the LHC would not have any
sensitivity to these benchmark points in early data. 

\begin{figure}[htpb!]
 	\centering
 	\subfloat[\MET distribution]{
 		\includegraphics[width=0.8\linewidth]{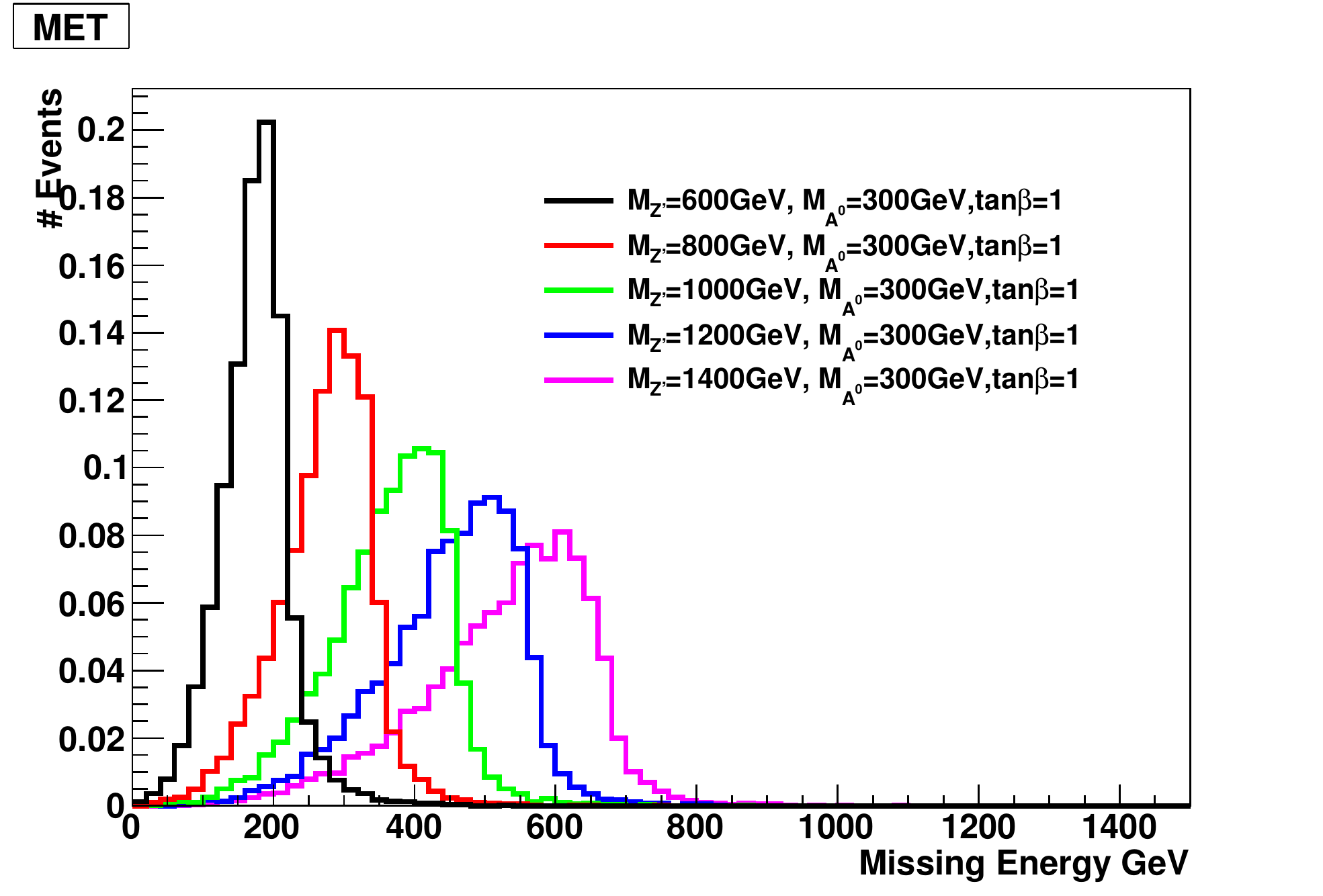}
 	}
 	\hfill
 	\subfloat[Leading $b-$jet $p_T$  distribution]{
 		\includegraphics[width=0.8\linewidth]{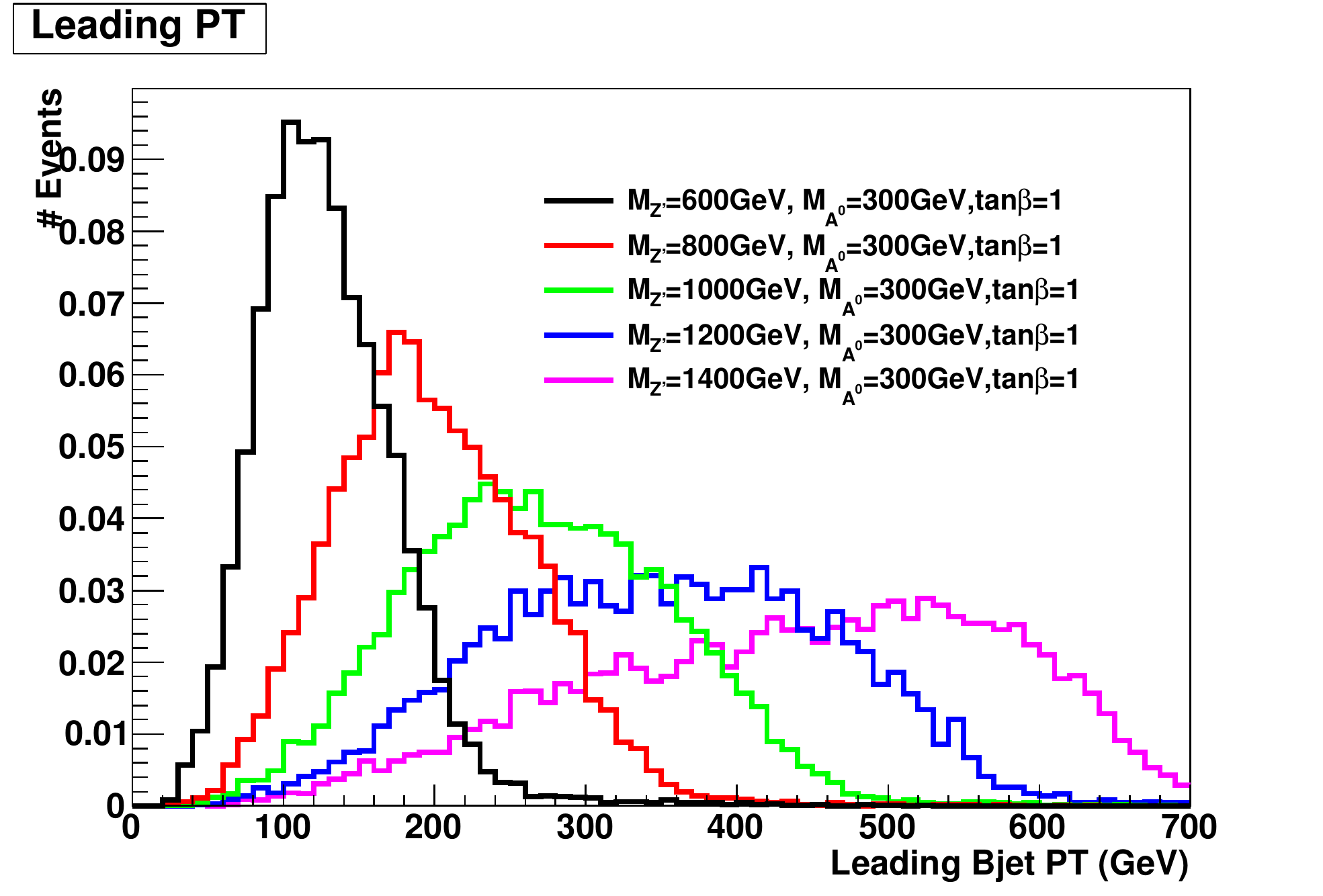}
 	}
 	\hfill
 	\subfloat[$\Delta\phi$ distance between the two $b-$ jets]{
 		\includegraphics[width=0.8\linewidth]{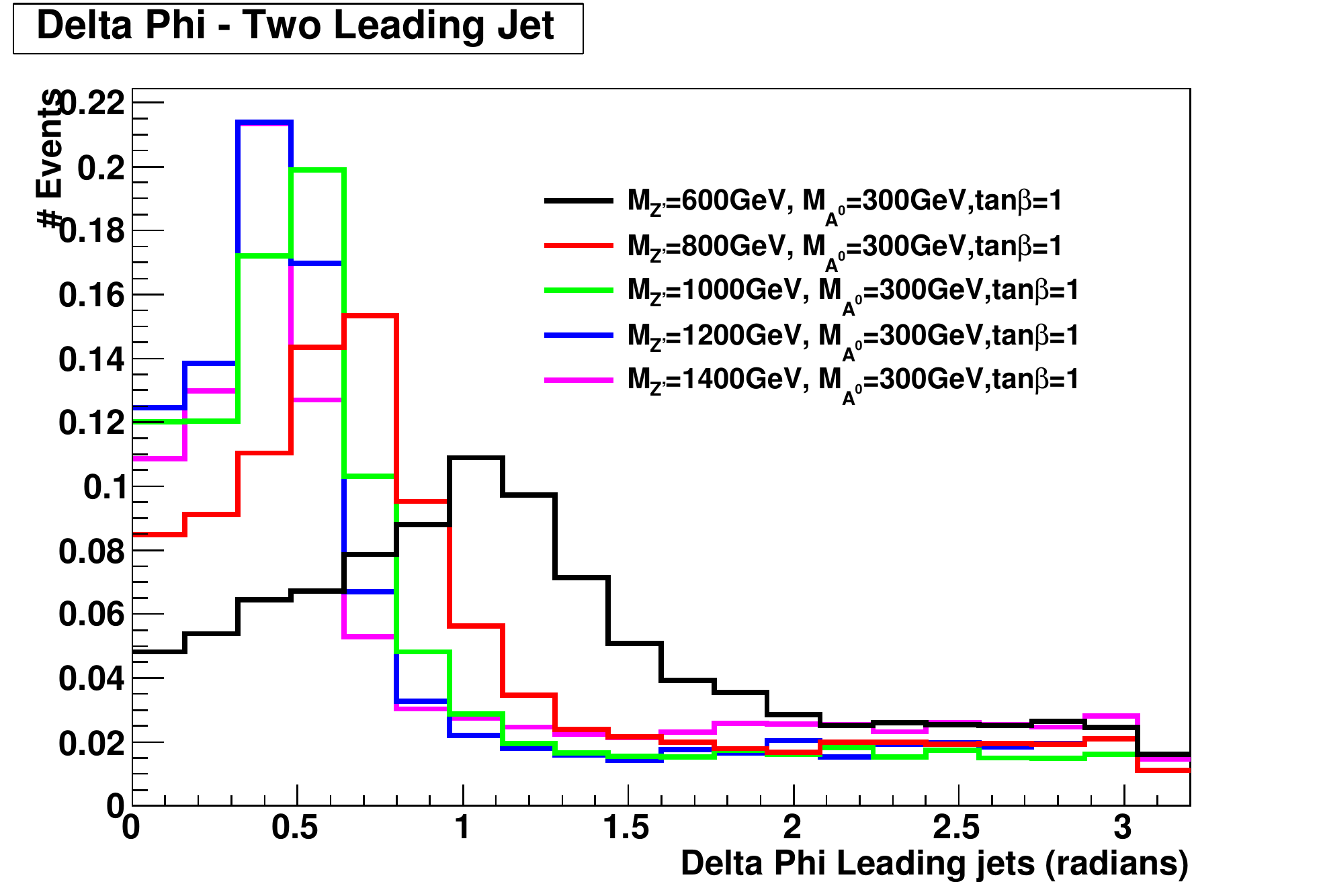}
 	}
 	
 	\caption{Kinematic distributions of the signal process varying $M_{\Zprime}$, for $\mDM=100$~\gev, $M_{A^0}=300$~\gev.}
 	\label{fig:DMH_mzp}
 \end{figure}

\begin{figure}[htpb!]
   	\centering
   	\subfloat[\MET distribution]{
   		\includegraphics[width=0.8\linewidth]{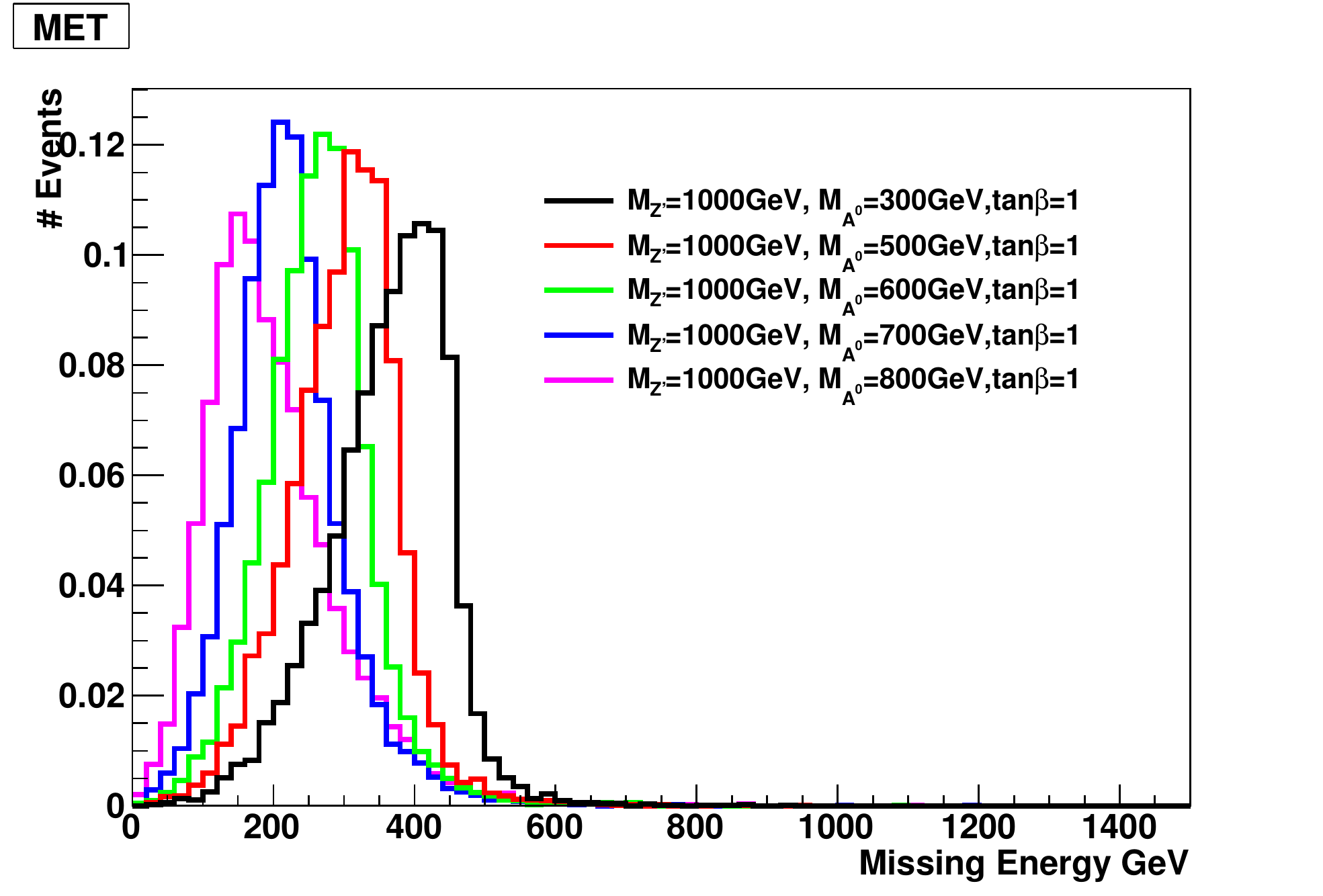}
   	}\hfill
   	\subfloat[Leading $b-$jet $p_T$ distribution]{
   		\includegraphics[width=0.8\linewidth]{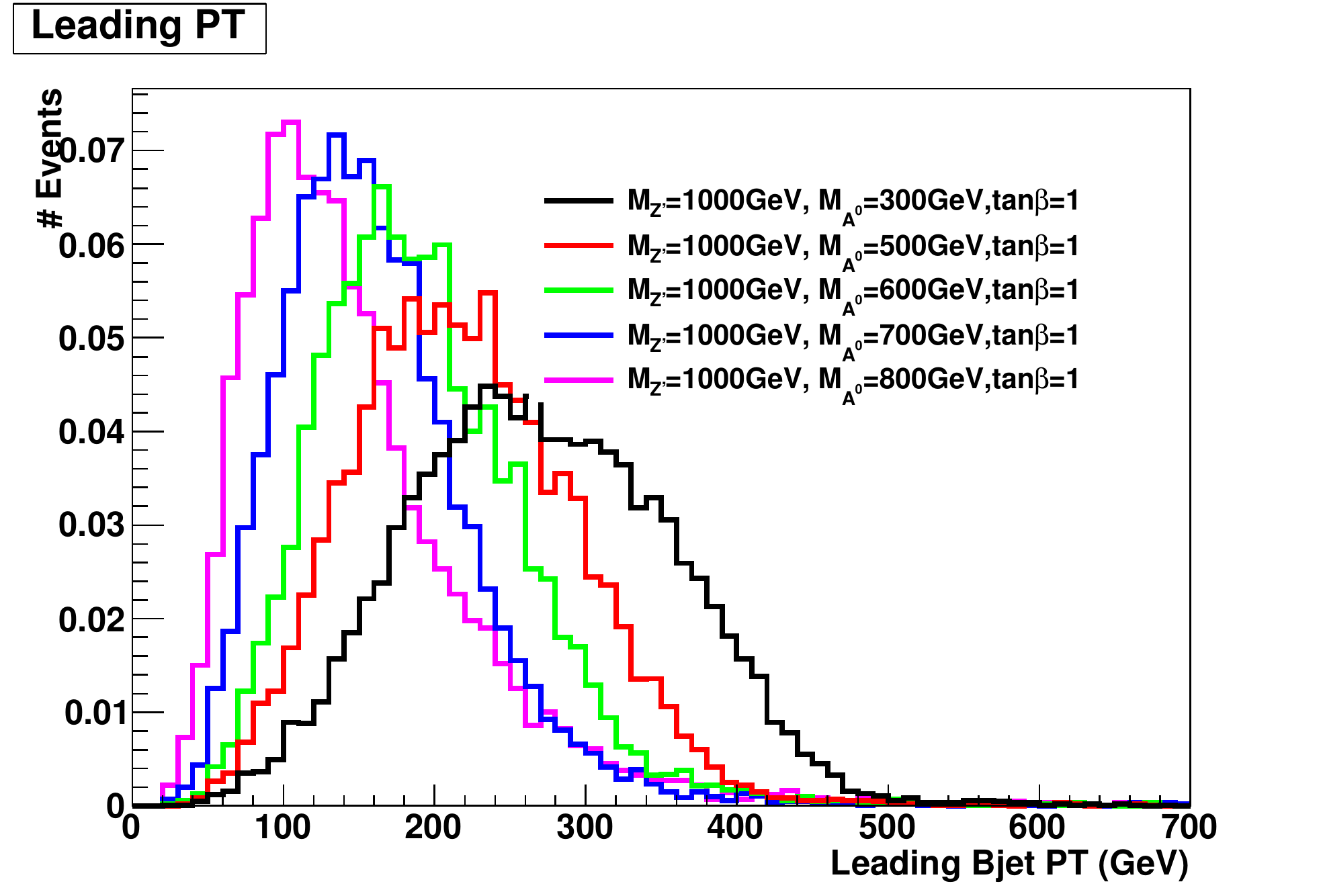}
   	}
   	\hfill
   	\subfloat[$\Delta\phi$ distance between the two $b-$ jets]{
   		\includegraphics[width=0.8\linewidth]{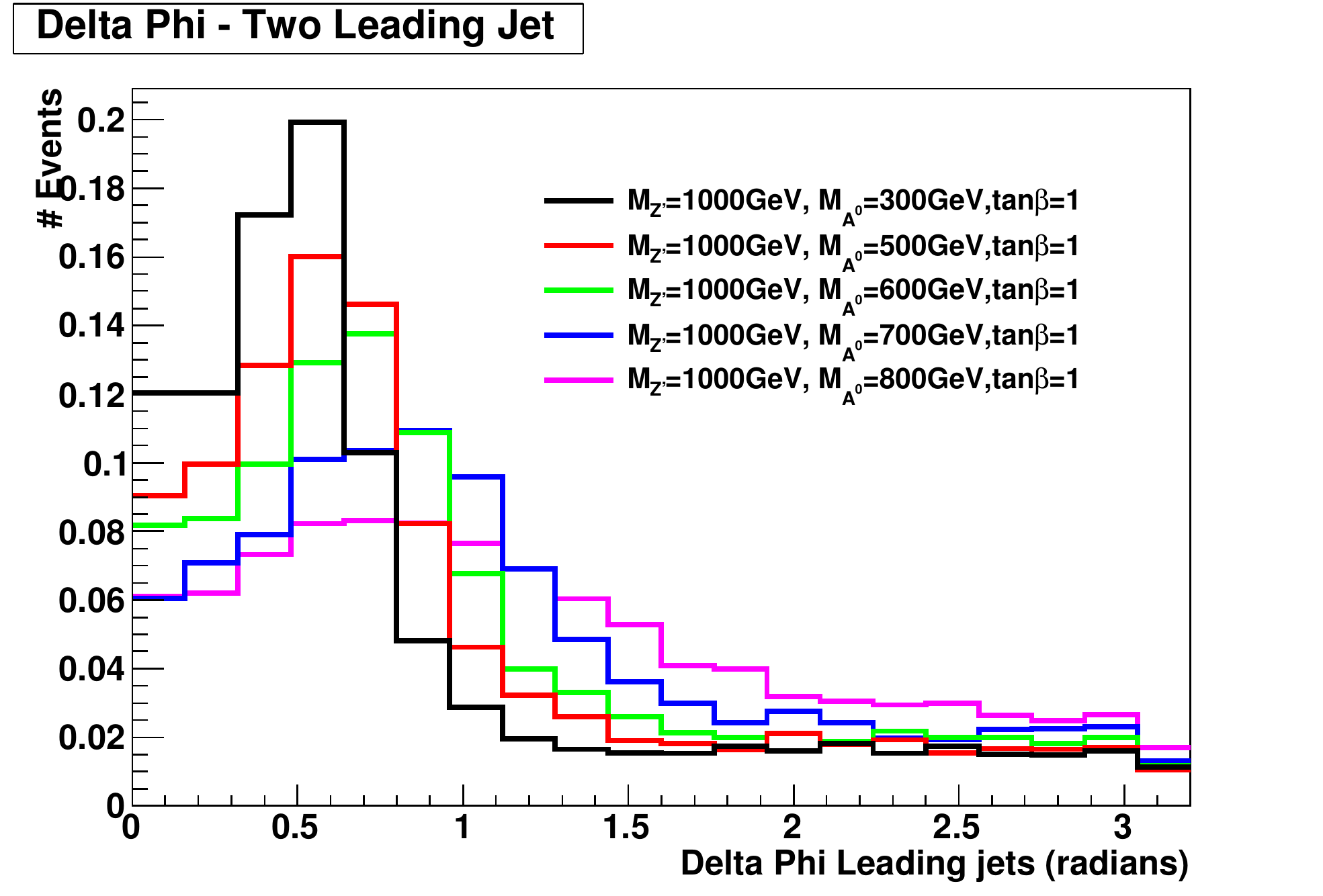}
   	}
   	\caption{Kinematic distributions of the signal process varying $M_{A^0}$, for $\mDM=100$~\gev, $M_{\Zprime}=1000$~\gev.}
   	\label{fig:DMH_ma0}
   \end{figure}

The kinematic distributions with varying $M_{\Zprime}$ for fixed $M_{A^0}$ are shown in Fig.~\ref{fig:DMH_mzp}, while the dependency on $M_{A^0}$ is shown in Fig.~\ref{fig:DMH_ma0}. 
  
\begin{figure}[htpb!]
  	\centering
  	\subfloat[\MET distribution]{
  		\includegraphics[width=0.8\linewidth]{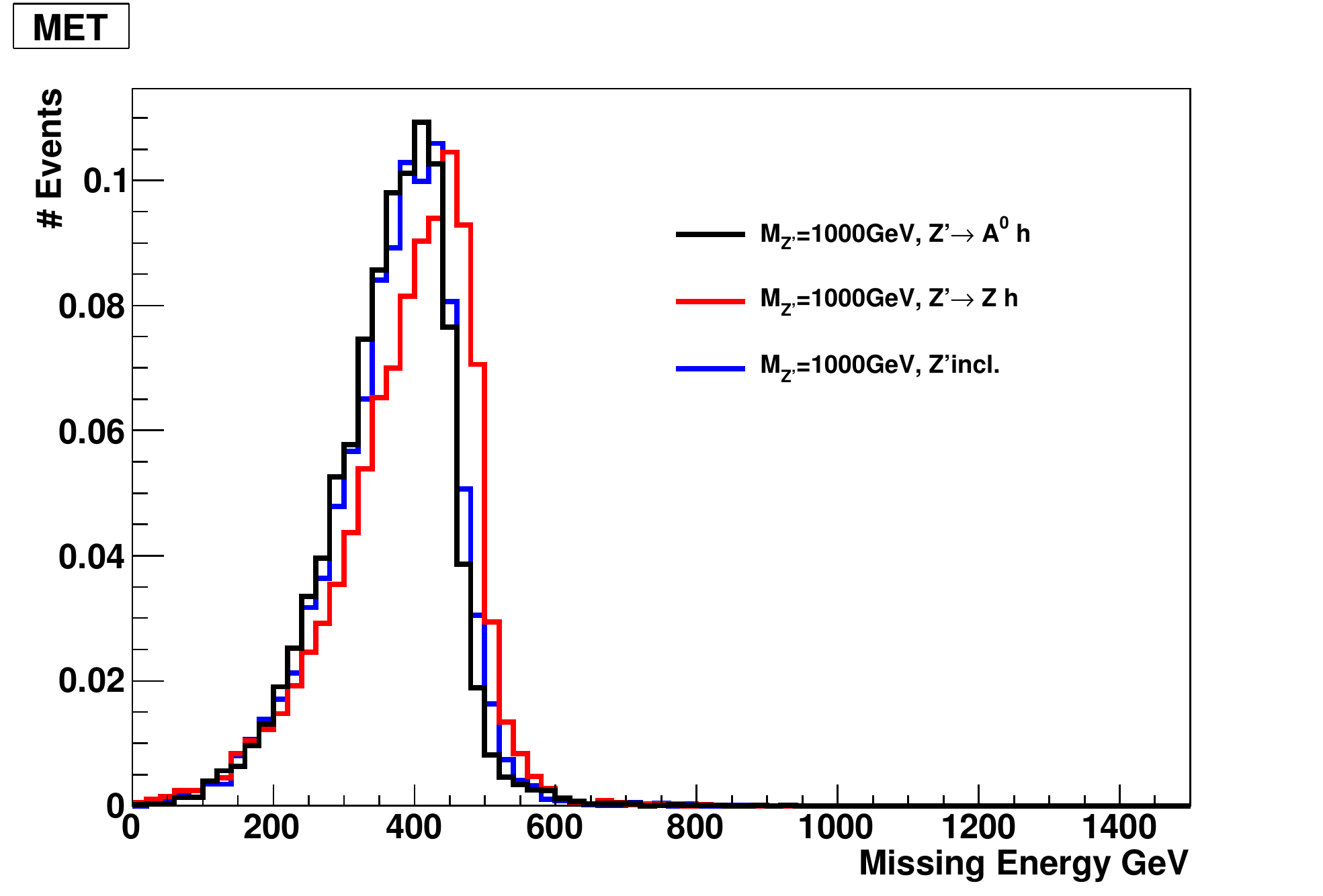}
  	}\hfill
  	\subfloat[Leading $b-$jet $p_T$ distribution]{
  		\includegraphics[width=0.8\linewidth]{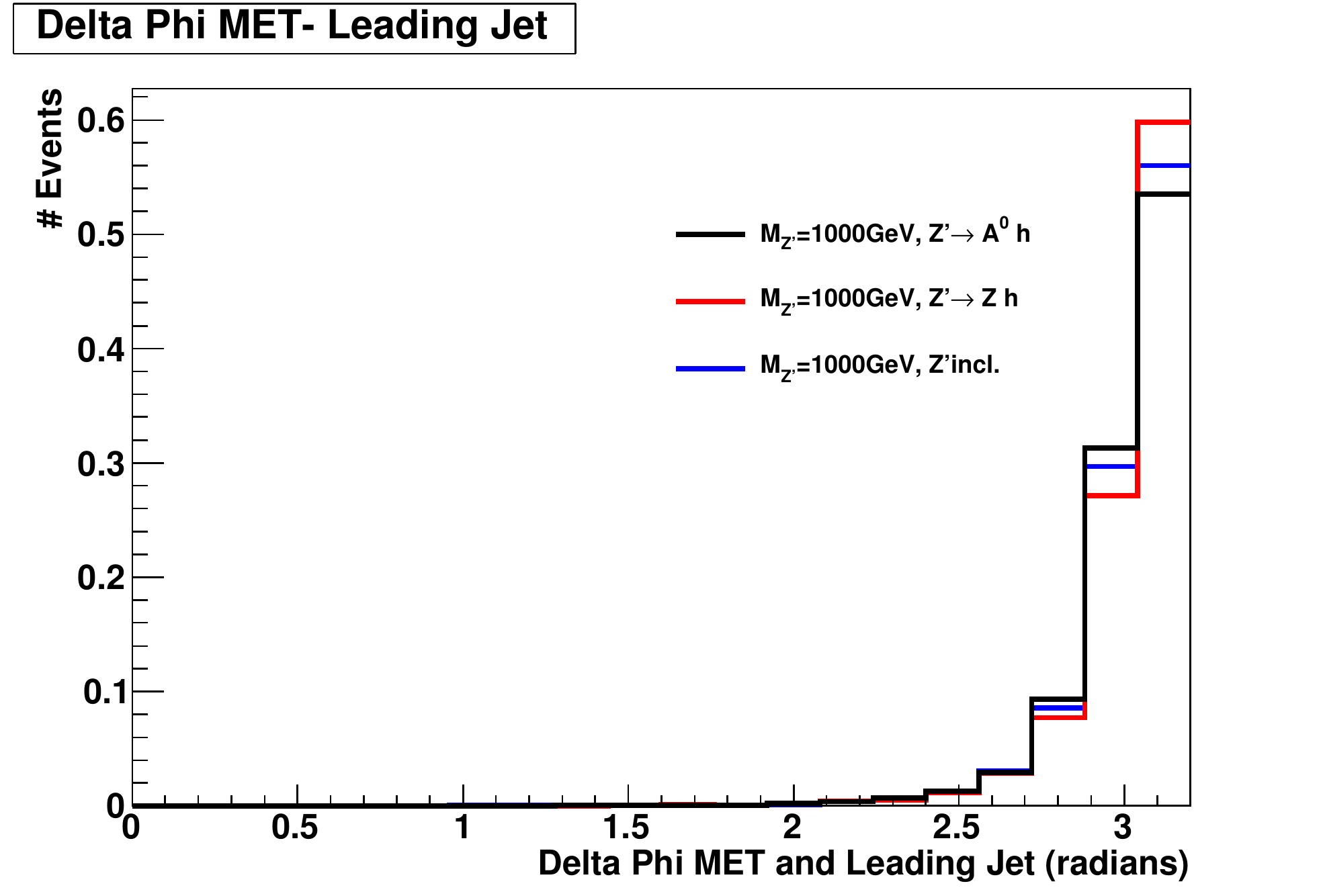}
  	}
  	\hfill
  	\subfloat[$\Delta\phi$ distance between the two $b-$ jets]{
  		\includegraphics[width=0.8\linewidth]{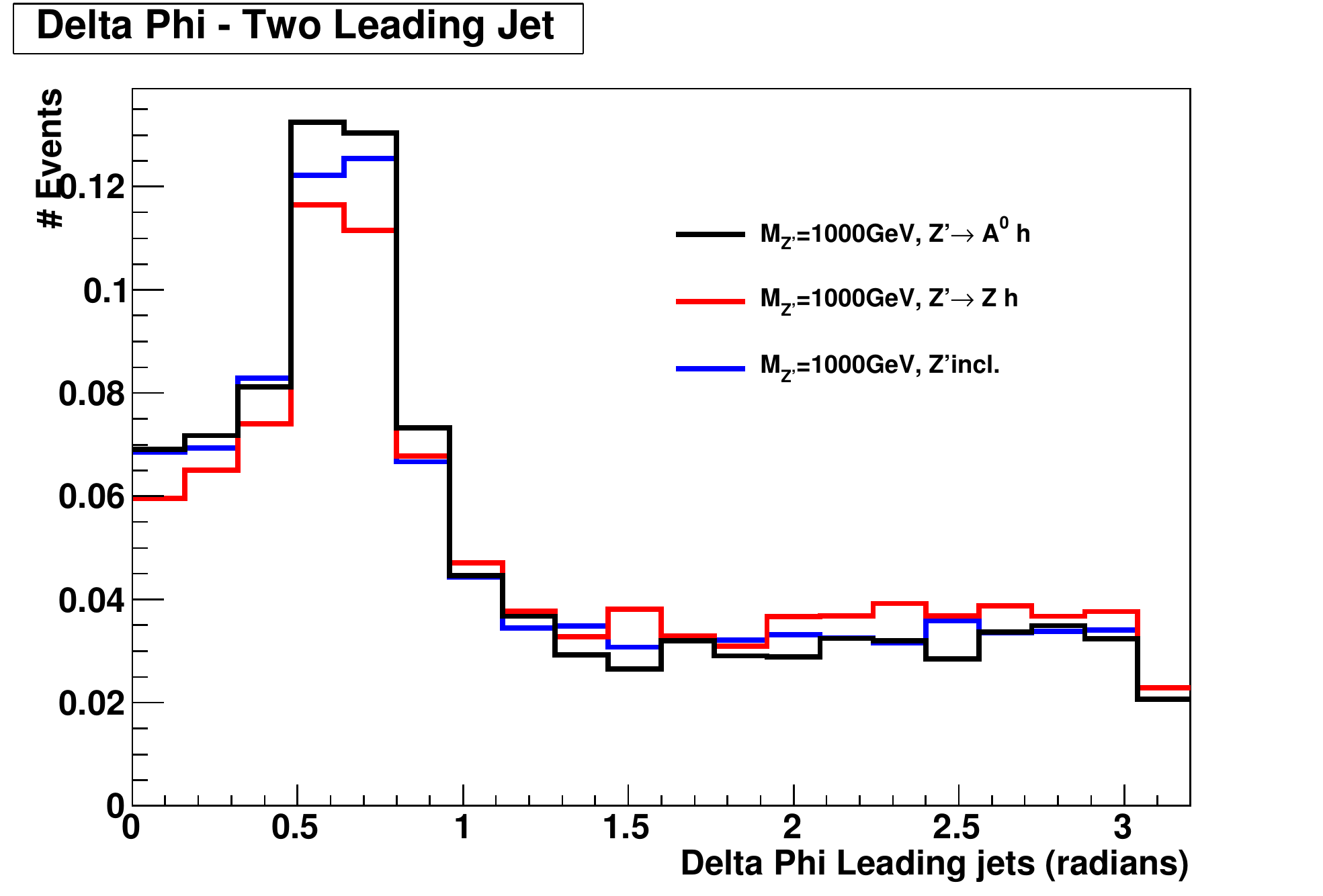}
  	}
  	\caption{Kinematic distributions of $\Zprime \to A^0\,h$ exclusive production, $\Zprime \to Zh$ exclusive production and \Zprime inclusive production for $M_{\Zprime}=1000$~\gev and $M_{A^0}=300$~\gev}
  	\label{fig:DMH_zpincl}
\end{figure}

 This model also allows for an additional source of Higgs plus \MET signal with a similar kinematics (Fig.~\ref{fig:DMH_zpincl}, shown with detector simulation 
 samples) to the signal process from the decay of $\Zprime \to h Z$, where the $Z$ decays invisibly. The partial decay width for the \Zprime is:

 \begin{equation}
 \Gamma_{\Zprime \to hZ}  = (g_z \cos \alpha \sin \beta)^2 \frac{|p|}{24 \pi} \left( \frac{ |p|^2 }{M_{\Zprime}^2} + 3 \frac{M_Z^2}{M_{\Zprime}^2} \right),
 \end{equation}
The values for the \Zprime masses scanned for those samples should follow those of the previous samples, 
namely values of $M_{\Zprime}=600, 800, 1000, 1200, 1400$~\gev.  This signal process has no $M_A$ dependence.
 
\section{EFT models with direct DM-boson couplings}
\label{sec:EFT_models_with_direct_DM_boson_couplings}


The EFT operators considered in this section do not have an implementation 
of a simplified model completion for Dirac fermion Dark Matter available to date. 
They provide kinematic distributions that 
are unique to mono-boson signatures, and that in most cases 
are not reproduced by an equivalent simplified model.\sidenote{Wherever this is
the case, for practical reasons one can only generation a simplified model result 
in the limiting EFT case, as the results can be rescaled and reinterpreted.}

A complete list of effective operators with direct DM/boson couplings for
Dirac DM, up to dimension 7, can be found in~\cite{Cotta:2012nj, Carpenter:2012rg, Crivellin:2015wva}. 
Higher dimensional operators, up to dimension 8, leading to Higgs+\MET signatures,
are mentioned in~\cite{Carpenter:2012rg, Berlin:2014cfa}. The first part of this Section outlines
the main characteristics for a limited number of these models that could be 
considered in early Run-2 searches. 
However, the EFT approximation made for these operators can be problematic, see Ref.~\cite{Berlin:2014cfa} for discussion.
For this reason, model-independent results as in Appendix~\ref{app:Presentation_Of_Experimental_Results} 
should be privileged over considering these operators as realistic benchmarks. 

However, the Forum discussion highlighted that the EFT approach allows
more model-independence when reinterpreting results, and that it is worth still considering
interpretation of the results available in terms of these operators. Furthermore, once simplified models are available
for those operators, EFT results can be used as a limiting case for consistency checks. 
We devote the end of this Section to a discussion on the presentation of results 
from this model, including an assessment of their reliability 
using a conservative procedure that is only dependent on EFT parameters.

The studies in this Section
have been performed using a UFO model within \madgraph v2.2.3, interfaced to \pythia 8 for the parton shower.  
The implementation of these models is discussed further in Section~\ref{sub:EFTModels}.


\subsection{Dimension 5 operators}
\label{sub:EW_EFT_Dim5}

\begin{figure}[!htb]
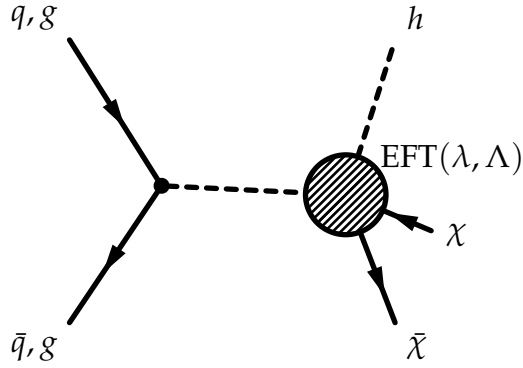

	\centering
	\unitlength=0.005\textwidth
    \vspace{3\baselineskip}
	\begin{feynmandiagram}[modelMonoHEFT]
		\fmfleft{i1,i2}
		\fmfright{o1,o2,ohidden,o3}
		\fmf{fermion}{i2,v1,i1}
		\fmflabel{\Large $q,g$}{i2}
		\fmflabel{\Large $\bar{q},g$}{i1}
		\fmf{dashes}{v1,v2}
		\fmfv{decor.shape=circle,decor.filled=shaded, decor.size=30,label={\Large $\text{EFT}(\lambda,,\Lambda)$},label.a=30,label.d=15}{v2}
		\fmf{dashes}{v2,o3}
		\fmflabel{\Large $h$}{o3}
		\fmf{fermion}{o2,v2,o1}
		\fmflabel{\Large ${\bar{\chiDM}}$}{o1}
		\fmflabel{\Large ${\chiDM}$}{o2}
		\fmfdot{v1}
	\end{feynmandiagram}
    \vspace{3\baselineskip}
	\caption{Diagram for EFT operators giving rise to a Higgs+\MET signature.}
	\label{fig:modelMonoHEFT}
\end{figure}

The lowest dimension benchmark operators we consider are effective dimension 5,
such as the one depicted in Figure~\ref{fig:modelMonoHEFT}.  

Following the notation of~\cite{Carpenter:2012rg},  models
from this category have a Lagrangian that, after electroweak symmetry breaking, 
includes terms such as:

\begin{eqnarray}
\frac{m_W^2}{\Lambda_5^3} ~\bar{\chiDM} \chiDM ~W^{+ \mu} W^{-}_\mu
+ \frac{m_Z^2}{2 \Lambda_5^3} ~ \bar{\chiDM} \chiDM ~ Z^\mu Z_\mu ~,
\end{eqnarray}
where $m_Z$ and $m_W$ are the masses of the $Z$ and $W$ boson, $W^{\mu}$ and $Z^{\mu}$
are the fields of the gauge bosons, $\chiDM$ denotes the Dark Matter fields
and $\Lambda_5$ is the effective field theory scale. Note that these operators are of true dimension 7, 
but reduce to effective dimension 5 once the Higgs vacuum expectation values, 
contained in the W and Z mass terms, are inserted.  
As such, one expects that these operators would naturally arise in UV complete models where Dark Matter 
interacts via a Higgs portal where heavy mediators couple to the Higgs or other fields in an extended Higgs sector. 
In such models the full theory may be expected to contain additional operators with Higgs-Dark Matter couplings~\cite{Djouadi:2012zc}.
The above operator also induces signatures with 
\MET in conjunction with Z and W bosons at tree level, as shown in Fig.~\ref{fig:VPlusMET_EFT},
while at loop level it induces couplings to photon pairs and $Z \gamma$ through W loops.
In these models, a clear relation exists between final states with photons, EW bosons
and Higgs boson. 

\begin{figure}
	\includegraphics[width=0.95\textwidth]{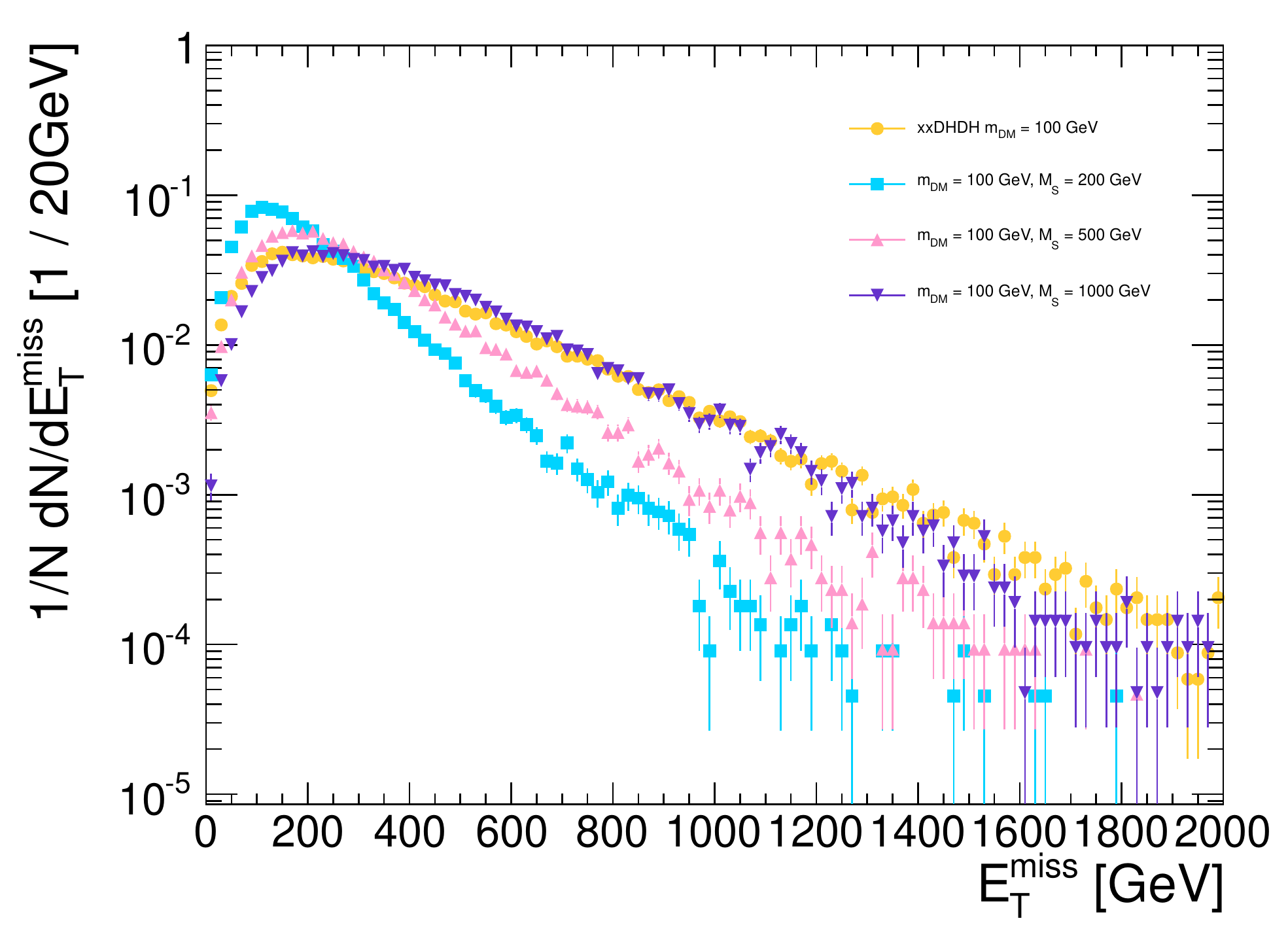}
	\caption{Comparison of the missing transverse momentum for the simplified model
		where a scalar mediator is exchanged in the \schannel and the model including 
		a dimension-5 scalar contact operator, in the leptonic Z+\MET final state. All figures in this Section
		have been performed using a UFO model within \madgraph v2.2.3, interfaced to \pythiaEight for the parton shower.  }
	\label{fig:EW_EFT5_Zlep_MET}
\end{figure}

As shown in Fig.~\ref{fig:EW_EFT5_Zlep_MET}, the 
kinematics of this model can be approximated by that of a simplified model including 
a high-mass scalar mediator exchanged in the \schannel described in Section~\ref{sub:EW_Scalar}. 
For this reason, the list of benchmark models with direct boson-DM couplings for photon, Z and W 
only includes dimension 7 operators: if the scalar model with initial state radiation of an EW boson
is already generated, then its results can be rescaled. 

The Higgs+\MET analysis,
however, will not consider the scalar simplified model as benchmark, due to the very low sensitivity 
in early LHC analyses, and will instead use this dimension 5 operator. 

\subsubsection{Parameter scan}

The two parameters of this model are the scale of new physics $\lambda$ 
and the DM particle mass. SM-DM coupling and new physics scale are related by 
$\gDM = {(246~\gev)}/{\lambda}$.

The initial value of the new physics scale $\lambda$ chosen 
for the sample generation is 3~\tev. This is a convention and does not affect the signal kinematics:
the cross-section of the samples can be rescaled when deriving the constraints on this scale. 
However, more care should be given when rescaling Higgs+\MET operators
of higher dimensions, as different diagrams have a different $\lambda$ dependence. 

The DM mass values for the benchmark points to be simulated are chosen to
span a sufficient range leading to different kinematics, 
that is within the LHC sensitivity for early searches and that is consistent across 
the various signatures and EFT operators. We therefore start the mass scan
at \mdm=1~\gev, where collider experiments are complementary to direct and indirect detection
and choose the last point corresponding to a DM mass of 1~\tev. 
We recommend a scan in seven mass points, namely:
$$
\mdm = { 1, 10, 50, 100, 200, 400, 800, 1300 } \gev. 	
$$

\begin{figure}[hbpt!]
 	\centering
 	\subfloat[Leading $b-$jet transverse momentum]{
 		\includegraphics[width=0.8\linewidth]{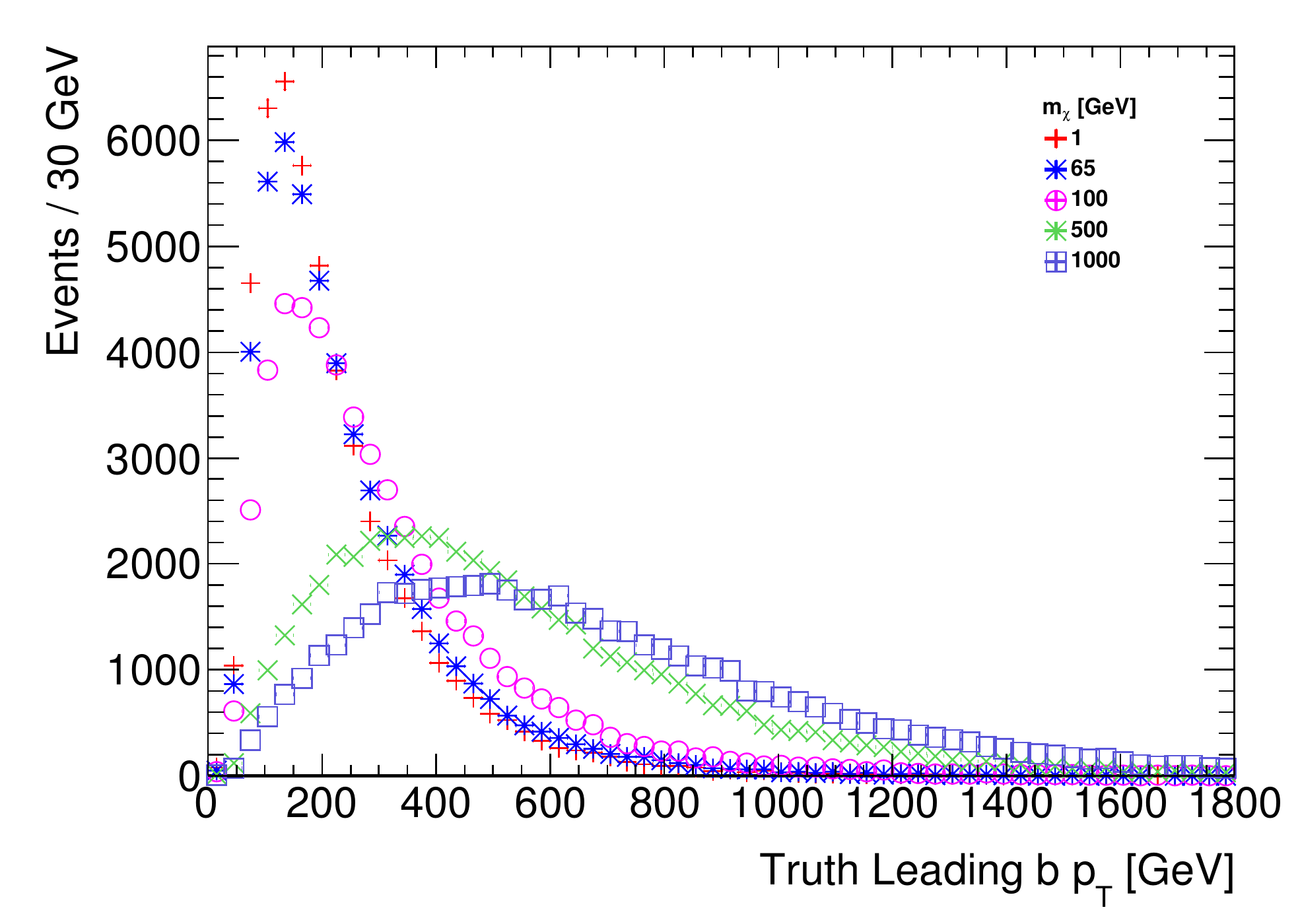} 
 	}
 	\hfill
 	\subfloat[Leading $b-$jet pseudorapidity]{
 		\includegraphics[width=0.8\linewidth]{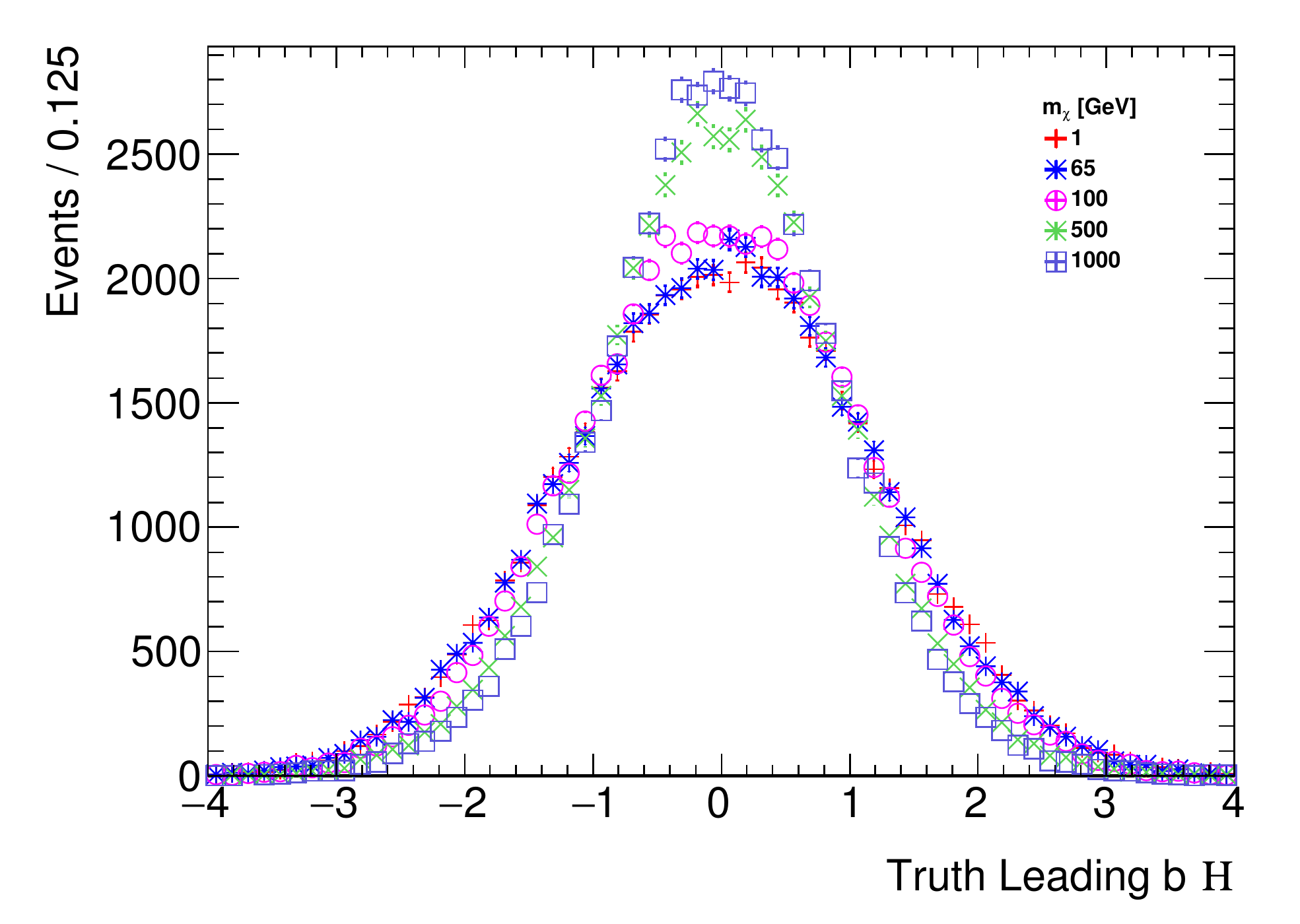} 
 	}
 	\hfill

 	\subfloat[Angular distance between the two leading $b-$jets]{
 		\includegraphics[width=0.8\linewidth]{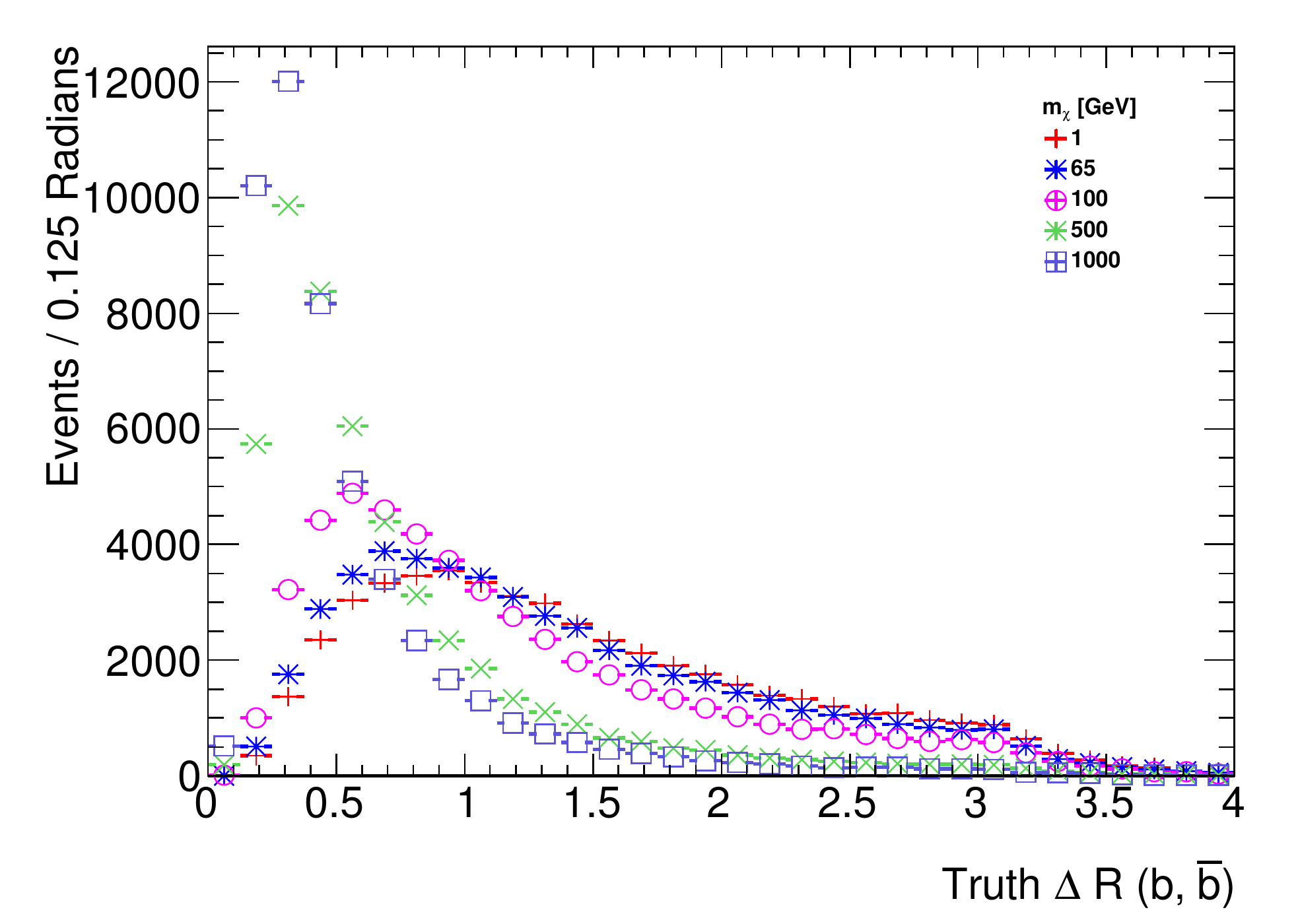} 
 	}
 	\caption{Comparison of the kinematic distributions for the two leading $b-$ jets (from the Higgs decay) in the model with direct interactions
 		between the Higgs boson and the DM particle, when varying the DM mass. 
 		\label{fig:Hbb_Dim5}}
 \end{figure}

A set of kinematic distributions from the Higgs+\MET signature where the Higgs decays 
into two $b-$quarks is shown in Fig.~\ref{fig:Hbb_Dim5}, for points similar to those of the grid scan proposed. 
  
\subsection{Dimension 7 operators}
\label{sub:EW_EFT_Dim7}


The dimension-7 benchmark models  contain the $SU(2)_L \times U(1)_Y$ gauge-invariant couplings between 
DM fields and the kinetic terms of the EW bosons. The CP-conserving scalar couplings of this type can be written as
\begin{equation} \label{eq:Lc1c2}
\frac{c_1}{\Lambda_S^3} \, \bar \chiDM \chiDM \, B_{\mu \nu} B^{\mu \nu }  + \frac{c_2}{\Lambda_S^3} \, \bar \chiDM \chiDM \, W_{\mu \nu}^i W^{i, \mu \nu }  \,.
\end{equation}
Here $B_{\mu \nu} = \partial_\mu B_\nu - \partial_\nu B_\mu$ and $W_{\mu \nu}^i =  \partial_\mu W_\nu^i - \partial_\nu W_\mu^i + g_2 \hspace{0.25mm} \epsilon^{ijk}  \hspace{0.25mm}  W_\mu^j \hspace{0.25mm} W_\mu^k$ are the $U(1)_Y$ and $SU(2)_L$ field strength tensor, respectively, and  $g_2$ denotes the weak coupling constant. In the case of the pseudoscalar couplings, one has instead
\begin{equation} \label{eq:Lc3c4}
\frac{c_1}{\Lambda_P^3} \, \bar \chiDM \gamma_5 \chiDM \, B_{\mu \nu} \tilde B^{\mu \nu }  + \frac{c_2}{\Lambda_P^3} \, \bar \chiDM \gamma_5 \chiDM \, W_{\mu \nu}^i \tilde W^{i, \mu \nu }  \,,
\end{equation}
where $\tilde B_{\mu \nu} = 1/2 \hspace{0.5mm} \epsilon_{\mu \nu  \lambda \rho}  \hspace{0.25mm}  B^{\lambda \rho}$ and $\tilde W_{\mu \nu}^i = 1/2 \hspace{0.5mm} \epsilon_{\mu \nu  \lambda \rho}  \hspace{0.25mm}  W^{i, \lambda \rho}$ are the dual  field strength tensors. In addition to the CP-conserving interactions (\ref{eq:Lc1c2}) and (\ref{eq:Lc3c4}), there are also four CP-violating couplings that are obtained from the above operators by the replacement $\bar \chiDM \chiDM \leftrightarrow \bar \chiDM \gamma_5 \chiDM$.

The effective interactions introduced in (\ref{eq:Lc1c2}) and  (\ref{eq:Lc3c4}) appear  in models of Rayleigh DM~\cite{Weiner:2012cb}. Ultraviolet completions where the operators are generated through loops of states charged under $U(1)_Y$ and/or $SU(2)_L$  have been proposed in \cite{Weiner:2012gm} and their LHC signatures have been studied in \cite{Liu:2013gba}. If these new charged particles  are  light, the high-$p_T$ gauge bosons that participate in  the \MET processes considered here are able to resolve the substructure of the loops. This generically suppresses the cross sections compared to the EFT predictions~\cite{Haisch:2012kf}, and thus will weaken the bounds on the interaction strengths of  DM and the EW gauge bosons  to some extent.  Furthermore, the light charged mediators may be produced  on-shell in $pp$ collisions, rendering direct LHC searches potentially more restrictive than \MET searches. Making the above statements precise would require further studies beyond the timescale of this forum.

Since for $\Lambda_S = \Lambda_P$ the effective interactions (\ref{eq:Lc1c2}) and (\ref{eq:Lc3c4}) predict essentially the same value of the mono-photon, mono-$Z$ and mono-$W$ cross section \cite{Carpenter:2012rg,Crivellin:2015wva}, we consider below only the former couplings. We emphasize however that measurements of the jet-jet azimuthal angle difference in \MET$+ 2 j$ events may be used to disentangle whether DM couples more strongly to the combination $B_{\mu \nu} B^{\mu \nu}$ ($W_{\mu \nu}^i W^{i, \mu \nu }$) or the product $B_{\mu \nu} \tilde B^{\mu \nu}$ ($W_{\mu \nu}^i \tilde W^{i, \mu \nu }$) of field strength tensors \cite{Cotta:2012nj,Crivellin:2015wva}.

After EW symmetry breaking the interactions (\ref{eq:Lc1c2}) induce direct couplings 
between pairs of DM particles and  gauge bosons.  The corresponding Feynman rule reads:

\begin{equation}  \label{eq:feynman}
\frac{4 \hspace{0.25mm} i}{\Lambda_S^3} \; g_{V_1 V_2} \, \big (  p_1^{\mu_2} \hspace{0.25mm} p_2^{\mu_1} - g^{\mu_1 \mu_2}  \, p_1 \cdot p_2 \big ) \,,
\end{equation}
where $p_i$ ($\mu_i$) denotes the momentum (Lorentz index) of the vector field $V_i$ and for simplicity the spinors associated with the DM fields have been dropped. The couplings $g_{V_i V_j}$ take the form:

\begin{equation} \label{eq:gViVj}
\begin{split}
g_{\gamma \gamma} & = c_w^2 \hspace{0.25mm} c_1+ s_w^2  \hspace{0.25mm} c_2 \,, \\[1mm]
g_{\gamma Z}   & = - s_w c_w \, \big (  c_1  - c_2  \big ) \,, \\[1mm]
g_{ZZ}  & = s_w^2 \hspace{0.25mm} c_1 + c_w^2  \hspace{0.25mm} c_2  \,, \\[1mm]
g_{WW} & = c_2 \,,
\end{split}
\end{equation}
with $s_w$ ($c_w$) the sine (cosine) of the weak mixing angle. Note that our coefficients $c_1$ and $c_2$ are identical to the coefficients $C_B$ and $C_W$ used in \cite{Crivellin:2015wva}, while they are related via $k_1 = {c_w}^2 c_1$ and $k_2 = {s_w}^2 c_2$ to the coefficients $k_1$ and $k_2$ introduced in \cite{Carpenter:2012rg}.

%
%

The coefficients $c_1$ and $c_2$ appearing in (\ref{eq:gViVj}) determine the relative importance of each of the \MET channels and their correlations. For example, one observes that:
\begin{itemize}
 \item Only $c_2$ enters the coupling between DM and $W$ bosons, meaning that only models with $c_2 \neq 0$ predict a mono-$W$ signal;
 \item If $c_1 = c_2$ the mono-photon (mono-$Z$) signal does not receive contributions from diagrams involving $Z$ (photon) exchange;
  \item Since numerically $c_w^2/s_w^2 \simeq 3.3$ the mono-photon channel is particularly sensitive to $c_1$.
\end{itemize}

\subsubsection{Parameter scan}

As stated above and shown in Ref.~\cite{Nelson:2013pqa}, 
the kinematic distributions for dimension-7 scalar and pseudoscalar operators
only shows small differences. This has been verified from a generator-level study:
the signal acceptance after a simplified analysis selection 
(\MET$>$350~\gev, leading jet $p_T > $ 40~\gev, minimum azimuthal difference between
either of the two jets and the \MET direction $>$ 0.4) is roughly 70\% for both models, independent from the coefficients $c_1$ and $c_2$. 
We therefore only suggest to generate one of the two models.


The differences in kinematics for the various signatures
are negligible when changing the coefficients $c_1$ and $c_2$, 
since these coefficient factorize in the matrix element. 
Only the case $c_1=c_2=1$ is generated as benchmark;
other cases are left for reinterpretation as they will only need a 
rescaling of the cross-sections. 

\begin{figure}[h!]
  \centering
  	\includegraphics[width=0.95\textwidth]{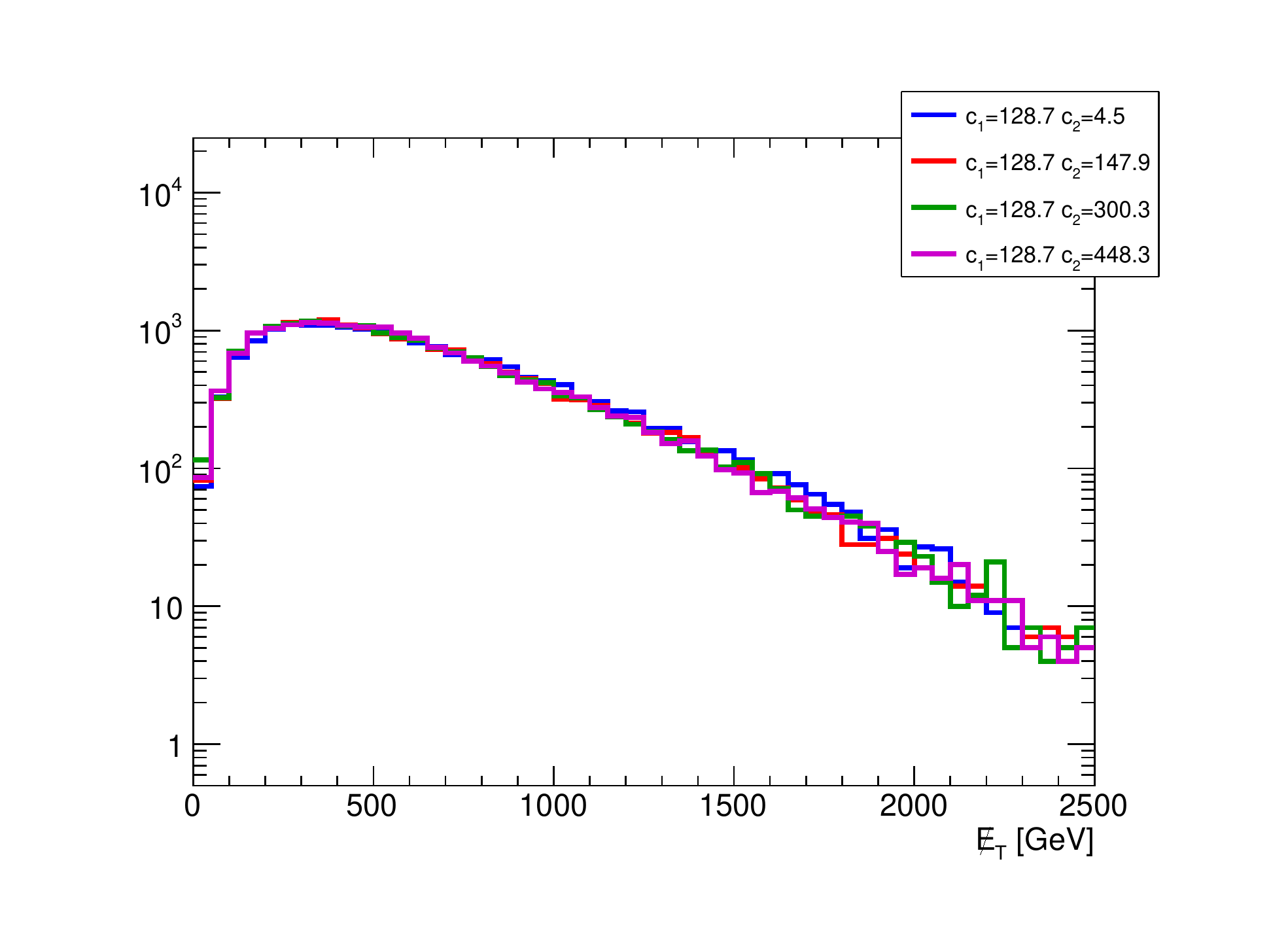}
    \caption{\MET distribution for the dimension-7 model with a hadronically decaying Z in the final state,
    for the scalar and pseudoscalar operators representing direct interactions between DM and bosons. The values of the coefficients in the legend are multiplied by 100.}
    \label{fig:EFTD7_EW_kinematics}
\end{figure}


\subsection{Higher dimensional operators}

Many higher dimensional operators can induce signals of photons or $W/Z/H$ bosons
in the final state. A complete list can be found in Refs.~\cite{Carpenter:2013xra, Berlin:2014cfa, Petrov:2013nia}
and references therein. 

Although with lower priority with respect to the operators above, 
a representative dimension-8 operators can be chosen as benchmark, with the form:
 
$$\frac{1}{\Lambda^4} \bar{\chiDM} \gamma^{\mu} \chiDM B_{\mu \nu} H^{\dagger} D^{\nu} H$$

In this case, the new physics scale is $\Lambda$ is related to the coupling
of the DM as $\displaystyle y_{\chiDM} = \frac{1}{\Lambda^{4}}$.
An advantage of this operator is that it includes all signatures with EW bosons,
allowing to assess the relative sensitivity of the various channels with the same model.  
The kinematics for this operator is different with respect to other operators,
leading to a harder \MET spectrum, 
as illustrated by comparing the leading $b-$jet distribution for the dimension 5 operator
to the dimension 8 operator. 
  
   \begin{figure}[hbpt!]
   	\centering
   	\subfloat[Dimension 5 operator]{
   		\includegraphics[width=0.95\linewidth]{figures/EW/monoH/xxhhg5/truth_leading_b_pt} 
   	}
   	\hfill
   	\subfloat[Dimension 7 operator]{
   		\includegraphics[width=0.95\linewidth]{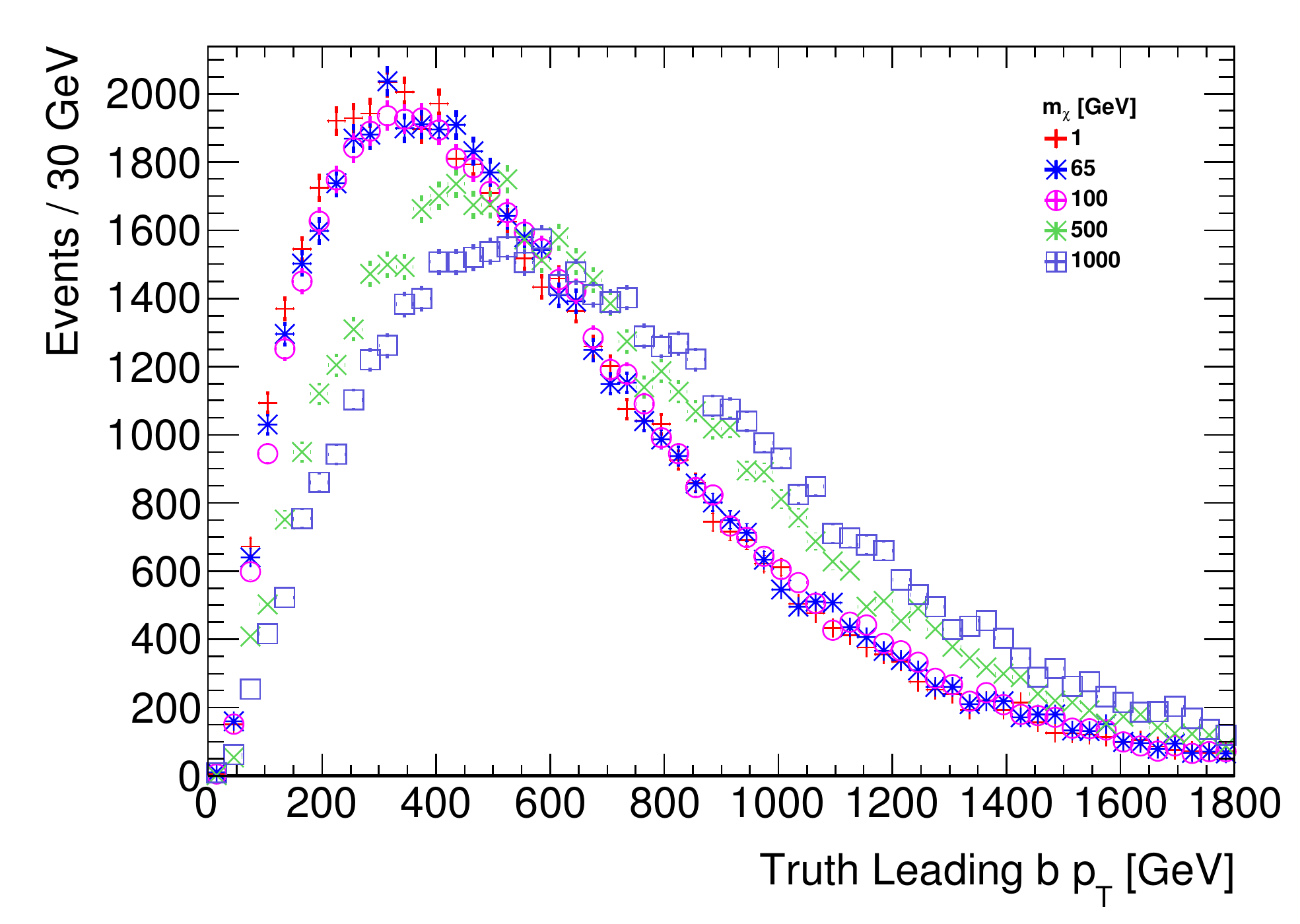} 
   	}
   	\caption{Comparison of the transverse momentum for the leading $b-$ jet from the Higgs decay for a dimension 5 and dimension 7 operator
   		with direct boson-DM couplings.}
   	\end{figure}

%
%
%
%

\subsection{Validity of EW contact operators and possible completions}
\label{sub:validityEWContact}

It is important to remember that the operators described in 
this section may present problems in terms of the validity of the contact interaction
approach for the energy scales reached at the LHC. 

As outlined in~\cite{Berlin:2014cfa}, designing very high \MET search signal regions
that are exclusively motivated by the hard \MET spectra of the dimension 7 and 8 operators
will mean that the momentum transfer in the selected events is larger. This in turn
means that processes at that energy scale (mediators, particles exchanged in loops)
are accessible, and a simple contact interaction will not be able to correctly
describe the kinematics of these signals. 

Contact interaction operators like the ones in this section 
remain useful tools for comparison of the sensitivity of different search channels, 
and for reinterpretation of other models under the correct assumptions. 
To date, while UV-complete models are known, their phenomenology has
not been studied in full detail as their completion involves loops~\sidenote{
An example case for the need of loop completions is a simplified model with an additional scalar exchanged at tree level.
The scalar couples to $WW$ and $ZZ$ in a gauge-invariant way, Integrating out the mediator 
does not lead to the Lorentz structure of a dimension-7 operator, so it is not possible
to generate dimension-7 operators that satisfy gauge and Lorentz invariance at the same time.
A model with a \spinone mediator cannot be considered as an candidate for completion either, since dimension-7 operators only have scalar or pseudoscalar couplings.}. 

However, this may be the focus of future theoretical exploration, as discussed in Ref.~\cite{Crivellin:2015wva}.
An example of a complete model 
for scalar DM corresponding to the dimension-5 operator 
is provided in the Appendix~\ref{app:EWSpecificModels_Appendix}.
Providing results for the pure EFT limit of these models will prove useful
to cross-check the implementation of future. 

Given these considerations, we recommend to present results 
for these models as follows: 

\begin{itemize}
\item Deliver fiducial limits on the cross section of any new physics events, without any model assumption, according to the guidelines in Appendix~\ref{app:Presentation_Of_Experimental_Results}.  
\item Assess the percentage of events that pass a condition of validity for the EFT approximation that does not
depend on a specific completion, and present results removing 
of the invalid events using the procedure in Section~\ref{sec:EFTValidity} alongside the raw EFT results.
\end{itemize}



\chapter{Implementation of Models}
\label{app:MonojetLikeModels_Appendix}
\section{\texorpdfstring{Implementation of \schannel and \tchannel models for \MET+X analyses}{Implementation of \schannel and \tchannel models for MET+X analyses}}

In the studies to date, a number of different Monte Carlo tools have
been used to simulate DM signals.  In this Chapter, we make recommendations 
on the accuracy at which simulations should be performed for different final states. We 
also provide explicit examples  of codes and implementations (including specific settings) 
that have been used to obtain the results in this report.
We stress that these recommendations are based on the current
status of publicly available codes and users should always check whether new results
at a better accuracy have appeared in the meantime. In that case, we recommend to update
the corresponding analyses  directly using the new releases and/or codes, and in case this would not be possible,
to at least take into account the new information in the analysis (e.g., via a MC comparison with the latest predictions, 
or by effectively using global/local $K$-factors). For all models included in this report, 
\pythiaEight has been used to provide the parton shower simulation. Nevertheless, 
we note that showering matrix element events with Herwig~\cite{Bahr:2008pv,Corcella:2002jc,Corcella:2000bw,Marchesini1992465} should be considered as 
an equally valid alternative.

\subsection{Implementation of  \schannel  models for mono-jet signature}
\label{sec:monojet_implementation}

These models include those discussed in Secs.~\ref{sec:monojet_V} and \ref{sec:monojet_scalar}.
In monojet analyses, i.e. when final states are selected with a few jets and \MET{}, observables and in particular the \MET{} spectrum depend upon the accuracy of the simulation of QCD radiation.
For the vector and axial vector models, the current state of the art is  NLO+PS. It is particularly simple to obtain simulations for these processes at NLO+PS and even for merged samples at NLO accuracy, starting from SM implementations.  We therefore recommend simulations to be performed at NLO+PS, and in case multi-jet observables are employed,  by merging samples with different multiplicities. 
Results at such accuracy can be obtained either in dedicated implementations, such as that of  \powheg \cite{Haisch:2013ata}, or via general purpose NLO tools like \madgraph employing available UFO models at NLO. A testing version of the full set of these UFO models has been made available
only in June 2015 \cite{NewMadgraphModels}. For this reason, it was not used as part of the studies of this Forum on initial Run-2 benchmark models. Nevertheless, we encourage further study of these UFO models by the ATLAS and CMS collaborations.

A study using POWHEG~\cite{Haisch:2013ata,Fox:2012ru} has shown that the NLO corrections result in a substantial reduction in the dependence on the choice of the renormalization and factorization scales and hence a reduced theoretical uncertainty on the signal prediction. For the central choice of renormalization and factorization scales, the NLO corrections also provide a minor enhancement in the cross section due to the jet veto that has been so far employed in Run-1 analyses.

For the scalar and pseudoscalar models, the lowest order process
already involves a one-loop amplitude in QCD.
Because of the complexity of performing NLO calculations for this class of processes and in particular the absence of general methods for computing two-loop virtual contributions, only LO predictions are currently available. These can be interfaced to shower programs exactly as usual tree-level Born computations, i.e. by considering one parton multiplicity at the time
or by merging different parton multiplicities via CKKW or MLM schemes to generate inclusive samples with jet rates at LO accuracy.  For spin-0 mediators in the mono-jet final state,
the top-quark loop is the most important consideration.
The matrix element implementation with exact top-loop dependence of the \schannel spin-0 mediated DM production is available in \mcfm~\cite{Fox:2012ru,Harris:2014hga}~\footnote{Only the scalar mediator is available in the public release.} at fixed order and in \powheg~\cite{Haisch:2015ioa} and \madgraph~\cite{NewMadgraphModels} for event generation at LO+PS level. The \powheg and \mcfm implementations include the finite
top quark mass dependence for DM pair production and one extra parton at LO. 
The same processes are available in \madgraph v2.3 and could be made
available in the future in codes like {\sc Sherpa+OpenLoops/GoSam}, including up to two extra partons in the final state. Samples can be merged employing CKKW, $K_T$-MLM procedures.

Most of the results that have been presented in this document for these processes have been obtained with \powheg interfaced to \pythiaEight, matching the state of the art calculation as of Spring 2015. For future reference, we document the specific settings needed to run the \powheg generation for the Dark Matter models so they can serve as nominal benchmarks for the early Run-2 ATLAS and CMS DM analyses. \powheg parameter cards for all models can be found on the Forum SVN repository~\cite{ForumSVN_DMA, ForumSVN_DMV, ForumSVN_DMS_tloop, ForumSVN_DMP_tloop}.

\subsubsection{\powheg configuration for \schannel DM models}

The latest \powheg release is available for download using the
instructions at
\url{http://powhegbox.mib.infn.it/}. The Forum recommends
using at least version \texttt{3059}.

\begin{itemize}

\item \powheg can generate either unweighted (uniformly--weighted) or
weighted events.  
The relevant keywords in the input card are \bornsuppfact and \bornktmin. 

\begin{enumerate}
\item unweighted events: 

\bornsuppfact: negative or absent\\
\bornktmin PT

This runs the program in the most straightforward way,
but it is likely not the more convenient choice, as will be
explained below. \powheg will generate unweighted events using a sharp
lower cut (with value \texttt{PT}) on the leading-jet \pT. Since this is a
generation cut, the user must check that the choice of \bornktmin
does not change the cross section for signal events passing analysis selections.
It is good practice to use as a value in the input card a
transverse momentum 10-20\% smaller than the final analysis selection
on \MET{}, and check that the final result is independent, by exploring an even
smaller value of \bornktmin. The drawback of using this mode is that
it is difficult to populate well, and in a single run, both the low-\pT
region as well as the high-\pT tail.

\item weighted events: 

\bornsuppfact PTS\\
\bornktmin PT

\powheg will now produce weighted events, thereby allowing to generate
a single sample that provides sufficient statistics in all signal
regions. Events are still generated with a sharp lower cut set by
\bornktmin, but the \bornsuppfact parameter is used to set the event
suppression factor according to

\begin{equation}
F(\kT)=\frac{\kT^2}{\kT^2+\bornsuppfact^2} \;.
\end{equation}

In this way, the events at, for instance, low \MET, are suppressed
but receive higher weight, which ensures at the same time higher
statistics at high \MET. We recommend to set \bornsuppfact to 1000.

The \bornktmin parameter can be used in conjunction with \bornsuppfact to suppress the low \MET region
even further.  It is recommended to set \bornktmin to one--half the value of
the lowest \MET selection. For instance,  for the event selection used in the
CMS/ATLAS monojet analyses, assuming the lowest \MET region being defined above 300\,GeV, the proposed value for
\bornktmin is 150.  However, this parameter should be set keeping in
mind the event selection of all the analyses that will use these
signal samples, and hence a threshold lower than 150 may be required.

\end{enumerate}


\item The \powheg monojet implementations can generate events using two expressions for the mediator propagators. The default setup (i.e if the keyword \runningwidth is absent, commented out or set to 0) is such that a normal Breit-Wigner function is used for the propagator: in this case, the expression
\begin{equation*}
  Q^2 - M^2 + \complexi\,M\,\Gamma
\end{equation*}
is used for the propagator's denominator, where Q is the virtuality of the mediator, and M and $\Gamma$ are its mass and width, respectively.
This is the more straightforward, simple and transparent option, and it was used for the Forum studies. It should be the method of choice, unless one approaches regions of parameter space where gamma/M starts to approach order 1 values. In those cases, a more accurate modelling (or at least a check of the validity of the fixed width approach) can be achieved by using a running width: by setting the \runningwidth token to 1, \powheg uses as the denominator of the mediator’s propagator the expression
\begin{equation*}
 Q^2 - M^2 + \complexi\,Q^2\,\frac{\Gamma}{M},
\end{equation*}
which is known to give a more realistic description. See Ref.~\cite{Bardin:1989qr} for a discussion.


  

\item Set the parameters defining the bounds on the invariant mass of the Dark Matter pair, \masslow and \masshigh, to -1. In this way, \powheg will assign values internally. 
\item The minimal values for \ncallOne, \itmxOne, \ncallTwo, \itmxTwo are 250000, 5, 1000000, 5 for the vector model, respectively.
\item The minimal values for \ncallOne, \itmxOne, \ncallTwo, \itmxTwo are 100000, 5, 100000, 5 for the scalar top-loop model, respectively.
\item When NLO corrections are included (as for instance in the
  vector model), negative-weighted events could happen and should
  be kept in the event sample, hence \withnegweights should be set to
  1. If needed, their fraction can be decreased by setting \foldsci
  and \foldy to bigger value (2 for instance). \foldphi can be kept to
  1.
\item One should use the automatic calculation of systematic uncertainties associated with the choice of hard scale and PDFs as described in Section\,\ref{sec:TheoryUncertainties}.

\item \texttt{idDM} is the integer that identifies the DM particle in the Monte Carlo event record.  This should be chosen so that other tools can process the \powheg output properly.

\end{itemize}

\powheg in itself is not an event generator and must be interfaced with a tool that provides parton showering, hadronization, \textit{etc.}   For some time, a
\pythiaEight \cite{Sjostrand:2014zea} interface
has existed for \powheg.  The \pythiaEight runtime configuration is
the following:

\begin{verbatim}
POWHEG:veto = 1
POWHEG:pTdef = 1
POWHEG:emitted = 0
POWHEG:pTemt = 0
POWHEG:pThard = 0
POWHEG:vetoCount = 100
SpaceShower:pTmaxMatch = 2
TimeShower:pTmaxMatch = 2
\end{verbatim}
As always, it is recommended to use the latest \pythiaEight release,
available at \url{http://home.thep.lu.se/~torbjorn/Pythia.html}.
At the time of this report, the latest version is \texttt{8.209}.

\subsection{Merging samples with different parton multiplicities}
\label{sec:monojet_parton_match}

For the models discussed in the previous section, it is important
to calculate the hard process as accurately as possible in QCD.
For many other signal models, the \MET{} signature depends more
upon the production and decay of the mediator. In some cases, observables
built in terms of the jets present in the final state are considered, something that assumes inclusive samples accurate in higher jet multiplicities are available.  In these cases, one can employ LO+PS simulations where different parton multiplicities are merged and then matched to parton shower, using schemes such as CKKW or MLM merging.

Here, we consider the example of an EFT model produced in association
with up to 2 additional QCD partons.   A Monte Carlo sample based on
this method could be used in alternative to a NLO+PS sample for describing shapes 
and jet distributions (but not for the overall normalisation which would still be at LO).
The methodology described here could also be used for the \tchannel model
discussed in Sec.~\ref{sec:monojet_t_channel}.

For the calculation of tree-level merged samples for DM signals, tools that can  
read UFO files and implement multi-parton merging should be employed, such 
that \madgraph (+\pythiaEight or {\sc HERWIG++}) and {\sc Sherpa}~\cite{Hoche:2014kca}.
In this report we have mostly employed \madgraph. 
\madgraph provides a flexible and easy--to--use framework for implementing
new models via the {\sc FeynRules} package.
\madgraph can perform both LO and NLO calculations in QCD, matched/merged to parton showers~\cite{Alwall:2008qv}. For NLO ones, dedicated UFO model implementations at NLO should be used. Several UFO models at NLO are publicly available that while not developed specifically for DM, are suitable to make mode independent simulations at NLO accuracy, including multiparton merging 
via the FxFx technique~\cite{Frederix:2012ps}. A dedicated DM UFO implementation 
has been developed and it has been released as a testing version~\cite{NewMadgraphModels}.

Merging events generated via matrix elements with different number of partons in the final state can be achieved by a judicious procedure that  avoids double counting of the partons from matrix elements and parton showering.
Several merging techniques are available. Based on some comparative studies \,\cite{Alwall:0706.2569}, there is some advantage to using the CKKW-L merging scheme \cite{Lonnblad:2011xx} implemented in \pythiaEight.  Alternatively, one can use the $k_T$-MLM scheme also available in \pythiaEight.

\subsubsection{Generation of the LHE file}
\label{sub:MadgraphParameters}

The example presented here is a D5 EFT model, and
includes tree-level diagrams with $\chi\bar\chi$+0,1,2 partons.
We stress that \madgraph, like \powheg, is not in itself and event generator, but must be interfaced 
with an event generator through an LHE file.  The production of the LHE file proceeds through setting the 
process parameters and the run parameters.

The process parameters are:
\begin{verbatim}
import model MODELNAME
generate p p > chi chi~ [QCD] @0
add process p p > chi chi~ j [QCD] @1
add process p p > chi chi~ j j [QCD] @2
\end{verbatim}

The runtime parameters are more numerous, and define the
collider properties, PDF sets, etc.   The specific parameters
needed for matching are, for the example of CKKW-L matching:
\begin{verbatim}
ickkw = 0
ktdurham = matching scale
dparameter = 0.4
dokt = T
ptj=20
drjj=0
mmjj=0
ptj1min=0
\end{verbatim}
For different kinds of matching, a different choice of \texttt{ickkw} and
related parameters would be made.

\subsubsection{Implementation of the CKKW-L merging}
\label{sec:match_implementation}
To illustrate the settings related to merging different multipliticities, the EFT D5 samples were generated with \madgraph version 2.2.2 and showered in \pythiaEight.201, using the Madgraph parameters in the previous section (Sec.~\ref{sub:MadgraphParameters}).

The \pythiaEight parameters for the CKKW-L $k_T$-merging scheme are:
\begin{verbatim}
Merging:ktType           = 1
Merging:TMS              = matching scale
1000022:all = chi chi~ 2 0 0 30.0 0.0 0.0 0.0 0.0 
1000022:isVisible = false
Merging:doKTMerging      = on
Merging:Process          = pp>{chi,1000022}{chi~, -1000022}
Merging:nJetMax          = 2
\end{verbatim}
The matching scales should be the same for the generation and parton showering.
In the model implementation, the particle data group ID \texttt{1000022} is used for weakly
interacting dark matter candidates.   Since this is a Majorana particle by default (with no
corresponding anti-particle), and the model produces a DM Dirac fermion, the particle properties
are changed accordingly.  Also, the DM mass is set to 30\,\gev.
The \texttt{Merging:Process} command specifies the lowest parton emission process generated in \madgraph and \texttt{Merging:nJetMax = 2} gives the maximum number of additional parton emissions with respect to the lowest parton emission process. 

 In general, it is desired to take the hard parton emissions from the matrix element generation in \madgraph and allow \pythiaEight to take care of soft emissions only. The transition between these two regimes is defined by the matching scale and its optimal value can be determined by studying the cross-section as a function of the number of jets (differential jet rates). The differential rates $\frac{dN_{i\to j}}{d \log_{10}(k_\textrm{cut})}$ give the number of events which pass from $i$ jets to $j$ jets as the $k_T$ value increases beyond $k_\textrm{cut}$. An optimal matching scale should lead to smooth differential jet rates.

\begin{figure*}[h!]
 	\centering  
 	\subfloat[$1\rightarrow2$ jets]{%
 		\includegraphics[width=0.45\linewidth]{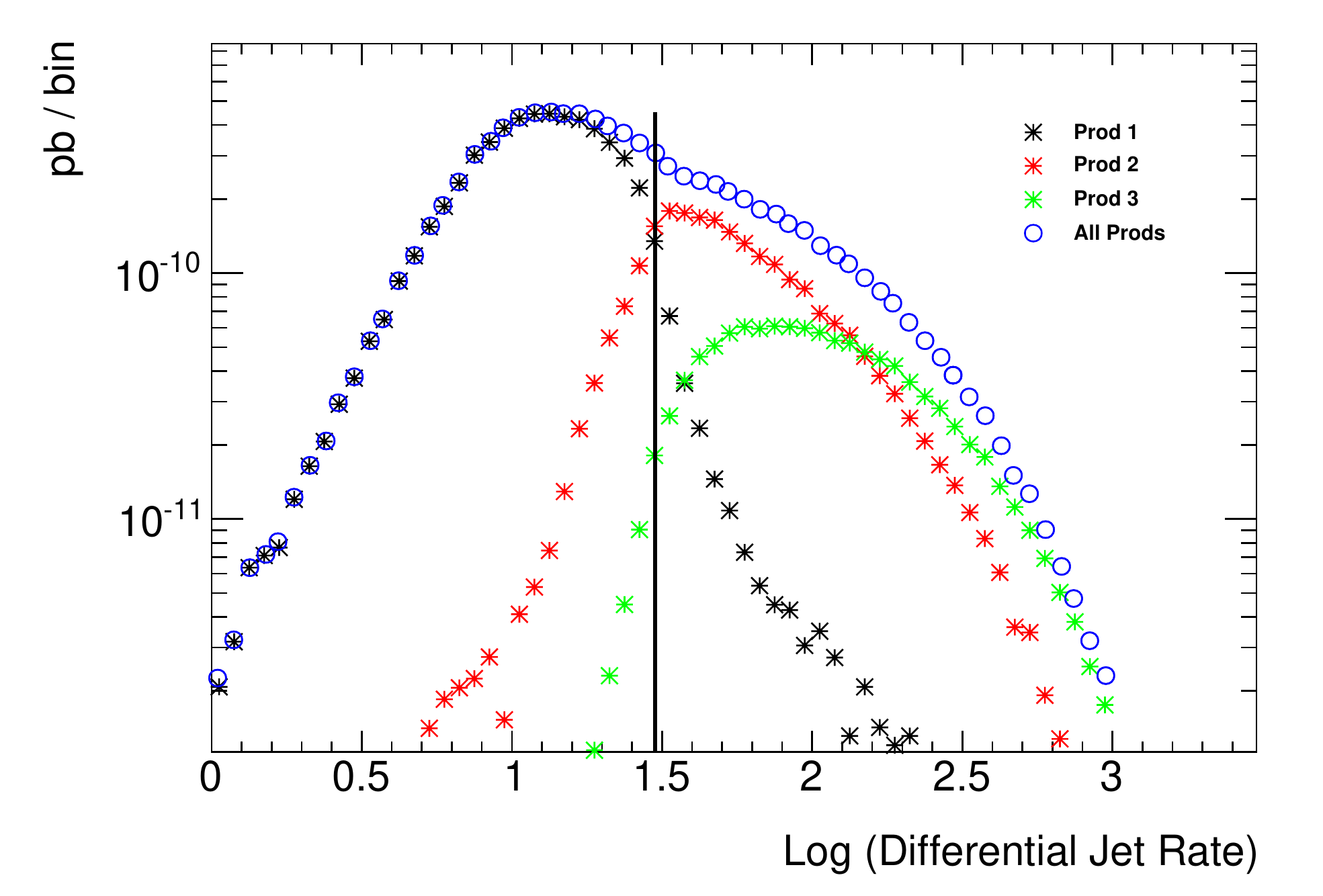}
 	}
 	\hfill
 	\subfloat[$2\rightarrow3$ jets]{%
 		\includegraphics[width=0.45\linewidth]{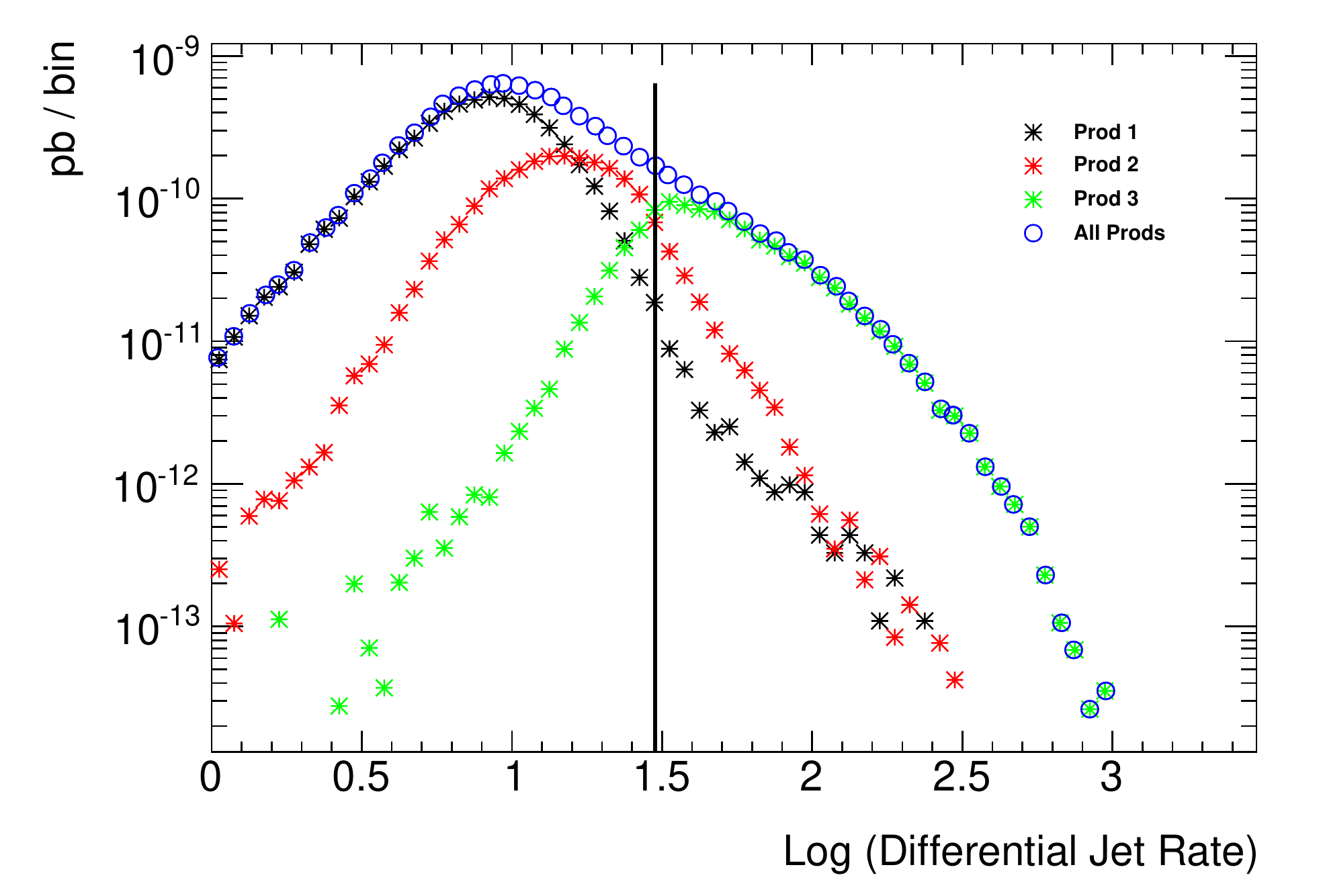}
 	}
 	\hfill
 	\subfloat[$3\rightarrow4$ jets]{%
 		\includegraphics[width=0.45\linewidth]{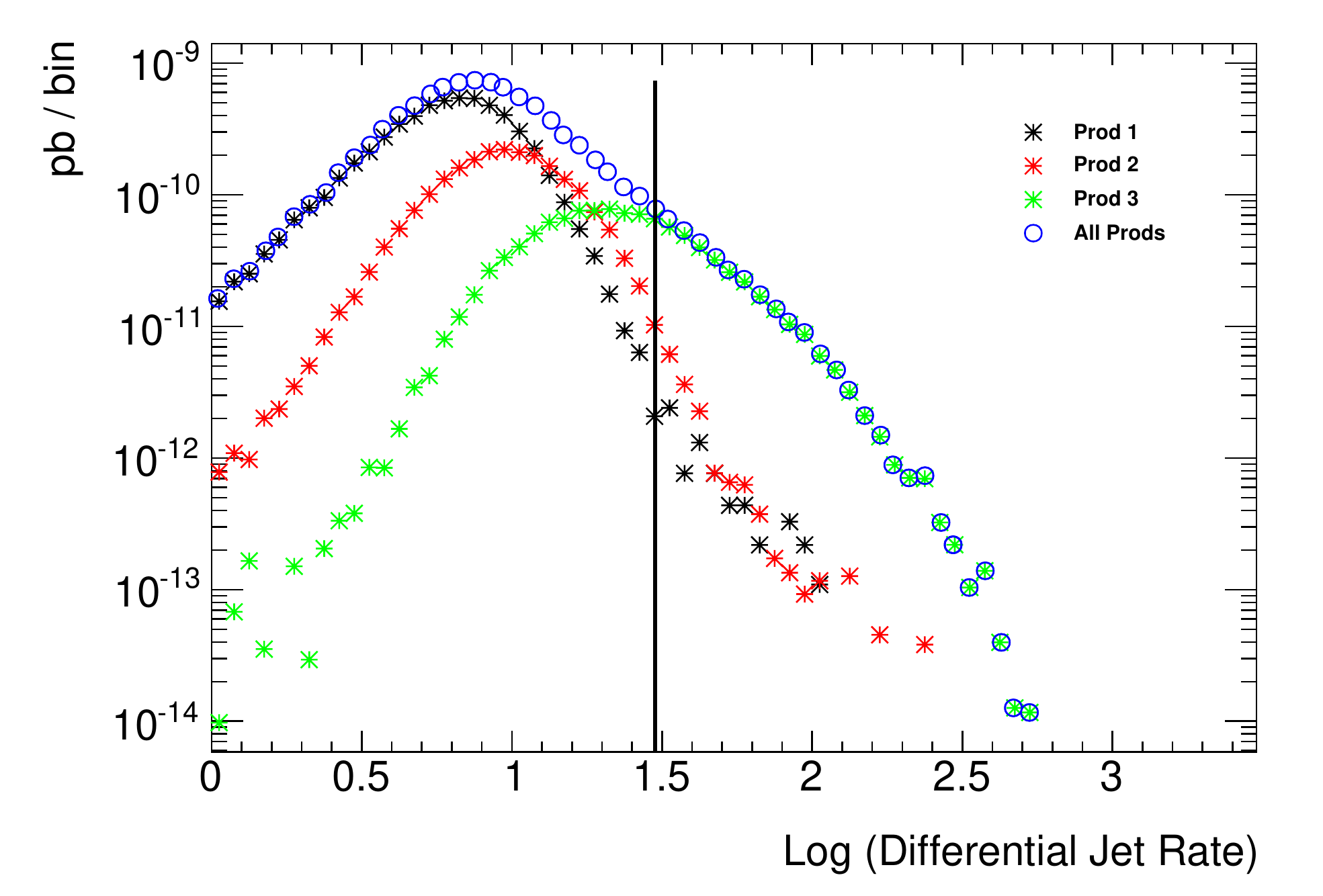}
 	}
 	\hfill
 	\subfloat[$4\rightarrow5$ jets]{%
 		\includegraphics[width=0.45\linewidth]{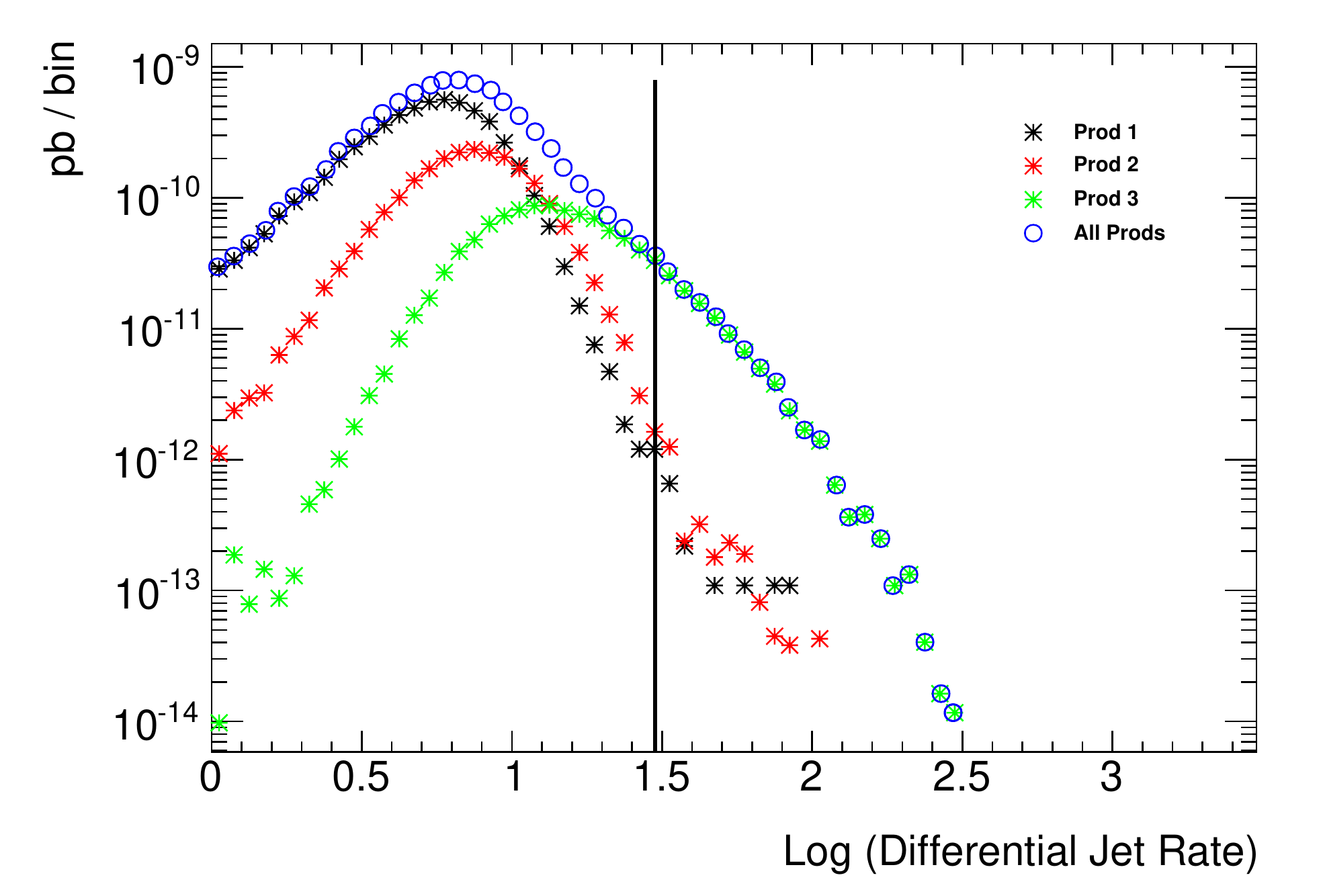}
 	}
 	\caption{Distributions of differential jet rates $\frac{dN_{i\to j}}{d \log_{10}(k_\textrm{cut})}$ for EFT D5 sample with CKKW-L matching scale at 30\,\gev. The 0-, 1- and 2-parton emission samples are generated separately and indicated in the plots as Prod 1, Prod 2 and Prod 3, respectively. A vertical line is drawn at the matching scale.}
 	\label{fig:CKKW_D5_30}
 \end{figure*}

\begin{figure*}[h!]
 	\centering  
 	\subfloat[$1\rightarrow2$ jets]{%
 		\includegraphics[width=0.45\linewidth]{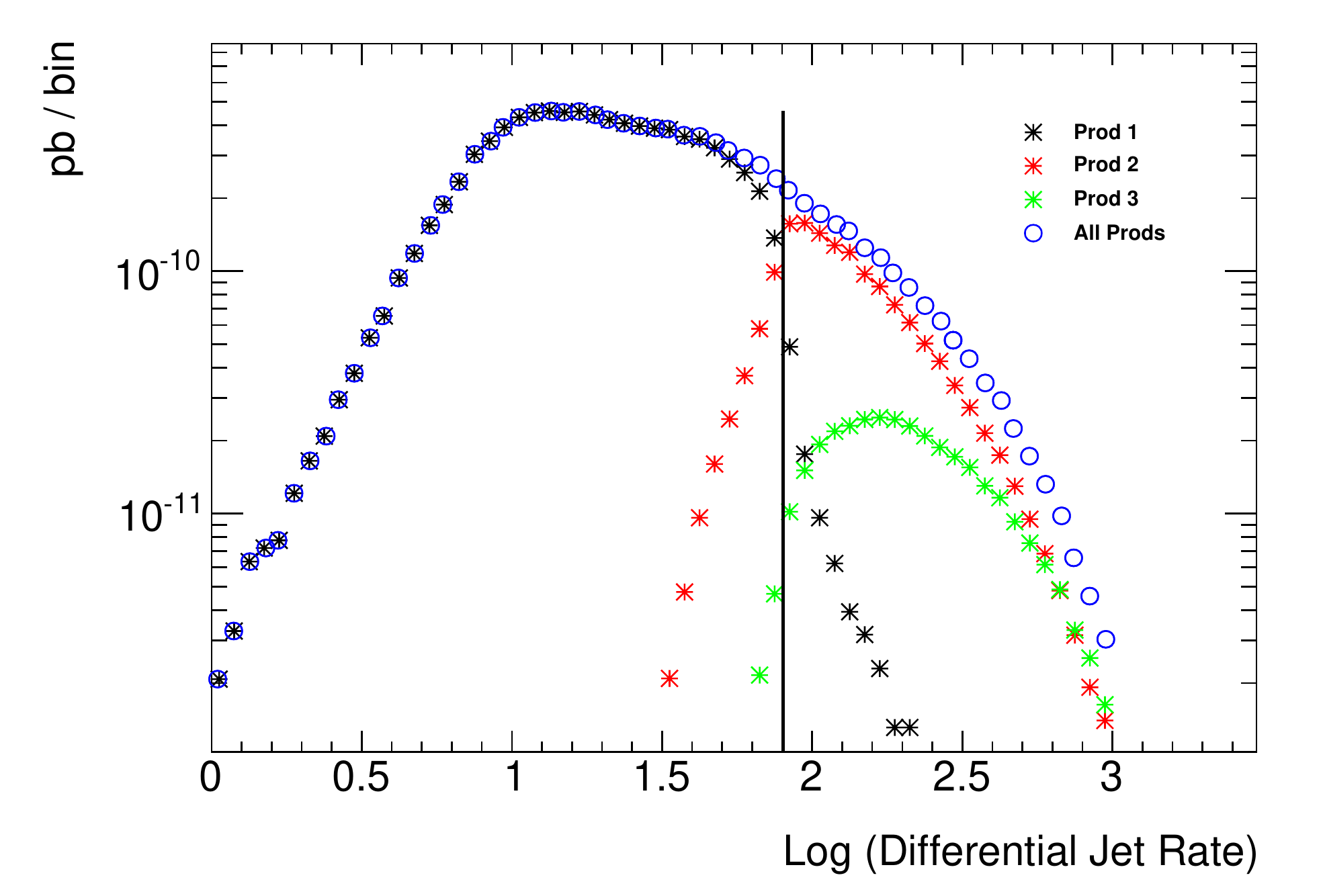}
 	}
 	\hfill
 	\subfloat[$2\rightarrow3$ jets]{%
 		\includegraphics[width=0.45\linewidth]{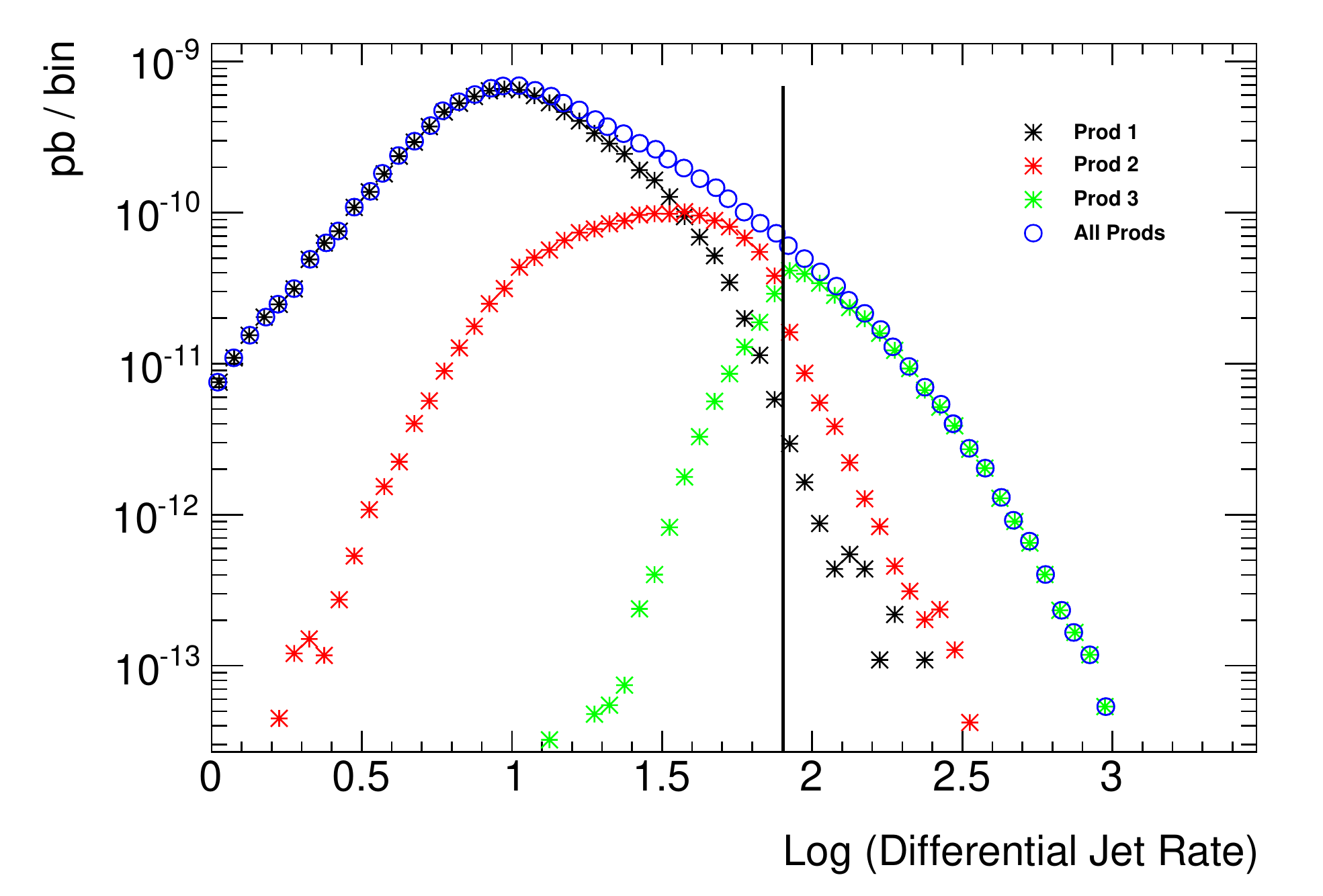}
 	}
 	\hfill
 	\subfloat[$3\rightarrow4$ jets]{%
     \includegraphics[width=0.45\linewidth]{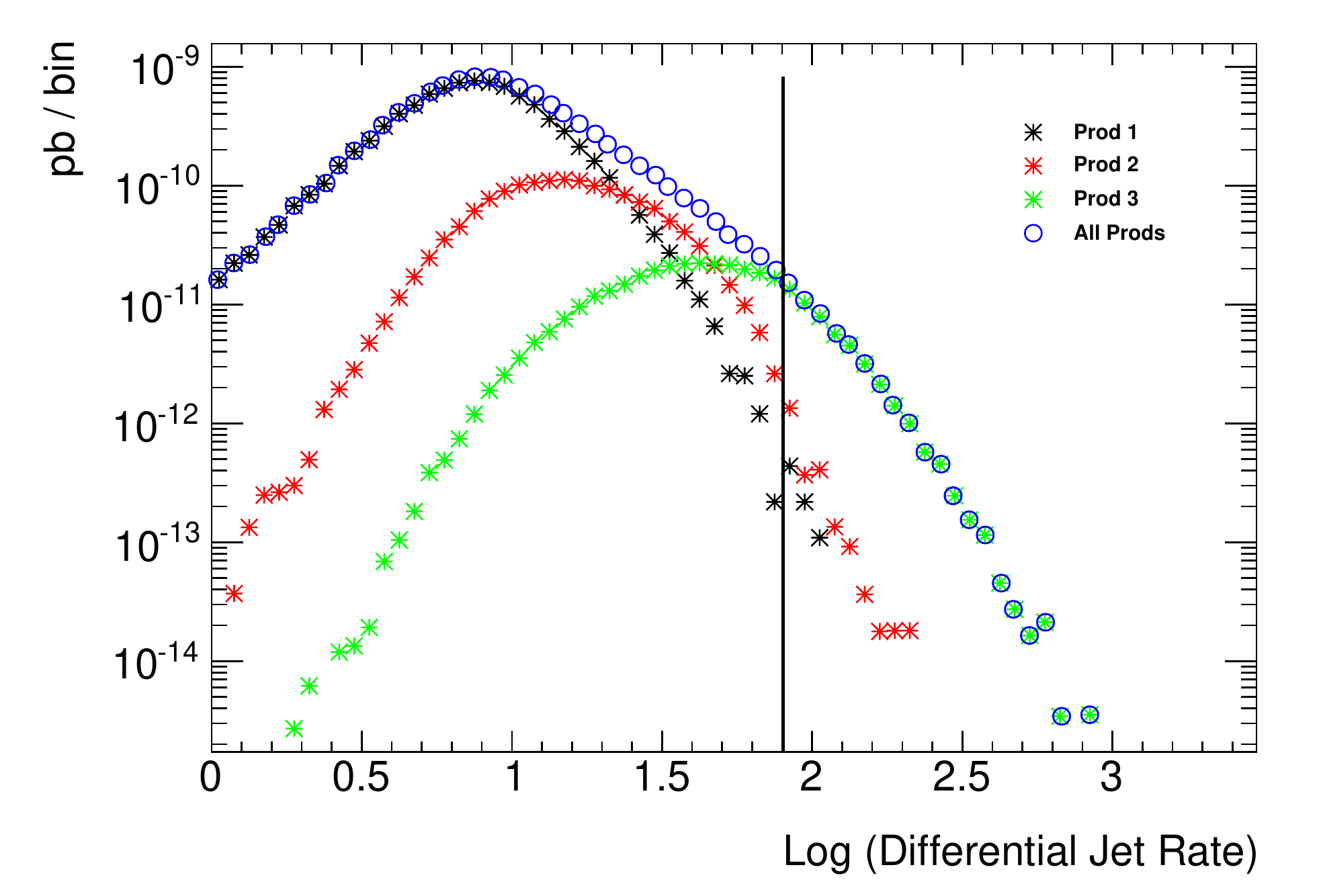}
 	}
 	\hfill
 	\subfloat[$4\rightarrow5$ jets]{%
     \includegraphics[width=0.45\linewidth]{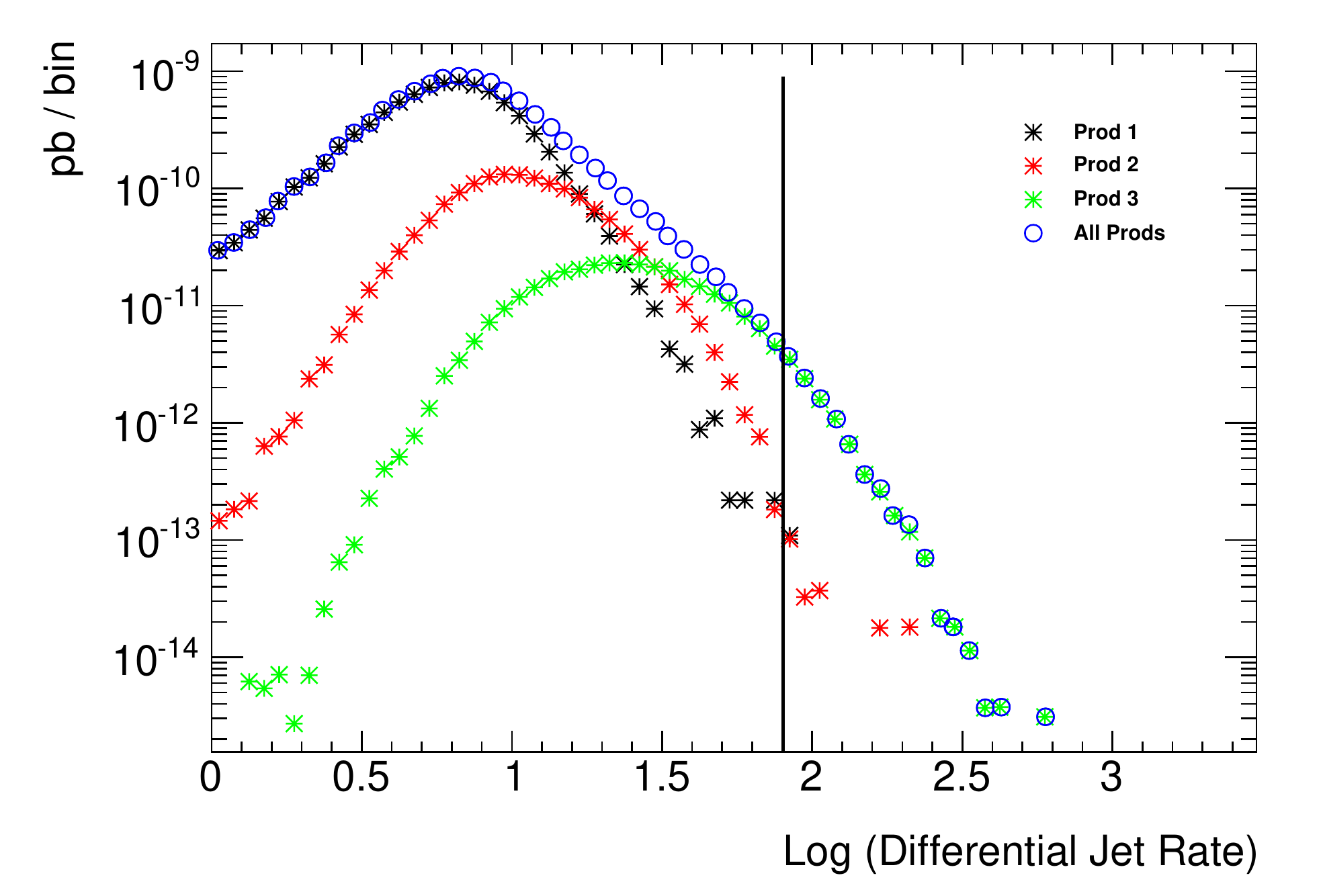}
 	}
   \caption{Distributions of differential jet rates $\frac{dN_{i\to j}}{d \log_{10}(k_\textrm{cut})}$ for EFT D5 sample with CKKW-L matching scale at 80\,\gev. The 0-, 1- and 2-parton emission samples are generated separately and indicated in the plots as Prod 1, Prod 2 and Prod 3, respectively. A vertical line is drawn at the matching scale.}
   \label{fig:CKKW_D5_80}
 \end{figure*}

\begin{figure*}[h!]
 	\centering  
 	\subfloat[$1\rightarrow2$ jets]{%
 		\includegraphics[width=0.45\linewidth]{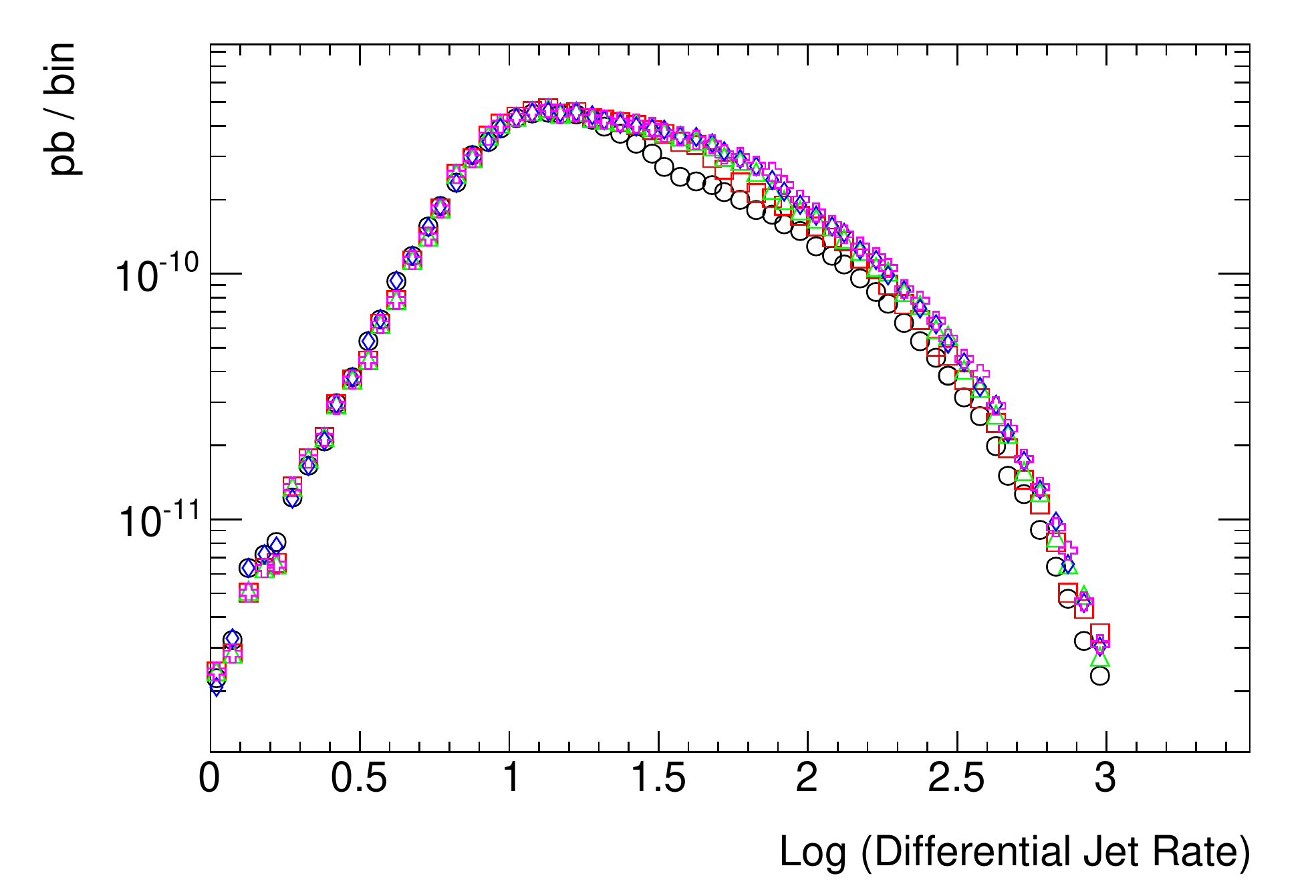}
 		\includegraphics[width=0.45\linewidth]{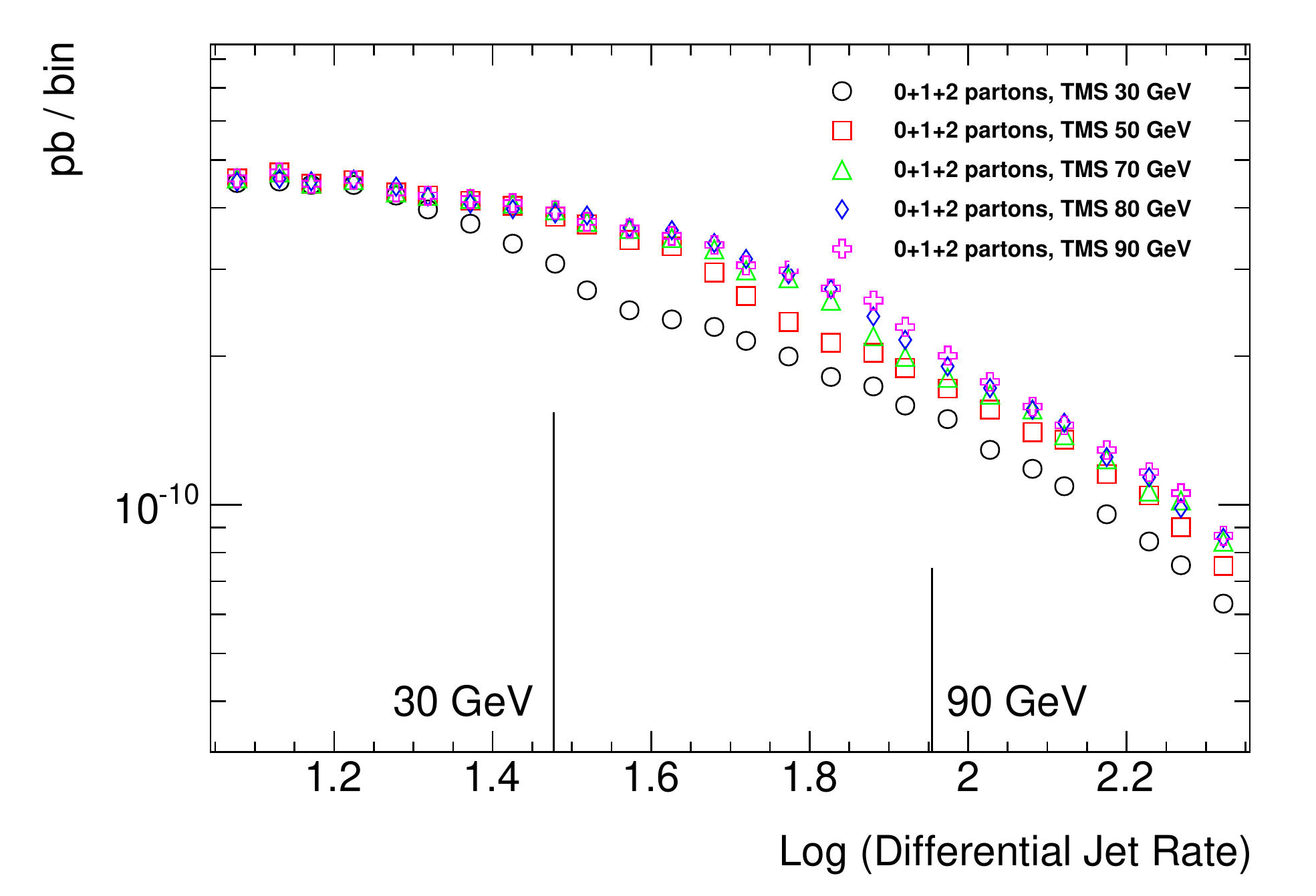}
 	}
 	\hfill
 	\subfloat[$2\rightarrow3$ jets]{%
 		\includegraphics[width=0.45\linewidth]{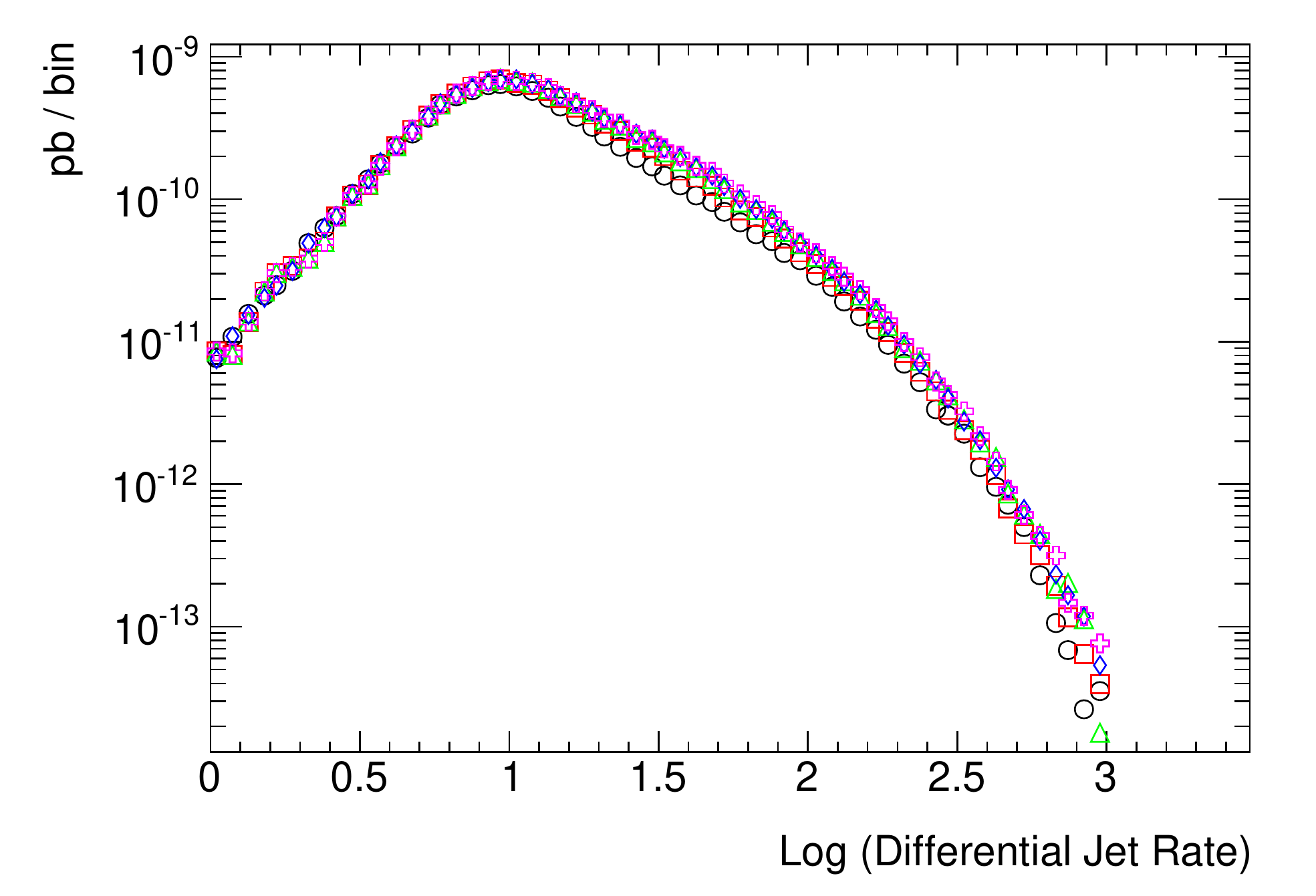}
 		\includegraphics[width=0.45\linewidth]{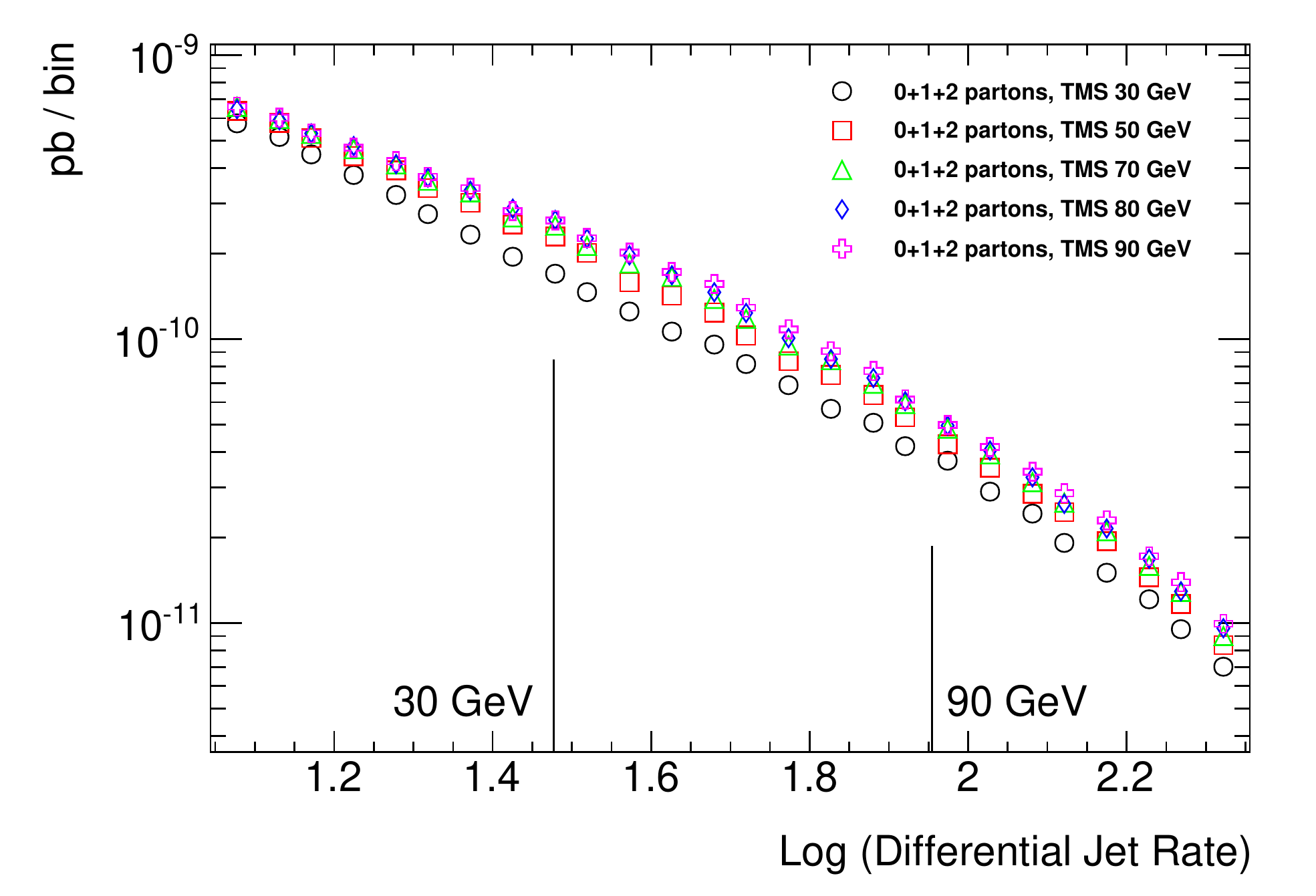}
 	}
 	\hfill
 	\subfloat[$3\rightarrow4$ jets]{%
 		\includegraphics[width=0.45\linewidth]{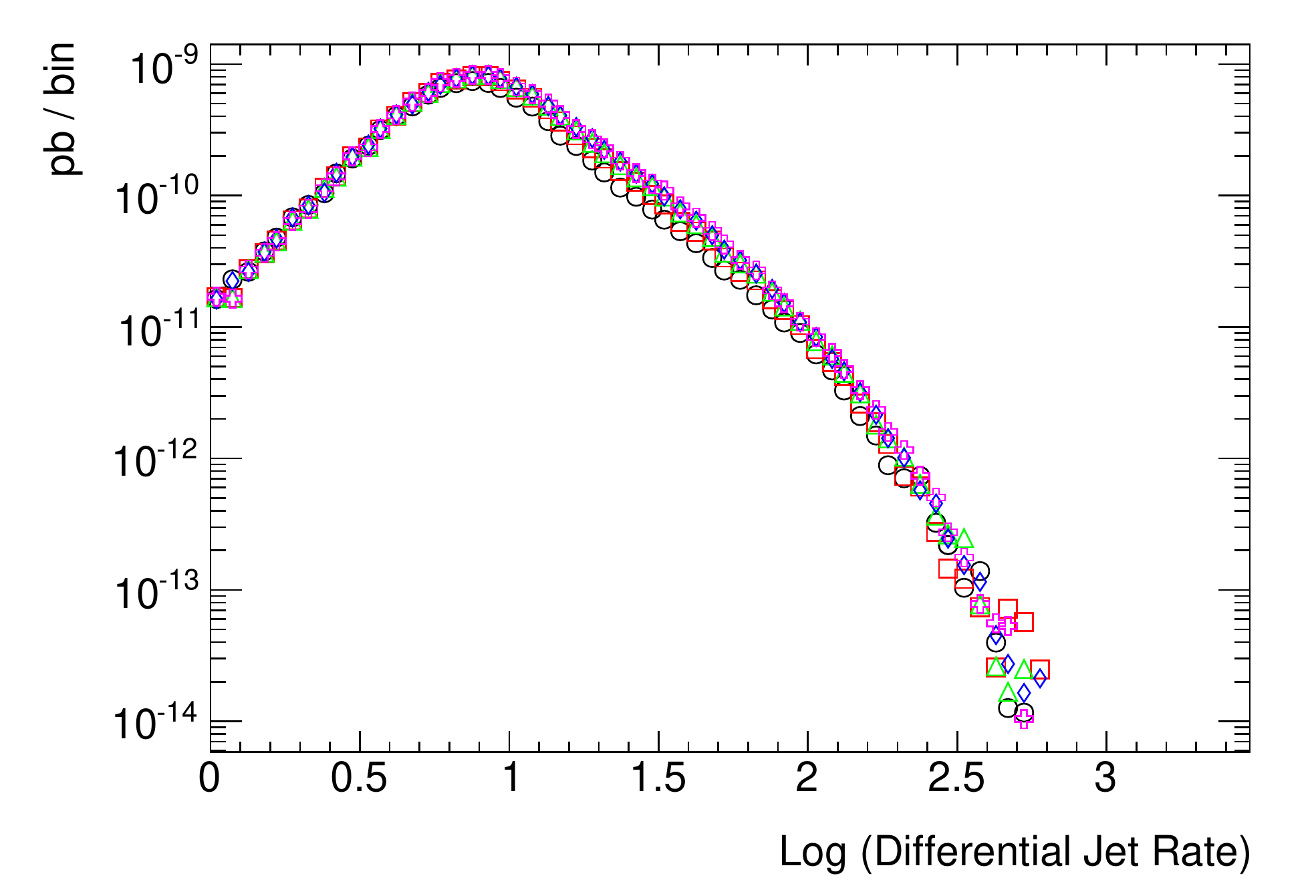}
 		\includegraphics[width=0.45\linewidth]{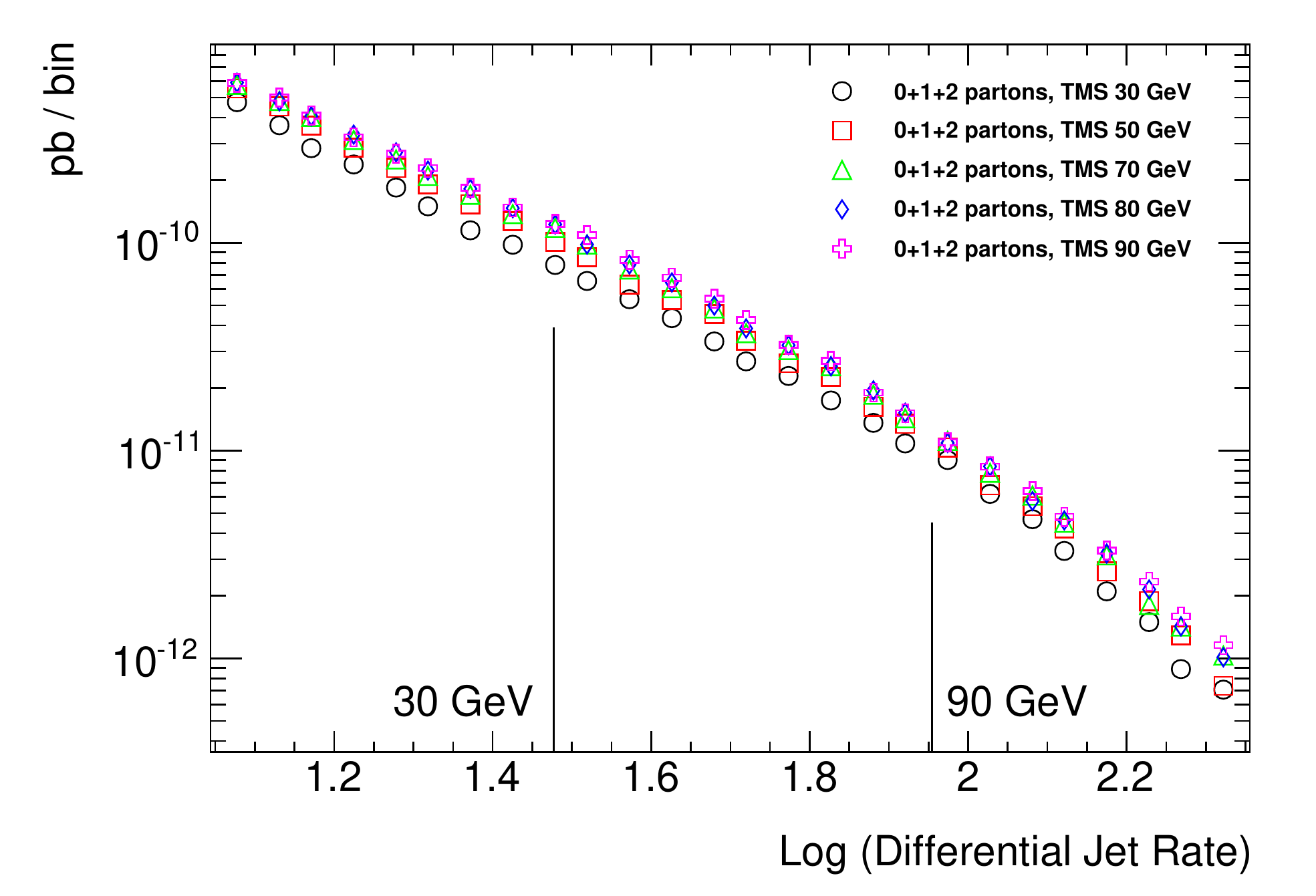}
 	}
 	\hfill
 	\subfloat[$4\rightarrow5$ jets]{%
 		\includegraphics[width=0.45\linewidth]{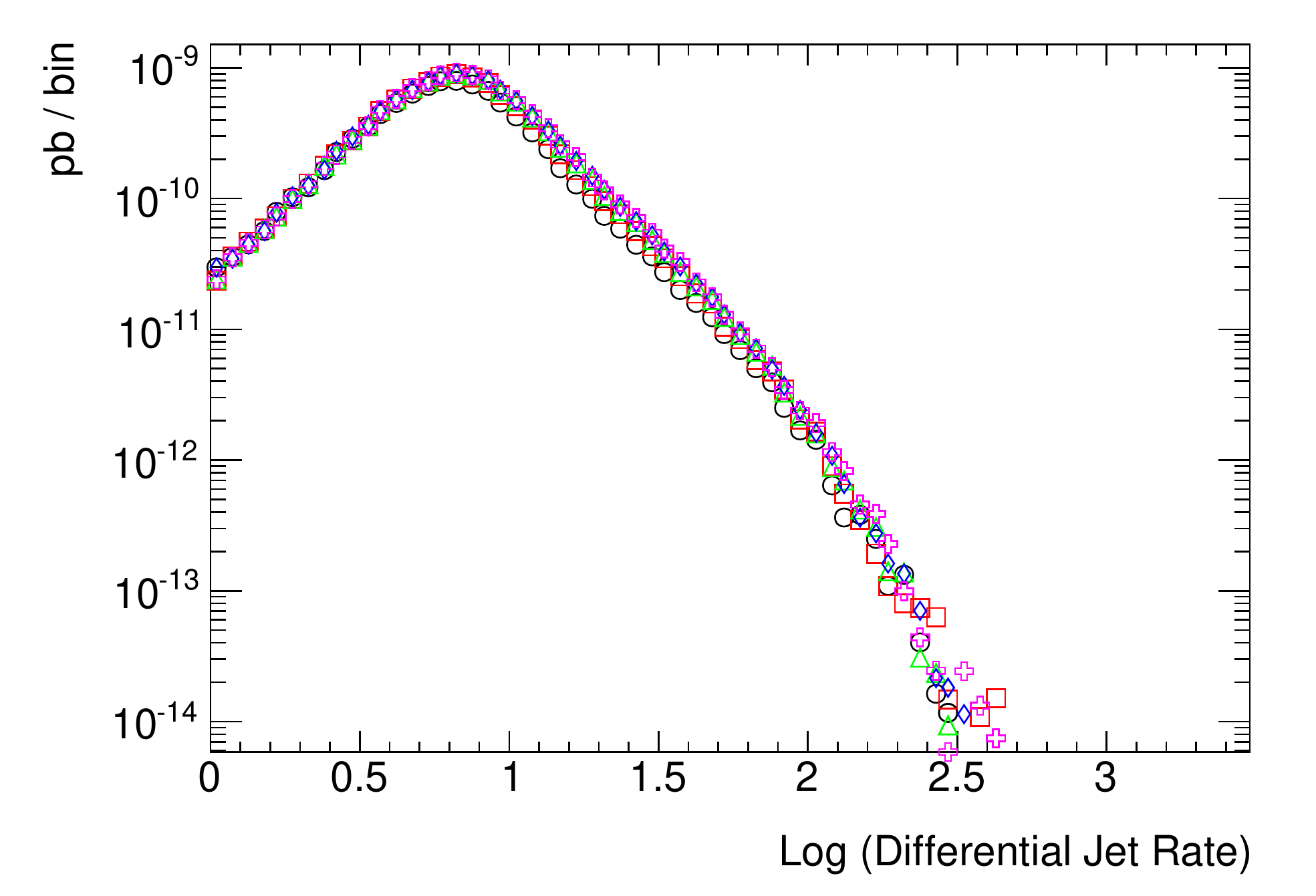}
 		\includegraphics[width=0.45\linewidth]{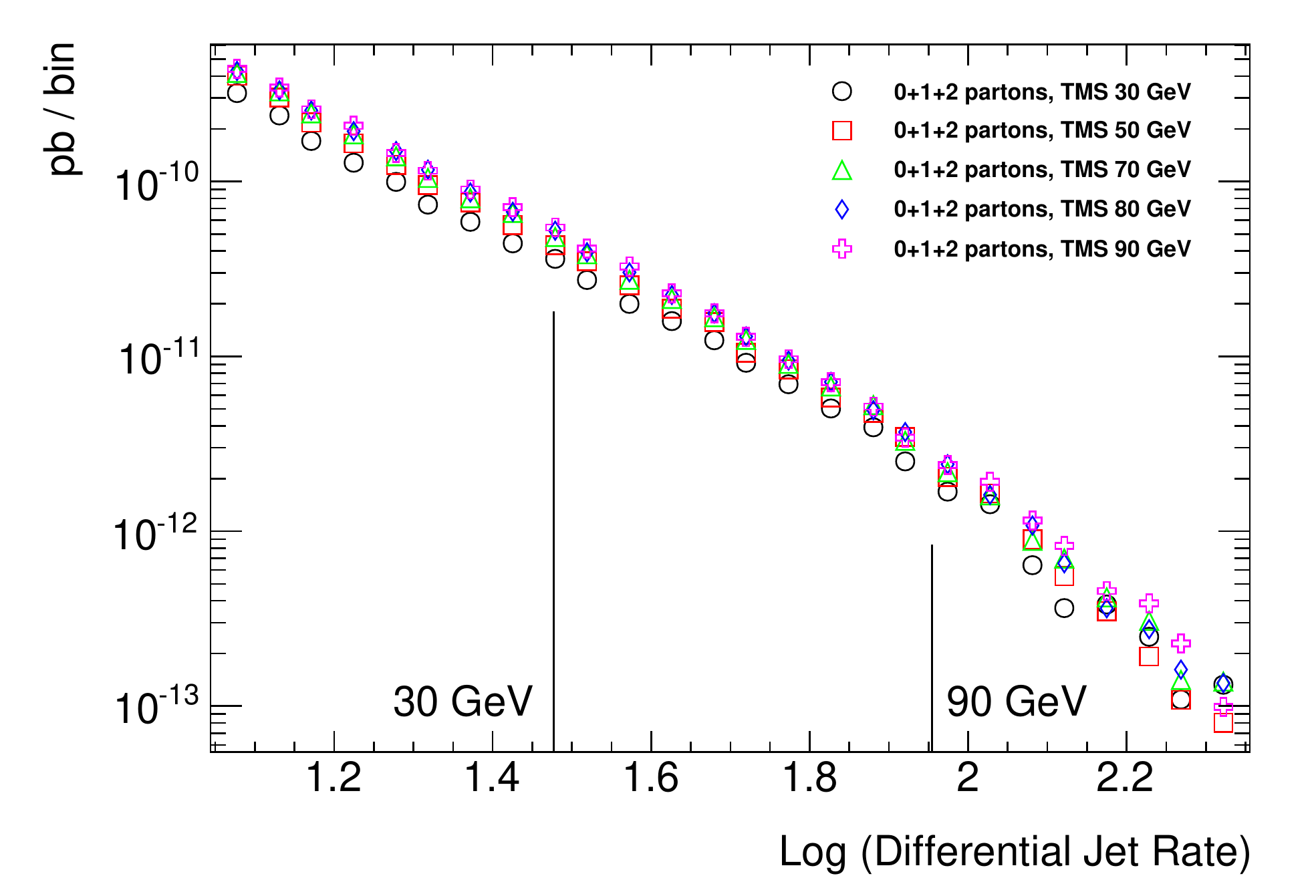}
 	}
   \caption{Distributions of differential jet rates $\frac{dN_{i\to j}}{d \log_{10}(k_\textrm{cut})}$ for EFT D5 sample with CKKW-L matching scale at 30, 50, 70, 80 and 90\,\gev. A zoom of the region around the matching scale values is shown on right.}
   \label{fig:CKKW_D5_zoom}
 \end{figure*}

 Two examples of differential jet rates, using matching scale 30\,\gev and 80\,\gev, from the EFT D5 sample generated as described in the previous section are given in Fig.\,\ref{fig:CKKW_D5_30} and \ref{fig:CKKW_D5_80}, respectively.
 Although a kink is visible around the matching scale value in both cases, the 80\,\gev scale leads to smoother distributions. 
 In order to find the optimal matching scale, additional samples with matching scale 50, 70, and 90\,\gev are generated as well and a detailed comparison of the differential jet rates close to the transition region is shown in Fig.\,\ref{fig:CKKW_D5_zoom}.
 The largest differences among the samples are visible for the $1\rightarrow2$ jets transition where the 30\,\gev and 50\,\gev scale lead to a drop of the rates around the matching scale values. On the contrary, there is a hint of an increased rate around the matching scale value in the sample generated with the 90\,\gev scale. Therefore, we recommend to use 80\,\gev as the baseline matching scale.

\begin{figure}[h!]
 	\centering  
     \includegraphics[width=0.95\linewidth]{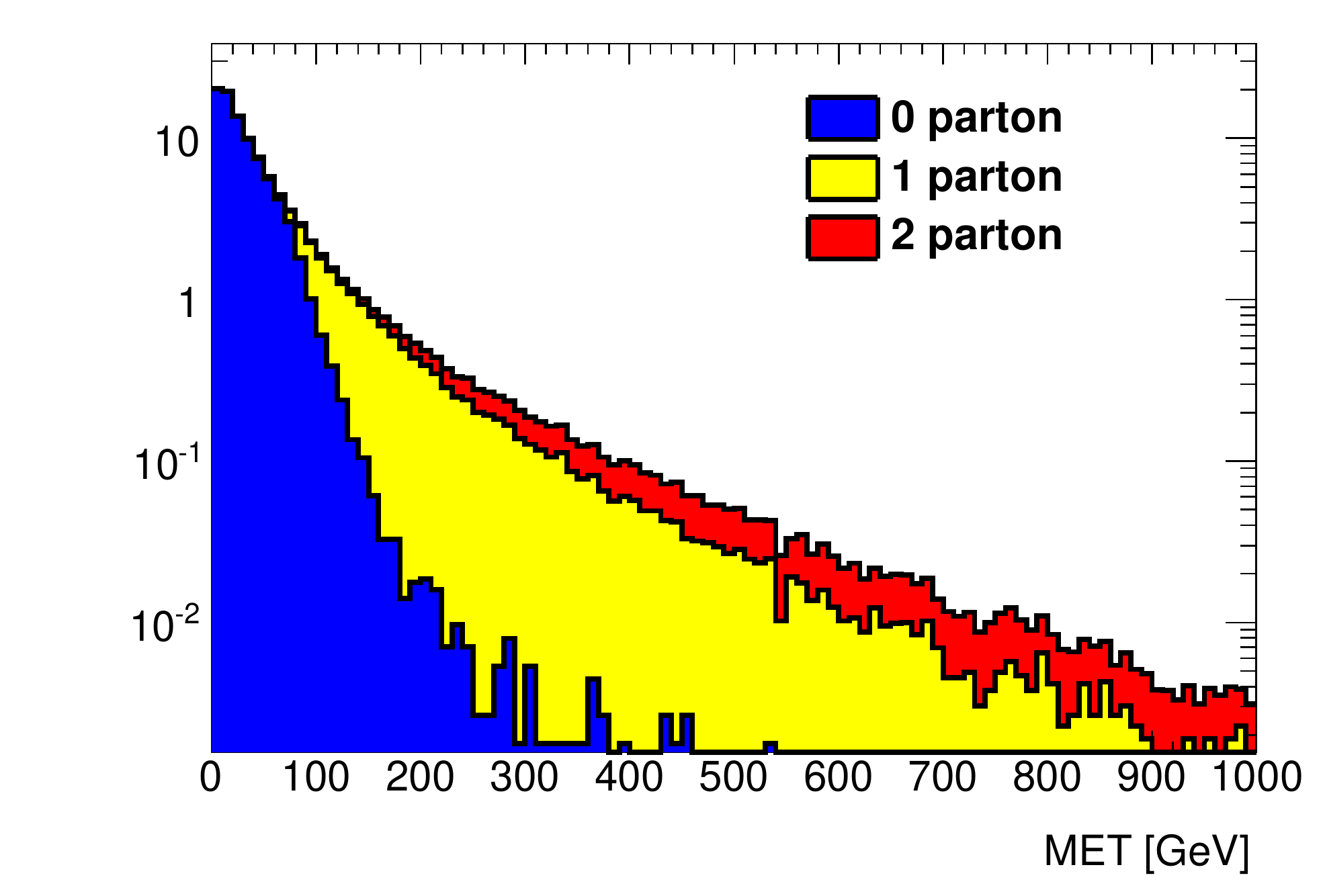}
 	\caption{Missing transverse momentum distributions for EFT D5 sample with CKKW-L matching scale at 80\,\gev. Individual contributions from the 0-, 1- and 2-parton emission samples are shown.}
 	\label{fig:Kine_D5_80}
 \end{figure}

 The prescription for the event generation given in Section\,\ref{sec:match_implementation} starts with the emission of 0 partons and ends with maxim 2 partons in addition. Producing the samples separately allows to investigate the relative composition of the individual samples in various parts of the phase space. Figure\,\ref{fig:Kine_D5_80} shows the \MET distribution of the EFT D5 sample with the matching scale at 80\,\gev. The plot reveals that the 0-parton sample gives the dominant contribution in the region below the matching scale value that rapidly decreases at higher \MET. Assuming the lowest analysis \MET cut in early Run-2 mono-jet analyses at 300\,\gev, the generation of the 0-parton emission sample can be safely omitted as it only gives $<1\%$ contribution at $\MET>300\,\gev$. For the 1- and 2-parton emission samples, one can use a generator cut on the leading parton $\pT$, \texttt{ptj1min}, in order to avoid generating low \MET events that are irrelevant for the analysis.

\begin{figure}[h!]
	\centering  
	\subfloat[No jet multiplicity cut]{%
		\includegraphics[width=0.95\linewidth]{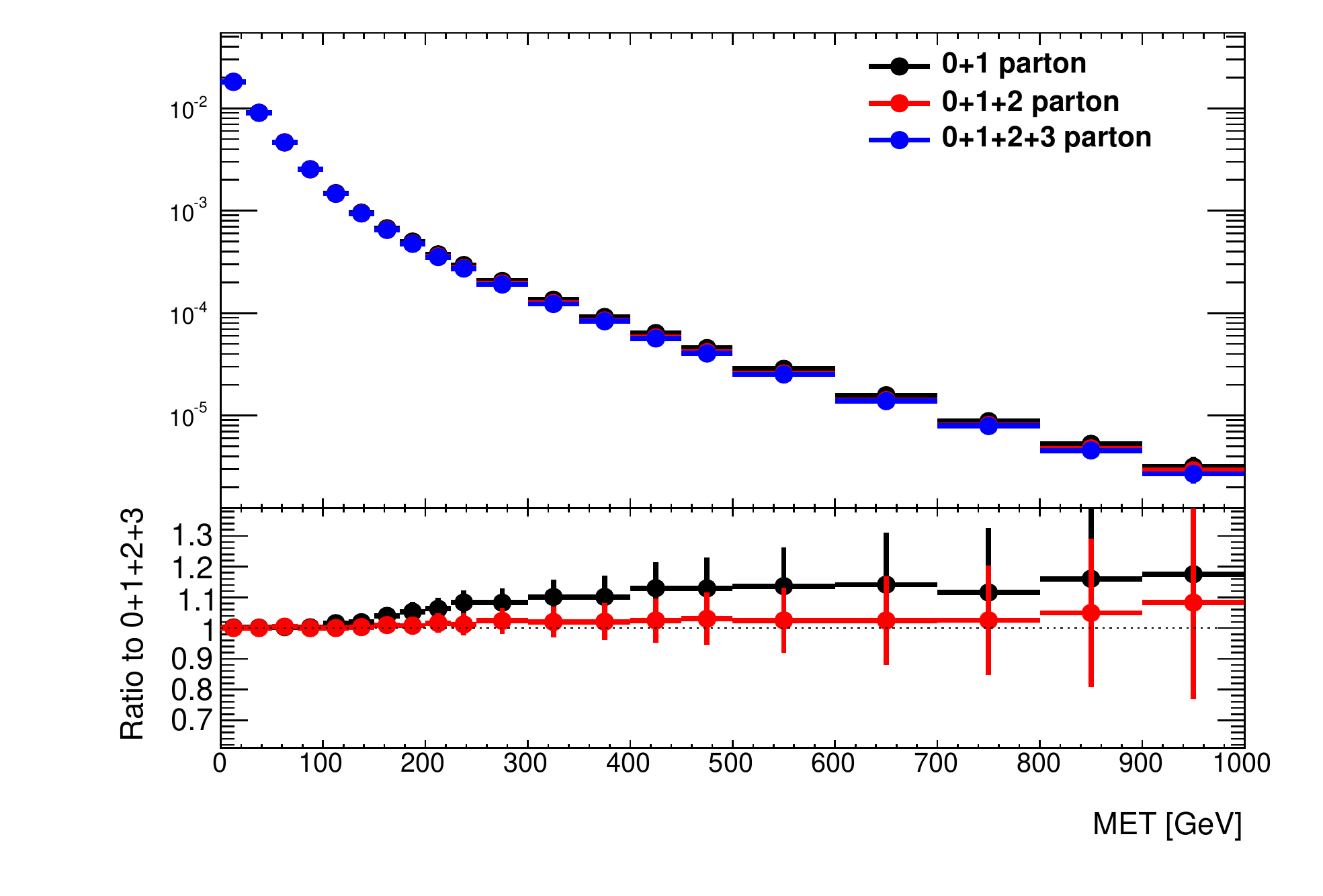}
	}
	\hfill
	\subfloat[$N_{\text{jet}}\leqslant$3]{%
		\includegraphics[width=0.95\linewidth]{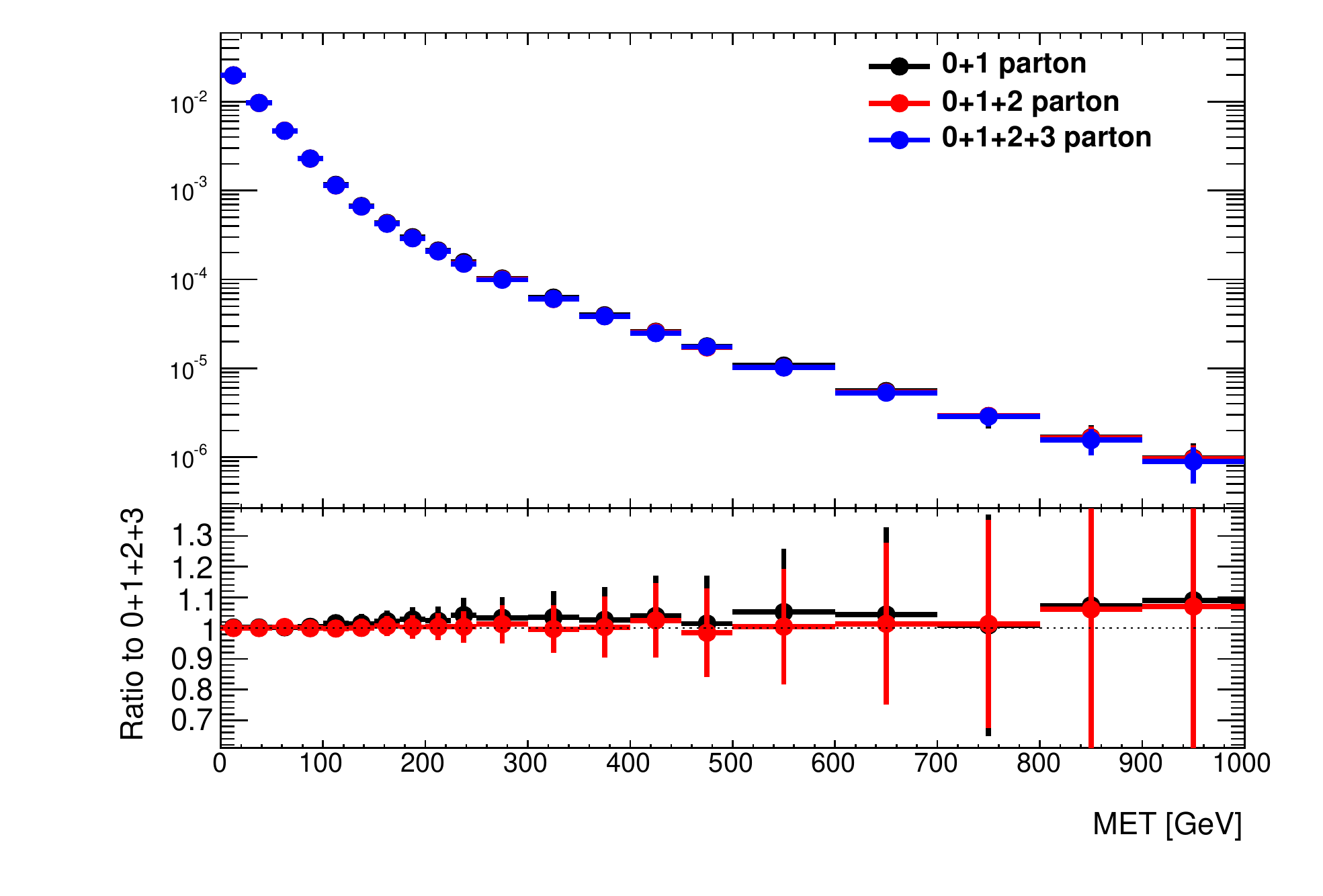}
	}
	\hfill
	\caption{Missing transverse momentum distributions for EFT D5 sample with CKKW-L matching scale at 80\,\gev produced with maximum 1 (black), 2 (red) and 3 (blue) partons emitted at the generator level. The ratios are shown with respect to the latter sample.}
	\label{fig:RatioKine_D5}
\end{figure}

In order to describe the signal kinematics correctly and save time during MC production, the parton emissions will only be generated up to a certain multiplicity. The higher multiplicity samples usually have small enough cross sections and the corresponding parts of the phase space can be sufficiently approximated by parton showering in \pythiaEight.
A dedicated study comparing samples generated with up to 1-, 2-, or 3-parton multiplicities was performed, using again the settings for the CKKW-L $k_T$-merging with the 80\,\gev matching scale and the \texttt{Merging:nJetMax} parameter adjusted accordingly.
Figure\,\ref{fig:RatioKine_D5} shows the \MET distribution of the samples at $\MET>250\,\gev$.

\begin{figure}[h!]
	\centering  
	\includegraphics[width=0.95\linewidth]{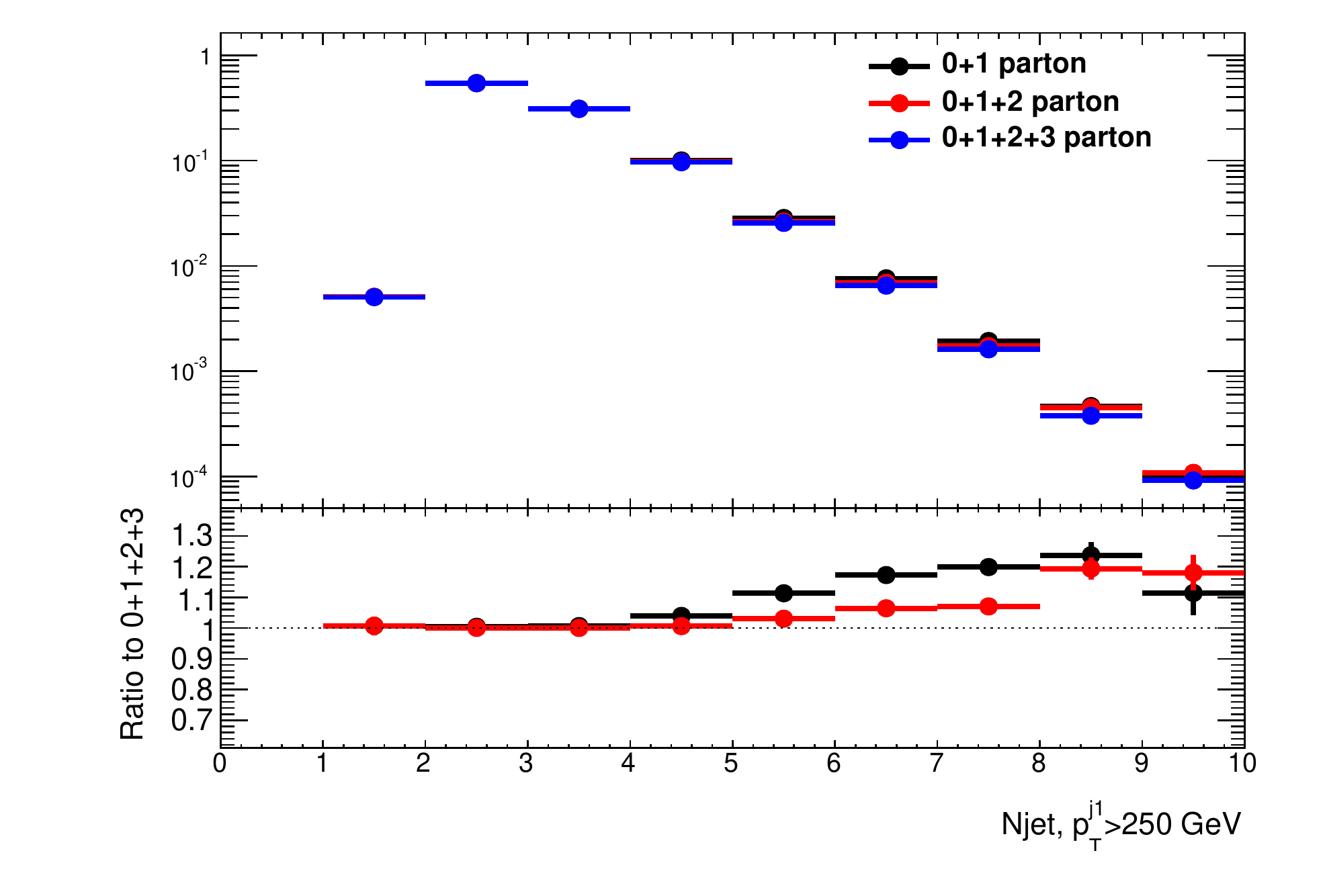}
	\caption{Multiplicity of jets with $\pT>30\,\gev$ and $|\eta|<2.8$ for EFT D5 sample with CKKW-L matching scale at 80\,\gev produced with maximum 1 (black), 2 (red) and 3 (blue) partons emitted at the generator level. The ratios are shown with respect to the latter sample. The leading jet $\pT$ is required to be larger than 250\,\gev.}
	\label{fig:RatioKine_D5_2}
\end{figure}

With an event selection requiring \MET and the leading jet \pT being larger than $250\,\gev$, the sample generated with up to 1 parton has 10.3\% larger yield compared to the sample with up to 3 partons, while the yield of the sample with up to 2 partons is only 2.3\% larger.
If an additional cut is applied allowing for up to 3 jets with $\pT>30\,\gev$, the agreement improves to 3.2\% larger for up to 1 parton and 0.7\% larger for up to 2 partons, compared with up to 3 partons.
A similar comparison is shown in Fig.\,\ref{fig:RatioKine_D5_2} for the jet multiplicity in the events with the leadning jet $\pT>250\,\gev$, where an agreement at the level of $\sim3\%$ between the samples with up to 2 and 3 parton emissions is observed for number of jets up to 7.
This justifies it is sufficient to produce samples with up to 2 parton emissions only at the generator level and ignore generating higher parton emissions.

\subsection{Implementation of \tchannel models for the jet+\MET{} final state}
\label{sec:tchannel_implementation}

The simulations for \tchannel models are available via LO UFO implementations, where events are generated at LO+PS accuracy. The UFO file and parameter cards for the \tchannel models with couplings to light quarks only~\cite{Papucci:2014iwa} can be found on the Forum SVN repository~\cite{ForumSVN_TChannel_PapucciVichiZurek}. The model files from Ref.~\cite{Bell:2012rg} can also be found on the repository~\cite{ForumSVN_TChannel_Amelia}. The latter is the implementation that has been used for the studies in this report: in the monojet case there are only cross section differences between this model and the model in~\cite{ForumSVN_TChannel_PapucciVichiZurek}. 
 
Multi-parton simulation and merging are necessary and require particular care for this model: this has not been a topic of detailed studies within the Forum, and we suggest to follow the procedure outlined in Ref.~\cite{Papucci:2014iwa}. 

\subsection{Implementation of \schannel and  \tchannel models with EW bosons in the final state}
\label{sec:EW_implementation}

Currently, simulations for most of these models are available via LO UFO implementations, allowing event generation at the LO+PS accuracy. We note, however, that inclusion of NLO corrections would be possible. In \madgraph, for example, this amounts to simply upgrading the currently employed UFO models to NLO, where the calculations exist for this class of processes. However, this was not available within the timescale of the Forum towards simulation of early Run-2 benchmarks. As a consequence, in this work we have used LO UFO implementations within \madgraph 2.2.3 interfaced to \pythiaEight for the parton shower. The corresponding parameter cards used for the Run-2 benchmark models can be found on the Forum SVN repository~\cite{ForumSVN_EW_DMV}. This is the implementation that will be used for early Run-2 LHC Dark Matter searches.

None of these models requires merging samples with different parton multiplicities  
since the visible signal comes from the production of a heavy SM boson whose transverse momentum distribution is sufficiently 
well described at LO+PS level.  As a result, no special runtime configuration is needed for \pythiaEight. 

\subsection{\texorpdfstring{Implementation of \schannel and \tchannel models with heavy flavor quark signatures}{Simulations for models with heavy flavor quark signatures}}
\label{sec:TTBar_implementation}

Dedicated implementations for DM signals in this final state are available at LO+PS accuracy. 
However, the state of the art of the simulations for $t \bar{t}$ and $b \bar{b}$ with a generic scalar and vector mediator is NLO+PS accuracy. For example, simulations for $t \bar{t}$ + scalar can be obtained via \powheg and {\sc sherpa} starting from the SM implementations. In \madgraph,  all final relevant final states, spin-0 (scalar and pseudo scalar) and spin-1, (vector and axial) are available  at NLO+PS via the dedicated NLO UFO for DM has been released in June 2015~\cite{NewMadgraphModels}).

In the work of this Forum, simulations for the $t \bar{t}$ and $b \bar{b}$ signatures of the scalar mediator model have been 
generated starting from a leading order UFO with \madgraph 2.2.2, using \pythiaEight for the parton shower. 
The UFO file and parameter cards that will be used as benchmarks 
for early Run-2 searches in these final states can be found on the Forum SVN repository~\cite{ForumSVN_DMTTBar}.
Multi-parton merging has been used for the $b \bar{b}$ case but it has not been studied in detail within this Forum. 
The b-flavored DM model of Section~\ref{sec:singleb} is simulated at LO+PS using \madgraph v2.2.3 and \pythiaEight for the parton shower.  
The corresponding UFO and parameter files can be found on the Forum SVN repository~\cite{ForumSVN_DMSingleB}.

\subsubsection{Quark flavor scheme and masses}

In the case of $b \bar{b}$ final state an additional care should be taken when choosing the flavor scheme 
generation and whether quarks should be treated as massive or massless.

The production of DM+$b\bar{b}$, Dark Matter in association with $b$ jets via a decay of a (pseudo) scalar boson, 
is dominated in simplified mediator models by the gluon-gluon initiated production, similar to the production of 
Z+$b\bar{b}$ at the LHC. The Z+$b\bar{b}$ process has been studied in detail in the Z(ll)+$b$-jets final state,  
which can be used to validate both the modeling of DM+bb and, its main background, Z(vv)+$b\bar{b}$. 
In this context, the $p_\textrm{T}$ of the Z boson is related to the observed MET, whereas the $b$-jet kinematics 
determines the ratio of mono-$b$/di-$b$ signatures in the detector.


For basic kinematic criteria applied to Z+$b\bar{b}$ production, 
this process leads in $\sim90\%$ of the events to a signature with only 1 $b$-jet in the acceptance (
'Z+1$b$-jet production') and only in  $\sim10\%$ of the events to a signature with 2 $b$-jets in the detector ('Z+2$b$-jets production). 
The production cross section of the Z+$b\bar{b}$ process can be calculated in the 'five-flavor scheme', 
where b quarks are assumed massless, and the 'four-flavor scheme', where massive b quarks are 
used~\cite{Campbell:2003dd,Maltoni:2005wd,Campbell:2005zv}.
Data slightly favour the cross-section predictions in the five-flavor scheme~\cite{Chatrchyan:2014dha} for the 1 $b$-jet signature. 
In this document we have preferred the 5-flavor scheme due to its simplicity and 
cross sections and models in the 5-flavor scheme are available in the repository. 
The PDF used to calculate these cross section is NNPDF3.0 (lhaid 263000). 

\begin{figure}[h!]
\begin{minipage}{0.49\textwidth}
	\centering 
	\includegraphics[scale=0.32]{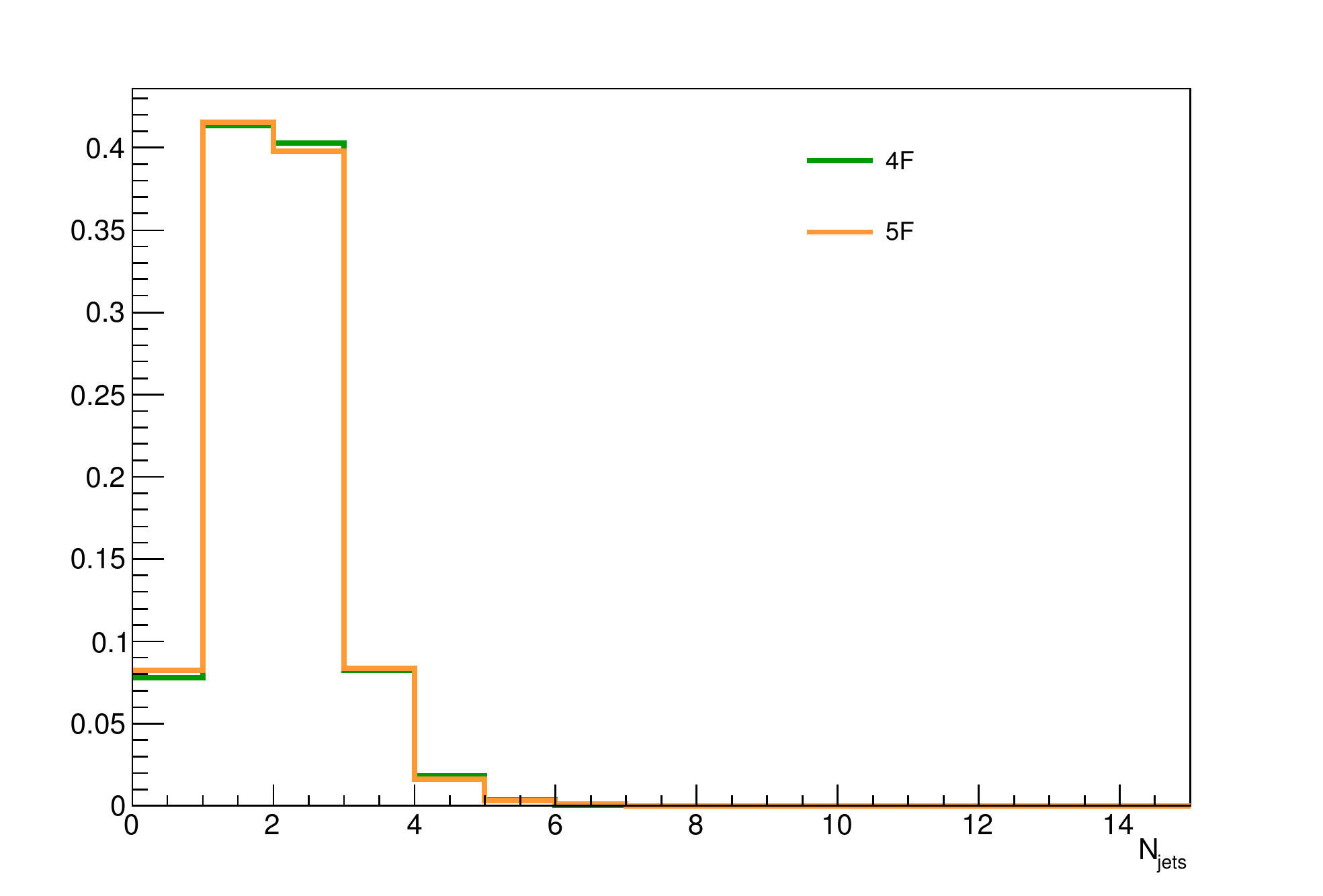}
\end{minipage}
\hfill
\begin{minipage}{0.49\textwidth}
	\centering 
	\includegraphics[scale=0.32]{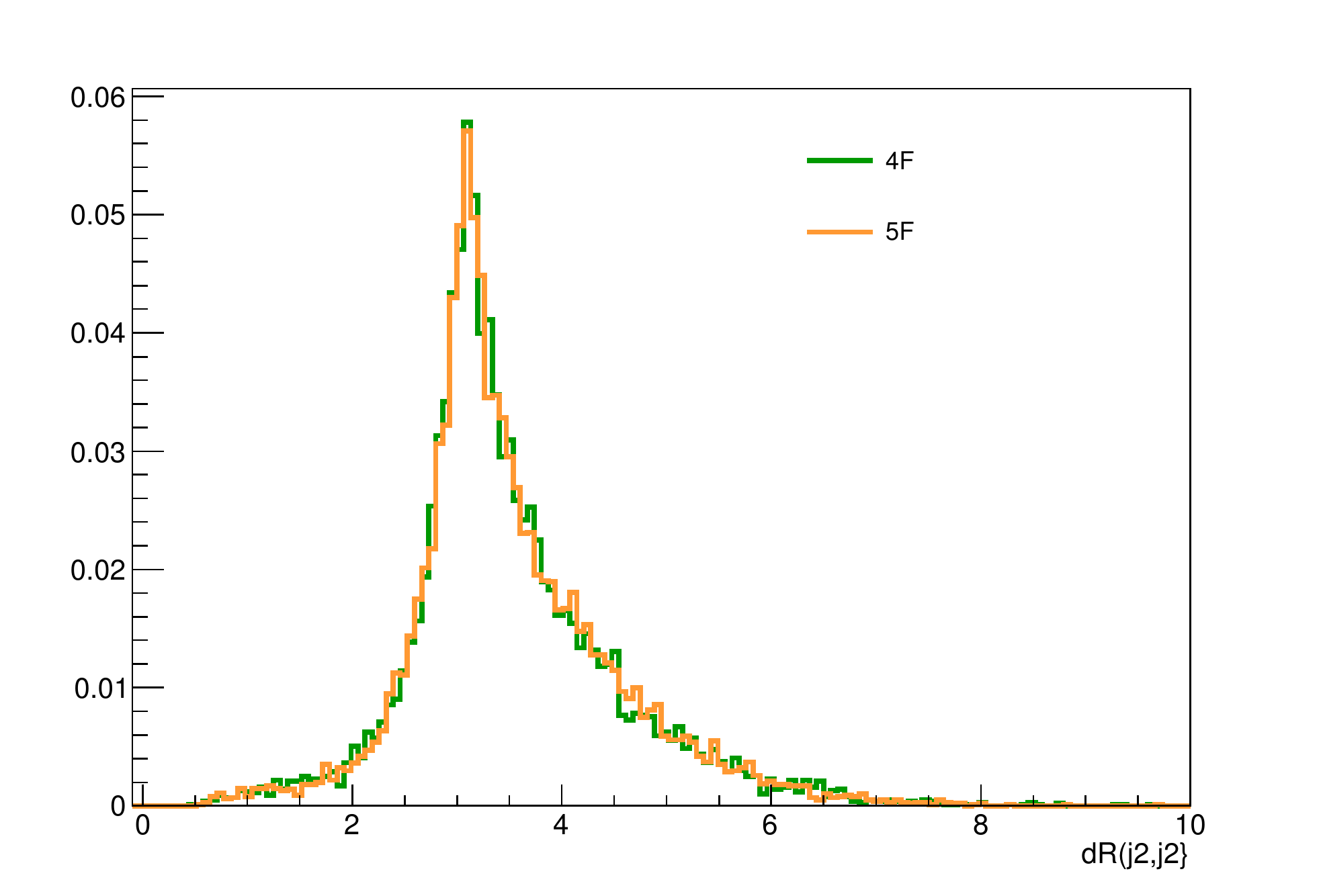}
\end{minipage}
\caption{Comparison of the jet multiplicity (left) and angular correction $\Delta R(j_1, j_2)$ (right) for the DM+$b\bar{b}$ scalar model generated in the 4-flavor and 5-scheme.
	The samples are generated for $m_\chi=1$~GeV and $m_\phi=10$~GeV.
	\label{fig:4Fvs5F}}
\end{figure}

On the other hand, both data~\cite{Chatrchyan:2014dha,Chatrchyan:2013zja,CMS:2015mba} 
and theoretical studies~\cite{Frederix:2011qg,Wiesemann:2014ioa} suggest that the best modelling of  an 
inclusive Z+$b\bar{b}$ sample especially for what concerns $b$-quark observables, is achieved at NLO+PS using a 4-flavor scheme 
and a massive treatment of the $b$-quarks.  
In Figure~\ref{fig:4Fvs5F} we show that, at LO, as expected, no appreciable difference is visible in the kinematics between 
either flavor scheme used for DM+$b\bar{b}$. In our generation we have used  NNPDF3.0 set (lhaid 263400).



\section{\texorpdfstring{Implementation of specific models for $V+\MET$ analyses}{Implementation of specific models for V+MET analyses}}

\subsection{Model implementation for mono-Higgs models}
\label{sub:monoHiggs}

Currently, simulations for most of these models are available via LO UFO implementations, allowing event generation at the 
LO+PS accuracy. We note, however, that the inclusion of NLO corrections would be possible but not available in time for the conclusion of these studies. In \madgraph, for example, this amounts to simply upgrading the currently employed UFO models to NLO. Simulation of  loop-induced associated production of DM and Higgs is also possible with the exact top-quark mass dependence. In \madgraph, for example, this can be obtained from the NLO UFO SM and 2HDM implementations. 

In this work all three Higgs+\MET models have been generated at leading
order with \madgraph 2.2.2, using \pythiaEight for the parton shower. No merging procedure has been employed.
The LO UFO implementations of the scalar and vector models that will be used as early Run-2 benchmarks can be found on the Forum SVN  
repository~\cite{ForumSVN_EWMonoHiggs}, while the 2HDM model can be found at this link~\cite{ForumSVN_EWMonoHiggs_2HDM}.

As a final technical remark, we suggest always to let the shower program handle the  $h$ decay 
(and therefore to generate a stable $h$ at the matrix element level).  
In so doing a much faster generation is achieved and the $h$ branching ratios are more accurately 
accounted for by the shower program. 

\subsubsection{\madgraph details for scalar mediator Higgs+MET model}

The case of the associated production of a Higgs and scalar mediator via a top-quark loop can be either considered 
exactly or via an effective Lagrangian where the top-quark is integrated out. While this latter model has 
been shown not to be reliable~\cite{Haisch:2012kf,Hespel:2014sla,Baur:1989cm}, for simplicity we have chosen to perform
the study in this tree-level effective formulation. A full study of the process including
finite top-quark mass and parton shower effects is possible yet left for future work.

\subsubsection{\madgraph details for 2HDM Higgs+MET model}
\label{sec:monoHImplementation}

While a 2HDM UFO implementation at NLO accuracy to be used with \madgraph has been made available at the end of the work
of the Forum~\cite{NewMadgraphModels}, in this work we have only considered LO simulations.
  
The two couplings that can be changed in the implemented model follow the nomenclature below:
 \begin{itemize}
 	\item \texttt{Tb} - $\tan \beta$
 	\item \texttt{gz} - $g_z$, gauge coupling of \Zprime to quarks
 \end{itemize}
 The other couplings are not changed, including \texttt{gx} (the $A \bar \chiDM \chiDM$ coupling) which has little impact on the signal. 
 $\sin \alpha$ is fixed internally such that $\cos (\beta-\alpha) = 0$. 
 The width of the \Zprime and $A$ can be computed automatically within \madgraph. 
 The couplings here don't affect the signal kinematics, so they can be fixed to default values  and then the signal rates can be scaled appropriately. 
 
The nomenclature for the masses in the implemented model is:
 \begin{itemize}
 	\item \texttt{MZp} - PDG ID 32 - \Zprime
 	\item \texttt{MA0} - PDG ID 28 - $A$
 	\item \texttt{MX} - PDG ID 1000022 - dark matter particle
 \end{itemize}
 
The other masses are unchanged and do not affect the result. 
 Both $\Zprime \to hZ(\bar \nu \nu)$ and  $\Zprime \to hA(\bar \chiDM \chiDM)$ contribute to the final state, scaling
 different with model parameters. We recommend to generate them separately, 
 and then add the two signal processes together weighted by cross sections.

\subsection{Implementation of EFT models for EW boson signatures}
\label{sub:EFTModels}

The state of the art for these models is LO+PS. NLO+PS can be achieved
as well, but the corresponding implementation is not yet available.
In our simulations we have implemented the models in the corresponding
UFO files and  generated events at LO via \madgraph 2.2.2, using \pythiaEight for the parton shower. 
UFO files and parameter cards that will be used as early Run-2 benchmarks can be found on the Forum SVN repository:~\cite{ForumSVN_EWMonoHiggs} for operators 
with Higgs+MET final states and ~\cite{ForumSVN_EWEFTD7} for $W/Z/\gamma$ final states.
These models do not require merging.

\chapter{Presentation of EFT results}
\label{sec:EFTValidity} 


Most of this report has focused on simplified models.
In this Chapter, we wish to emphasize the applicability of
Effective Field Theories (EFTs) 
in the interpretation of DM searches at the LHC.
Given our current lack of knowledge about the nature of a DM particle and
its interactions, it appears mandatory to provide the necessary information
for a model independent interpretation of the collider bounds.
This approach should be complemented with
an interpretation within a choice of simplified models.
We note that, even though EFT benchmarks are only valid in given conditions,
the results provided by the current list of simplified models cannot always
characterize the breadth of SM-DM interactions.
In at least one case, composite WIMPs~\cite{Nussinov:1985xr,Kaplan:1991ah,Banks:2010eh}, 
the contact interaction framework is the correct one to
constrain new confinement scales. 


Ideally, experimental constraints should be shown as bounds of allowed signal events in the kinematic regions considered for 
the search, as detailed in Appendix~\ref{app:Presentation_Of_Experimental_Results}. 
A problematic situation is the attempt to derive a limit on
nucleon-dark matter scattering cross sections from EFT results
based on collider data~\footnote{Comparisons between constraints from different experiments 
	meant to highlight their complementarity should be expressed as 
	a function of the model parameters rather than on derived observables;
	however this is a point that should be developed further after the conclusion of the work of this Forum.}. 
Experiments that directly probe the nucleon-dark matter scattering cross section 
are testing the regime of small momentum transfers, where the EFT approximation typically holds.  
Collider experiments, though, are sensitive to large momentum transfers: 
We first illustrate the complications
that can arise with EFTs at colliders by considering an effective interaction
$$ {\cal L}_\textrm{int} = {(\bar q\gamma_\mu q)(\bar\chiDM\gamma^\mu\chiDM) \over \Mstar^2}
= (\bar q\gamma_\mu q)(\bar\chiDM\gamma^\mu\chiDM) {g\over \Lambda^2}$$
that couples quarks and DM $\chi$ fields.\footnote{The \textit{exact} operator chosen is not important:
	as detailed in the following, statements concerning the applicability of an EFT can also be made without a specific relation to simplified models.}  
The strength of this interaction is
parametrized by $\displaystyle {1\over \Mstar^2} = {g\over \Lambda^2}$.
A monojet signature can be generated from this operator
by applying perturbation theory in the QCD coupling.
An experimental search will place a limit on \Mstar.   
For a fixed \Mstar, a small value of $g$ will correspond
to a small value of $\Lambda$.   The EFT approximation breaks down
if $Q>\Lambda$, where $Q$ is a typical hard scale of the process.
The limit on small $g$ can only be reliable if the
kinematic region $Q>\Lambda$ is removed from the event generation.
However, if a fraction of events is removed from the prediction,
the corresponding value of $g$ must increase to match the experimental
limit on \Mstar.
On the other hand, if, for the same value of \Mstar, a large $\Lambda$
is assumed so that the full set of events fulfill the EFT validity condition,
a larger value of g is required.  For large enough g, computations based on perturbation theory become unreliable.

In the first part of this Chapter, we summarize two methods that
have been advocated to truncate events that 
do not fulfill the condition necessary for the use of an EFT.
These methods are described in detail in Refs.~\cite{Busoni:2013lha,Busoni:2014sya,Busoni:2014haa,Aad:2015zva,Racco:2015dxa,Berlin:2014cfa}. 
We then propose a recommendation for the presentation of EFT results for early Run-2 LHC searches.

\section{Procedures for the truncation of EFT benchmark models}

\subsection{EFT truncation using the momentum transfer and information on UV completion}

\label{sec:TruncationWithQTr}

In the approach described in Ref.~\cite{Busoni:2014sya},
the EFT prediction is modified to incorporate the effect of a
propagator for a relatively light mediator.
For a tree-level interaction between DM and
the SM via some mediator with mass \mMed, 
the EFT approximation corresponds to expanding the propagator
for the mediator
in powers of $\Qtr^2/\mMed^2$, truncating at lowest order, and combining the remaining parameters into a single parameter \Mstar 
(connected to the scale of the interaction $\Lambda$ in the literature).
For an example scenario with a \Zprime-type mediator (leading to some combination of operators D5 to D8 in the notation of~\cite{Goodman:2010ku} for the EFT limit),
this corresponds to setting
\be
\frac{\gDM \gq}{Q_{\rm tr}^2-\mMed^2}=-\frac{\gDM \gq}{\mMed^2}\left(1+\frac{Q^2_{\rm tr}}{\mMed^2}+ \mathcal{O} \left(\frac{Q^4_{\rm tr}}{\mMed^4}\right)\right) \simeq -\frac{1}{{\Mstar^2}},
\ee
where $\Qtr$ is the momentum carried by the mediator, and $\gDM$,
$\gq$ are the DM-mediator and quark-mediator couplings
respectively.\footnote{Here, we ignore potential complications from
the mediator width when the couplings are large.}
A minimal condition that must be satisfied for this approximation to be valid is that $\Qtr^2 < \mMed^2 = \gDM \gq {\Mstar^2}$.
This requirement avoids the regions:
$\Qtr^2 \sim \mMed^2$, in which case the EFT misses a resonant enhancement, and it is conservative to ignore this enhancement;
and $\Qtr^2 \gg \mMed^2$, in which case the signal cross section
should fall according to a power of $\Qtr^{-1}$ instead of $\mMed^{-1}$.   The latter is the problematic kinematic region.

The condition $\Qtr^2 < \mMed^2 = \gDM \gq \Mstar^2$ was applied
to restrict the
kinematics of the signal and remove events for which the high-mediator-mass approximation made in the EFT would not be reliable.
This leads to a smaller effective cross-section, after imposing the event selection of the analysis.  This truncated signal was then used
to derive a new, more conservative limit on
$\Mstar$ as a function of $(\mDM, \gDM \gq)$.

For the example D5-like operator,
where the cross section $\sigma$ scales as $\Mstar^{-4}$,
there is a simple rule for converting a rescaled cross section into a rescaled constraint on ${\Mstar}$.
if the original limit is based on a simple cut-and-count procedure.
Defining $\sigma_{\rm EFT}^{\rm cut}$ as the cross section truncated such that all events pass the condition $\sqrt{\gDM \gq} \Mstar^{\rm rescaled} > \Qtr$, we have
\be
\Mstar^{\rm rescaled} = \left(\frac{\sigma_{\rm EFT}}{\sigma_{\rm EFT}^{\rm cut}(\Mstar^{\rm rescaled})}\right)^{1/4} \Mstar^{\rm original},
\ee
which can be solved for $\Mstar^{\rm rescaled}$ via either iteration or a scan.
Similar relations exist for a given UV completion of each operator. 

This procedure has been proposed in Ref.~\cite{Busoni:2014sya} 
and its application to ATLAS results can
be found in Ref.~\cite{Aad:2015zva} for a range of operators.
We reiterate: knowledge of the UV completion for a given
EFT operator was necessary for this procedure;
this introduces a model-dependence that was not present
in the non-truncated EFT results.

Currently, simplified models (including the full effect
of the mediator propagator) are available for comparison with
the data, and since knowledge of the simplified models is needed
for the truncation procedure,
there is no reason to apply this prescription.   Instead, the
simplified model limit for large \Mstar can be presented for
interpretation in terms of EFT operators.

\subsection{EFT truncation using the center of mass energy}
\label{sec:TruncationWithSHat}

The procedure presented in the previous section was predicated on
some knowledge of the simplified model.  This led to the identification
of the mass of the DM pair as the relevant kinematic quantity to use
in a truncation procedure.
In general, if no assumption is made about the underlying dynamics,
it is more conservative to place a limit on the total center
of mass energy $E_\text{cm}$ of the DM production process.
Furthermore, the direct connection between the mass scale of
the EFT validity, \Mcut, and the
mass scale that normalizes the EFT operator, \Mstar, is unknown.
For such cases,~Refs.\cite{Racco:2015dxa,Berlin:2014cfa} proposed
a procedure to extract model independent and consistent bounds within the EFT
that can be applied to any effective Lagrangian describing the interactions between the DM and the SM.
This procedure provides conservative limits that can be directly reinterpreted in any completion of the EFT.
The condition ensuring that the EFT approximation is appropriate is:
\begin{equation}
\label{Ecm<Mcut}
E_\text{cm}<M_\text{cut}\,.
\end{equation}



%
The relationship between \Mcut and \Mstar can be parameterized
by an \textit{effective coupling strength} \gstar, such that
$\Mcut=\gstar \, \Mstar\,.$
A scan over values of \gstar provides an indication of the
sensitivity of the prediction to the truncation procedure.
In the \Zprime-type model considered above, \gstar is equal to $\sqrt{\gDM\gq}$.
The resulting plots are shown in \cite{Racco:2015dxa} for a particular effective operator. 

The advantage of this procedure is that the obtained bounds can be directly and easily recast in any  completion of the EFT, by computing the parameters \Mstar, \Mcut in the full model as functions of the parameters of the complete theory. On the other hand, the resulting limits will be weaker than those obtained using \Qtr and a specific UV completion.

\subsection{Truncation at the generator level}

The conditions on the momentum transfer can also be applied directly at the generator level, by discarding 
events that are invalid and calculating the limits from this truncated shape. 
This provides the necessary rescaling of the cross section while keeping the information on the change in the kinematic distributions due to the removal of the invalid events. This procedure is more general with 
respect to rescaling the limit in the two sections above, and it should be followed if
a search is not simply a counting experiment and exploits the shapes of kinematic distributions.

\subsection{Sample results of EFT truncation procedures}

An example of the application of the two procedures to the limit on $M_*$ from Ref.~\cite{ATL-PHYS-PUB-2014-007} as a function of the product of the couplings is shown in Figure~\ref{Ecm<Mcut}. Only the region between the dashed and the solid line is excluded. It can be seen that the procedure from~\cite{Racco:2015dxa} outlined in Section~\ref{sec:TruncationWithSHat}, shown in blue, is more conservative than the procedure from Refs.~\cite{Busoni:2014sya,Aad:2015zva}, described in Section~\ref{sec:TruncationWithQTr}.

\begin{figure}
	\centering
	\includegraphics[width=0.95\textwidth]{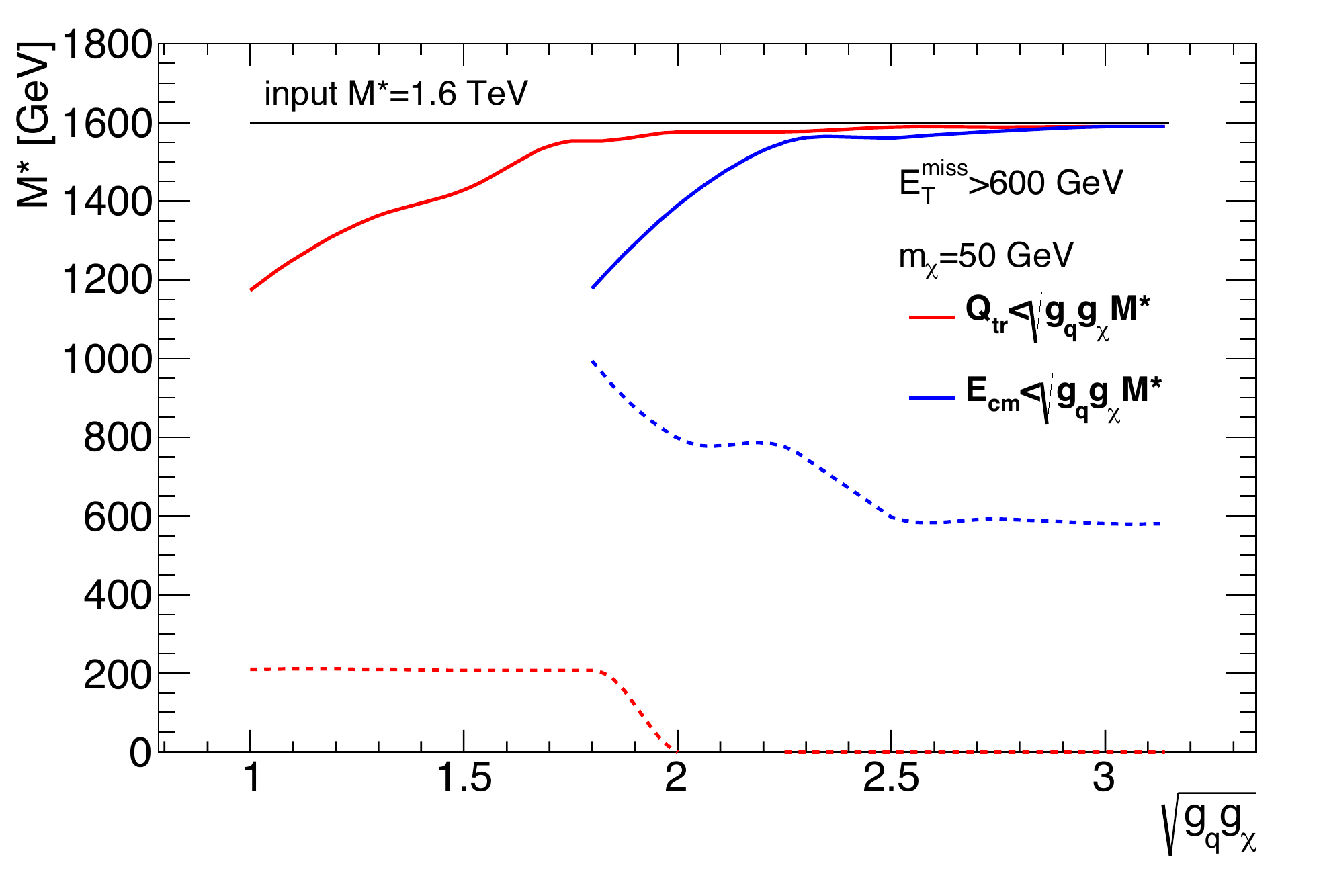}
	\caption{95\% CL lower limits on the scale of the interaction of the D5 operator at 14~\tev, after the two truncation procedures. 
		The procedure from~\cite{Racco:2015dxa} outlined in Section~\ref{sec:TruncationWithSHat} is shown in blue, while the procedure from Refs.~\cite{Busoni:2014sya,Aad:2015zva}, described in Section~\ref{sec:TruncationWithQTr} is shown in red. Only the region between the dashed and the solid lines is excluded. Even though the intersection between the two lines is not shown in this plot, it should be noted that no limit can be set anymore for sufficiently low couplings, whatever truncation method is used.}
	\label{fig:monojet_MstarMmed}
\end{figure}


\subsection{Comments on unitarity considerations}

A further consideration applicable to EFT operators at hadron colliders
is the potential violation of unitarity.  An analysis of the operator
$\displaystyle {\bar{q}\gamma_\mu q\bar\chi\gamma^\mu\chi \over \Mstar^2}$
provides the limit:
\be
\Mstar > \beta(s) \sqrt{s}  \sqrt{ {\sqrt{3} \over 4 \pi}  },
\ee
where $\sqrt{s}$ is (maximally) the collider energy and $\beta(s)$ is
the DM velocity \cite{Shoemaker:2011vi}.
Constraints for other operators have also been derived \cite{Endo:2014mja}.
This constraint on \Mstar still is open to interpretation, since the
relation to \Mcut is not resolved, except for a specific simplified model.
Derived limits on \Mstar should be compared to this unitarity bound to
check for consistency.

\section{Recommendation for presentation of EFT results} 
\label{sec:RecommendationEFTResults}

In this report,
we make two recommendations for the presentation of collider results
in terms of Effective Field Theories for the upcoming Run-2 searches. 
A full discussion of the presentation of
collider results in relation to other experiments
is left to work beyond this Forum, where ATLAS, CMS, the theory community
and the Direct and Indirect Detection communities are to be involved. 

We divide the EFT operators in two categories: 
those that can be mapped to one or more UV-complete simplified models, such as those
commonly used in LHC searches so far and detailed in~\cite{Goodman:2010ku}, and those
for which no UV completion is available to LHC experiments, such as those outlined in Section~\ref{sec:EFT_models_with_direct_DM_boson_couplings}.

\subsection{EFT benchmarks with corresponding simplified models}
\label{sub:EFT_withSimp}

If a simplified model can be mapped to a given EFT, then the model's high-mediator-mass limit  will converge to the EFT.

\begin{figure*}[p!]
	\centering
	\subfloat[\MET{}]{
		\includegraphics[width=0.5\linewidth]{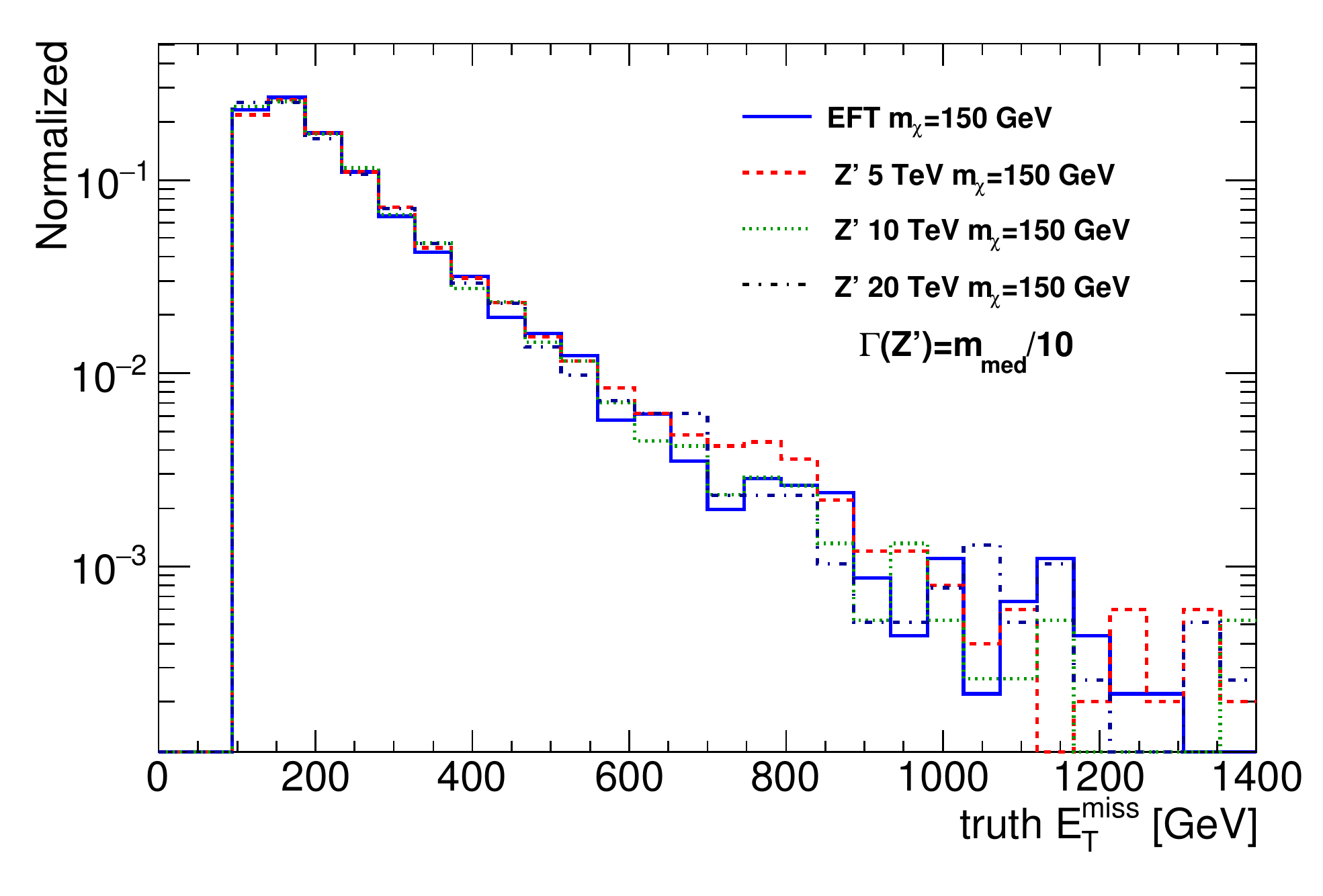}
	}
		\hfill
	\subfloat[Center of mass energy $E_{cm}$]{
	   \includegraphics[width=0.47\linewidth]{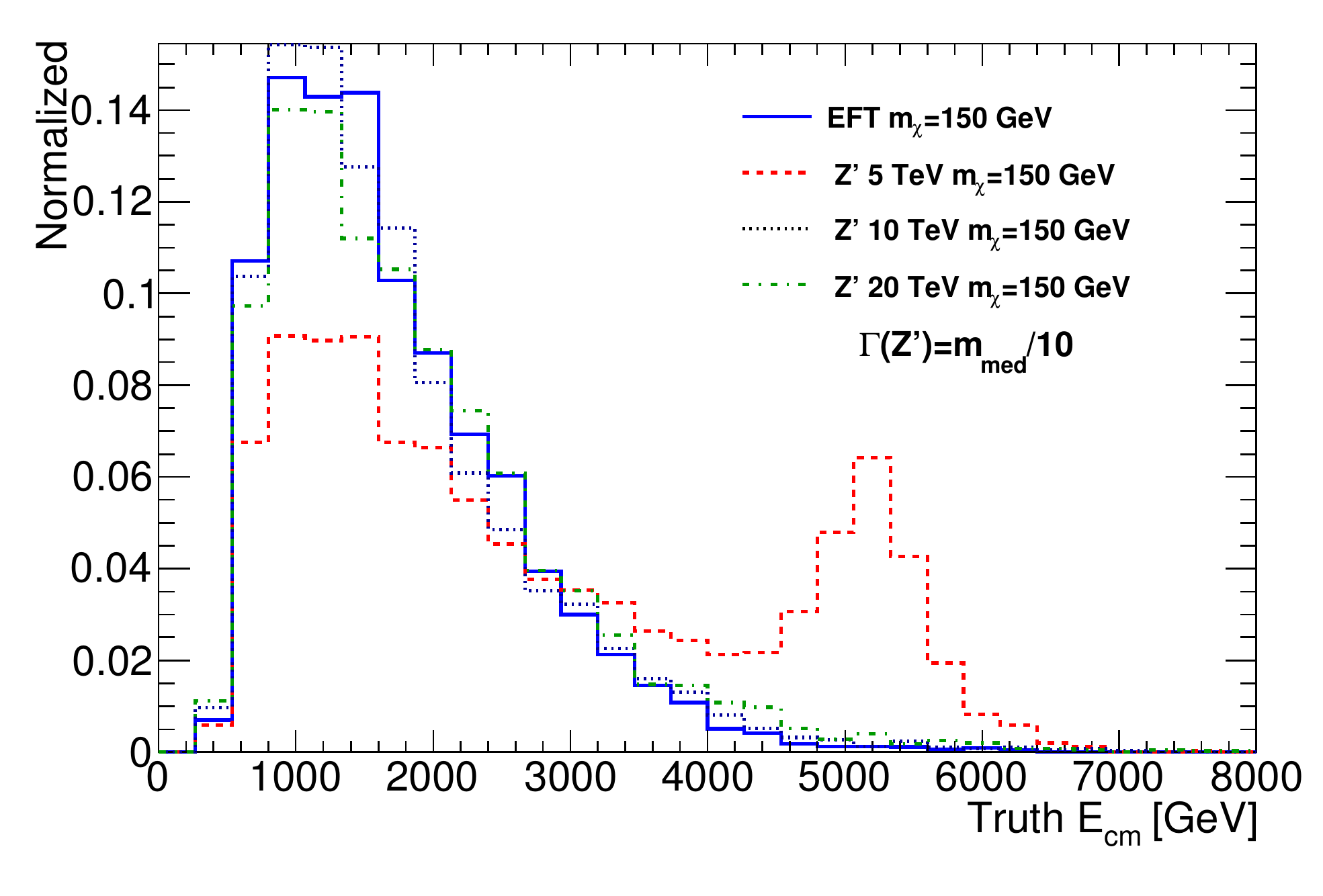} 
	}		
	\subfloat[Mediator transverse momentum]{
		\includegraphics[width=0.47\linewidth]{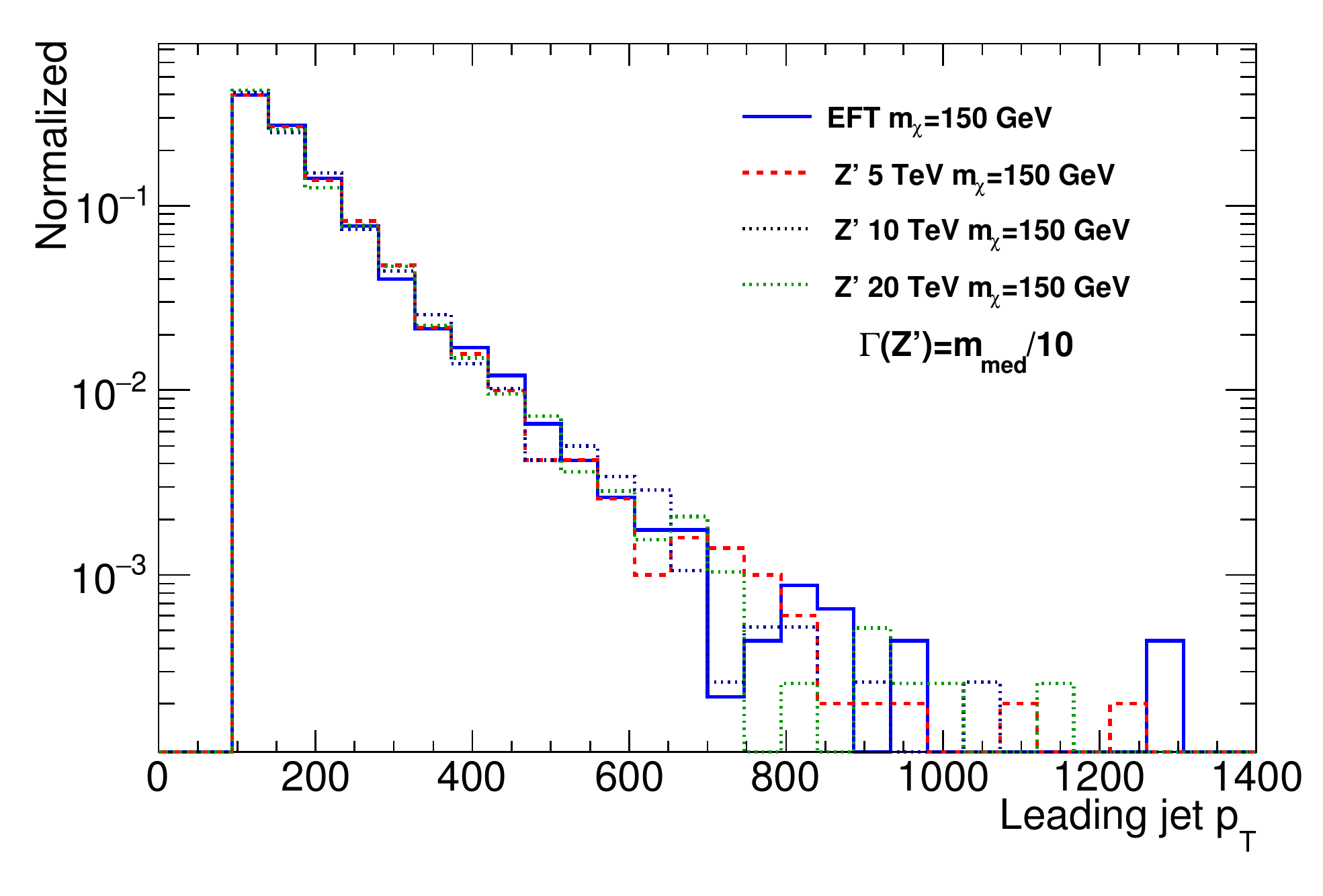}
	}
		\hfill
	\subfloat[Leading DM transverse momentum]{
		\includegraphics[width=0.47\linewidth]{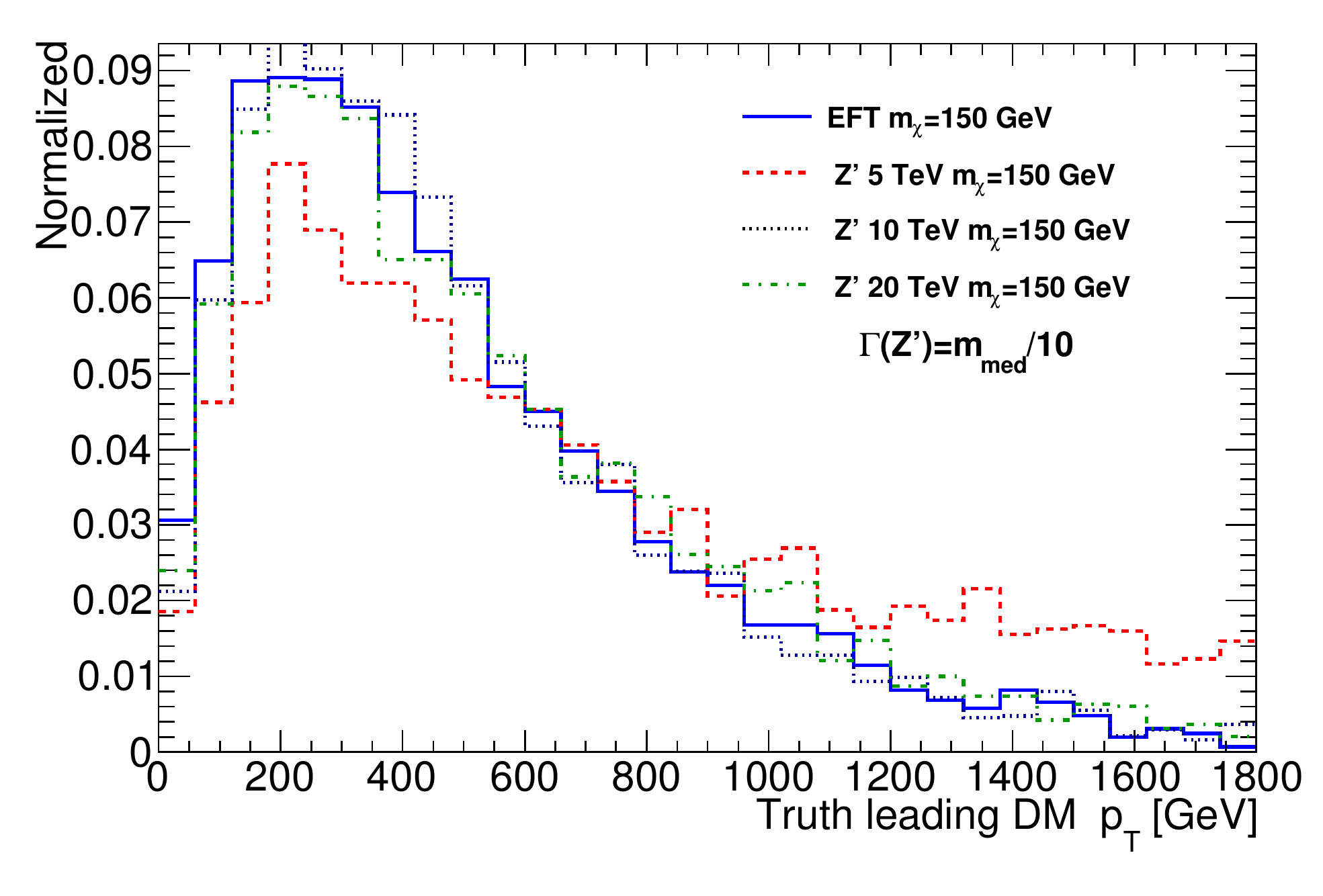} 
		}
	\subfloat[DM transverse momentum (sub-leading)]{
		\includegraphics[width=0.47\linewidth]{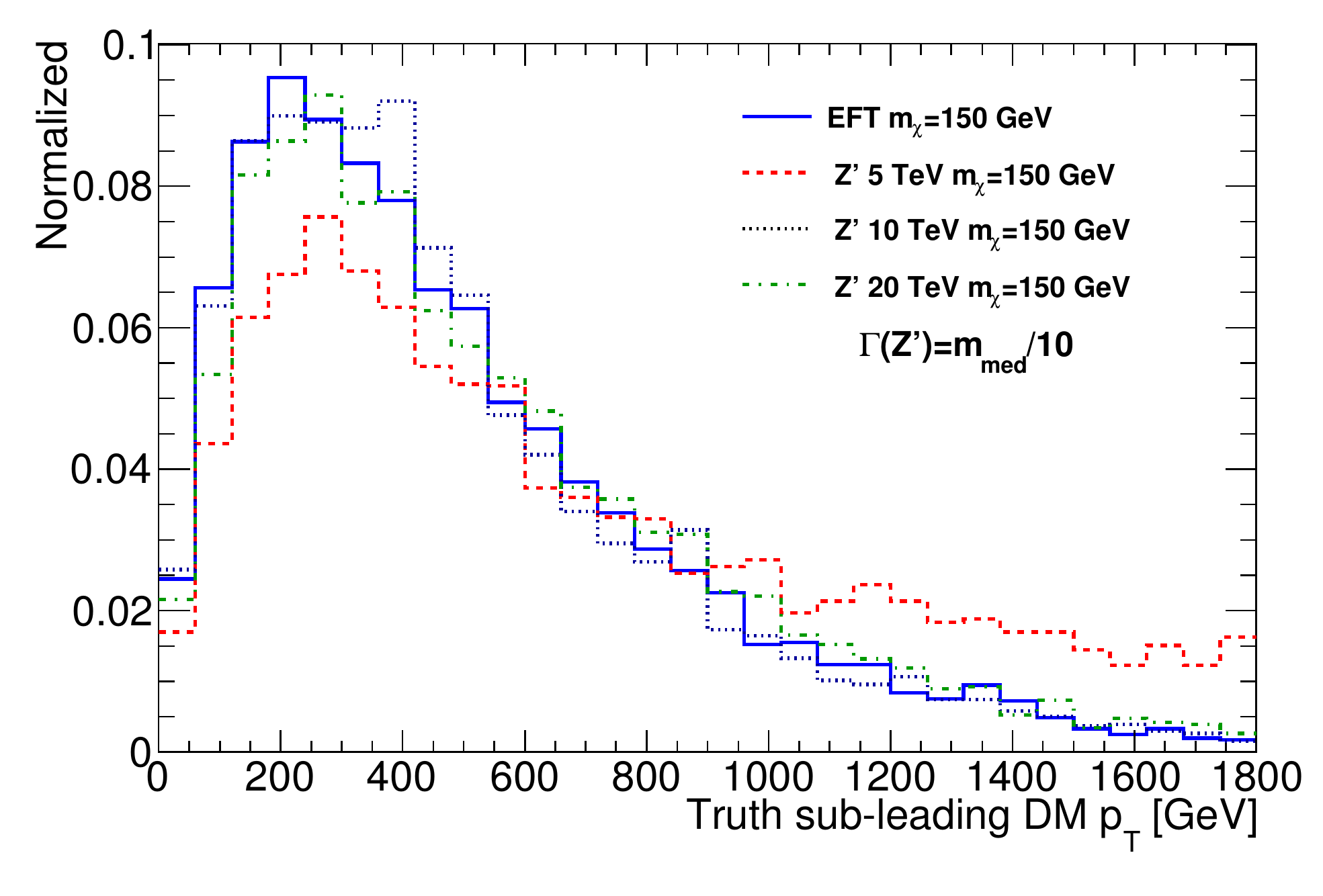} 
	}
	\vskip20pt
		\caption{Comparison of the kinematic distributions at 14~\tev between a narrow \schannel mediator and the
			corresponding D5 contact operator, at generator level for a jet+\MET{} signature. 
			\label{fig:EFT_kinematics}}
		
\end{figure*}

A study of 14 TeV benchmarks for narrow resonances with \gq = 0.25 and \gDM = 1 (see Section~\ref{sub:parameter_scan_monojet})
shows that a mediator with a mass of at least 10 TeV fully reproduces the kinematics of a contact
interaction and has no remaining dependence on the presence of a resonance. 
A comparison of the main kinematic variables for the \schannel vector 
mediator model with a width of 0.1 \mMed
is shown in Fig.~\ref{fig:EFT_kinematics}.\footnote{The use of a fixed width rather than the minimal width is exclusive of these plots.} 


\begin{figure*}[p!]
	\centering
	\subfloat[\MET{}, D5 operator]{
		\includegraphics[width=0.47\linewidth]{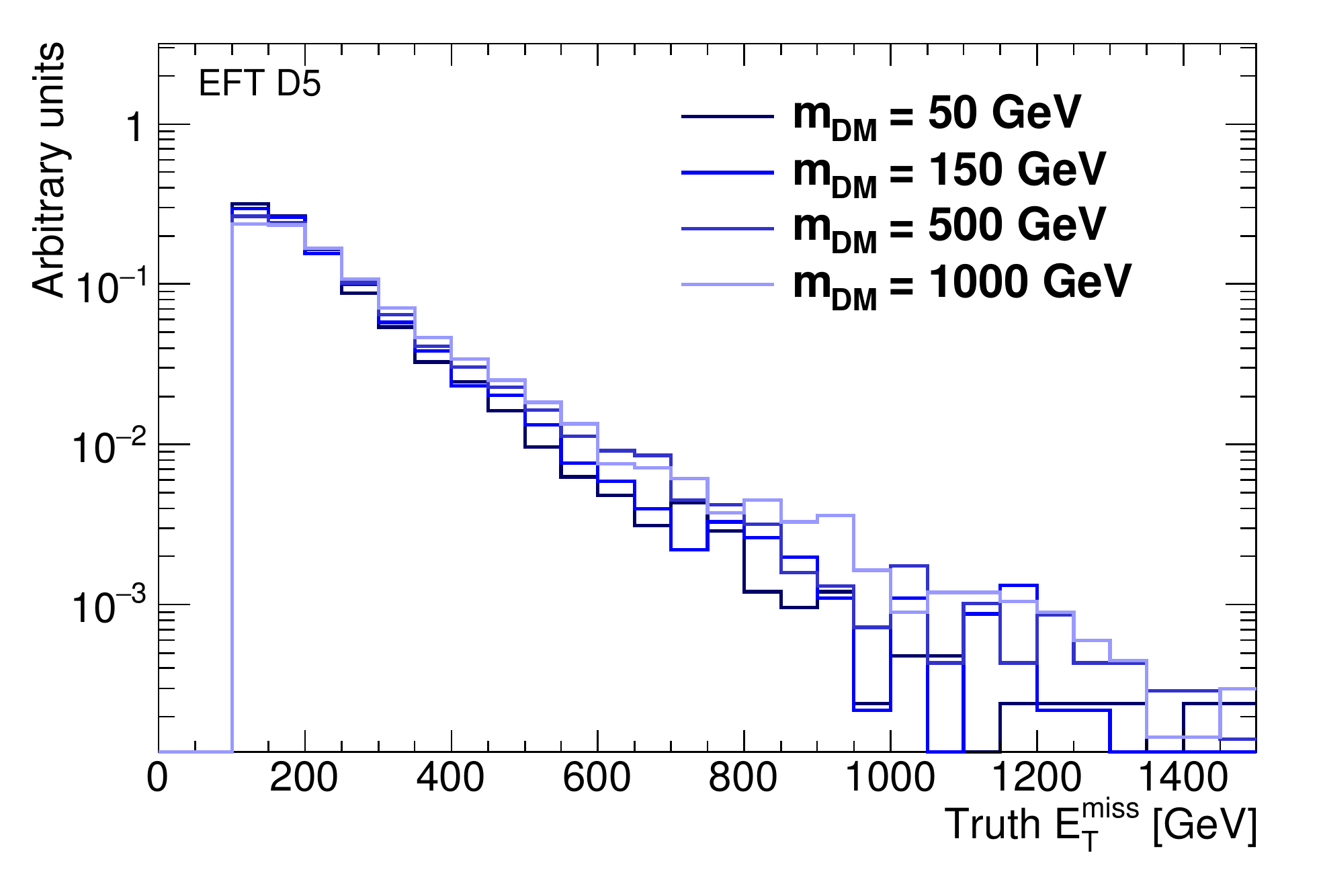}
	}
	\subfloat[Invariant mass of the two WIMPs $m_{\chi \chi}$, D5 operator]{
		\includegraphics[width=0.47\linewidth]{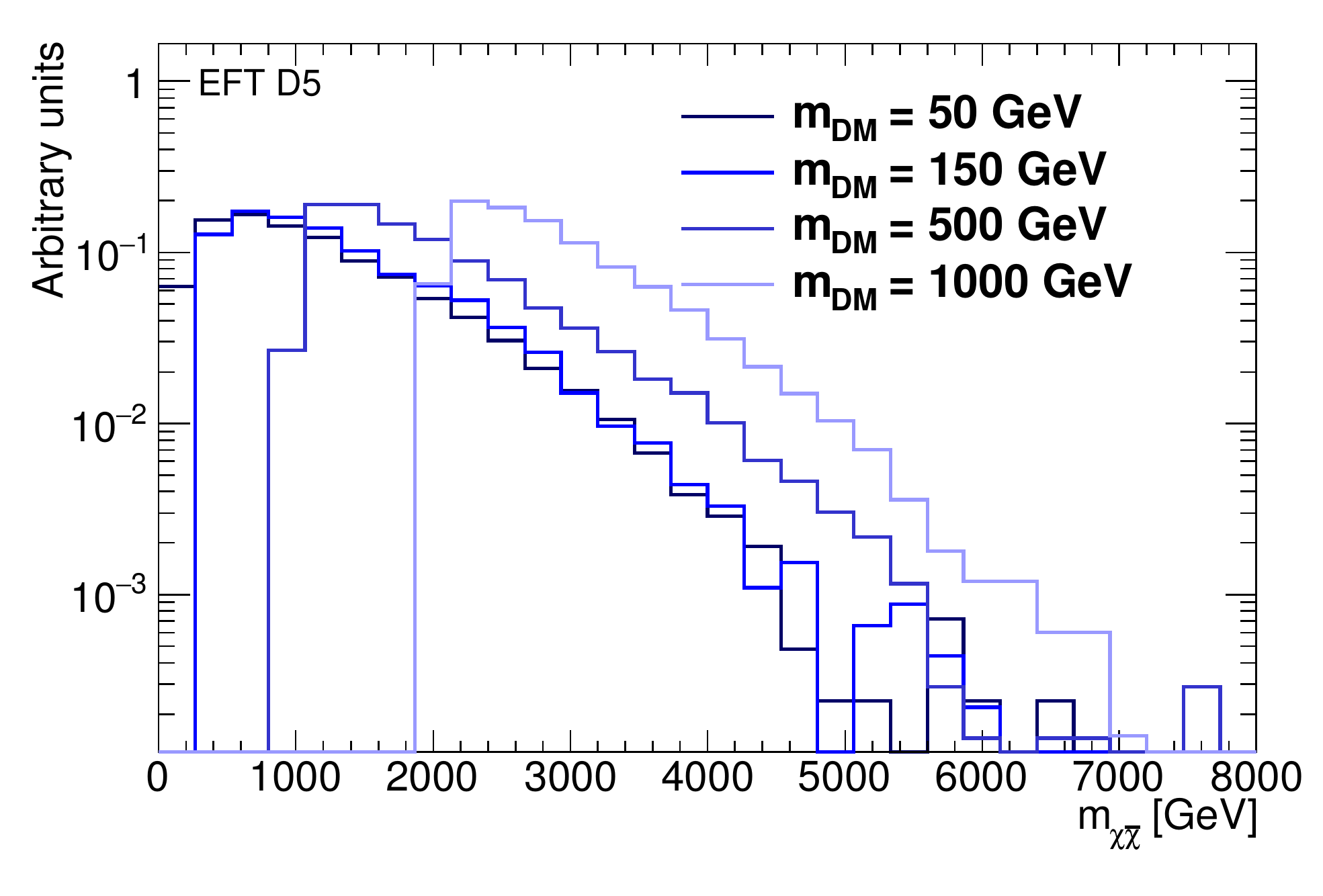} 
	}		
	\hfill
	\subfloat[\MET{}, simplified model]{
		\includegraphics[width=0.47\linewidth]{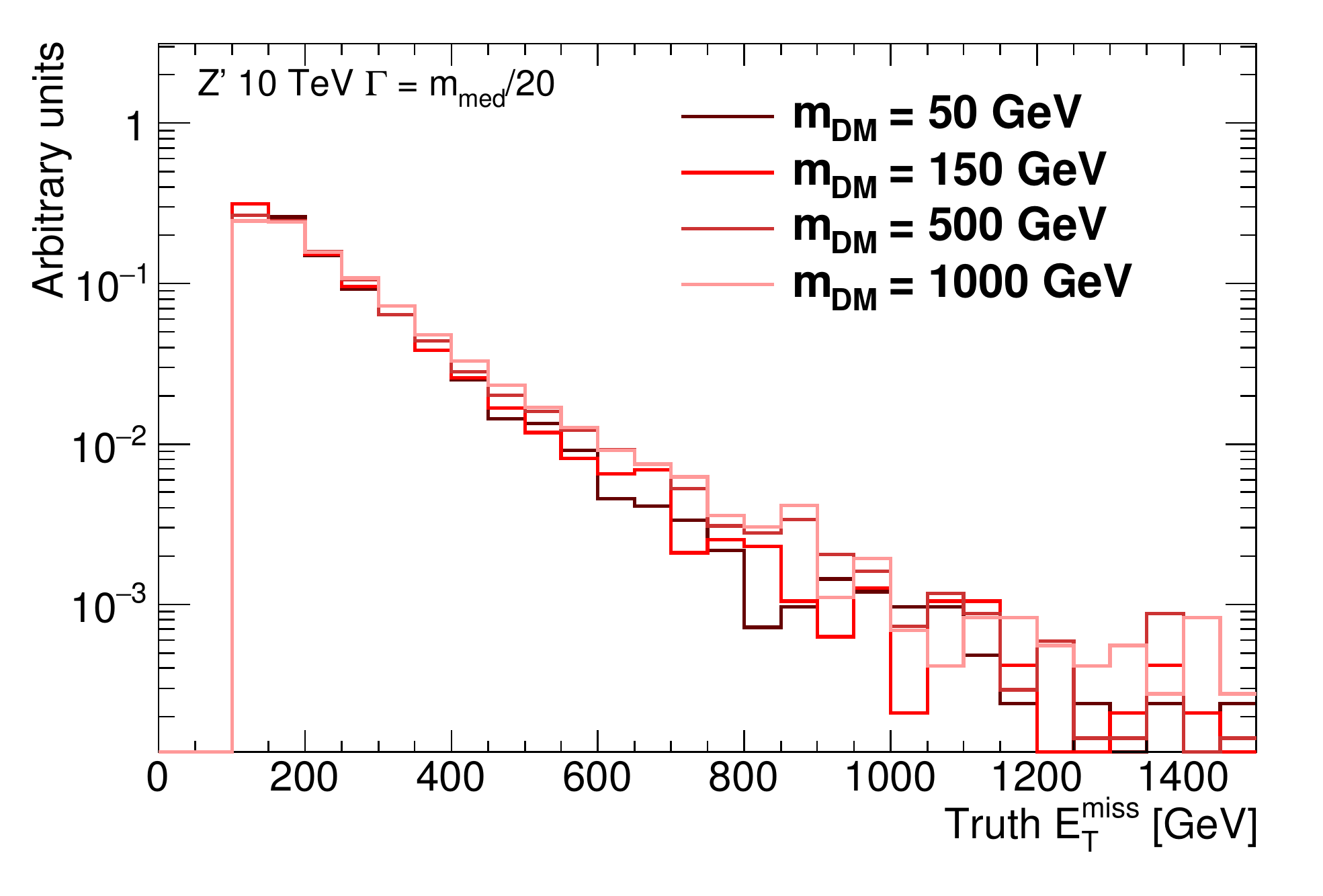}
	}
	\subfloat[Invariant mass of the two WIMPs $m_{\chi \chi}$, simplified model]{
		\includegraphics[width=0.47\linewidth]{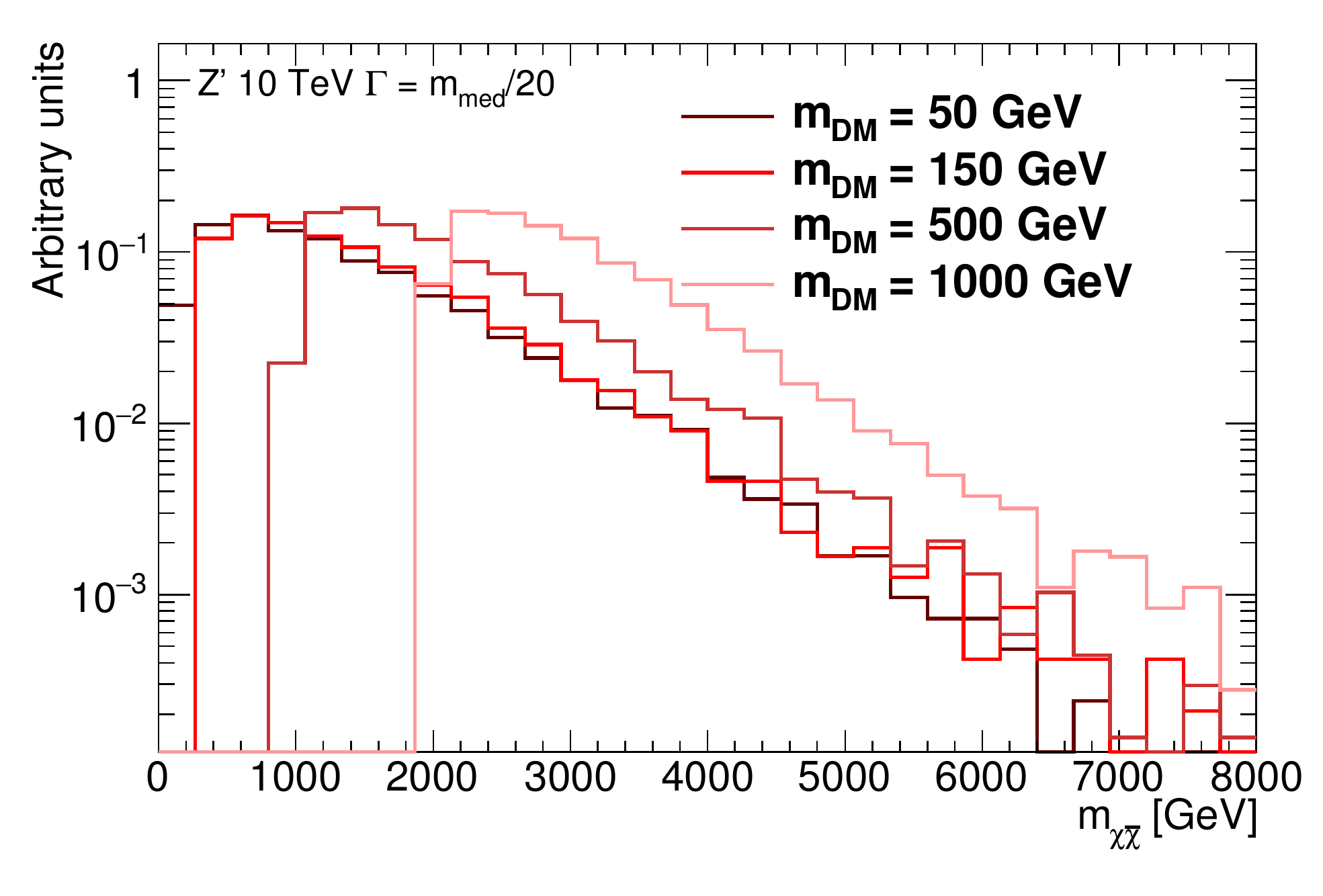} 
	}		
	\caption[][28pt]{Comparison of the kinematic distributions for a narrow \schannel mediator, 
		at generator level for a jet+\MET{} signature, for varying DM masses. 
		\label{fig:EFT_kinematics_mDM}}
\end{figure*}

As already observed in Section~\ref{sub:parameter_scan_monojet}, varying the DM mass changes the kinematics,
both in the simplified model and in the EFT case. This can be seen in Fig.~\ref{fig:EFT_kinematics_mDM}. 

\vskip20pt
	
Based on these studies, the Forum recommends experimental collaborations to 
add one grid scan point at very high mediator mass (10 \tev) to the scan, 
for each of the DM masses for the \schannel simplified models described in Section~\ref{subsec:MonojetLikeModels}. 
This will allow to reproduce the results of an equivalent contact interaction
as a simple extension of the existing parameter scan. 

It should be checked that the high-mass mediator case for the simplified model is correctly implemented 


\subsection{EFT benchmarks with no corresponding simplified models}


Whenever a UV completion is not available, an EFT still
captures a range of possible theories beyond the simplified models that we already consider. 
However, in the case of the dimension-7 operators detailed in Section~\ref{sec:EFT_models_with_direct_DM_boson_couplings}
we can only roughly control how well the EFT approximation holds, as described in Section~\ref{sub:validityEWContact}.
Despite the fact that a propagator was introduced to motivate
the truncation procedure for \schannel models, the prescription from Sec.~\ref{sub:EFT_withSimp}
depends upon the simplified model to derive the
energy scaling that is used for the comparison with the momentum transfer. 
The simple fact remains that the effective
coupling of the operator -- $g/\Lambda^n$ -- should not allow
momentum flow $Q>\Lambda$ or $g>4\pi$.  Given our ignorance of
the actual kinematics, 
the truncation procedure recommended for this purpose
is the one described in Section~\ref{sec:TruncationWithSHat},
as it is independent from any UV completion details. 

Because there is no UV completion,
the parameter \Mcut can be treated more freely than
an explicit function of $g$ and $\Lambda$.
It makes sense to choose \Mcut such that we 
identify the transition region where the EFT stops being
a good description of UV complete 
theories. This can be done using the ratio \Reft, which is defined
as the fraction of events for which $\hat{s} > \Mcut^2$. 
For large values of \Mcut, no events are thrown away in the truncation 
procedure, and \Reft = 1. As \Mcut becomes smaller, eventually all events are thrown 
away in the truncation procedure, i.e. \Reft = 0, and the EFT gives no 
exclusion limits for the chosen acceptance.  

We propose a rough scan over \Mcut, such that we find the values of \Mcut 
for which \Reft ranges from 0.1 to 1. The analysis can then perform a scan over 
several values of \Mcut, and show the truncated limit 
for each one of them. 


\chapter{Evaluation of signal theoretical uncertainties}
\label{sec:TheoryUncertainties} 

A comprehensive and careful assessment of signal theoretical uncertainties
plays in general a more important role for the background estimations
(especially when their evaluation is non-entirely data-driven) than it
does for signal simulations. Nevertheless, also for signal samples theoretical 
uncertainties are relevant, and may become even dominant in certain regions of phase space.

The uncertainties on the factorization and renormalization scales are assessed by the experimental collaborations by varying
the original scales of the process by factors of 0.5 and 2. The evaluation of the uncertainty on the choice of PDF follows
the PDF4LHC recommendation ~\cite{PDF4LHC} of considering the envelope of different PDF error sets, in order to account
for the uncertainty on the various PDFs as well as the uncertainty on the choice of the central value PDF. 
The Forum has not discussed the uncertainties related to the merging of different samples, nor the uncertainty
due to the choice of the modeling of the parton shower.  
This Chapter provides technical details on how scale and PDF uncertainties can be assessed 
for events generated with \powheg and \madgraph.


\section{POWHEG}

When using \powheg~\cite{Frixione:2007vw,Alioli:2010xd,Nason:2004rx}, it is
possible to study scale and PDF errors for the dark matter signals. A
fast reweighting machinery is available in \namecaps{POWHEG-BOX} that
allows one to add, after each event, new weights according to
different scale or PDF choices, without the need to regenerate all the
events from scratch. 

To enable this possibility, the variable \texttt{storeinfo\_rwgt} should be set 
to 1 in the \powheg input file when the events are generated for the 
first time\sidenote{Notice that even if the variable is not present, by 
default it is set to 1.}. After each event, a line starting with 
\begin{verbatim}
   #rwgt 
\end{verbatim}
is appended, containing the necessary information to generate extra 
weights. In order to obtain new weights, corresponding to different 
PDFs or scale choice, after an event file has been generated, a line 
\begin{verbatim}
   compute_rwgt 1 
\end{verbatim}
should be added in the input file along with the change in parameters
that is desired. For instance, \texttt{renscfact} and
\texttt{facscfact} allow one to study scale variations on the
renormalization and factorization scales around a central value. By
running the program again, a new event file will be generated, named
\texttt{<OriginalName>-rwgt.lhe}, with one more line at the end of each event of the form
\begin{verbatim}
   #new weight,renfact,facfact,pdf1,pdf2 
\end{verbatim}
followed by five numbers and a character string. The first of these 
numbers is the weight of that event with the new parameters chosen. By 
running in sequence the program in the reweighting mode, several 
weights can be added on the same file. Two remarks are in order.

\begin{itemize} 

\item The file with new weights is always named \\
\texttt{<OriginalName>-rwgt.lhe}\\
hence care has to be taken to save it as \\
\texttt{<OriginalName>.lhe}\\
before each iteration of the reweighting procedure. 

\item Due to the complexity of the environment where the program is 
likely to be run, it is strongly suggested as a self-consistency check 
that the first reweighting is done keeping the initial parameters. If 
the new weights are not exactly the same as the original ones, then 
some inconsistency must have happened, or some file was probably 
corrupted. 

\end{itemize} 

\noindent It is possible to also have weights written in the version 3 Les Houches format. 
To do so, in the original run, at least the token\\
\texttt{lhrwgt\_id 'ID'}\\
\noindent must be present. The reweighting procedure is the same as described 
above, but now each new run can be tagged by using a different value 
for the \texttt{lhrwgt\_id} keyword. After each event, the following lines will 
appear: 
\small{
\begin{verbatim}
  <rwgt> 
  <wgt id='ID'>
  <wgt id='ID1'>
  </rwgt> 
\end{verbatim}}
\normalsize

A more detailed explanation of what went into the computation of every 
single weight can be included in the \texttt{<header>} section of the event 
file by adding/changing the line \\
\texttt{lhrwgt\_descr 'some info'}\\
\noindent in the input card, before each ``reweighting'' run is performed. Other 
useful keywords to group together different weights are 
\texttt{lhrwgt\_group\_name} and \texttt{lhrwgt\_group\_combine}. 

More detailed information can be obtained by inspecting the document in 
\texttt{/Docs/V2-paper.pdf} under the common \namecaps{POWHEG-BOX-V2} directory. 

\section{The SysCalc package in \madgraph}

\syscalc is a post-processing package for parton-level events as obtained from leading-order calculations in \madgraph.  It can associate to each event a series of weights corresponding to the evaluation of a certain class of theoretical uncertainties. The event files in input and output are compliant with  the Les Houches v3 format.
For NLO calculations, PDF and scale uncertainties are instead evaluated automatically by setting corresponding instructions in the \texttt{run\_card.dat} and no post-processing is needed (or possible).

The requirements of the package as inputs are : 
\begin{itemize}
\item A systematics file (which can be generated by MadGraph 5 v. 1.6.0 or later) \cite{Alwall:2014hca,Alwall:2011uj}.
\item The Pythia-PGS package (v. 2.2.0 or later) \cite{Sjostrand:2006za}. This is needed only in the case of matching scales variations.
\item The availability of LHAPDF5 \cite{Whalley:2005nh}.
\item A configuration file (i.e. a text file) specifying the parameters to be varied. 
\end{itemize}

\syscalc supports all leading order computations generated in \madgraph including fixed-order computation and matched-merged computation performed in the MLM scheme~\cite{Mangano:2006rw}.
\madgraph stores additional information inside the event in order to have access to all the information required to compute the convolution of the PDFs with the matrix element for the various supported systematics.

Below follows an example configuration file which could serve as an example:\\[2mm]

{\footnotesize
\texttt{\# Central scale factors \\
scalefact: \\
0.5 1 2\\
\# Scale correlation \\
\# Special value -1: all combination (N**2)\\
\# Special value -2: only correlated variation\\
\# Otherwise list of index N*fac\_index + ren\_index\\
\#\ \ \ \ index starts at 0\\
scalecorrelation: \\
-1 \\
\#   $\alpha_s$ emission scale factors \\
alpsfact:\\
0.5 1 2\\
\#   matching scales\\
matchscale:\\
30 60 120\\
\# PDF sets and number of members (optional)\\
PDF:\\
CT10.LHgrid 53\\
MSTW2008nlo68cl.LHgrid\\[3mm]}
}

Without matching/merging, \syscalc\ is able to compute the variation of renormalisation and factorisation scale (parameter \texttt{scalefact}) and the change of PDFs.
The variation of the scales can be done in a correlated and/or uncorrelated way, basically following the value of the \texttt{scalecorrelation} parameter which can take the following values:
\begin{itemize}
\item  -1 : to account for all $N^2$ combinations.
\item  -2 : to account only for the correlated variations.
\item A set of positive values corresponding to the following entries (assuming \emph{0.5, 1, 2} for the  \texttt{scalefact} entry):
\begin{enumerate}
{\footnotesize
\item[0:] $\mu_F =\mu^{\rm orig}_F /2,\, \mu_R =\mu^{\rm orig}_R /2$
\item[1:] $\mu_F =\mu^{\rm orig}_F /2,\, \mu_R =\mu^{\rm orig}_R $
\item[2:] $\mu_F =\mu^{\rm orig}_F /2,\, \mu_R =\mu^{\rm orig}_R * 2$
\item[3:] $\mu_F =\mu^{\rm orig}_F, \phantom{/2}\, \mu_R =\mu^{\rm orig}_R /2$
\item[4:] $\mu_F =\mu^{\rm orig}_F, \phantom{/2}\, \mu_R =\mu^{\rm orig}_R$
\item[5:] $\mu_F =\mu^{\rm orig}_F, \phantom{/2}\, \mu_R =\mu^{\rm orig}_R * 2$
\item[6:] $\mu_F =\mu^{\rm orig}_F *2, \mu_R =\mu^{\rm orig}_R /2$
\item[7:] $\mu_F =\mu^{\rm orig}_F *2, \mu_R =\mu^{\rm orig}_R$
\item[8:] $\mu_F =\mu^{\rm orig}_F *2, \mu_R =\mu^{\rm orig}_R * 2$
}
\end{enumerate}
\end{itemize}

Without correlation, the weight associated to the renormalisation scale is the following:
\begin{equation}
\mathcal{W}^{\mu_R}_{\rm new} =  \frac{\alpha_S^{N}(\Delta*\mu_ R)}{\alpha_S^{N}(\mu_R)} * \mathcal{W}_{\rm orig}, 
\end{equation}
where $\Delta$ is the scale variation considered, $\mathcal{W_{\rm{orig}}}$ and $\mathcal{W_{\rm{new}}}$ are respectively the original/new weights associated to the event. $N$ is the power in the strong coupling for the associated event (interference is not taken account on an event by event basis).
The weight associated to the scaling of the factorisation scale is:
\begin{equation}
\mathcal{W}^{\mu_F}_{\rm {new}} =   \frac{f_{\rm 1,orig} (x_1, \Delta*\mu_F) * f_{\rm 2,orig} (x_2, \Delta*\mu_F)}{f_{\rm 1,orig}(x_1, \mu_F) * f_{\rm 2,orig}(x_2, \mu_F)} * \mathcal{W}_{\rm orig}, 
\end{equation}
where $f_{\rm i,orig}$ are the probabilities from the original PDF set associated to the incoming partons, which hold a proton momentum fraction $x_1$ and $x_2$ for the first and second beam respectively.

The variations for the PDF are given by the corresponding weights associated to the new PDF sets:
\begin{equation}
\mathcal{W}^{\rm PDF}_{\rm new} =  \frac{f_{\rm 1,new} (x_1, \mu_F) * f_{\rm 2,new} (x_2, \mu_F)}{f_{\rm 1,orig} (x_1, \mu_F) * f_{\rm 2,orig} (x_2, \mu_F)}* \mathcal{W}_{\rm orig},
\end{equation}
where $f_{\rm i,new}$ is the new PDF probability associated to parton $i$.

In presence of matching, \madgraph associates one history of radiation (initial and/or final state radiation) obtained by a $k_T$ clustering algorithm, and calculates $\alpha_s$ at each vertex of the history to a scale given by the aforementioned clustering algorithm. Furthermore, \madgraph reweights the PDF in a fashion similar to what a parton shower would do. \syscalc can perform the associated re-weighting (parameter \texttt{alpsfact}) by dividing and multiplying by the associated factor.

For each step in the history of the radiation (associated to a scale $\mu_i = k_{T,i}$), this corresponds to the following expression for a Final State Radiation (FSR):
\begin{equation}
\mathcal{W}^{\rm FSR}_{\rm new} = \frac{ \alpha_s (\Delta* \mu_i)} { \alpha_s (\mu_i)}  * \mathcal{W}_{\rm orig},
\end{equation}
and to the following expression for Initial State Radiation  (ISR), associated to a scale $\mu_i$ and fraction of energy $x_i$:
\begin{equation}
\mathcal{W}^{\rm ISR}_{\rm new} = \frac{ \alpha_s (\Delta* \mu_i)} { \alpha_s (\mu_i)} \frac{ \frac{f_a(x_i,\Delta*\mu_i)}{f_b(x_i,\Delta*\mu_{i+1})}} { \frac{f_a(x_i, \mu_i)}{f_b(x_i, \mu_{i+1})} }
  *\mathcal{W}_{\rm orig},
\end{equation}
where $\mu_{i+1}$ is the scale of the next step in the (initial state) history of radiation.

\syscalc can include the weight associated to different merging scales in the MLM matching/merging mechanism (for output of the \texttt{pythia6} package or \texttt{pythia-pgs} package). 

In that case, the parton shower does not veto any event according to the MLM algorithm, although in the output file the scale of the first emission is retained. Having this information, \syscalc can test each value of the specified matching scales under the \texttt{matchscale} parameter block. \syscalc\ will then test for each of the values specified in the parameter \texttt{matchscale} if the event passes the MLM criteria or not. If it does not, then a zero weight is associated to the event, while if it does, then a weight 1 is kept. As a reminder,  those weights are the equivalent of having a (approximate) Sudakov form-factor and removing at the same time the double counting between the events belonging to different multiplicities.\\

Finally, we give an example of the \syscalc output which follows the LHEF v3 format. The following block appears in the header of the output file:

\footnotesize{
\begin{verbatim}
<header>
  <initrwgt>
    <weightgroup type="Central scale variation" combine="envelope">
      <weight id="1"> mur=0.5 muf=0.5 </weight>
      <weight id="2"> mur=1 muf=0.5 </weight>
      <weight id="3"> mur=2 muf=0.5 </weight>
      <weight id="4"> mur=0.5 muf=1 </weight>
      <weight id="5"> mur=1 muf=1 </weight>
      <weight id="6"> mur=2 muf=1 </weight>
      <weight id="7"> mur=0.5 muf=2 </weight>
      <weight id="8"> mur=1 muf=2 </weight>
      <weight id="9"> mur=2 muf=2 </weight>
    </weightgroup>
    <weightgroup type="Emission scale variation" combine="envelope">
      <weight id="10"> alpsfact=0.5</weight>
      <weight id="11"> alpsfact=1</weight>
      <weight id="12"> alpsfact=2</weight>
    </weightgroup>
    <weightgroup type="CT10nlo.LHgrid" combine="hessian">
      <weight id="13">Member 0</weight>
      <weight id="14">Member 1</weight>
      <weight id="15">Member 2</weight>
      <weight id="16">Member 3</weight>
      ...
      <weight id="65">Member 52</weight>
    </weightgroup>
  </initrwgt>
</header>
\end{verbatim}}

\noindent For each event, the weights are then written as follows:
\footnotesize{
\begin{verbatim}
<rwgt>
  <wgt id="1">83214.7</wgt>
  <wgt id="2">61460</wgt>
  <wgt id="3">47241.9</wgt>
  <wgt id="4">101374</wgt>
  ...
  <wgt id="64">34893.5</wgt>
  <wgt id="65">41277</wgt>
</rwgt>
\end{verbatim}}
\normalsize

\chapter{Conclusions}
\label{chapter:conclusions}

The ATLAS/CMS Dark Matter Forum concluded its work in June 2015. Its mandate was focused on identifying a prioritized, compact set of simplified model benchmarks
to be used for the design of the early Run-2 LHC searches for \MET+X final states. Its participants included many of the experimenters from both collaborations that are involved in these searches, as well as many of the theorists working actively on these models.
This report has documented this basis set of models, as well as studies of the kinematically-distinct regions of the parameter space of the models, to aid the design of the searches.
Table 6.1 summarizes the state of the art of the calculations, event generators, and tools that are available to the two LHC collaborations to simulate these models at the start of Run-2. It also describes some that are known to be under development as the report was finalized.


\begin{footnotesize}

\begin{table*}[!p]
	\centering\scriptsize
\noindent
\begin{tabular}{llrl} \toprule \multicolumn{4}{c}{\textbf{Benchmark models for ATLAS and CMS Run-2 DM searches}}\\

	\cmidrule(r){1-4} 
	\multicolumn{4}{c}{vector/axial vector mediator, \schannel (Sec.~\ref{sec:monojet_V})}\\
	\cmidrule(r){1-4} 

	Signature & State of the art calculation and tools & Implementation & References \\ 
	\cmidrule(r){1-4} 
    jet + \MET{} & \textbf{NLO+PS (\powheg, SVN r3059)} & \cite{ForumSVN_DMA, ForumSVN_DMV} & \parbox{3.5cm} {\cite{Haisch:2013ata,Haisch:2015ioa,Alioli:2010xd,Nason:2004rx,Frixione:2007vw} }\\ 
	& NLO+PS (\textit{DMsimp} UFO + \madgraph v2.3.0) & \cite{NewMadgraphModels} & \parbox{3.5cm} {\cite{Alwall:2014hca,Alloul:2013bka,Degrande:2011ua} } \\ 		
	& NLO (\mcfm v7.0) & Upon request & \parbox{3.5cm} {\cite{Fox:2012ru,Harris:2014hga} }\\[5pt] 
	$W/Z/\gamma$ + \MET{} & \textbf{LO+PS (UFO + \madgraph v2.2.3)} & \cite{ForumSVN_EW_DMV} & 
	\parbox{3.5cm} {\cite{Alwall:2014hca,Alloul:2013bka,Degrande:2011ua} }  \\ 
	 & NLO+PS (\textit{DMsimp} UFO + \madgraph v2.3.0) & \cite{NewMadgraphModels} & 
	\parbox{3.5cm} {\cite{Alwall:2014hca,Alloul:2013bka,Degrande:2011ua} }  \\ 

	\cmidrule(r){1-4} 
	\multicolumn{4}{c}{scalar/pseudoscalar mediator, \schannel (Sec.~\ref{sec:monojet_scalar})}\\
	\cmidrule(r){1-4} 
	Signature & State of the art calculation and tools& Implementation & References \\ 
		\cmidrule(r){1-4} 
		
	jet + \MET{} & \textbf{LO+PS, top loop (\powheg, r3059)} &  \cite{ForumSVN_DMS_tloop, ForumSVN_DMP_tloop} &  \parbox{3.5cm} {\cite{Haisch:2013ata,Haisch:2015ioa,Alioli:2010xd,Nason:2004rx,Frixione:2007vw} } \\ 
	& LO+PS, top loop (\textit{DMsimp} UFO + \madgraph v.2.3.0) & \cite{NewMadgraphModels} & \parbox{3.5cm} {\cite{Alwall:2014hca,Hirschi:2011pa,Alloul:2013bka,Degrande:2011ua} }\\ 		
	& LO, top loop (\mcfm v7.0) & Upon request & \cite{Fox:2012ru,Harris:2014hga} \\ [5pt] 
	$W/Z/\gamma$ + \MET{} & \textbf{LO+PS (UFO + \madgraph v2.2.3}) & & \parbox{3.5cm} { \cite{Alwall:2014hca,Alloul:2013bka,Degrande:2011ua} }\\ [5pt] 
    $t\bar{t},b\bar{b}$+ \MET{} & \textbf{LO+PS (UFO + \madgraph v2.2.3}) &  \cite{ForumSVN_DMTTBar} & \parbox{3.5cm} { \cite{Alwall:2014hca,Alloul:2013bka,Degrande:2011ua} }\\ 
    	& NLO+PS (\textit{DMsimp} UFO + \madgraph v2.3.0) & \cite{NewMadgraphModels} & \parbox{3.5cm} {\cite{Alwall:2014hca,Alloul:2013bka,Degrande:2011ua} } \\

%
	\cmidrule(r){1-4} 
	\multicolumn{4}{c}{scalar mediator, \tchannel (Sec.~\ref{sec:monojet_t_channel})}\\
	\cmidrule(r){1-4} 
	Signature & State of the art calculation and tools & Implementation & References \\ 
		\cmidrule(r){1-4} 
		
	jet(s) + \MET{} (2-quark gens.) & LO+PS (UFO + \madgraph v2.2.3) & \cite{ForumSVN_TChannel_PapucciVichiZurek}&  \parbox{3.5cm} {\cite{Papucci:2014iwa,Alwall:2014hca,Alloul:2013bka,Degrande:2011ua} }\\ [5pt] 
	jet(s) + \MET{} (3-quark gens.) & \textbf{LO+PS (UFO + \madgraph v2.2.3)} & \cite{ForumSVN_TChannel_Amelia}&  \parbox{3.5cm} {\cite{Bell:2012rg,Alwall:2014hca,Alloul:2013bka,Degrande:2011ua} } \\ [5pt] 
	$W/Z/\gamma$ + \MET{} & \textbf{LO+PS (UFO + \madgraph v2.2.3)} & TBC  & \parbox{3.5cm} {\cite{Bell:2012rg, Alwall:2014hca,Alloul:2013bka,Degrande:2011ua}}\\ [5pt] 
    $b$ + \MET{} & \textbf{LO+PS (UFO + \madgraph v2.2.3)} & \cite{ForumSVN_DMSingleB}  & \parbox{3.5cm} {\cite{Lin:2013sca,Agrawal:2014una, Alwall:2014hca,Alloul:2013bka,Degrande:2011ua}}\\ 
			
	\cmidrule(r){1-4} 
	\multicolumn{4}{c}{Specific simplified models with EW bosons (Sec.~\ref{sec:monoHiggs})}\\
	\cmidrule(r){1-4} 
	Signature and model & State of the art calculation and tools & Implementation & References \\ 
	\cmidrule(r){1-4} 
	
	Higgs + \MET{}, vector med. & \textbf{LO+PS (UFO + \madgraph v2.2.3)} & \cite{ForumSVN_EWMonoHiggs}& \parbox{3.5cm} {\cite{Carpenter:2013xra,Berlin:2014cfa,Alwall:2014hca,Alloul:2013bka,Degrande:2011ua} }\\ 
	Higgs + \MET{}, scalar med. & \textbf{LO+PS (UFO + \madgraph v2.2.3)} & \cite{ForumSVN_EWMonoHiggs}& \parbox{3.5cm} {\cite{Carpenter:2013xra,Berlin:2014cfa,Alwall:2014hca,Alloul:2013bka,Degrande:2011ua} }\\ 
	Higgs + \MET{}, 2HDM & \textbf{LO+PS (UFO + \madgraph v2.2.3)} & \cite{ForumSVN_EWMonoHiggs_2HDM}& \parbox{3.5cm} {\cite{Berlin:2014cfa,Alwall:2014hca,Alloul:2013bka,Degrande:2011ua}} \\

	\cmidrule(r){1-4} 
	\multicolumn{4}{c}{Contact interaction operators with EW bosons (Sec.~\ref{sec:monoHiggs})}\\
	\cmidrule(r){1-4} 
	Signature and model & State of the art calculation and tools & Implementation & References \\ 
	\cmidrule(r){1-4} 
	
	 $W/Z/\gamma$ + \MET{}, dim-7& \textbf{LO+PS (UFO + \madgraph v2.2.3)} & \cite{ForumSVN_EWEFTD7}& \parbox{3.5cm} {\cite{Cotta:2012nj, Carpenter:2012rg, Crivellin:2015wva,Berlin:2014cfa,Alwall:2014hca,Alloul:2013bka,Degrande:2011ua} }\\ 
	 Higgs + \MET{}, dim-4/dim-5 & \textbf{LO+PS (UFO + \madgraph v2.2.3)} & \cite{ForumSVN_monoHEFTD5}& \parbox{3.5cm} {\cite{Carpenter:2013xra,Petrov:2013nia,Berlin:2014cfa,Alwall:2014hca,Alloul:2013bka,Degrande:2011ua} }\\ 
	 Higgs + \MET{}, dim-8 & \textbf{LO+PS (UFO + \madgraph v2.2.3)} & \cite{ForumSVN_EWMonoHiggs}&
	 \parbox{3.5cm} {\cite{Carpenter:2013xra,Petrov:2013nia,Berlin:2014cfa,Alwall:2014hca,Alloul:2013bka,Degrande:2011ua} }\\

	\bottomrule 
	\end{tabular}
	
	\begin{center}
		\normalsize 
		Table 6.1: 
		Summary table for available benchmark models considered within the works of this Forum. 
		The results in this document have been obtained with the implementations in bold. 	
		\label{tab:summaryModels}
	\end{center}
	
\end{table*}

\end{footnotesize}

.

This document primarily presents studies related to simplified models.
The presentation of results for EFT benchmark models is also discussed.
The studies contained in this report are meant to highlight the use of
EFTs as a benchmark that is complementary to simplified models,
and to demonstrate how that collider results could be presented a function of the
fraction of events that are valid within the contact interaction approximation.

A number of points remain to be developed beyond the scope of this Forum,
in order to fully benefit from LHC searches in the global quest for Dark Matter.
First and foremost, to accomodate the urgent need of a basis set of simplified models, 
this work has made many grounding assumptions, as stated in the introduction. Departures from these assumptions have not been fully explored.
As a consequence, the list of models and implementations employed by the ATLAS and CMS collaborations 
for early LHC Run-2 searches is not meant to exhaust the range of possibilities for mediating processes, 
let alone cover all plausible mdoels of collider dark matter production. 
Rather, it is hoped that others will continue the systematic exploration of the most generic possibilites 
for collider dark matter production, building upon the framework used in this report just as this report has 
relied heavily on the work of many others. 
This also applies to models that exist in literature but do not have an implementation yet: we
hope that this work will further encourage the theory and generator community to improve the 
implementation of new models as well as the precision of the calculations of existing ones. 
The role of constraints on the mediator particles
from direct past and present collider searches should also be developed further.

Furthermore, we see the need for broader discussion on the comparison
of experimental results amongst collider and non-collider searches for particle dark matter. 
This point will have to be addressed before the presentation of Run-2 results:
The uncertainties in the comparisons between experiments should be discussed and conveyed, so that the different results
can be placed in their correct context, and so we can collectively build a fair and comprehensive picture of our understanding of particle Dark Matter.

\chapter{Acknowledgements}

The authors would like to thank Daniel Whiteson for helping in the review of this document. 
This research was supported by the Munich Institute for Astro- and Particle Physics (MIAPP) of the DFG cluster of excellence "Origin and Structure of the Universe".
The authors would like to express a special thanks to the Mainz Institute for Theoretical Physics (MITP) for its hospitality and support. 
P. Pani wishes to thank the support of the Computing Infrastructure of
Nikhef.

\appendix


\chapter{Appendix: Additional models for Dark Matter searches}
\label{app:EWSpecificModels_Appendix}
\section{\texorpdfstring{Models with a single $top-$quark + \MET}{Models with a single top-quark + MET}}
\label{sec:singletop}

Many different theories predict final states with a single top and associated missing 
transverse momentum (monotop), some of them including dark matter candidates. 
A simplified model encompassing the processes leading to this phenomenology is described in Refs.~\cite{AndreaFuksMaltoni,Agram:2013wda,Boucheneb:2014wza},
and is adopted as one of the benchmarks for Run 2 LHC searches. 

\begin{figure}[!h!tpd]
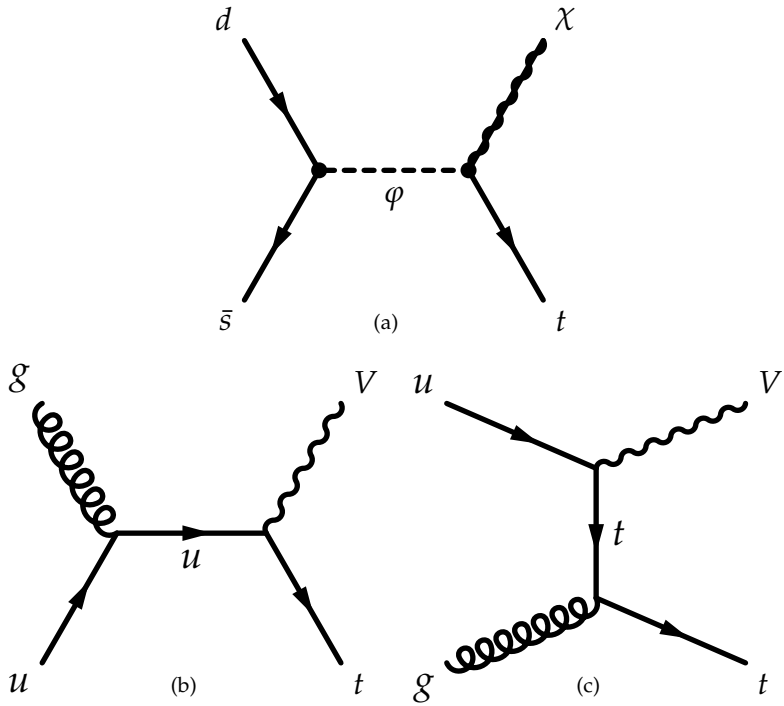

\centering
\unitlength=0.0046\textwidth
\subfloat[\label{subfig:S1}]{
  \begin{feynmandiagram}[modelS1]
    \fmfleft{i1,i2}
    \fmfright{o1,o2}
    \fmf{dashes,label={\Large $\varphi$}}{v1,v2}
    \fmf{fermion}{i2,v1,i1}
    \fmf{fermion}{v2,o1}
    \fmf{plain,tension=0}{v2,o2}
    \fmf{wiggly}{v2,o2}
    \fmfdot{v1,v2}
    \fmflabel{\Large ${\bar{s}}$}{i1}
    \fmflabel{\Large ${d}$}{i2}
    \fmflabel{\Large ${t}$}{o1}
    \fmflabel{\Large ${\chiDM}$}{o2}
  \end{feynmandiagram}
}\\\vspace{\baselineskip}
\subfloat[\label{subfig:S4s}]{
  \begin{feynmandiagram}[modelS4s]
    \fmfleft{i1,i2}
    \fmfright{o1,o2}
    \fmf{fermion,label={\LARGE $u$}}{v1,v2}
    \fmf{gluon}{i2,v1}
    \fmflabel{\LARGE $g$}{i2}
    \fmf{fermion}{i1,v1}
    \fmflabel{\LARGE $u$}{i1}
    \fmf{wiggly}{v2,o2}
    \fmf{fermion}{v2,o1}
    \fmflabel{\Large $V$}{o2}
    \fmflabel{\Large $t$}{o1}
  \end{feynmandiagram}
}
\subfloat[\label{subfig:S4t}]{
  \begin{feynmandiagram}[modelS4t]
    \fmfleft{i1,i2}
    \fmfright{o1,o2}
    \fmf{fermion}{i2,vup}
    \fmflabel{\LARGE $u$}{i2}
    \fmf{gluon}{i1,vdown}
    \fmflabel{\LARGE $g$}{i1}
    \fmf{fermion, label={\LARGE $t$}}{vup,vdown}
    \fmf{fermion}{vdown,o1}
    \fmflabel{\Large $t$}{o1}
    \fmf{wiggly}{vup,o2}
    \fmflabel{\Large $V$}{o2}
  \end{feynmandiagram}
  }
\caption
{
Feynman diagrams of leading order processes leading to monotop events: production of
a colored scalar resonance $\varphi$ decaying into a top quark and a spin-$1/2$ fermion $\chiDM$ (a),
$s-$ (b) and \tchannel (c) non resonant production of a top quark in association with
a spin-1 boson $V$ decaying invisibly.
}
\label{fig:feyn_prod}
\end{figure}

The simplified model is constructed by imposing that the model Lagrangian
respects the electroweak $SU(2)_L \times U(1)_Y$ gauge symmetry and by
requiring minimality in terms of new states to supplement to the Standard
Model fields. As a result, two monotop production mechanisms are possible.
In the first case, the monotop system is constituted by an invisible (or
long-lived with respect to detector distances) fermion $\chi$ and a top quark.
It is produced as shown in the diagram of \ref{fig:feyn_prod}~(a) where a colored
resonance $\varphi$ lying in the triplet representation of $SU(3)_C$ decays
into a top quark and a $\chi$ particle. In the second production mode, the
monotop state is made of a top quark and a vector state $V$ connected to a
hidden sector so that it could decay invisibly into, e.g., a pair of dark
matter particles as studied in~\cite{Boucheneb:2014wza}. The production proceeds via
flavor-changing neutral interactions of the top quark with a quark of the
first or second generation and the invisible $V$ boson (see the diagrams of
\ref{fig:feyn_prod}~(b) and (c)).

\newthought{Resonant production}
\label{sec:ResonantProd}

In this case, a colored $2/3$-charged scalar ($\varphi$) is produced and decays into a top quark and a spin-$1/2$ invisible particle, $\chiDM$.  The dynamics of the new sector is described by the following Lagrangian:
\be\label{eq:lagrangianResonant}\bsp
\lag  =
  \bigg[
    \varphi \bar d^c \Big[a^q_{SR} + b^q_{SR} \gamma_5 \Big] d +
    \varphi \bar u \Big[a^{1/2}_{SR} + b^{1/2}_{SR} \gamma_5 \Big] \chiDM
    +  {\rm h.c.} 
  \bigg] ,
\esp\ee
where $u$ ($d$) stands for any up-type (down-type) quark, the notation $SR$
refers to the monotop production mechanism via a scalar resonance and all
flavor and color indices are understood for clarity.

In the notation of~\cite{Agram:2013wda}, 
the couplings of the new colored fields to down-type quarks are
embedded into the $3\times 3$ antisymmetric matrices
$a^q_{SR}$ (scalar couplings) and $b^q_{SR}$ (pseudoscalar couplings)
while those to the new fermion $\chiDM$ and one
single up-type quark are given by the three-component vectors
$a^{1/2}_{S R}$ and $b^{1/2}_{S R}$
in flavor space.

Under the form of Eq.~\eqref{eq:lagrangianResonant}, the Lagrangian is the one
introduced in the original monotop search proposal~\cite{AndreaFuksMaltoni}. It has been
used by the CMS collaboration for Run I analyses after neglecting all pseudoscalar components
of the couplings and adding the vector resonance case for which minimality
requirements are difficult to accommodate~\cite{Khachatryan:2014uma}. In contrast, the
study of Ref.~\cite{Boucheneb:2014wza} has imposed electroweak gauge invariance and
required minimality. This enforces all new couplings to be right-handed so that
\begin{equation}
a^{1/2}_{SR} = b^{1/2}_{SR} = \frac12 y_s^*
\qquad\text{and}\qquad
a^q_{SR} = b^q_{SR} = \frac12 \lambda_s \ ,
\end{equation}
where the objects $y_s$ and $\lambda_s$ are
a tridimensional vector and a $3\times 3$ matrix
in flavor space respectively. 
This class of scenarios is the one that has been adopted by the ATLAS collaboration for its
Run~I monotop searches~\cite{Aad:2014wza} and will be considered by both
collaborations for Run~II analyses.

The resulting model can be likened to the MSSM with an $R$-parity violating of
a top squark to the Standard Model down-type quarks and an $R$-parity
conserving interaction of a top quark and a top-squark to a neutralino.

\newthought{Non-Resonant production}
\label{sec:NonResonantProd}

For non-resonant monotop production, the monotop state is produced via
flavor-changing neutral interactions of the top quark, a lighter up-type
quark and a new invisible vector particle $V$. 
This is the only case considered, as having a new scalar 
would involve in particular a mixing with the SM Higgs boson and therefore a larger number of free parameters. 
The Lagrangian describing the
dynamics of this non-resonant monotop production case is:
\be\label{eq:lagrangianNonResonantVector}\bsp
\lag =
  \bigg[
    V_\mu \bar u \gmu \Big[a^1_{FC} \!+\! b^1_{FC} \gamma_5 \Big] u  
    + \rm h.c. 
  \bigg] \ ,
\esp\ee
where the flavor and color indices are
again understood for clarity.
The strength of the interactions among these two states and a pair
of up-type quarks is modeled via two $3\times 3$ matrices in flavor space $a^{1}_{FC}$ for the vector couplings
and $b^{1}_{FC}$ for the axial vector couplings, the $FC$ subscript referring to the flavor-changing neutral
monotop production mode and the $(1)$ superscript to the vectorial nature of the invisible particle.

As for the resonant case, the Lagrangian of Eq.~\eqref{eq:lagrangianNonResonantVector} is the one that
has been used by CMS after reintroducing the scalar option for the invisible
state and neglecting all pseudoscalar interactions~\cite{Khachatryan:2014uma}. As
already mentioned, a simplified setup motivated by gauge invariance and
minimality has been preferred so that, as shown in Ref.~\cite{Boucheneb:2014wza}, we
impose all interactions to involve right-handed quarks only,
\begin{equation}
a^1_{FC} = b^1_{FC} = \frac12 a_R
\end{equation}
where $a_R$ denotes a $3\times 3$ matrix in flavor space.
This implies the vector field to be an $SU(2)_L$
singlet.

\newthought{Model parameters and assumptions}
 
The models considered as benchmarks for the first LHC searches
contain further assumptions in terms of the flavor structure of the model
with respect to the Lagrangians of the previous subsection.
In order to have an observable monotop signature at the LHC, the Lagrangians
introduced above must include not too small couplings of the new particles to
first and second generation quarks. For simplicity, we assumed that only
channels enhanced by parton density effects will be considered, so that we fix
\begin{equation}\bsp
(a_R)_{13} = (a_R)_{31} = a \ , \\
(\lambda_s)_{12} = - (\lambda_s)_{21} = \lambda
\qquad\text{and}\qquad
(y_s)_3 = y \ ,
\esp\end{equation}
all other elements of the matrices and vectors above being set to zero.


\newthought{Implementation}
     In order to allow one for the Monte Carlo simulation of events relevant for
     the monotop production cases described above, we consider the Lagrangian
     \be
       \lag =
         \bigg[
             a V_\mu \bar u \gmu P_R t
           + \lambda \varphi \bar d^c P_R s
           + y \varphi \bar \chiDM P_R t
           +  {\rm h.c.} \ ,
         \bigg]
     \ee
     where $P_R$ stands for the right-handed chirality projector and the new
     physics couplings are defined by the three parameters $a$, $\lambda$ and
     $y$. We additionally include a coupling of the invisible vector boson $V$
     to a dark sector (represented by a fermion $\psi$) whose strength can be
     controlled through a parameter $g_{DM}$,
     \be
       \lag = g_{DM} V_\mu \bar\psi \gamma^\mu\psi \ .
     \ee
     This ensures the option to make the $V$-boson effectively invisible by
     tuning $g_{DM}$ respectively to $a$. We implement the entire model in the
     {\sc FeynRules} package~\cite{Alloul:2013bka} so that the model can be
     exported to a UFO library~\cite{Degrande:2011ua} to be linked to
     {\sc MadGraph5\_aMC@NLO}~\cite{Alwall:2014hca} for event generation,
     following the approach outlined in~\cite{Christensen:2009jx}.

\subsection{Parameter scan}

Under all the assumptions of the previous sections, the parameter space of
the resonant model is defined by four quantities, namely the mass of the
new scalar field $\varphi$, the mass of the invisible fermion $\chi$ and
the strengths of the interactions of the scalar resonance with the monotop
system $y$ and with down-type quarks $\lambda$. One of both coupling
parameters could however be traded with the width of the resonance.

The parameter space of the non-resonant model is defined by two
parameters, namely the mass of the invisible state $V$ and its
flavor-changing neutral coupling to the up-type quarks $a_R$.

In the case of the non-resonant model, the invisible vector is connected to
a hidden sector that could be, in its simplest form, parameterized by a new
fermion~\cite{Boucheneb:2014wza}. This has effects on the width of the invisible $V$ state.

A consensus between the ATLAS and CMS collaborations has been reached in
the case of non-resonant monotop production. The results have been
described above. In contrast, discussions in the context of resonant
monotop production are still on-going. The related parameter space
contains four parameters and must thus be further simplified for practical
purposes. Several options are possible and a choice necessitates additional
studies that will be achieved in a near future.

\begin{figure*}[!htb]
	\centering
	\includegraphics[width=0.3\textwidth]{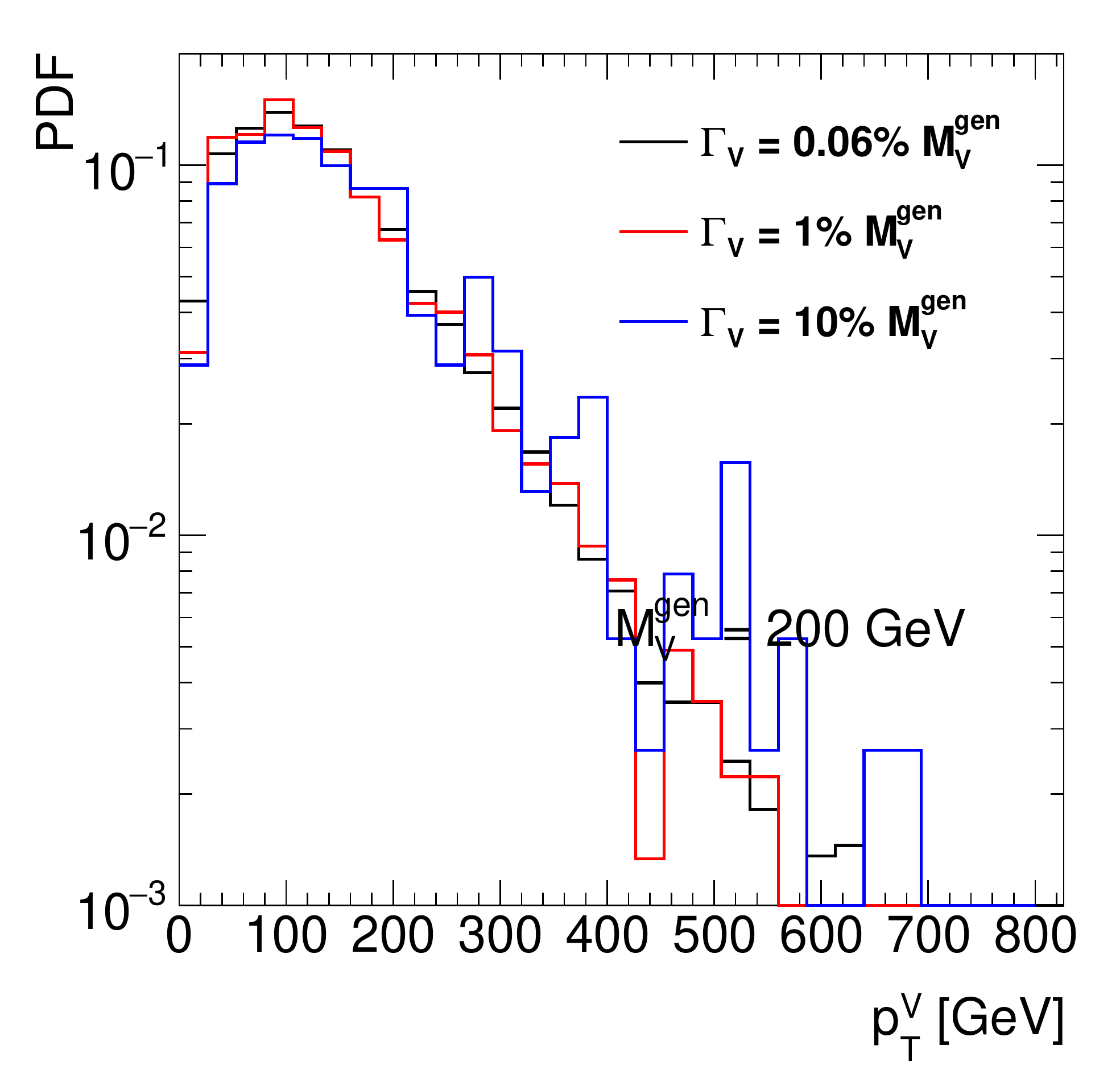}
	\includegraphics[width=0.3\textwidth]{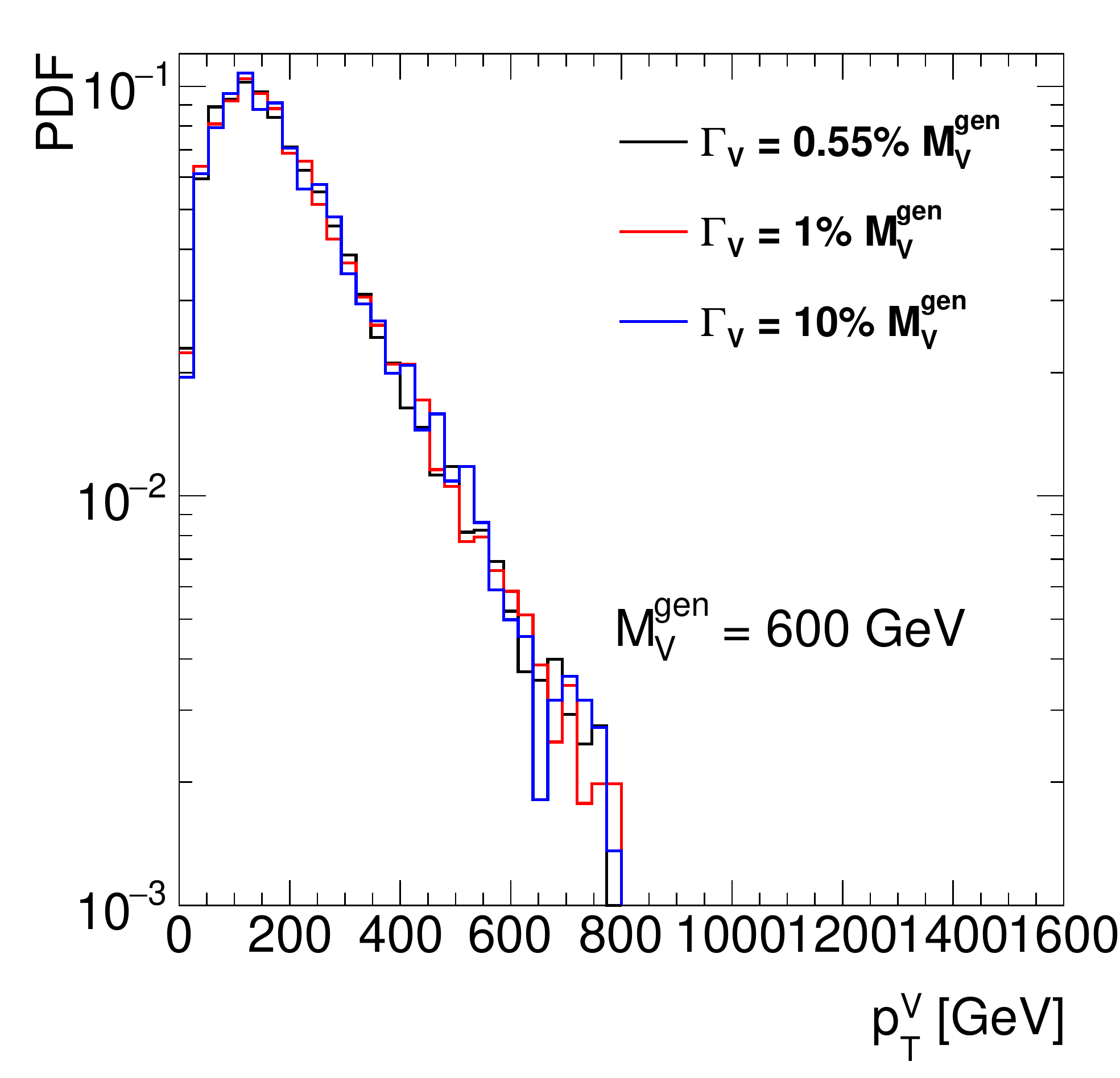}
	\includegraphics[width=0.3\textwidth]{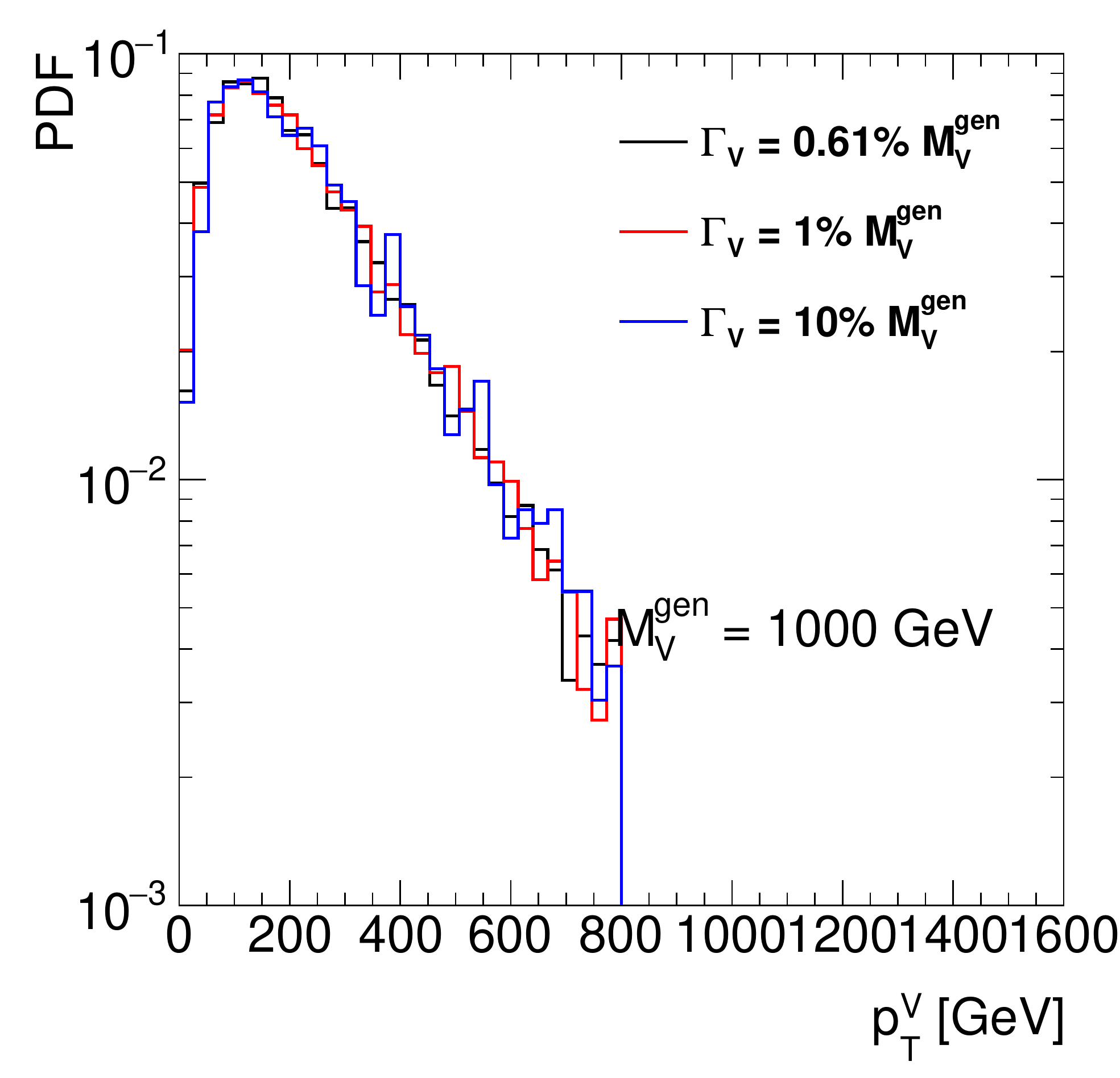}
	\caption{

Distributions of the transverse momentum of the $V$ boson in the case of
the process $p p \to tV \to t (t\bar u + {\rm c.c.})$. We have imposed
that the $V$-boson is produced on-shell and have chosen its mass to be
$m_V = 200$, 600 and 1000~GeV (left, central and right panels).
We have considered three possible cases for the total width
of the $V$-boson, which has been fixed to 0.61\%, 0.1\% and 10\% of the
mass.
}
	\label{fig:appB:pTV}
\end{figure*}

It has been verified that the kinematics do not depend on the width of the
invisible state in the case where this width is at most 10\% of the
$V$-mass. This is illustrated in Fig.~\ref{fig:appB:pTV},
where we show the transverse-momentum
spectra of the $V$-boson when it decays into a top-up final state and for
different $V$-boson masses. The results are independent of the
visible or invisible decay modes as we are only concerned with the
kinematic properties of the invisible state.


\subsection{Single Top Model implementation}
  Card files for \madgraph are provided on the Forum SVN
  repository~\cite{ForumSVN_EWMonoTop} and correspond to the Lagrangian that has
  been implemented in {\sc FeynRules}. Each coupling constant of the model can
  be set via the block \texttt{COUPX} of the parameter card. Its entries 1, 2
  and 3 respectively correspond to the monotop-relevant parameters $a$,
  $\lambda$ and $y$, while the width (and in particular the invisible partial
  width) of the $V$-boson can be tuned {\it via} the $g_{DM}$ parameter to given
  in the entry 10 of the \texttt{COUPX} block.

  The masses of the particles are set in the \texttt{MASS} block of the
  parameter card, the PDG codes of the new states being 32 (the vector state
  $V$), 1000006 (the $\varphi$ colored resonance), 1000022 (the invisible fermion
  $\chiDM$) and 1000023 (the fermion $\psi$ connecting the $V$ state to the dark
  sector). The width of the new vector has to be computed from all open
  tree-level decays (after fixing $g_{DM}$ to a large value and setting the
  relevant entry to \texttt{Auto} in the \texttt{DECAY} block of the parameter
  card), while the way to calculate the width of the resonance $\phi$ is under
  discussion by both the ATLAS and CMS collaborations. The $chi$ and $psi$
  fermions are taken stable so that their width vanishes.

\section{\texorpdfstring{Further W+\MET models with possible cross-section enhancements}{Further W+MET models with possible cross-section enhancements}}
\label{app:monoWExtramodel}

As pointed out in Ref.~\cite{Bell:2015sza}, the mono-$W$ signature can probe the iso-spin violating interactions of dark matter with quarks. The relevant operator after the electroweak symmetry breaking is 
\begin{equation}
\frac{1}{\Lambda^2}\overline{\chiDM} \gamma_\mu \chiDM \left( \overline{u}_L \gamma^\mu u_L + \xi \bar{d}_L \gamma^\mu d_L \right) \,.
\end{equation}
Here, we only keep the left-handed quarks because the right-handed quarks do not radiate a $W$-gauge boson from the weak interaction. As the LHC constrains the cutoff to higher values, it is also important to know the corresponding operators before the electroweak symmetry. At the dimension-six level, the following operator
\begin{equation}
\frac{c_6}{\Lambda^2}\overline{\chiDM} \gamma_\mu \chiDM \,\overline{Q}_L \gamma^\mu Q_L 
\end{equation}
conserves iso-spin and provides us $\xi=1$~\cite{Bell:2015sza}. At the dimension-eight level, new operators appear to induce iso-spin violation and can be
\begin{equation}
\frac{c^d_8}{\Lambda^4}\overline{\chiDM} \gamma_\mu \chiDM \,(H\overline{Q}_L) \gamma^\mu (Q_L H^\dagger) 
+ \frac{c^u_8}{\Lambda^4}\overline{\chiDM} \gamma_\mu \chiDM \,(\tilde{H}\overline{Q}_L) \gamma^\mu (Q_L \tilde{H}^\dagger)  \,.
\end{equation}
After inputting the vacuum expectation value of the Higgs field, we have 
\begin{equation}
\xi = \frac{c_6 \,+\, c_8^d\,v_{\rm EW}^2/2\Lambda^2}{c_6 \,+\, c_8^u \,v_{\rm EW}^2/2\Lambda^2} \,.
\end{equation}
For a nonzero $c_6$ and $v_{\rm EW} \ll \Lambda$, the iso-spin violation effects are suppressed. On the other hand, the values of $c_6$, $c^d_8$ and $c^u_8$ depend on the UV-models. 

There is one possible UV-model to obtain a zero value for $c_6$ and non-zero values for $c^d_8$ and $c^u_8$. One can have the dark matter and the SM Higgs field charged under a new $U(1)^\prime$ symmetry. There is a small mass mixing between SM $Z$-boson and the new \Zprime with a mixing angle of ${\cal O}(v_{\rm EW}^2/M^2_{\Zprime})$. After integrating out \Zprime, one has different effective dark matter couplings to $u_L$ and $d_L$ fields, which are proportional to their couplings to the $Z$ boson. For this model, we have $c_6=0$ and 
\begin{equation}
\xi = \frac{-\frac{1}{2} + \frac{1}{3} \sin^2{\theta_W} }{ \frac{1}{2} - \frac{2}{3} \sin^2{\theta_W}} \approx  -2.7 
\end{equation}
and order of unity. 

\section{Simplified model corresponding to dimension-5 EFT operator}


As an example of a simplified model corresponding to the dimension-5 EFT operator 
described in Section~\ref{sec:EFT_models_with_direct_DM_boson_couplings}, 
we consider a Higgs portal with a scalar mediator. Models of this kind
are among the most concise versions of simplified models that produce 
couplings of Dark Matter to pairs of gauge-bosons.  Scalar fields may couple directly to pairs of electroweak gauge bosons, 
but must carry part of the electroweak vacuum expectation value.  One may thus consider a simple model where Dark Matter couples to a a scalar 
singlet mediator, which mixes with the fields in the Higgs sector.
\begin{equation}
L\subset \frac{1}{2} m_s S^2 + \lambda S^2|H|^2 +\lambda^{'} S |H|^2 + y S \chiDM \overline{\chiDM}
\end{equation}
Where H is a field in the Higgs sector that contains part of the electroweak vacuum expectation value, 
S is a heavy scalar singlet and $\chiDM$ is a Dark Matter field. 
There is then an \schannel diagram where DM pairs couple to the singlet field S, 
which then mixes with a Higgs-sector field, and couples to W and Z bosons. 
This diagram contains 2 insertions of EW symmetry breaking fields, 
corresponding in form to the effective dimension-5 operator in Section~\ref{sub:EW_EFT_Dim5}.

\section{Inert two-Higgs Doublet Model (IDM)}\label{sec:i2HDM}

For most of the simplified models included in this report, the mass of
the mediator and couplings/width are non-trivial parameters of the
model. In these scenarios, we remain agnostic about the theory behind
the dark matter sector and try to parameterize it in simple terms.

We have not addressed how to extend the simplified models to realistic
and viable models which are consistent with the symmetries of the
Standard Model. Simplified models often violate gauge invariance which
is a crucial principle for building a consistent BSM model which
incorporates SM together with new physics. For example, with a new
heavy gauge vector boson mediating DM interactions, one needs not just
the dark matter and its mediator, but also a mechanism which provides
mass to this mediator in a gauge invariant way.

Considering both the simplified model and other elements necessary for a consistent theory is a next logical step. The authors of \cite{Belyaev:2015tap} term these Minimal
Consistent Dark Matter (MCDM) models. MCDM models are at the same time still toy models that can be 
easily incorporated into a bigger BSM model and explored via
complementary constraints from collider and direct/indirect DM search
experiments as well as relic density constraints. 



The idea of an inert Two-Higgs Doublet Model (IDM) was introduced
more than 30 years ago in Ref~\cite{Deshpande:1977rw}. The IDM was
first proposed as a Dark Matter model in Ref.~\cite{IDMnaturalness} and its
phenomenology further studied in Refs.~\cite{LopezHonorez:2006gr,ScalarMultiplet,IDMnewviable,IDMgammalines,Dolle:2009fn,ATL-PHYS-PUB-2014-007,IDMnu1,IDMnu2,IDMpos,IDMVIB,Goudelis:2013uca,Belyaev:2015tap}. It is an extension of the SM with a second scalar 
doublet $\phi_2$ with no direct coupling to fermions.  This doublet has a discrete $Z_2$ symmetry, 
under which $\phi_2$ is odd and all the other fields are even. 
The Lagrangian of the odd sector is,
\begin{equation}
  \mathcal{L} = \frac{1}{2}(D_{\mu}\phi_2)^2 -V(\phi_1,\phi_2)
\end{equation}
with the  potential $V$  containing mass terms and $\phi_1 - \phi_2$
interactions:
\begin{fullwidth}
  \begin{eqnarray}
    V &=& -m_1^2 (\phi_1^{\dagger}\phi_1) - m_2^2 (\phi_2^{\dagger}\phi_2) + \lambda_1 (\phi_1^{\dagger}\phi_1)^2 + \lambda_2 (\phi_2^{\dagger}\phi_2)^2    \nonumber  \\
      &+&  \lambda_3(\phi_2^{\dagger}\phi_2)(\phi_1^{\dagger}\phi_1)  + \lambda_4(\phi_2^{\dagger}\phi_1)(\phi_1^{\dagger}\phi_2) + 
          \frac{\lambda_5}{2}\left[(\phi_1^{\dagger}\phi_2)^2 + (\phi_2^{\dagger}\phi_1)^2 \right],
  \end{eqnarray}
\end{fullwidth}

where $\phi_1$ and  $\phi_2$ are SM and inert Higgs doublets respectively carrying the same hypercharge. These doublets can be parameterized as
\begin{equation}
  \phi_1=\frac{1}{\sqrt{2}}
  \begin{pmatrix}
    0\\
    v+H 
  \end{pmatrix}
  \qquad
  \phi_2= \frac{1}{\sqrt{2}}
  \begin{pmatrix}
    \sqrt{2}{h^+} \\
    h_1 + ih_2
  \end{pmatrix}
\end{equation}

\begin{figure}
\includegraphics[width=\textwidth]{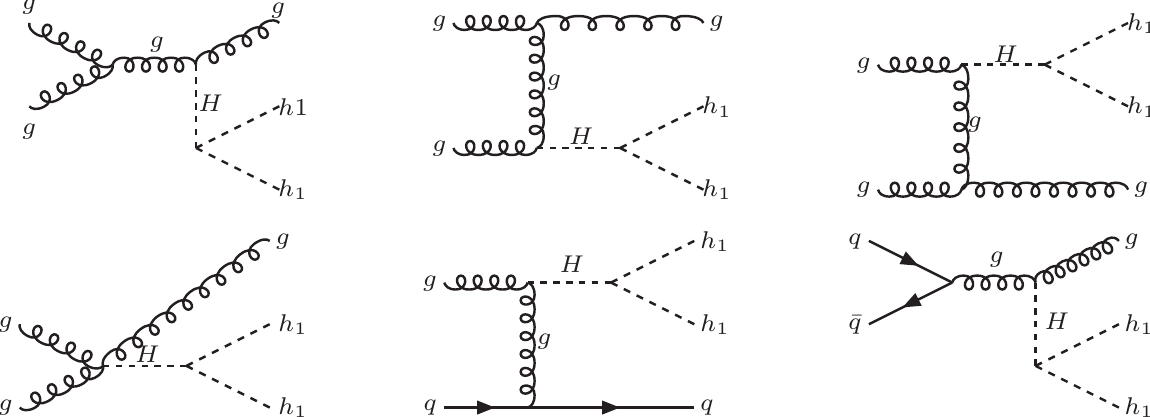} 
\caption{Feynman diagrams for $gg\to h_1 h_1+g$ process
contributing to mono-jet signature, adapted from \cite{Belyaev:2015tap}.}
\label{fig:fdmonojet1}
\end{figure}

\begin{figure}[htb]
\includegraphics[width=\textwidth]{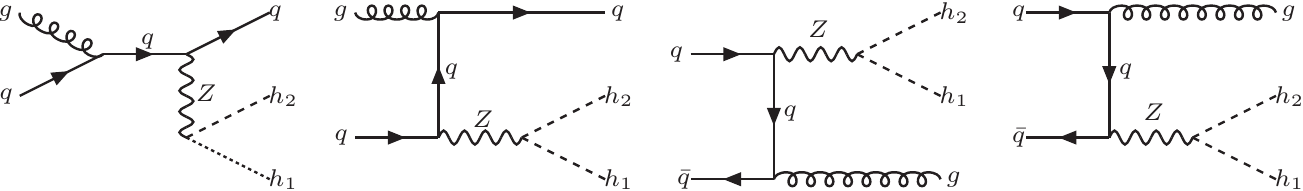} 
\caption{Feynman diagrams for $q\bar{q}\to h_1 h_2+g$ ($gq\to h_1 h_2+q$) process 
contributing to mono-jet signature, adapted from \cite{Belyaev:2015tap}.}
\label{fig:fdmonojet2}
\end{figure}

\begin{figure}[htb]
\includegraphics[width=\textwidth]{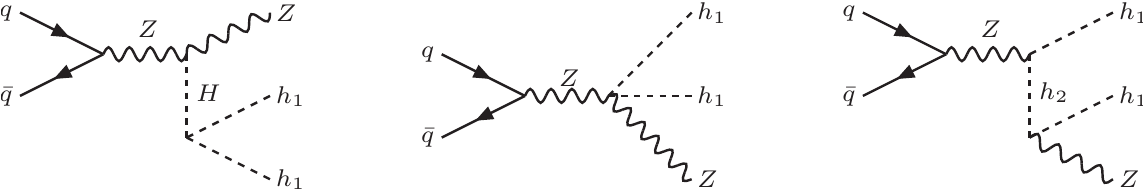} 
\caption{Feynman diagrams for $q\bar{q}\to h_1 h_1+Z$  process 
contributing to mono-Z signature, adapted from \cite{Belyaev:2015tap}.}
\label{fig:fdmonoZ}
\end{figure}

\begin{figure}[htb]
\includegraphics[width=\textwidth]{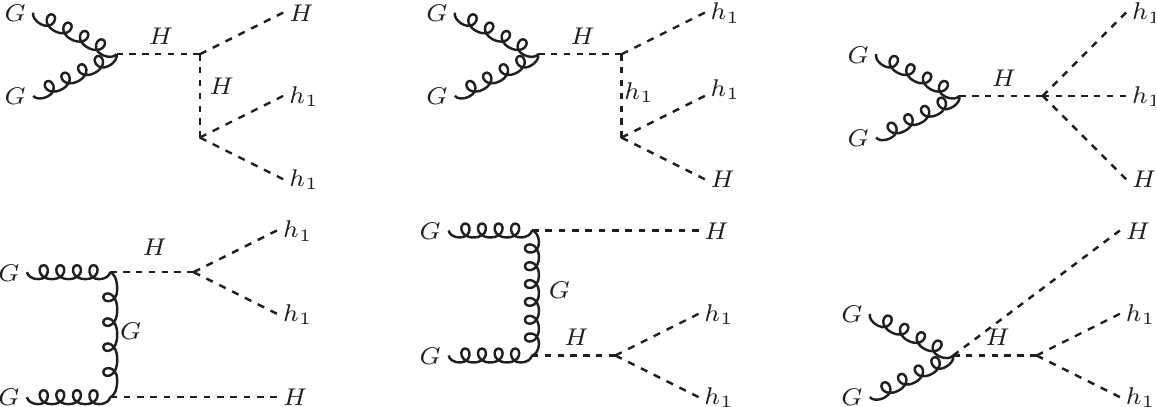} 
\caption{Feynman diagrams for $gg\to h_1 h_1+H$  process 
contributing to mono-Higgs signature, adapted from \cite{Belyaev:2015tap}.}
\label{fig:fdmonoH1}
\end{figure}

\begin{figure}[htb]
\includegraphics[width=\textwidth]{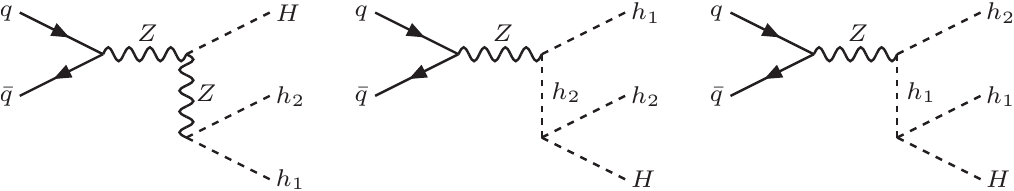} 
\caption{Feynman diagrams for $q\bar{q}\to h_1 h_2+H$  process 
contributing to mono-Higgs signature, adapted from \cite{Belyaev:2015tap}.}
\label{fig:fdmonoH2}
\end{figure}

\begin{figure}[htb]
\includegraphics[width=\textwidth]{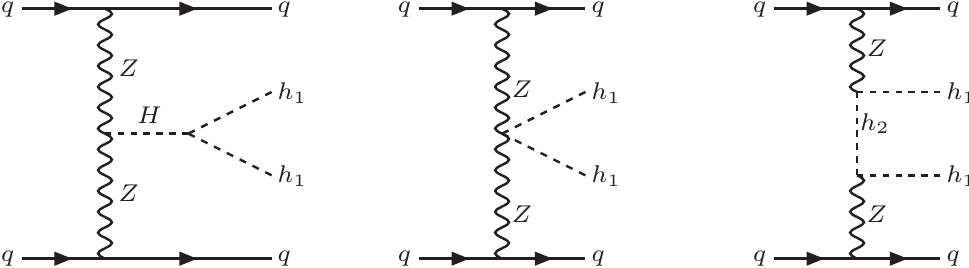} 
\caption{Diagrams for $qq\to qq h_1 h_1$ DM production in vector boson
fusion process, adapted from \cite{Belyaev:2015tap}.}
\label{fig:fdvbf}
\end{figure}

In addition to the SM, the IDM introduces four more degrees of freedom coming from the inert doublet in the form of a $Z_2$-odd charged scalar $h^\pm$ and two neutral $Z_2$-odd scalars $h_1$ and $h_2$. The lightest neutral scalar, $h_1$ is identified as the dark matter candidate. Aspects of the IDM collider phenomenology have been studied in \cite{Burgess:2000yq, Andreas:2008xy, Arhrib:2013ela, Belyaev:2015tap,IDMnaturalness,IDMLEPII,IDMLHChinvfirst,IDMdileptons1,IDMtrileptons,IDMmultileptons,IDMhgaga1,IDMhgaga2,IDMposthiggs,IDMdileptonsII}. Its LHC signatures include dileptons \cite{IDMdileptons1,IDMdileptonsII}, trileptons \cite{IDMtrileptons} and multileptons \cite{IDMmultileptons} along with missing transverse energy, modifications of the Higgs branching ratios \cite{IDMhgaga1,IDMhgaga2,Goudelis:2013uca}, as well as $\slashed{E}_T + \rm{jet}$, Z, and Higgs and $\slashed{E}_T + \rm{VBF}$ signals (see Figs.~\ref{fig:fdmonojet1}--\ref{fig:fdvbf}).


Based on the various LHC search channels, DM phenomenology issues and theoretical considerations, numerous works have proposed benchmark scenarios for the IDM, see e.g. \cite{IDMmultileptons,Goudelis:2013uca} while a FeynRules implementation (including MadGraph, CalcHEP and micrOMEGAs model files) was provided in \cite{IDMmultileptons}. An updated analysis of the parameter space has recently been performed in Ref.~\cite{Belyaev:2015tap}. 

\begin{figure}[htb]
	\includegraphics[width=\textwidth]{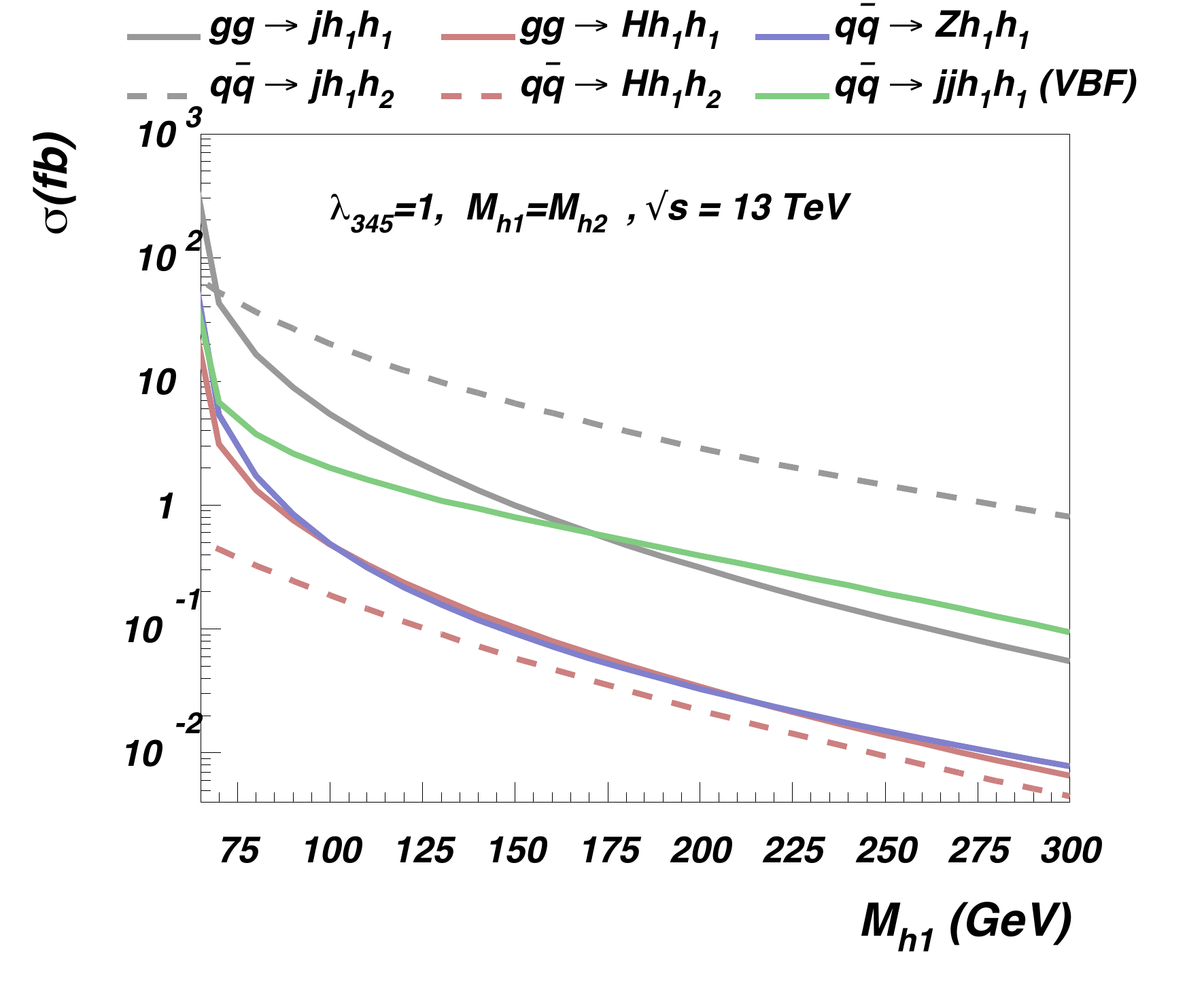} 
	\caption{LHC cross section at 13~\tev for various signatures, from \cite{Belyaev:2015tap}.}
	\label{fig:IDM_xsecs}
\end{figure}

The authors suggested to study mono-X  signatures that are relevant to model-independent collider DM searches,
and evaluated their rates presented below. They have implemented and cross-checked the IDM model into CalcHEP and micrOMEGAs,
with an implementation publicly available on the  ~\href{http://hepmdb.soton.ac.uk/hepmdb:0615.0189}{HEPMDB database}, including loop-induced $HHG$ and $\gamma\gamma H$ models. They propose an additional set of benchmark points, mostly inspired by mono-$X$ and VBF searches (Table.~\ref{tab:IDMbenchMarks}). Though the overall parameter space of IDM is 5-dimensional, once all relavant constraints are applied the parameter space relevant to a specific LHC signature typically reduces to 1-2 dimensional. In the mono-jet case, one can use two separate simplified models, a $gg \rightarrow h_1 h_1 + g$ process (via Higgs mediator) and a $qq \rightarrow h_1 h_2 + g (gq \rightarrow h_1 h_2 + q)$ process (through a Z-boson mediator) to capture the physics relevant to the search.
The cross sections for the various mono-$X$ and VBF signatures produced by this model are displayed in Fig.~\ref{fig:IDM_xsecs}.

\begin{table}[htb]
	\centering
	\begin{tabular}{|c||c|c|c|c|c|c|}
		\hline
		{\bf BM}                       &  {\bf 1}  & {\bf 2}  & {\bf 3}  & {\bf 4}  &  {\bf 5}  \\
		\hline\hline 
		$M_{h_{1}}$ (GeV)     & 48      	& 53 		& 70 		& 82 	&120 \\
		\hline
		$M_{h_{2}}$ (GeV)     & 55      	& 189 		& 77  		&  89  & 140 \\
		\hline
		$M_{h_{\pm}}$ (GeV)   & 130     	& 182 		& 200  	&  150  &  200 \\
		\hline
		$\lambda_{2}$         &  0.8    	& 1.0 		& 1.1 		& 0.9 	& 1.0 \\ 
		\hline
		$\lambda_{345}$       & $-0.010$ 	& $-0.024$  	& $+0.022$ 	& $-0.090$  & $-0.100$      \\
		\hline
		$\Omega h^2$          & $3.4 \times 10^{-2}$ & $8.1 \times 10^{-2}$  & $9.63 \times 10^{-2}$  & $1.5 \times 10^{-2}$  &  $2.1 \times 10^{-3}$ \\
		\hline 
		$\sigma_{SI}$ (pb)   & $2.3 \times 10^{-10}$ &  $7.9 \times 10^{-10}$  & $5.1 \times 10^{-10}$  & $4.5 \times 10^{-10}$  &  $2.6 \times 10^{-9}$ \\
		\hline 
		$\sigma_{LHC}$ (fb)     & $1.7 \times 10^{2}$ &  $7.7 \times 10^{2}$  & $4.3 \times 10^{-2}$  & $1.2 \times 10^{-1}$  &  $2.3 \times 10^{-2}$ \\
		\hline\hline
	\end{tabular}
	\caption{Five benchmarks for IDM in  ($M_{h_{1}},M_{h_{2}},M_{h_{\pm}},\lambda_{2},\lambda_{345}$) parameter space.
		We also present the corresponding relic density ($\Omega h^2$), the spin-independent cross section for DM scattering on the proton ($\sigma_{SI}$),
		and the LHC cross section at 13 TeV for mono-jet process $pp\to h_1,h_1+jet$ for $p_T^{jet}>100$~GeV cut ($\sigma_{LHC}$).}
	\label{tab:IDMbenchMarks}
\end{table}

\clearpage

\chapter{Appendix: Presentation of experimental results for reinterpretation}
\label{app:Presentation_Of_Experimental_Results}
 When collider searches present results with the recommended benchmarks, we suggest the following:
 \begin{itemize}
 \item Provide limits in collider language, on fundamental parameters of
 the interaction: the couplings and masses of particles in simplified model.
 \item Translate limits to non-collider language, for a range of
 assumptions, in order to convey a rough idea of the range of
 possibilities. The details of this point are left for work beyond the scope of this Forum. 
 \item Provide all necessary material for theorists to reinterpret simplified
 model results as building blocks for more complete models (e.g. signal cutflows,
 acceptances, etc). This point is detailed further in this appendix.
\item Provide model-independent results in terms of limits on
  cross-section times efficiency times acceptance of new phenomena for all cases, but
  especially when EFTs are employed as benchmarks. This recommendation has been issued before: see
  Ref.~\cite{Kraml:2012sg} for detailed suggestions.
 \item Provide easily usable and clearly labeled results in a digitized format, e.g.~\cite{HEPData} entries, ROOT histograms and macros
 or tables available on analysis public pages.
 \end{itemize}

This appendix describes further considerations for reinterpretation and reimplementation of the analyses, 
as well as for the use of simplified model results directly given by the collaborations. 

\section{Reinterpretation of analyses}

In the case of reinterpretation for models different than those provided by the experimental collaborations,
the information needed primarily includes expected and observed exclusion lines along with their $\pm 1 \sigma$ uncertainty, 
expected and observed upper limits in case of simplified models, efficiency maps and kinematic distributions
as reported in the analysis. If the kinematics of the new model to be tested in the reinterpretation is similar to that 
of the original model provided by the collaboration, it
will be straight-forward to rescale the results provided to match the new model cross-section
using this information. 

\section{Reimplementation of analyses}

One of the important developments in recent years is an active development of software codes~\cite{Dumont:2014tja, Conte:2014zja, Kim:2015wza,Cranmer:2010hk,ATOM,Barducci:2014ila} necessary for recasting analyses. The aim of these codes is to provide a public library of LHC analyses that have been reimplemented and validated, often by the collaborations themselves. Such libraries can then be used to analyze validity of a BSM scenario in a systematic and effective manner. The availability of public libraries further facilitates a unified framework and can lead to an organized and central structure to preserve LHC information long term.
The reimplementation of an analysis consists of several stages. Typically, the analysis note is used as a basis for the implementation of the preselection and event selection cuts in the user analysis code within the recasting frameworks. Signal events are generated, and passed through a parameterized detector simulation using software such as Delphes or PGS~\cite{deFavereau:2013fsa,PGS}. The reconstructed objects are then analyzed using the code written in the previous step, and the results in terms of number of events are passed through a statistical analysis framework to compare with the backgrounds provided by the collaborations. 

In order to be able to effectively use such codes, it is important to get a complete set of information from the collaborations. 

For what concerns the generation of the models, it is desirable to have the following items as used by the collaborations:
\begin {itemize}
	\item Monte Carlo generators: Monte Carlo generators along with the exact versions used to produce the event files should be listed. 
	\item Production cross sections: The order of production cross sections (e.g. LO,NLO,NLL) as well as the codes which were used to compute them should be provided. Tables of reference cross sections for several values of particle masses are useful as well. 
	\item Process Generation: Details of the generated process, detailing number of additional partons generated. 
	\item LHE files: selected LHE files (detailing at least a few events if not the entire file) corresponding to the benchmarks listed in the analysis could also be made available in order to cross check process generation. Experimental collaborations may generate events on-the-fly without saving the intermediate LHE file; we advocate that the cross-check of process generation is straight-forward if this information is present, so we encourage the generation of a few selected benchmark points allowing for a LHE file to be saved. Special attention should be paid to list the parameters which change the production cross section or kinematics of the process e.g. mixing angles. 
	\item Process cards: Process cards including  PDF choices, details of matching algorithms and scales and details of process generation. If process cards are not available, the above items should be clearly identified. 
	\item Model files: For models which are not already implemented in \madgraph, the availability of the corresponding model files in the UFO format~\cite{Degrande:2011ua} is highly desired. This format details the exact notation used in the model and hence sets up a complete framework. In case \madgraph is not used, enough information should be provided in order to clearly identify the underlying model used for interpretations and reproduce the generation. 
\end{itemize}
The ATLAS/CMS Dark Matter Forum provides most of the information needed within its SVN repository~\cite{ForumSVN} and on a dedicated HEPData~\cite{HEPData} page dedicated to the results in this report.

Efficiency maps and relevant kinematic distributions as reported in the analysis should be provided, in a digitized format with clearly specified units.
If selection criteria cannot be easily simulated through parameterized detector simulation, the collaborations should provide the efficiency of such cuts. 
Overall reconstruction and identification efficiencies of physics objects are given as an input to the detector simulation software. 
It is thus very useful to get parametrized efficiencies for reconstructed objects (as a function of the rapidity $\eta$ and/or transverse momentum $p_T$), 
along with the working points at which they were evaluated (e.g. loose, tight selection). Object definitions should be clearly identifiable. 
Digitized kinematic distributions are often necessary for the validation of the analysis so that the results from the collaboration are obtained, 
and so are tables containing the events passing each of the cuts. 

The availability of digitized data and backgrounds is one of the primary requirements for fast and efficient recasting. 
Platforms such as HepData~\cite{HEPData} can be used as a centralized repository; alternatively, analysis public pages and tables can be used
for dissemination of results. Both data and Standard Model backgrounds should be provided in the form of binned histogram that can be interpolated if needed. 

A detailed description of the likelihood used in order to derive the limits from the comparison of data to signal plus background should be given. 
This can be inferred from the analysis documentation itself, however direct availability of the limit setting code as a workspace in RooStats or HistFitter~\cite{Baak:2014wma} is highly desirable. 

Finally, the collaborations can also provide an analysis code directly implemented in one of the public recasting codes detailed above. 
Such codes can be published via INSPIRE~\cite{INSPIRE} in order to track versioning and citations. 

\section{Simplified model interpretations}

Dark Matter searches at the LHC will include simplified model interpretations in their search results. These interpretations are simple and can be used for a survey of viability of parameter space. Codes such as~\cite{Kraml:2013mwa, Kraml:2014sna, Papucci:2014rja} can make use of the simplified model results given in the form of 95\% Confidence Level (CLs) upper limit  or efficiency maps in order to test Beyond the Standard Model parameter space. As mentioned above, it will thus be extremely useful if the results are given in a digitized form that is easily usable by the theory community. 

The parameter space of these models should be clearly specified. For example, for a simplified model containing dark matter mass \mDM, mediator mass \Mmed and couplings \gDM, \gq it will be very useful to have upper limits on the product of couplings $\sqrt{\gDM\gq}$ or cross section times branching ratio as a function of \mDM, \Mmed. Limits on visible cross sections of the simplified models considered for interpretations should be made available.

The usage of simplified model results relies on interpolating between upper limit values. In order to facilitate the interpolation, regions where large variation of upper limits is observed should contain denser grid, if a uniform grid over the entire plane is not possible. For simplified model involving more than three parameters (two masses and product of couplings), slices of upper limits in the additional dimensions will be necessary for reinterpretation. 

As already mentioned in the introduction to this Chapter, acceptance and efficiency maps for all the signal regions involved in the analysis should be made available. These results are not only useful for model testing using simplified models but also to validate implementation of the analysis. Information about the most sensitive signal regions as a function of new particle masses is also useful in order to determine the validity of approximate limit setting procedures commonly used by theorists.

\chapter{Appendix: Additional details and studies within the Forum}
\label{app:Additional_details}
\section*{\texorpdfstring{Further information for baryonic \Zprime Model}{Further information for baryonic Z' Model}}

\subsection*{Cross-section scaling}

\begin{figure}[hbpt!]
	\includegraphics[width=0.99\linewidth]{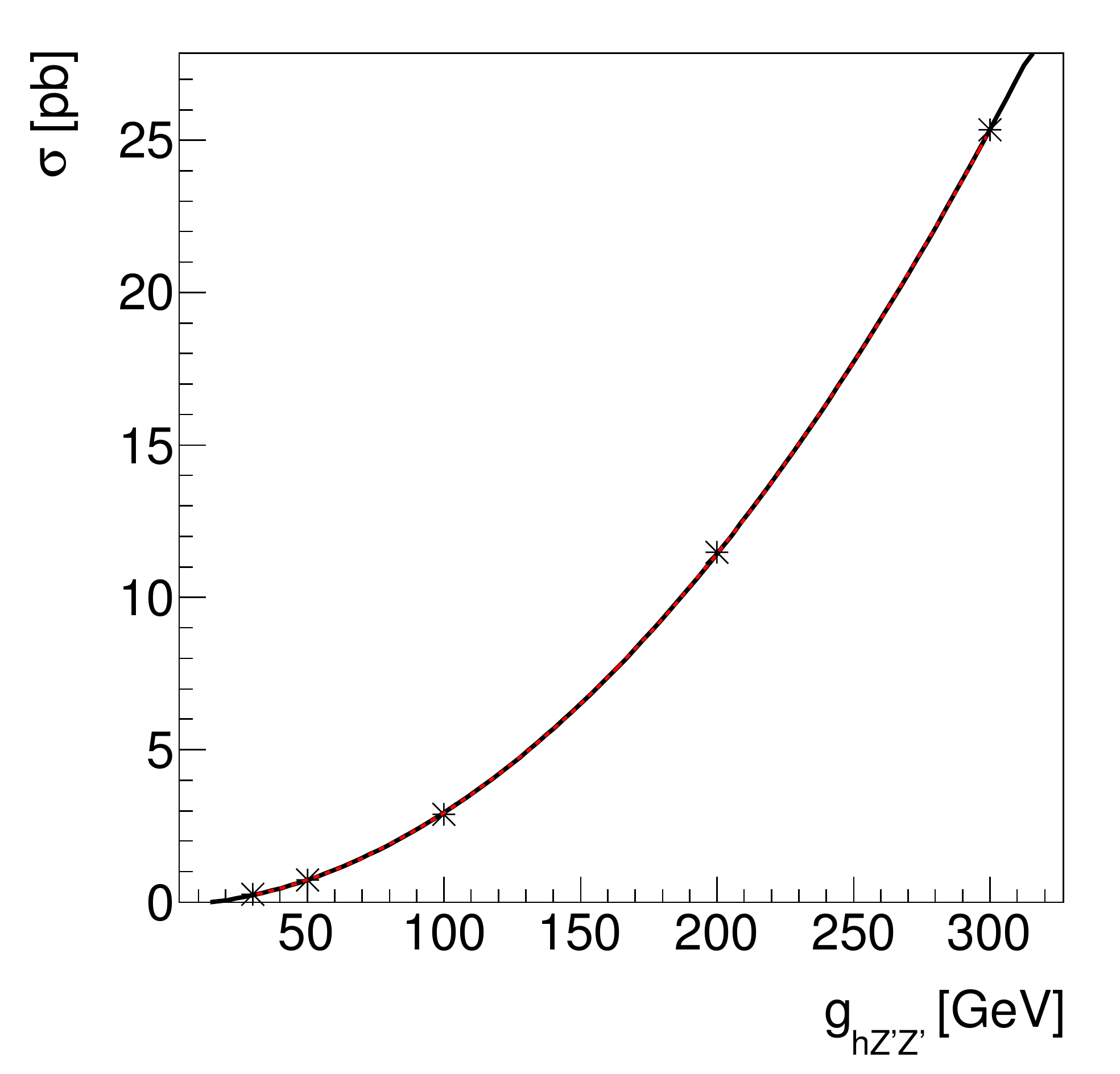}\\
	\includegraphics[width=0.99\linewidth]{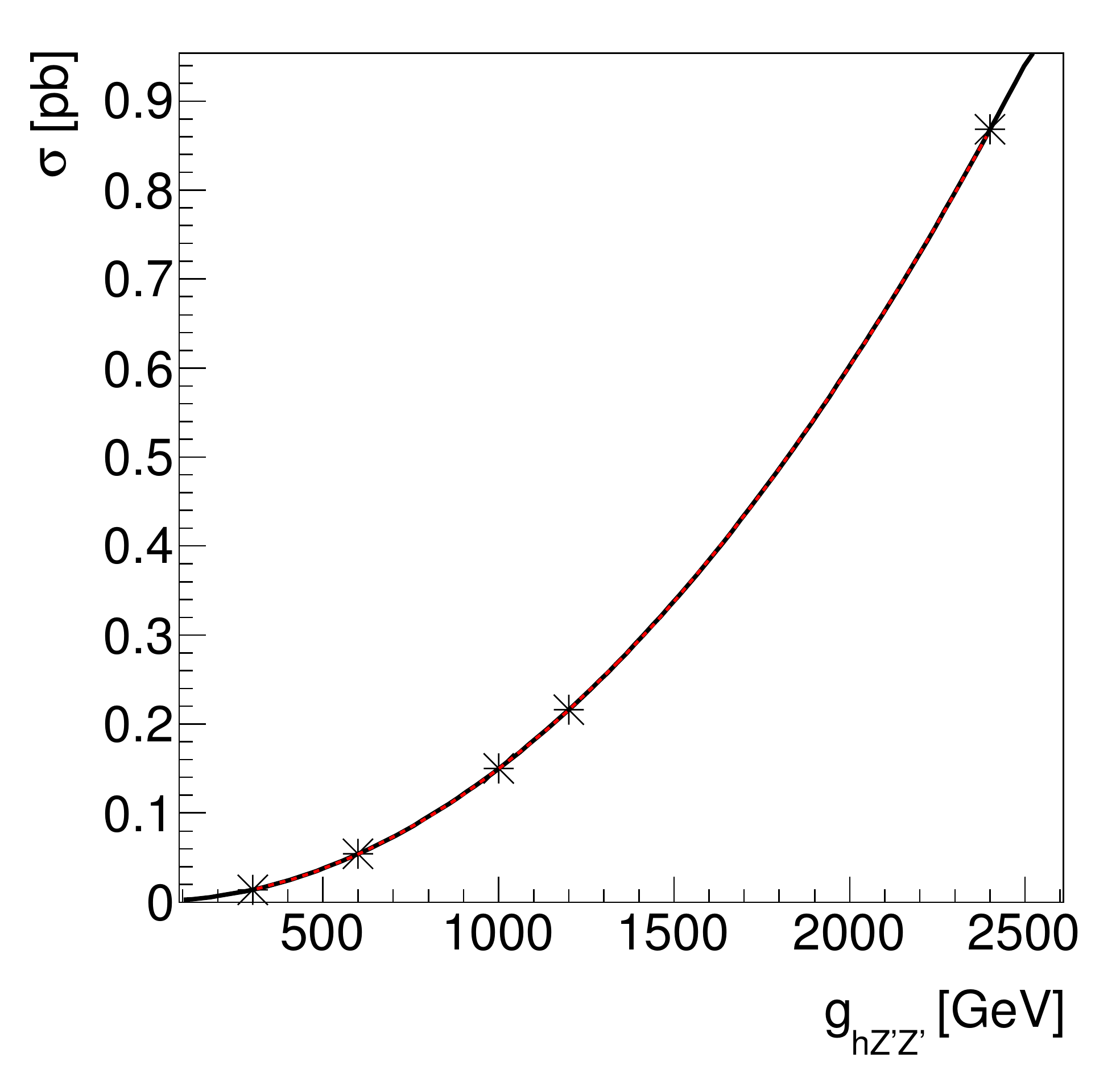}
	\caption{Cross section of the $pp \rightarrow H\chiDM\bar{\chiDM}$ process as a function of 
		$g_{h \Zprime \Zprime}$ for $m_{\Zprime} = 100$~\gev (left) 
		and $m_{\Zprime} = 1$~\tev (right). The fit functions are shown in the text. 
		\label{fig:vectorXSdeps}}
\end{figure}

The dependence of the cross section of the $pp \rightarrow H\chiDM\bar{\chiDM}+X$ process 
on $g_{h \Zprime \Zprime}$ is shown in Figure~\ref{fig:vectorXSdeps}. 
The curves have been fit to second-order polynomials, where $y$ is the cross-section
and $x$ is the coupling $g_{h \Zprime \Zprime}$. 

For $m_{med} = 100$~\gev, the fit function is 
$$y = -0.12 - 3.4\times10^{-3}x + 2.7\times10^{-4}x^2$$.
For $m_{med} = 1$~\tev, the fit function is 
is $$y = 0.0012 - 2.4\times10^{-7}x + 1.5\times10^{-7}x^2$$,

\be
y = -0.12 - 3.4\times10^{-3}x + 2.7\times10^{-4}x^2.
\ee
For $\mMed = 1$ TeV, the fit function is 
is:

\be
y = 0.0012 - 2.4\times10^{-7}x + 1.5\times10^{-7}x^2.
\ee

\printbibliography

\end{document}